\documentclass[a4paper,fleqn,dvipsnames,web,table]{cas-dc}

\usepackage{framed,multirow}
\usepackage{amsmath,amssymb, mathrsfs, stmaryrd}
\usepackage{xcolor}
\usepackage{hyperref}
\usepackage{tikz}
\usetikzlibrary{arrows.meta, spy, calc, arrows, patterns, patterns.meta}
\usepackage{adjustbox}
\usepackage{floatrow}
\usepackage{array}
\usepackage{graphicx}
\usepackage{array}
\usepackage{makecell}
\usepackage{hhline}
\usepackage{colortbl}
\usepackage[round,authoryear]{natbib}

\newcolumntype{P}[1]{>{\centering\arraybackslash}p{#1}}
\newcolumntype{M}[1]{>{\centering\arraybackslash}m{#1}}

\newcommand{\norm}[1]{\left\lVert#1\right\rVert}

\begin{document}
\shorttitle{Generalizable multi-task, multi-domain deep segmentation of sparse pediatric imaging datasets}
\shortauthors{A. Boutillon et~al.}

\title[mode=title]{Generalizable multi-task, multi-domain deep segmentation of sparse pediatric imaging datasets via multi-scale contrastive regularization and multi-joint anatomical priors}

\author[1,2]{Arnaud Boutillon}[orcid=0000-0001-5855-2770]
\cormark[1]

\author[1,2]{Pierre-Henri Conze}[orcid=0000-0003-2214-3654]

\author[2,3,4]{Christelle Pons}[orcid=0000-0003-3924-6035]

\author[1,2]{Valérie Burdin}[orcid=0000-0001-6012-9883]

\author[2,3,5]{Bhushan Borotikar}[orcid=0000-0002-3404-6547]

\address[1]{IMT Atlantique, Brest, France}
\address[2]{LaTIM UMR 1101, Inserm, Brest, France}
\address[3]{Centre Hospitalier Régional et Universitaire (CHRU) de Brest, Brest, France}
\address[4]{Fondation ILDYS, Brest, France}
\address[5]{Symbiosis Center for Medical Image Analysis, Symbiosis International University, Pune, India}

\cortext[cor1]{Corresponding author: \href{mailto:arnaud.boutillon@imt-atlantique.fr}{\nolinkurl{arnaud.boutillon@imt-atlantique.fr}} (A. Boutillon)}

\begin{abstract}
Clinical diagnosis of the pediatric musculoskeletal system relies on the analysis of medical imaging examinations. In the medical image processing pipeline, semantic segmentation using deep learning algorithms enables an automatic generation of patient-specific three-dimensional anatomical models which are crucial for morphological evaluation. However, the scarcity of pediatric imaging resources may result in reduced accuracy and generalization performance of individual deep segmentation models. In this study, we propose to design a novel multi-task, multi-domain learning framework in which a single segmentation network is optimized over the union of multiple datasets arising from distinct parts of the anatomy. Unlike previous approaches, we simultaneously consider multiple intensity domains and segmentation tasks to overcome the inherent scarcity of pediatric data while leveraging shared features between imaging datasets. To further improve generalization capabilities, we employ a transfer learning scheme from natural image classification, along with a multi-scale contrastive regularization aimed at promoting domain-specific clusters in the shared representations, and multi-joint anatomical priors to enforce anatomically consistent predictions. We evaluate our contributions for performing bone segmentation using three scarce and pediatric imaging datasets of the ankle, knee, and shoulder joints. Our results demonstrate that the proposed approach outperforms individual, transfer, and shared segmentation schemes in Dice metric with statistically sufficient margins. The proposed model brings new perspectives towards intelligent use of imaging resources and better management of pediatric musculoskeletal disorders.
\end{abstract}


\maketitle

\section{Introduction}
\label{sec:introduction}

\begin{figure}[ht!]
\centering
\begin{adjustbox}{width=\textwidth}
\begin{tikzpicture}[every node/.style={inner sep=0,outer sep=0}]

\draw[line width=0.1mm, rounded corners=2.5, fill=white!50!lightgray] (-1.1,-2.4) -- (5.1,-2.4) -- (5.1,1.5) -- (-1.1,1.5) -- cycle;
\draw[line width=0.1mm, fill=white!20!gray, rounded corners=2.5] (-.95,-1.15) -- (.95,-1.15) -- (.95,.95) -- (-.95,.95) -- cycle;
\draw[line width=0.1mm, fill=white!20!gray, rounded corners=2.5] (1.05,-1.15) -- (2.95,-1.15) -- (2.95,.95) -- (1.05,.95) -- cycle;
\draw[line width=0.1mm, fill=white!20!gray, rounded corners=2.5] (3.05,-1.15) -- (4.95,-1.15) -- (4.95,.95) -- (3.05,.95) -- cycle;
\draw[line width=0.1mm, fill=white!40!gray, rounded corners=2.5] (-.95,-1.7) -- (.95,-1.7) -- (.95,-2.25) -- (-.95,-2.25) -- cycle;
\draw[line width=0.1mm, fill=white!40!gray, rounded corners=2.5] (1.05,-1.7) -- (2.95,-1.7) -- (2.95,-2.25) -- (1.05,-2.25) -- cycle;
\draw[line width=0.1mm, fill=white!40!gray, rounded corners=2.5] (3.05,-1.7) -- (4.95,-1.7) -- (4.95,-2.25) -- (3.05,-2.25) -- cycle;

\node[inner sep=0pt] (mri_a) at (0,0)
    {\includegraphics[width=.21\textwidth]{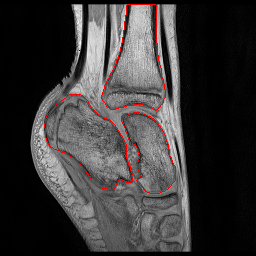}};
\draw [white!20!gray, rounded corners=2.5, line width=2.5] (mri_a.north west) -- (mri_a.north east) -- (mri_a.south east) -- (mri_a.south west) -- cycle;

\node[inner sep=0pt] (mri_k) at (2,0)
    {\includegraphics[width=.21\textwidth]{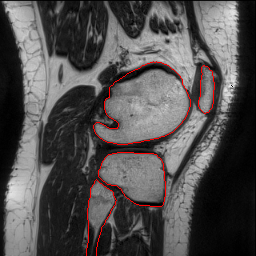}};
\draw [white!20!gray, rounded corners=2.5, line width=2.5] (mri_k.north west) -- (mri_k.north east) -- (mri_k.south east) -- (mri_k.south west) -- cycle;

\node[inner sep=0pt] (mri_s) at (4,0)
    {\includegraphics[width=.21\textwidth]{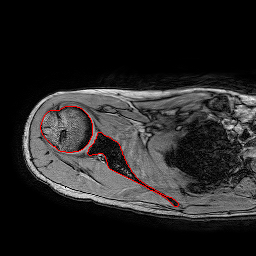}};
\draw [white!20!gray, rounded corners=2.5, line width=2.5] (mri_s.north west) -- (mri_s.north east) -- (mri_s.south east) -- (mri_s.south west) -- cycle;

\node at (2,1.225) {\scalebox{.8}{Pediatric Imaging Domains}};
\node at (0,-1) {\scalebox{.7}{Ankle}};
\node at (2,-1) {\scalebox{.7}{Knee}};
\node at (4,-1) {\scalebox{.7}{Shoulder}};

\node at (2,-1.425) {\scalebox{.8}{Segmentation Tasks}};
\node at (0,-1.85) {\scalebox{.7}{Calcaneus, Talus,}};
\node at (0,-2.1) {\scalebox{.7}{Tibia}};
\node at (2,-1.85) {\scalebox{.7}{Femur, Fibula,}};
\node at (2,-2.1) {\scalebox{.7}{Patella, Tibia}};
\node at (4,-2) {\scalebox{.7}{Humerus, Scapula}};

\end{tikzpicture}
\end{adjustbox}
  \caption{Pediatric ankle, knee, and shoulder joint imaging domains and their respective segmentation tasks consisting of the following bones: \texttt{[calcaneus, talus, tibia (distal)]}, \texttt{[femur (distal), fibula (proximal), patella, tibia (proximal)]}, and \texttt{[humerus, scapula]}. Ground truth delineations are in red (\textcolor{red}{\raisebox{1.8pt}{\rule{5pt}{1pt}}}).}
  \label{fig:introduction}
\end{figure}

Semantic segmentation allows the identification and localization of meaningful anatomical structures in a medical image by extracting their boundaries and consequently acts as a crucial pre-processing step in the medical image analysis workflow guiding clinical decisions. For the management of pediatric musculoskeletal disorders, three-dimensional (3D) solid or surface models of muscles, bones, cartilages, and ligaments generated from imaging examinations can significantly assist clinicians in diagnosing pathologies, assessing morphological evolution over time, and optimally guiding treatment strategies  \citep{hirschmann_artificial_2019}. An accurate understanding of the pediatric anatomy is especially needed as clinical verdict requires precise knowledge of the morphological deformity and associated joint dysfunction \citep{balassy_role_2008}. This study focuses specifically on ankle, knee and shoulder joints (Fig. \ref{fig:introduction}) for which kinematic and dynamic analyses are based on patient-specific 3D bone models \citep{hirschmann_artificial_2019}. However, the process of segmenting magnetic resonance (MR) images typically relies on manual delineation which is tedious, time-consuming, and suffers from intra- and inter-observer variability \citep{jaramillo_pediatric_2008, meyer_musculoskeletal_2008}. Moreover, the pediatric musculoskeletal system may be more challenging to segment than its adult counterpart due to thinner structures, the ongoing bone ossification process, and the existence of higher anatomical variability between age groups \citep{balassy_role_2008, jaramillo_pediatric_2008, meyer_musculoskeletal_2008}. Developing robust and fully-automated segmentation techniques becomes therefore a necessity to reduce analysis time and increase the reliability of morphological assessment.

Deep learning approaches have demonstrated promising results for solving medical imaging-based tasks, including classification, detection, reconstruction, registration, and segmentation \citep{litjens_survey_2017, lundervold_overview_2019}. Specifically, convolutional neural networks (CNN) have become state-of-the-art methods in numerous medical imaging-based applications due to their ability to learn hierarchical representations of image features in a purely data-driven manner \citep{litjens_survey_2017, lundervold_overview_2019}. For medical image segmentation, most deep learning models are designed based on UNet \citep{ronneberger_u-net_2015} and its 3D counterpart VNet \citep{milletari_v-net_2016} due to their impressive performances compared to other CNN architectures. Numerous refinements to the UNet convolutional encoder-decoder architecture have been proposed, including models which embed encoders pre-trained on large non-medical imaging databases (e.g. ImageNet) \citep{russakovsky_imagenet_2015} to leverage low-level features typically shared between different image types \citep{conze_healthy_2020}. Alternatively, one can mention attention models such as Attention UNet (Att-UNet) \citep{oktay_attention_2018} which integrates attention gates on the long range skip connections (between encoder and decoder) to focus on salient features. Furthermore, UNet and VNet models have already been employed for segmenting musculoskeletal structures in MR images, including adult knee bones, muscles, and cartilages \citep{ambellan_automated_2019, zhou_deep_2018}, adult shoulder bones \citep{he_effective_2019} as well as pediatric shoulder muscles \citep{conze_healthy_2020}. However, to the best of our knowledge, the literature on fully-automated pediatric bone segmentation and multi-joint learning schemes remains limited.

While the implementation and optimization of supervised CNN typically requires large amount of annotated data, the conception of imaging datasets is a slow and onerous process \citep{kohli_medical_2017} that is even more challenging for pediatric databases \citep{hirschmann_artificial_2019}. Hence, the inherent scarcity of pediatric imaging resources can induce limited generalization capabilities in neural networks and reduce their performance on unseen images, which in turn may restrict their integration into regular clinical applications. Several strategies have been reported to address this generalizability issue and avoid over-fitting. These include employing multi-task \citep{chen_multi-task_2019, le_multitask_2019, murugesan_psi-net_2019, song_end--end_2020} or multi-domain learning \citep{kamnitsas_unsupervised_2017, karani_lifelong_2018, valindria_multi-modal_2018, chang_domain-specific_2019, dou_unpaired_2020, liu_ms-net_2020}, as well as incorporating regularization terms during optimization \citep{nosrati_incorporating_2016, ravishankar_learning_2017, dalca_anatomical_2018, oktay_anatomically_2018, myronenko_3d_2019, boutillon_multi-structure_2021}. Intuitively, multi-task and multi-domain models benefit from parameter sharing to learn more robust and generic representations than their individual counterparts \citep{zhou_domain_2021, wang_generalizing_2021}. Regularization schemes, for their part, leverage prior knowledge to prevent model over-fitting and have therefore proven to be effective in achieving more accurate and consistent outcomes for medical image segmentation \citep{nosrati_incorporating_2016}. Regularization constraints can originate from different prior information such as boundaries, shape, atlas, or topology \citep{nosrati_incorporating_2016}. Since the common goal of these approaches is to reduce over-fitting, it could be beneficial to combine them, as well as to design regularization terms specific to multi-task, multi-domain learning to further improve performance and to build more generalizable models.

\subsection{Multi-task and multi-domain learning}
\label{sec:multi-task_and_multi-domain_learning}

For medical image analysis, multi-task learning aims at leveraging heterogeneous forms of annotations, from global image labels (e.g. healthy versus impaired musculoskeletal joint) to finer-grained and pixel-level segmentation, to improve the performance of deep models \citep{le_multitask_2019}. An additional advantage of these approaches is that a variety of tasks (e.g. classification, detection, regression, segmentation, etc.) can be solved simultaneously to provide a more complete clinical diagnosis \citep{song_end--end_2020}. Certain frameworks have also proposed to incorporate supplementary sub-tasks (e.g. contour prediction or distance map estimation) to refine coarse, non-smooth, and discontinuous segmentation predictions from convolutional models \citep{murugesan_psi-net_2019}. Additionally, \citep{chen_multi-task_2019} designed an attention based reconstruction task to leverage unlabeled medical images in a semi-supervised segmentation framework. Hence, two types of multi-task strategies emerge in the literature: cascade of task-specific sub-networks \citep{song_end--end_2020}, or networks with shared encoder and task-specific decoders \citep{chen_multi-task_2019, le_multitask_2019, murugesan_psi-net_2019}. The former is characterized by sub-models dedicated for each task that can leverage the output of the previous network as an input, while the latter defines models with partial parameters sharing between tasks. Both approaches have been reported to perform better than traditional independent models by enabling a better cooperation between tasks \citep{chen_multi-task_2019, le_multitask_2019, murugesan_psi-net_2019, song_end--end_2020}. However, the developed pipelines remain specific to a given intensity domain.

In parallel, recent contributions have proposed to train models over multiple intensity domains (e.g. multi-modal, multi-scanner, multi-center, multi-protocol, etc.) with the same segmentation task, in order to leverage a greater amount of training data \citep{kamnitsas_unsupervised_2017, karani_lifelong_2018, valindria_multi-modal_2018, chang_domain-specific_2019, dou_unpaired_2020, liu_ms-net_2020}. These architectures aim at benefiting from the correlation between intensity domains to learn more robust domain-invariant feature representations and prove to be particularly useful when dealing with datasets containing a limited number of samples \citep{karani_lifelong_2018}. Numerous multi-domain schemes have been implemented and reported to achieve better performance than individual approaches. In particular, one can mention models exploiting transfer learning and fine-tuning between domains \citep{karani_lifelong_2018}, models integrating adversarial networks to learn domain-in\-variant features \citep{kamnitsas_unsupervised_2017}, models that share their latent space only \citep{valindria_multi-modal_2018, dou_unpaired_2020}, and models composed of domain-specific encoders and a shared decoder \citep{valindria_multi-modal_2018}. Following this trend to re-use and share an increasing number of parameters, \citep{dou_unpaired_2020} developed a single encoder-decoder segmentation network using shared convolutional kernels and domain- specific internal feature normalization parameters, i.e. batch normalization. While this highly compact architecture reaches superior performance for multi-modal segmentation, their methodology is specific to a given anatomical region of interest (e.g. abdomen or cardiac) and the segmentation task involved the same organs of interest across various intensity domains. 

Furthermore, multi-task, multi-domain learning frameworks have been concurrently developed for natural image analysis. In the context of semantic scene labeling, \citep{fourure_semantic_2016} proposed to train a single network over the union of multiple datasets to address the limited amount of annotated data. In their approach, each dataset is characterized by its own task (segmentation label set) and domain (intensity distribution). Hence, this framework is more generic than traditional multi-task approaches which usually focus on multiple tasks in the same domain or, traditional multi-domain techniques which consider domains containing the same set of objects. Following this, studies on universal representations in computer vision proposed to employ a single model with agnostic kernels, as visual primitives may be shared across tasks and domains, and dataset-specific layers which enable task and domain specialization \citep{bilen_universal_2017, rebuffi_learning_2017, rebuffi_efficient_2018}. These approaches, based on shared representations, have been reported to perform at par or superior to traditional independent models. However, to the best of our knowledge, multi-task, multi-domain learning has rarely been applied to medical image analysis, with the exception of the work of \citep{moeskops_deep_2016} which demonstrated that a single neural network can segment multiple anatomies (i.e. brain, breast, and cardiac) simultaneously. Nevertheless, instead of generating pixel-wise segmentation masks, their model relied on a triplanar patches-based approach that predicted the class of a single pixel per input patch, which proved to be computationally expensive. In particular, their architecture did not comprise a decoder and associated skip connections as in UNet \citep{ronneberger_u-net_2015}, to directly provide whole image segmentation leveraging the global context. Most importantly, patch-wise training lacks the efficiency of fully convolutional training to provide dense output predictions \citep{long_fully_2015}. Their methodology also failed to account for the difference in intensity distribution between domains by, for instance, integrating internal domain-specific feature normalization.

\subsection{Improved generalizability via regularization}
\label{sec:improved_generalizability_via_regularization}

Even though multi-task and multi-domain models can integrate task- and domain-specific information through specialized layers, task and domain prior knowledge could be further exploited to improve the generalizability of learnt shared representations. For instance, \citep{dou_unpaired_2020} introduced a knowledge distillation regularization loss whose goal is to constrain the prediction distributions of their multi-modal segmentation model to be similar across domains. Similarly, \citep{zhu_cross-domain_2020} imposed a Gaussian mixture distribution on the shared latent representation of their image translation network to preserve fine structures between domains. However, such a hypothesis may be too restrictive. Indeed, in representation learning, a good representation can be characterized by the presence of natural clusters corresponding to the classes of the problem (i.e. disentangled representation) \citep{bengio_representation_2013}. Hence, a number of self-supervised representation learning techniques focus on pulling together data points from the same class and pushing apart negative samples in embedded space using a contrastive metric \citep{hadsell_dimensionality_2006, chen_simple_2020}. A recent contribution extended this idea to fully-supervised image classification setting by leveraging the label information and considering many positive anchors simultaneously \citep{khosla_supervised_2020}. Thus, the contrastive regularization maximizes the performance of the classifier by imposing intra-class cohesion and inter-class separation in latent space. In the context of semi-supervised medical image segmentation, \citep{hu_semi-supervised_2021} exploited unannotated data by designing a contrastive loss forcing pixels from the same class to assemble in embedded space. Unlike \citep{zhu_cross-domain_2020}, in these non-parametric contrastive approaches, it is not necessary to define a prior distribution (e.g. Gaussian, Poisson, etc.) for the latent variables. Hence, contrastive regularization techniques appear more generic and appropriate to impose domain-specific clusters in the shared representations of deep multi-task, multi-domain models.

In addition, regularization schemes can arise from other forms of prior information such as shape models of the targeted anatomical structures. For deep learning based medical image segmentation, incorporating shape or anatomical priors has already proven to be useful in reducing the effect of noise, low contrast, and artefacts \citep{ravishankar_learning_2017, dalca_anatomical_2018, oktay_anatomically_2018, myronenko_3d_2019, boutillon_multi-structure_2021}. Indeed, recent works have proposed to learn a representation of the anatomy from ground truth annotations using a deep auto-encoder. Data-driven models such as auto-encoders are suitable for learning anatomical prior information due to the constrained nature of anatomical structures. Hence, anatomical priors arising from round truth annotations usually integrate position, orientation, size, and shape information of the targeted structures. The learnt non-linear anatomical representation is then integrated in the segmentation network during optimization, by enforcing the predicted segmentation to be close to the ground truth in anatomical space using a regularization term based on Euclidean distance \citep{ravishankar_learning_2017, oktay_anatomically_2018}. One can also employ a shape code discriminator to guide the segmentation network towards more consistent and plausible shape delineations \citep{boutillon_multi-structure_2021}. However, to the best of our knowledge, none of these studies on anatomical priors have proposed to simultaneously encode multiple anatomical regions in order to leverage position, orientation, size, and shape correlations between similar anatomical objects, such as pediatric bones across distinct musculoskeletal joints.

\subsection{Contributions}
\label{sec:contributions}

In this study, we propose to implement and optimize a single segmentation network over the union of multiple pediatric imaging datasets arising from separate regions of the anatomy. Unlike previous methods that operate on individual pediatric musculoskeletal joint, our framework simultaneously learns multiple intensity domains and segmentation tasks emerging from distinct anatomical joints (Fig. \ref{fig:introduction}). This approach allows to overcome the inherent scarcity of pediatric data while benefiting from more robust shared representations. Our main contributions are summarized as follows:
\begin{itemize}
    \item We formalize a segmentation model which incorporates a pre-trained Efficient encoder, shared convolutional filters, multi-domain attention gates, domain-specific batch normalization, and domain-specific output layers.
    \item We integrate a multi-scale contrastive regularization during optimization  to improve the generalization capabilities of neural networks. As opposed to classical contrastive approaches that operate on image classes, we leverage dataset label information to enhance intra-domain similarity and impose inter-domain margins.
    \item We extend the multi-task, multi-domain segmentation learning framework by incorporating multi-joint anatomical priors which encode the anatomical characteristics of multiple joints and further constrain the delineation tasks.
    \item We design a multi-joint auto-encoder whose architecture consists of shared convolutional kernels and domain-specific layers while its learning scheme also relies on the proposed multi-scale contrastive regularization to impose clusterization constraint on anatomical representations.
    \item We illustrate the effectiveness of our approach for multi-task, multi-domain segmentation on three sparse, unpaired (from different patient cohorts), and heterogeneous pediatric musculoskeletal MR imaging datasets.
\end{itemize}

This paper constitutes an extended version of our earlier work \citep{boutillon_multi-task_2021} with respect to the following points: (a) Incorporation of an additional pediatric knee joint MR imaging dataset, which corresponds to a novel intensity domain and segmentation task. (b) Modifications to the segmentation network with a new design based on an Efficient encoder that leverages transfer learning from natural images classification task \citep{tan_efficientnet_2019}. (c) Extension of the contrastive regularization from single-scale to multi-scale and integration of multi-joint anatomical priors to further improve generalizability. (d) Finally, we perform extensive experiments including visualization and quantitative analysis of the learnt representations through t-SNE \citep{maaten_visualizing_2008} algorithm and cosine similarity metric to thoroughly assess our contributions.

\section{Methodology}
\label{sec:methodology}

\begin{figure*}[ht!]
\centering
\begin{adjustbox}{width=\textwidth}
\tikzstyle{dashed}=[dash pattern=on .85pt off .85pt]
\begin{tikzpicture}[every node/.style={inner sep=0,outer sep=0}]

\draw[line width=0.1mm, color=darkgray, rounded corners=1] (-.35, .35) -- (-.35,-2) -- (3.45,-2) -- (3.45,1.375) -- (2.8,1.375) -- (2.8,.35) -- cycle;
\draw[line width=0.1mm, color=darkgray, rounded corners=1] (-.35, .35) -- (-.35,1.375) -- (2.8,1.375) -- (2.8,.35) -- cycle;

\draw[line width=0.1mm, color=RedViolet!80, rounded corners=1, dashed] (-.325,-1.375) -- (1.545,-1.375) -- (1.545,-1.975) -- (-.325, -1.975) --  cycle;
\draw[line width=0.1mm, color=BrickRed!80, rounded corners=1, dashed] (3.425,-1.375) -- (1.555,-1.375) -- (1.555,-1.975) -- (3.425, -1.975) -- cycle;
\draw[line width=0.1mm, color=MidnightBlue!80, rounded corners=1, dashed] (3.425,-1.365) -- (2.825,-1.365) -- (2.825,-0.0175) -- (3.425, -0.0175) -- cycle;
\draw[line width=0.1mm, color=BurntOrange!80, rounded corners=1, dashed] (3.425,1.35) -- (2.825,1.35) -- (2.825,-0.0075) -- (3.425, -0.0075) -- cycle;

\node[inner sep=0pt] (mri_a) at (-.15,.125)
    {\includegraphics[width=.0125\textwidth]{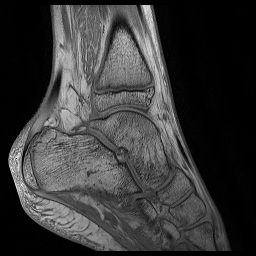}};
\draw [CornflowerBlue!40, rounded corners=.5, line width=.5] (mri_a.north west) -- (mri_a.north east) -- (mri_a.south east) -- (mri_a.south west) -- cycle;

\node[inner sep=0pt] (mri_k) at (0,-.125)
    {\includegraphics[width=.0125\textwidth]{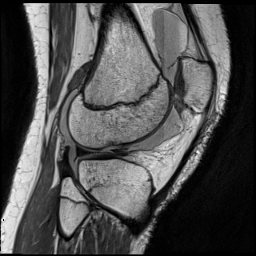}};
\draw [Orchid!40, rounded corners=.5, line width=.5] (mri_k.north west) -- (mri_k.north east) -- (mri_k.south east) -- (mri_k.south west) -- cycle;
    
\node[inner sep=0pt] (mri_s) at (.15,.125)
    {\includegraphics[width=.0125\textwidth]{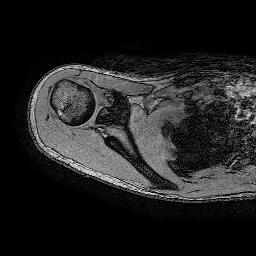}};
\draw [ForestGreen!40, rounded corners=.5, line width=.5] (mri_s.north west) -- (mri_s.north east) -- (mri_s.south east) -- (mri_s.south west) -- cycle;

\node[inner sep=0pt] (pred_a) at (1.35,.125)
    {\includegraphics[width=.0125\textwidth]{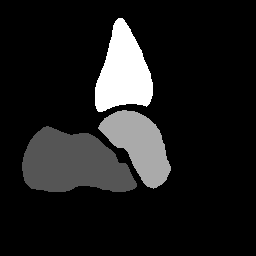}};  
\draw [CornflowerBlue!40, rounded corners=.5, line width=.5] (pred_a.north west) -- (pred_a.north east) -- (pred_a.south east) -- (pred_a.south west) -- cycle;

\node[inner sep=0pt] (pred_k) at (1.5,-.125)
    {\includegraphics[width=.0125\textwidth]{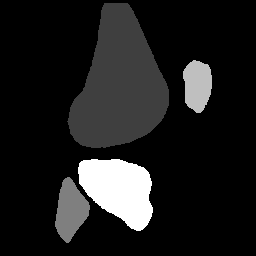}};
\draw [Orchid!40, rounded corners=.5, line width=.5] (pred_k.north west) -- (pred_k.north east) -- (pred_k.south east) -- (pred_k.south west) -- cycle;

\node[inner sep=0pt] (pred_s) at (1.65,.125)
    {\includegraphics[width=.0125\textwidth]{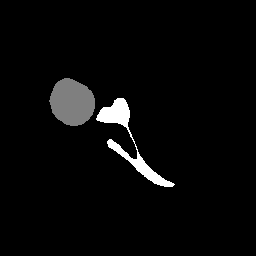}};
\draw [ForestGreen!40, rounded corners=.5, line width=.5] (pred_s.north west) -- (pred_s.north east) -- (pred_s.south east) -- (pred_s.south west) -- cycle;

\node[inner sep=0pt] (gt_a) at (1.35,-.875)
    {\includegraphics[width=.0125\textwidth]{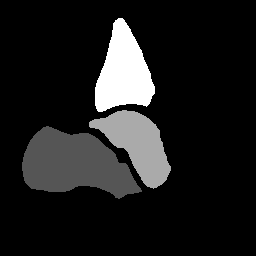}};   
\draw [CornflowerBlue!40, rounded corners=.5, line width=.5] (gt_a.north west) -- (gt_a.north east) -- (gt_a.south east) -- (gt_a.south west) -- cycle;

\node[inner sep=0pt] (gt_k) at (1.5,-1.125)
    {\includegraphics[width=.0125\textwidth]{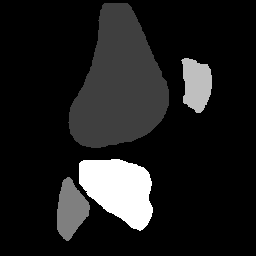}};
\draw [Orchid!40, rounded corners=.5, line width=.5] (gt_k.north west) -- (gt_k.north east) -- (gt_k.south east) -- (gt_k.south west) -- cycle;

\node[inner sep=0pt] (gt_s) at (1.65,-.875)
    {\includegraphics[width=.0125\textwidth]{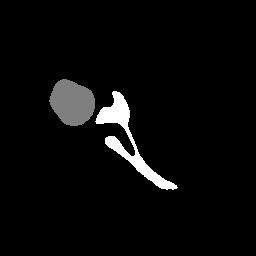}};
\draw [ForestGreen!40, rounded corners=.5, line width=.5] (gt_s.north west) -- (gt_s.north east) -- (gt_s.south east) -- (gt_s.south west) -- cycle;

\draw[line width=0.1mm, rounded corners = 1, -{Latex[length=1.5pt, width=1.5pt]}] (.3,0) -- (.45,0);
\draw[line width=0.1mm, rounded corners = 1, -{Latex[length=1.5pt, width=1.5pt]}] (1.05,0) -- (1.2,0);
\draw[line width=0.1mm, rounded corners = 1, -{Latex[length=1.5pt, width=1.5pt]}] (1.8,0) -- (1.95,0);
\draw[line width=0.1mm, rounded corners = 1, -{Latex[length=1.5pt, width=1.5pt]}] (1.8,-1) -- (1.95,-1);

\node at (0,-.285) {\scalebox{.2}{MR images}};
\node at (1.5,-.285) {\scalebox{.2}{Predictions}};
\node at (1.5,-.715) {\scalebox{.2}{Ground truths}};

\node at (.75,.315) {\scalebox{.2}{Efficient-UNet $S$}};
\draw[line width=0.1mm, fill=BurntOrange!60, rounded corners = 1] (.985,.2229) -- (.75,.125) -- (.45,.25) -- (.45,-.25) -- (.75,-.125) --  (.985,-.2229);
\draw[line width=0.1mm, fill=CarnationPink!70, rounded corners = 1] (.985,.2229) -- (1.05,.25) -- (1.05,-.25) --  (.985,-.2229);
\draw[line width=0.01mm, fill=BrickRed!60] (.485,.2354) -- (.515,.2229) -- (.515,-.2229) -- (.485,-.2354) -- cycle;
\draw[line width=0.01mm, pattern={Lines[angle=45,distance={3pt/sqrt(2)}]}, pattern color=black, solid] (.485,.2354) -- (.515,.2229) -- (.515,-.2229) -- (.485,-.2354) -- cycle;
\draw[line width=0.01mm, fill=BrickRed!60] (.585,.1937) -- (.615,.1812) -- (.615,-.1812) -- (.585,-.1937) -- cycle;
\draw[line width=0.01mm, pattern={Lines[angle=45,distance={3pt/sqrt(2)}]}, pattern color=black] (.585,.1937) -- (.615,.1812) -- (.615,-.1812) -- (.585,-.1937) -- cycle;
\draw[line width=0.01mm, fill=BrickRed!60] (.685,.1520) -- (.715,.1395) -- (.715,-.1395) -- (.685,-.1520) -- cycle;
\draw[line width=0.01mm, pattern={Lines[angle=45,distance={3pt/sqrt(2)}]}, pattern color=black] (.685,.1520) -- (.715,.1395) -- (.715,-.1395) -- (.685,-.1520) -- cycle;
\draw[line width=0.01mm, fill=BrickRed!60] (.785,.1395) -- (.815,.1520) -- (.815,-.1520) -- (.785,-.1395) -- cycle;
\draw[line width=0.01mm,  pattern={Lines[angle=45,distance={3pt/sqrt(2)}]}, pattern color=black] (.785,.1395) -- (.815,.1520) -- (.815,-.1520) -- (.785,-.1395) -- cycle;
\draw[line width=0.01mm, fill=BrickRed!60] (.885,.1812) -- (.915,.1937) -- (.915,-.1937) -- (.885,-.1812) -- cycle;
\draw[line width=0.01mm, pattern={Lines[angle=45,distance={3pt/sqrt(2)}]}, pattern color=black] (.885,.1812) -- (.915,.1937) -- (.915,-.1937) -- (.885,-.1812) -- cycle;
\draw[line width=0.01mm, fill=BrickRed!60] (.985,.2229) -- (1.015,.2354) -- (1.015,-.2354) -- (.985,-.2229) -- cycle;
\draw[line width=0.01mm, pattern={Lines[angle=45,distance={3pt/sqrt(2)}]}, pattern color=black] (.985,.2229) -- (1.015,.2354) -- (1.015,-.2354) -- (.985,-.2229) -- cycle;
\draw[line width=0.1mm, rounded corners = 1] (.45,.25) -- (.75,.125) -- (1.05,.25) -- (1.05,-.25) -- (.75,-.125) -- (.45,-.25) -- cycle;

\draw[line width=0.1mm, rounded corners = .5, -{Latex[length=1pt, width=1pt]}] (.55,.2083) -- (.55,.2583) -- (.825,.2583);
\draw[line width=0.1mm,  -{Latex[length=1pt, width=1pt]}] (.85,.1666) -- (.85,.2333);
\draw[line width=0.1mm, rounded corners = .5, , -{Latex[length=1pt, width=1pt]}] (.875,.2583) -- (.95,.2583) -- (.95,.2083);
\draw[RedViolet!80, line width=0.1mm] (.85, .2583) circle (.025);
\draw[RedViolet!80, line width=0.1mm, rounded corners = 1] (.8677,.276) -- (.8323,.2406);
\draw[RedViolet!80, line width=0.1mm, rounded corners = 1] (.8323,.276) -- (.8677,.2406);

\draw[line width=0.1mm, rounded corners = 1, -{Latex[length=1.5pt, width=1.5pt]}] (.75,-.13) -- (.75,-.375);
\draw[line width=0.1mm, rounded corners = 1, -{Latex[length=1.5pt, width=1.5pt]}] (.65,-.1666) -- (.65,-.275) -- (.75,-.275) -- (.75,-.375);
\draw[line width=0.1mm, rounded corners = 1, -{Latex[length=1.5pt, width=1.5pt]}] (.85,-.1666) -- (.85,-.275) -- (.75,-.275) -- (.75,-.375);
\draw[line width=0.1mm, rounded corners = 1, -{Latex[length=1.5pt, width=1.5pt]}] (.55,-.2083) -- (.55,-.275) -- (.75,-.275) -- (.75,-.375);
\draw[line width=0.1mm, rounded corners = 1, -{Latex[length=1.5pt, width=1.5pt]}] (.95,-.2083) -- (.95,-.275) -- (.75,-.275) -- (.75,-.375);

\draw[line width=0.1mm, rounded corners = 1] (.625,-.375) -- (.875,-.375) -- (.875,-.475) -- (.625,-.475) -- cycle;
\node at (.75,-.425) {\scalebox{.2}{$\mathcal{L}_{\text{MSC}}$}};

\draw[line width=0.1mm, rounded corners = 1, -{Latex[length=1.5pt, width=1.5pt]}] (1.5,-.325) -- (1.5,-.45);
\draw[line width=0.1mm, rounded corners = 1, -{Latex[length=1.5pt, width=1.5pt]}] (1.5,-.675) -- (1.5,-.55);

\draw[line width=0.1mm, rounded corners = 1] (1.375,-.45) -- (1.625,-.45) -- (1.625,-.55) -- (1.375,-.55) -- cycle;
\node at (1.5,-.5) {\scalebox{.2}{$\mathcal{L}_{\text{CE}}$}};

\draw[line width=0.1mm, fill=BurntOrange!60, rounded corners = 1] (2.25,.125) -- (1.95,.25) -- (1.95,-.25) -- (2.25,-.125) -- cycle;
\draw[line width=0.1mm, fill=CarnationPink!70, rounded corners = 1] (2.015,.2229) -- (1.95,.25) -- (1.95,-.25) --  (2.015,-.2229);
\draw[line width=0.01mm, fill=BrickRed!60] (1.985,.2354) -- (2.015,.2229) -- (2.015,-.2229) -- (1.985,-.2354) -- cycle;
\draw[line width=0.01mm, pattern={Lines[angle=45,distance={3pt/sqrt(2)}]}, pattern color=black, solid] (1.985,.2354) -- (2.015,.2229) -- (2.015,-.2229) -- (1.985,-.2354) -- cycle;
\draw[line width=0.01mm, fill=BrickRed!60] (2.085,.1937) -- (2.115,.1812) -- (2.115,-.1812) -- (2.085,-.1937) -- cycle;
\draw[line width=0.01mm, pattern={Lines[angle=45,distance={3pt/sqrt(2)}]}, pattern color=black] (2.085,.1937) -- (2.115,.1812) -- (2.115,-.1812) -- (2.085,-.1937) -- cycle;
\draw[line width=0.01mm, fill=BrickRed!60] (2.185,.1520) -- (2.215,.1395) -- (2.215,-.1395) -- (2.185,-.1520) -- cycle;
\draw[line width=0.01mm, pattern={Lines[angle=45,distance={3pt/sqrt(2)}]}, pattern color=black] (2.185,.1520) -- (2.215,.1395) -- (2.215,-.1395) -- (2.185,-.1520) -- cycle;
\draw[line width=0.1mm, rounded corners = 1] (1.95,.25) -- (2.25,.125) -- (2.25,-.125) -- (1.95,-.25) -- cycle;

\draw[line width=0.1mm, fill=BurntOrange!60, rounded corners = 1] (2.25,-.875) -- (1.95,-.75) -- (1.95,-1.25) -- (2.25,-1.125) -- cycle;
\draw[line width=0.1mm, fill=CarnationPink!70, rounded corners = 1] (2.015,-.7771) -- (1.95,-.75) -- (1.95,-1.25) --  (2.015,-1.2229);
\draw[line width=0.01mm, fill=BrickRed!60] (1.985,-.7646) -- (2.015,-.7771) -- (2.015,-1.2229) -- (1.985,-1.2354) -- cycle;
\draw[line width=0.01mm, pattern={Lines[angle=45,distance={3pt/sqrt(2)}]}, pattern color=black, solid] (1.985,-.7646) -- (2.015,-.7771) -- (2.015,-1.2229) -- (1.985,-1.2354) -- cycle;
\draw[line width=0.01mm, fill=BrickRed!60] (2.085,-.8063) -- (2.115,-.8188) -- (2.115,-1.1812) -- (2.085,-1.1937) -- cycle;
\draw[line width=0.01mm, pattern={Lines[angle=45,distance={3pt/sqrt(2)}]}, pattern color=black] (2.085,-.8063) -- (2.115,-.8188) -- (2.115,-1.1812) -- (2.085,-1.1937) -- cycle;
\draw[line width=0.01mm, fill=BrickRed!60] (2.185,-.8480) -- (2.215,-.8605) -- (2.215,-1.1395) -- (2.185,-1.1520) -- cycle;
\draw[line width=0.01mm, pattern={Lines[angle=45,distance={3pt/sqrt(2)}]}, pattern color=black] (2.185,-.8480) -- (2.215,-.8605) -- (2.215,-1.1395) -- (2.185,-1.1520) -- cycle;
\draw[line width=0.1mm, rounded corners = 1] (1.95,-.75) -- (2.25,-.875) -- (2.25,-1.125) -- (1.95,-1.25) -- cycle;

\node at (2.1,-.425) {\scalebox{.2}{Multi-joint}};
\node at (2.1,-.5) {\scalebox{.2}{Anatomical Encoder $F$}};
\node at (2.1,-.575) {\scalebox{.2}{(Fixed Weights)}};
\draw[line width=0.1mm, color=MidnightBlue!80, dashed, {Latex[length=1.5pt, width=1.5pt]}-] (2.1, -.385) -- (2.1,-.18745);  
\draw[line width=0.1mm, color=MidnightBlue!80, dashed, -{Latex[length=1.5pt, width=1.5pt]}] (2.1, -.615) -- (2.1,-.81255); 

\draw[line width=0.1mm, rounded corners = 1, -{Latex[length=1.5pt, width=1.5pt]}] (2.25,0) -- (2.6,0) -- (2.6,-.45);
\draw[line width=0.1mm, rounded corners = 1, -{Latex[length=1.5pt, width=1.5pt]}] (2.25,-1) -- (2.6,-1) -- (2.6,-.55);

\draw[line width=0.1mm, rounded corners = 1] (2.475,-.45) -- (2.725,-.45) -- (2.725,-.55) -- (2.475,-.55) -- cycle;
\node at (2.6,-.5) {\scalebox{.2}{$\mathcal{L}_{\text{MJAP}}$}};

\node[inner sep=0pt] (gt_a) at (.325,1.125)
    {\includegraphics[width=.0125\textwidth]{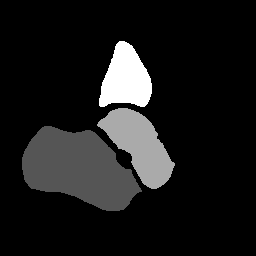}};
\draw [CornflowerBlue!40, rounded corners=.5, line width=.5] (gt_a.north west) -- (gt_a.north east) -- (gt_a.south east) -- (gt_a.south west) -- cycle;

\node[inner sep=0pt] (gt_k) at (.475,.875)
    {\includegraphics[width=.0125\textwidth]{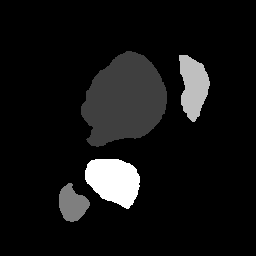}};
\draw [Orchid!40, rounded corners=.5, line width=.5] (gt_k.north west) -- (gt_k.north east) -- (gt_k.south east) -- (gt_k.south west) -- cycle;

\node[inner sep=0pt] (gt_s) at (.625,1.125)
    {\includegraphics[width=.0125\textwidth]{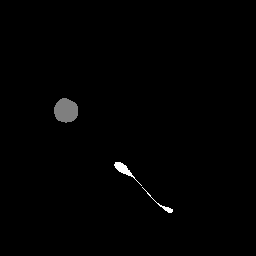}};
\draw [ForestGreen!40, rounded corners=.5, line width=.5] (gt_s.north west) -- (gt_s.north east) -- (gt_s.south east) -- (gt_s.south west) -- cycle;

\node[inner sep=0pt] (pred_a) at (1.825,1.125)
    {\includegraphics[width=.0125\textwidth]{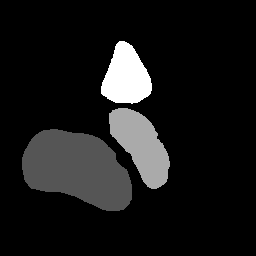}};   
\draw [CornflowerBlue!40, rounded corners=.5, line width=.5] (pred_a.north west) -- (pred_a.north east) -- (pred_a.south east) -- (pred_a.south west) -- cycle;

\node[inner sep=0pt] (pred_k) at (1.975,.875)
    {\includegraphics[width=.0125\textwidth]{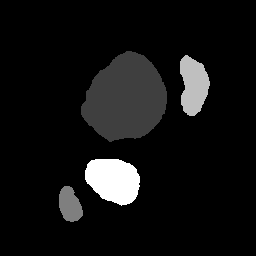}};
\draw [Orchid!40, rounded corners=.5, line width=.5] (pred_k.north west) -- (pred_k.north east) -- (pred_k.south east) -- (pred_k.south west) -- cycle;

\node[inner sep=0pt] (pred_s) at (2.125,1.125)
    {\includegraphics[width=.0125\textwidth]{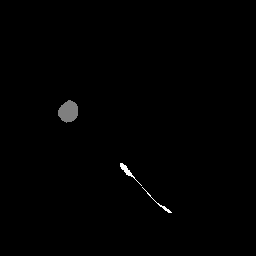}};
\draw [ForestGreen!40, rounded corners=.5, line width=.5] (pred_s.north west) -- (pred_s.north east) -- (pred_s.south east) -- (pred_s.south west) -- cycle;

\draw[line width=0.1mm, rounded corners = 1, -{Latex[length=1.5pt, width=1.5pt]}] (.775,1) -- (.925,1);
\draw[line width=0.1mm, rounded corners = 1, -{Latex[length=1.5pt, width=1.5pt]}] (1.525,1) -- (1.675,1);

\node at (1.975,.715) {\scalebox{.2}{Predictions}};
\node at (.475,.715) {\scalebox{.2}{Ground truths}};

\node[anchor=west] at (-.3,0.6) {\scalebox{.22}{\textbf{1. Learning multi-joint}}};
\node[anchor=west] at (-.3,0.5125) {\scalebox{.22}{\textbf{anatomical priors}}};

\node at (1.225,1.325) {\scalebox{.2}{Multi-joint Auto-encoder}};
\draw[line width=0.1mm, fill=BurntOrange!60, rounded corners = 1] (1.46,1.2229) -- (1.225,1.125) -- (.925,1.25) -- (.925,.75) -- (1.225,.875) --  (1.46,.7771);
\draw[line width=0.1mm, fill=CarnationPink!70, rounded corners = 1] (1.46,1.2229) -- (1.525,1.25) -- (1.525,.75) --  (1.46,.7771);
\draw[line width=0.1mm, fill=CarnationPink!70, rounded corners = 1] (.99,1.2229) -- (.925,1.25) -- (.925,.75) --  (.99,.7771);
\draw[line width=0.01mm, fill=BrickRed!60] (.96,1.2354) -- (.99,1.2229) -- (.99,.7771) -- (.96,.7646) -- cycle;
\draw[line width=0.01mm, pattern={Lines[angle=45,distance={3pt/sqrt(2)}]}, pattern color=black, solid] (.96,1.2354) -- (.99,1.2229) -- (.99,.7771) -- (.96,.7646) -- cycle;
\draw[line width=0.01mm, fill=BrickRed!60] (1.06,1.1937) -- (1.09,1.1812) -- (1.09,.8188) -- (1.06,.8063) -- cycle;
\draw[line width=0.01mm, pattern={Lines[angle=45,distance={3pt/sqrt(2)}]}, pattern color=black] (1.06,1.1937) -- (1.09,1.1812) -- (1.09,.8188) -- (1.06,.8063) -- cycle;
\draw[line width=0.01mm, fill=BrickRed!60] (1.16,1.1520) -- (1.19,1.1395) -- (1.19,.8605) -- (1.16,.848) -- cycle;
\draw[line width=0.01mm, pattern={Lines[angle=45,distance={3pt/sqrt(2)}]}, pattern color=black] (1.16,1.1520) -- (1.19,1.1395) -- (1.19,.8605) -- (1.16,.848) -- cycle;
\draw[line width=0.01mm, fill=BrickRed!60] (1.26,1.1395) -- (1.29,1.1520) -- (1.29,.848) -- (1.26,.8605) -- cycle;
\draw[line width=0.01mm,  pattern={Lines[angle=45,distance={3pt/sqrt(2)}]}, pattern color=black] (1.26,1.1395) -- (1.29,1.1520) -- (1.29,.848) -- (1.26,.8605) -- cycle;
\draw[line width=0.01mm, fill=BrickRed!60] (1.36,1.1812) -- (1.39,1.1937) -- (1.39,.8063) -- (1.36,.8188) -- cycle;
\draw[line width=0.01mm, pattern={Lines[angle=45,distance={3pt/sqrt(2)}]}, pattern color=black] (1.36,1.1812) -- (1.39,1.1937) -- (1.39,.8063) -- (1.36,.8188) -- cycle;
\draw[line width=0.01mm, fill=BrickRed!60] (1.46,1.2229) -- (1.49,1.2354) -- (1.49,.7646) -- (1.46,.7771) -- cycle;
\draw[line width=0.01mm, pattern={Lines[angle=45,distance={3pt/sqrt(2)}]}, pattern color=black] (1.46,1.2229) -- (1.49,1.2354) -- (1.49,.7646) -- (1.46,.7771) -- cycle;
\draw[line width=0.1mm, rounded corners = 1] (.925,1.25) -- (1.225,1.125) -- (1.525,1.25) -- (1.525,.75) -- (1.225,.875) -- (.925,.75) -- cycle;

\draw[line width=0.1mm, rounded corners = 1, -{Latex[length=1.5pt, width=1.5pt]}] (1.225,.87) -- (1.225,.625);
\draw[line width=0.1mm, rounded corners = 1, -{Latex[length=1.5pt, width=1.5pt]}] (1.125,.8334) -- (1.125,.725) -- (1.225,.725) -- (1.225,.625);
\draw[line width=0.1mm, rounded corners = 1, -{Latex[length=1.5pt, width=1.5pt]}] (1.325,.8334) -- (1.325,.725) -- (1.225,.725) -- (1.225,.625);
\draw[line width=0.1mm, rounded corners = 1, -{Latex[length=1.5pt, width=1.5pt]}] (1.025,.7917) -- (1.025,.725) -- (1.225,.725) -- (1.225,.625);
\draw[line width=0.1mm, rounded corners = 1, -{Latex[length=1.5pt, width=1.5pt]}] (1.425,.7917) -- (1.425,.725) -- (1.225,.725) -- (1.225,.625);

\draw[line width=0.1mm, rounded corners = 1] (1.1,.625) -- (1.35,.625) -- (1.35,.525) -- (1.1,.525) -- cycle;
\node at (1.225,.575) {\scalebox{.2}{$\mathcal{L}_{\text{MSC}}$}};

\draw[line width=0.1mm, rounded corners = 1, -{Latex[length=1.5pt, width=1.5pt]}] (1.975,.675) -- (1.975,.45) -- (1.35,.45);
\draw[line width=0.1mm, rounded corners = 1, -{Latex[length=1.5pt, width=1.5pt]}] (.475,.675) -- (.475,.45) -- (1.1,.45);
    
\draw[line width=0.1mm, rounded corners = 1] (1.1,.4) -- (1.35,.4) -- (1.35,.5) -- (1.1,.5) -- cycle;
\node at (1.225,.45) {\scalebox{.2}{$\mathcal{L}_{\text{CE}}$}};

\draw[line width=0.1mm, color=MidnightBlue!80, rounded corners=1, dashed] (1.225,  1.15) -- (1.225, .85) -- (.9, .715) -- (.9, 1.285) -- cycle;  
\draw[line width=0.1mm, color=MidnightBlue!80, rounded corners=1, dashed, -{Latex[length=1.5pt, width=1.5pt]}] (1.075, 1.2077) -- (1.075,1.275) -- (2.325,1.275) -- (2.325, .3) -- (2.1, .3) -- (2.1, .18745);  

\draw[line width=0.05mm, fill=BurntOrange!60, rounded corners =.5] (-.3,-.85) -- (-.3,-.8) -- (-.1,-.8) -- (-.1,-.85) -- cycle;
\draw[line width=0.05mm, fill=BrickRed!60, rounded corners =.5] (-.3,-.925) -- (-.3,-.975) -- (-.1,-.975) -- (-.1,-.925) -- cycle;
\draw[line width=0.05mm, pattern={Lines[angle=45,distance={3pt/sqrt(2)}]}, pattern color=black, rounded corners =.5] (-.3,-.925) -- (-.3,-.975) -- (-.1,-.975) -- (-.1,-.925) -- cycle;
\draw[line width=0.05mm, fill=CarnationPink!70, rounded corners =.5] (-.3,-1.0875) -- (-.3,-1.1375) -- (-.1,-1.1375) -- (-.1,-1.0875) -- cycle;

\draw[line width=0.1mm,  -{Latex[length=1pt, width=1pt]}] (-.3,-1.2325) -- (-0.225,-1.2325);
\draw[line width=0.1mm,  -{Latex[length=1pt, width=1pt]}] (-.175,-1.2325) -- (-.1,-1.2325);
\draw[line width=0.1mm,  -{Latex[length=1pt, width=1pt]}] (-.2,-1.3325) -- (-.2,-1.2575);
\draw[RedViolet!80, line width=0.1mm] (-.2, .-1.2325) circle (.025);
\draw[RedViolet!80, line width=0.1mm] (-0.2177,-1.2148) -- (-0.1823,-1.2502);
\draw[RedViolet!80, line width=0.1mm] (-0.1823,-1.2148) -- (-0.2177,-1.2502);

\node[anchor=west] at (-.3,-.6125) {\scalebox{.22}{\textbf{2. Multi-task, multi-domain}}};
\node[anchor=west] at (-.3,-.7) {\scalebox{.22}{\textbf{segmentation framework}}};
\node[anchor=west] at (-.05,-.825) {\scalebox{.2}{Shared convolutional filters}};
\node[anchor=west] at (-.05,-.95) {\scalebox{.2}{Domain-specific batch normalization}};
\node[anchor=west] at (-.05,-1.075) {\scalebox{.2}{Domain-specific input/output}};
\node[anchor=west] at (-.05,-1.15) {\scalebox{.2}{segmentation layers}};
\node[anchor=west] at (-.05,-1.275) {\scalebox{.2}{Multi-domain attention gate}};

\node at (2.49,-1.45) {\scalebox{.21}{\textbf{2.c. Domain-specific batch normalization (DSBN)}}};
\draw[line width=0.1mm, color=BrickRed!80, rounded corners=1, dashed, -{Latex[length=1.5pt, width=1.5pt]}] (1.025, -.95) -- (1.1, -.95) -- (1.1,-1.275) -- (2.1875,-1.275) -- (2.1875, -1.375);  

\draw[line width=0.05mm, -{Latex[length=1pt, width=1pt]}] (1.73, -1.8959) -- (2.5625, -1.6168);
\draw[line width=0.05mm, -{Latex[length=1pt, width=1pt]}] (1.73, -1.8501) -- (2.2625, -1.8501);
\draw[line width=0.05mm, -{Latex[length=1pt, width=1pt]}] (1.9375, -1.7751) -- (2.4875, -1.7751);
\draw[line width=0.05mm, -{Latex[length=1pt, width=1pt]}] (2.1625, -1.7001) -- (2.7125, -1.7001);
\draw[line width=0.05mm, -{Latex[length=1pt, width=1pt]}] (1.8625, -1.8501) -- (1.8625, -1.6751);
\draw[line width=0.05mm, -{Latex[length=1pt, width=1pt]}] (2.0875, -1.7751) -- (2.0875, -1.6001);
\draw[line width=0.05mm, -{Latex[length=1pt, width=1pt]}] (2.3125, -1.7001) -- (2.3125, -1.5251);

\draw[line width=0.1mm, domain=-.15:.15, smooth, variable=\x, CornflowerBlue] plot ({\x + 2.3375} , {exp(-(\x*\x)/(2*.0036)) * .09 - 1.7001});
\draw[line width=0.1mm, domain=-.15:.15, smooth, variable=\x, Orchid] plot ({\x + 2.3875} , {exp(-(\x*\x)/(2*.0009)) * .15 - 1.7001});
\draw[line width=0.1mm, domain=-.15:.15, smooth, variable=\x, ForestGreen] plot ({\x + 2.4375} , {exp(-(\x*\x)/(2*.0025)) * .12 - 1.7001});

\draw[line width=0.1mm, domain=-.15:.15, smooth, variable=\x, CornflowerBlue] plot ({\x +2.1375} , {exp(-(\x*\x)/(2*.0009)) * .16 - 1.7751});
\draw[line width=0.1mm, domain=-.15:.15, smooth, variable=\x, Orchid] plot ({\x +2.2375} , {exp(-(\x*\x)/(2*.0036)) * .10 - 1.7751});
\draw[line width=0.1mm, domain=-.15:.15, smooth, variable=\x, ForestGreen] plot ({\x + 2.0375} , {exp(-(\x*\x)/(2*.0016)) * .14 - 1.7751});

\draw[line width=0.1mm, domain=-.15:.15, smooth, variable=\x, CornflowerBlue] plot ({\x + 1.8875} , {exp(-(\x*\x)/(2*.0009)) * .15 - 1.8501});
\draw[line width=0.1mm, domain=-.15:.15, smooth, variable=\x, Orchid] plot ({\x + 1.8375} , {exp(-(\x*\x)/(2*.0007)) * .16 - 1.8501});
\draw[line width=0.1mm, domain=-.15:.15, smooth, variable=\x, ForestGreen] plot ({\x + 1.9375} , {exp(-(\x*\x)/(2*.0004)) * .17 - 1.8501});

\node[anchor=west] at (1.8825,-1.9) {\scalebox{.2}{Features}};
\node at (2.6925, -1.55) {\scalebox{.2}{Network}};
\node at (2.6925, -1.625) {\scalebox{.2}{depth}};
\draw[line width=0.1mm, CornflowerBlue] (2.82, -1.7) -- (2.97, -1.7);
\node[anchor=west] at (2.995, -1.7) {\scalebox{.2}{Ankle}};
\draw[line width=0.1mm, Orchid] (2.82, -1.8) -- (2.97, -1.8);
\node[anchor=west] at (2.995, -1.8) {\scalebox{.2}{Knee}};
\draw[line width=0.1mm, ForestGreen] (2.82, -1.9) -- (2.97, -1.9);
\node[anchor=west] at (2.995, -1.9) {\scalebox{.2}{Shoulder}};

\node at (3.125,1.275) {\scalebox{.21}{\textbf{2.a. Shared}}};
\node at (3.125,1.2) {\scalebox{.21}{\textbf{representation}}};
\draw[line width=0.1mm, color=BurntOrange!80, rounded corners=1, dashed, -{Latex[length=1.5pt, width=1.5pt]}] (2.15, .1666) -- (2.15, .2266) -- (2.825, .2266); 

\fill [CornflowerBlue!20, opacity=.75] plot [smooth cycle] coordinates {(2.86, 1.0325) (2.99,1.1325) (3.15, 1.0225) (3.04, 0.9325)};
\fill[CornflowerBlue] (3.02,1.0125) circle (.02);
\fill[CornflowerBlue] (2.92,1.0225) circle (.02);
\fill[CornflowerBlue] (2.99,0.9725) circle (.02);
\fill[CornflowerBlue] (3.01,1.0925) circle (.02);
\fill[CornflowerBlue] (2.94,1.0725) circle (.02);
\fill[CornflowerBlue] (3.09,1.0325) circle (.02);
\fill[CornflowerBlue] (3.05,1.0725) circle (.02);

\fill [Orchid!20, opacity=.75] plot [smooth cycle] coordinates {(3.0, 0.9025) (3.06,1.0325) (3.24, 0.9125) (3.17, 0.8425)};
\fill[Orchid] (3.11,0.9925) circle (.02);
\fill[Orchid] (3.09,0.9125) circle (.02);
\fill[Orchid] (3.03,0.9325) circle (.02);
\fill[Orchid] (3.18,0.9525) circle (.02);
\fill[Orchid] (3.14,0.8725) circle (.02);
\fill[Orchid] (3.07,0.9725) circle (.02);
\fill[Orchid] (3.18,0.9125) circle (.02);

\fill [ForestGreen!20, opacity=.75] plot [smooth cycle] coordinates {(3.06, 0.9825) (3.24,1.1025) (3.38, 0.9525) (3.29, 0.8925) (3.22, 0.9025)};
\fill[ForestGreen] (3.25,0.9925) circle (.02);
\fill[ForestGreen] (3.23,1.0525) circle (.02);
\fill[ForestGreen] (3.14,0.9725) circle (.02);
\fill[ForestGreen] (3.16,1.0225) circle (.02);
\fill[ForestGreen] (3.21,0.9425) circle (.02);
\fill[ForestGreen] (3.33,0.9725) circle (.02);
\fill[ForestGreen] (3.29,0.9225) circle (.02);

\draw[line width=0.15mm, -{Latex[length=2pt, width=2pt]}] (3.125, 0.8125) -- (3.125, 0.6625);
\node at (3.125, 0.6175) {\scalebox{.2}{$\mathcal{L}_{\text{MSC}}$}};

\fill [CornflowerBlue!20, opacity=.75] plot [smooth cycle] coordinates {(2.85, 0.4825) (2.96,0.5925) (3.11, 0.4625) (2.99, 0.3625)};
\fill[CornflowerBlue] (2.97,0.4625) circle (.02);
\fill[CornflowerBlue] (3.01,0.5225) circle (.02);
\fill[CornflowerBlue] (2.89,0.4725) circle (.02);
\fill[CornflowerBlue] (2.99,0.4025) circle (.02);
\fill[CornflowerBlue] (2.96,0.5425) circle (.02);
\fill[CornflowerBlue] (2.94,0.4225) circle (.02);
\fill[CornflowerBlue] (3.05,0.4725) circle (.02);

\fill [Orchid!20, opacity=.75] plot [smooth cycle] coordinates {(2.97, 0.2725) (3.11,0.3825) (3.22, 0.2725) (3.15, 0.2025)};
\fill[Orchid] (3.11,0.2825) circle (.02);
\fill[Orchid] (3.02,0.2725) circle (.02);
\fill[Orchid] (3.16,0.3025) circle (.02);
\fill[Orchid] (3.09,0.3425) circle (.02);
\fill[Orchid] (3.17,0.2425) circle (.02);
\fill[Orchid] (3.09,0.2425) circle (.02);
\fill[Orchid] (3.04,0.3125) circle (.02);

\fill [ForestGreen!20, opacity=.75] plot [smooth cycle] coordinates {(3.12, 0.4225) (3.26,0.5525) (3.4, 0.4325) (3.29, 0.3025)};
\fill[ForestGreen] (3.27,0.4425) circle (.02);
\fill[ForestGreen] (3.180,.4225) circle (.02);
\fill[ForestGreen] (3.3,0.4825) circle (.02);
\fill[ForestGreen] (3.21,0.3725) circle (.02);
\fill[ForestGreen] (3.33,0.4225) circle (.02);
\fill[ForestGreen] (3.21,0.4625) circle (.02);
\fill[ForestGreen] (3.29,0.3625) circle (.02);

\fill[CornflowerBlue] (2.9,0.1625) circle (.02);
\node[anchor=west] at (2.95, 0.1625) {\scalebox{.2}{Ankle}};
\fill[Orchid] (3.2,0.1625) circle (.02);
\node[anchor=west] at (3.25, 0.1625) {\scalebox{.2}{Knee}};
\fill[ForestGreen] (3,0.0625) circle (.02);
\node[anchor=west] at (3.05, 0.0625) {\scalebox{.2}{Shoulder}};

\node at (0.61,-1.45) {\scalebox{.21}{\textbf{2.d. Multi-domain attention gate}}};
\draw[line width=0.1mm, color=RedViolet!80, rounded corners=1, dashed, -{Latex[length=1.5pt, width=1.5pt]}] (.78,-1.275) -- (0.835, -1.275) -- (0.835,-1.375);  

\draw[line width=0.1mm, rounded corners=1, -{Latex[length=1pt, width=1pt]}] (-.275,-1.55) -- (-.175,-1.55);
\draw[line width=0.1mm, rounded corners=1, fill=BurntOrange!60] (-.175,-1.6) -- (-.175,-1.5) -- (-.025,-1.5) -- (-.025,-1.6) -- cycle; 
\node at (-.1,-1.55) {\scalebox{.2}{1$\times$1}};

\draw[line width=0.1mm, rounded corners=1, -{Latex[length=1pt, width=1pt]}] (-.025,-1.55) -- (.025,-1.55);
\draw[line width=0.1mm, rounded corners=1, fill=BrickRed!60] (.025,-1.6) -- (.025,-1.5) -- (.175,-1.5) -- (.175,-1.6) -- cycle; 
\draw[line width=0.05mm, pattern={Lines[angle=45,distance={3pt/sqrt(2)}]}, pattern color=black, rounded corners =1] (.025,-1.6) -- (.025,-1.5) -- (.175,-1.5) -- (.175,-1.6) -- cycle; 
\node at (.1,-1.55) {\scalebox{.14}{DSBN}};

\draw[line width=0.1mm, rounded corners=1, -{Latex[length=1pt, width=1pt]}] (.175,-1.55) -- (.275,-1.55) -- (.275,-1.65);

\draw[line width=0.1mm] (.275, .-1.7) circle (.05);
\draw[line width=0.1mm] (.275,-1.66) -- (.275,-1.74);
\draw[line width=0.1mm] (.235,-1.7) -- (.315,-1.7);

\draw[line width=0.1mm, rounded corners=1, -{Latex[length=1pt, width=1pt]}] (-.275,-1.85) -- (-.175,-1.85);
\draw[line width=0.1mm, rounded corners=1, fill=BurntOrange!60] (-.175,-1.9) -- (-.175,-1.8) -- (-.025,-1.8) -- (-.025,-1.9) -- cycle; 
\node at (-.1,-1.85) {\scalebox{.2}{1$\times$1}};

\draw[line width=0.1mm, rounded corners=1, -{Latex[length=1pt, width=1pt]}] (-.025,-1.85) -- (.025,-1.85);
\draw[line width=0.1mm, rounded corners=1, fill=BrickRed!60] (.025,-1.9) -- (.025,-1.8) -- (.175,-1.8) -- (.175,-1.9) -- cycle; 
\draw[line width=0.05mm, pattern={Lines[angle=45,distance={3pt/sqrt(2)}]}, pattern color=black, rounded corners =1] (.025,-1.9) -- (.025,-1.8) -- (.175,-1.8) -- (.175,-1.9) -- cycle; 
\node at (.1,-1.85) {\scalebox{.14}{DSBN}};

\draw[line width=0.1mm, rounded corners=1, -{Latex[length=1pt, width=1pt]}] (0.175,-1.85) -- (.275,-1.85) -- (.275,-1.75);

\draw[line width=0.1mm, rounded corners=1, -{Latex[length=1pt, width=1pt]}] (.325,-1.7) -- (.425,-1.7);
\draw[line width=0.1mm, rounded corners=1] (.425,-1.65) -- (.575,-1.65) -- (.575,-1.75) -- (.425,-1.75) -- cycle; 
\draw[line width=0.1mm] (.45,-1.725) -- (.5,-1.725) -- (.55,-1.675) ; 

\draw[line width=0.1mm, rounded corners=1, -{Latex[length=1pt, width=1pt]}] (.575,-1.7) -- (.625,-1.7);
\draw[line width=0.1mm, rounded corners=1, fill=BurntOrange!60] (.625,-1.65) -- (.775,-1.65) -- (.775,-1.75) -- (.625,-1.75) -- cycle; 
\node at (.7,-1.7) {\scalebox{.2}{1$\times$1}};

\draw[line width=0.1mm, rounded corners=1, -{Latex[length=1pt, width=1pt]}] (.775,-1.7) -- (.825,-1.7);
\draw[line width=0.1mm, rounded corners=1, fill=BrickRed!60] (.825,-1.65) -- (.975,-1.65) -- (.975,-1.75) -- (.825,-1.75) -- cycle; 
\draw[line width=0.05mm, pattern={Lines[angle=45,distance={3pt/sqrt(2)}]}, pattern color=black, rounded corners =1] (.825,-1.65) -- (.975,-1.65) -- (.975,-1.75) -- (.825,-1.75) -- cycle; 
\node at (.9,-1.7) {\scalebox{.14}{DSBN}};

\draw[line width=0.1mm, rounded corners=1, -{Latex[length=1pt, width=1pt]}] (.975,-1.7) -- (1.025,-1.7);
\draw[line width=0.1mm, rounded corners=1] (1.025,-1.65) -- (1.175,-1.65) -- (1.175,-1.75) -- (1.025,-1.75) -- cycle; 
\draw[line width=0.1mm, rounded corners=.75] (1.05,-1.725) -- (1.1,-1.725) -- (1.1,-1.675) -- (1.15,-1.675); 

\draw[line width=0.1mm, rounded corners=1, -{Latex[length=1pt, width=1pt]}] (1.175,-1.7) -- (1.275,-1.7);

\draw[line width=0.1mm] (1.325, -1.7) circle (.05);
\draw[line width=0.1mm] (1.3533,-1.6717) -- (1.2967,-1.7283);
\draw[line width=0.1mm] (1.3533,-1.7283) -- (1.2967,-1.6717);

\draw[line width=0.1mm, rounded corners=1, -{Latex[length=1pt, width=1pt]}] (1.375,-1.7) -- (1.475,-1.7);

\draw[line width=0.1mm, rounded corners=1, -{Latex[length=1pt, width=1pt]}]  (-.235,-1.85) -- (-.235,-1.935) -- (1.325,-1.935) -- (1.325,-1.75);

\node at (.5,-1.8) {\scalebox{.2}{ReLU}};
\node at (1.1,-1.8) {\scalebox{.2}{Sigmoid}};

\node at (3.125,-0.0925) {\scalebox{.21}{\textbf{2.b. Multi-joint}}};
\node at (3.125,-0.1675) {\scalebox{.21}{\textbf{anatomical priors}}};
\draw[line width=0.1mm, color=MidnightBlue!80, rounded corners=1, dashed, -{Latex[length=1.5pt, width=1.5pt]}] (2.725,-.5) -- (2.825, -.5); 

\node[inner sep=0pt] (mri_a) at (3.125,-0.4425)
    {\includegraphics[width=.025\textwidth]{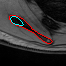}};
\draw[ForestGreen!40, rounded corners=.5, line width=.5] (mri_a.north west) -- (mri_a.north east) -- (mri_a.south east) -- (mri_a.south west) -- cycle;

\draw[line width=0.15mm, -{Latex[length=2pt, width=2pt]}] (3.125, -.6975) -- (3.125, -.8625);
\node at (3.275, -.78) {\scalebox{.2}{$\mathcal{L}_{\text{MJAP}}$}};

\node[inner sep=0pt] (mri_a) at (3.125,-1.1175)
    {\includegraphics[width=.025\textwidth]{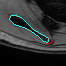}};
\draw [ForestGreen!40, rounded corners=.5, line width=.5] (mri_a.north west) -- (mri_a.north east) -- (mri_a.south east) -- (mri_a.south west) -- cycle;

\end{tikzpicture}
\end{adjustbox}
\caption{The first step of the proposed method involves a multi-joint auto-encoder that learns multi-joint anatomical priors (1) arising from ground truth segmentation of each joint. As a second step, we optimize a segmentation network $S$ based on Efficient-UNet \citep{ronneberger_u-net_2015, tan_efficientnet_2019} in a multi-task, multi-domain framework (2) defined by imaging datasets of three pediatric joints (ankle, knee, and shoulder). The auto-encoder and segmentation networks comprise shared convolutional filters, domain-specific batch normalization (DSBN) calibrating the internal features statistics (2.c) and domain-specific input/output segmentation layers delineating distinct anatomical regions. Their training procedures rely on the cross-entropy loss function $\mathcal{L}_{\text{CE}}$ and integrate a multi-scale contrastive regularization $\mathcal{L}_{\text{MSC}}$ to promote inter-domain separation in the shared representations (2.a). In addition, Efficient-UNet incorporates multi-domain attention gates attached to its skip connections (2.d) and its optimization scheme is combined with multi-joint anatomical priors $\mathcal{L}_{\text{MJAP}}$ computed by the multi-joint anatomical encoder $F$ with fixed weights to enforce anatomically consistent predictions (2.b).}
  \label{fig:framework}
\end{figure*}
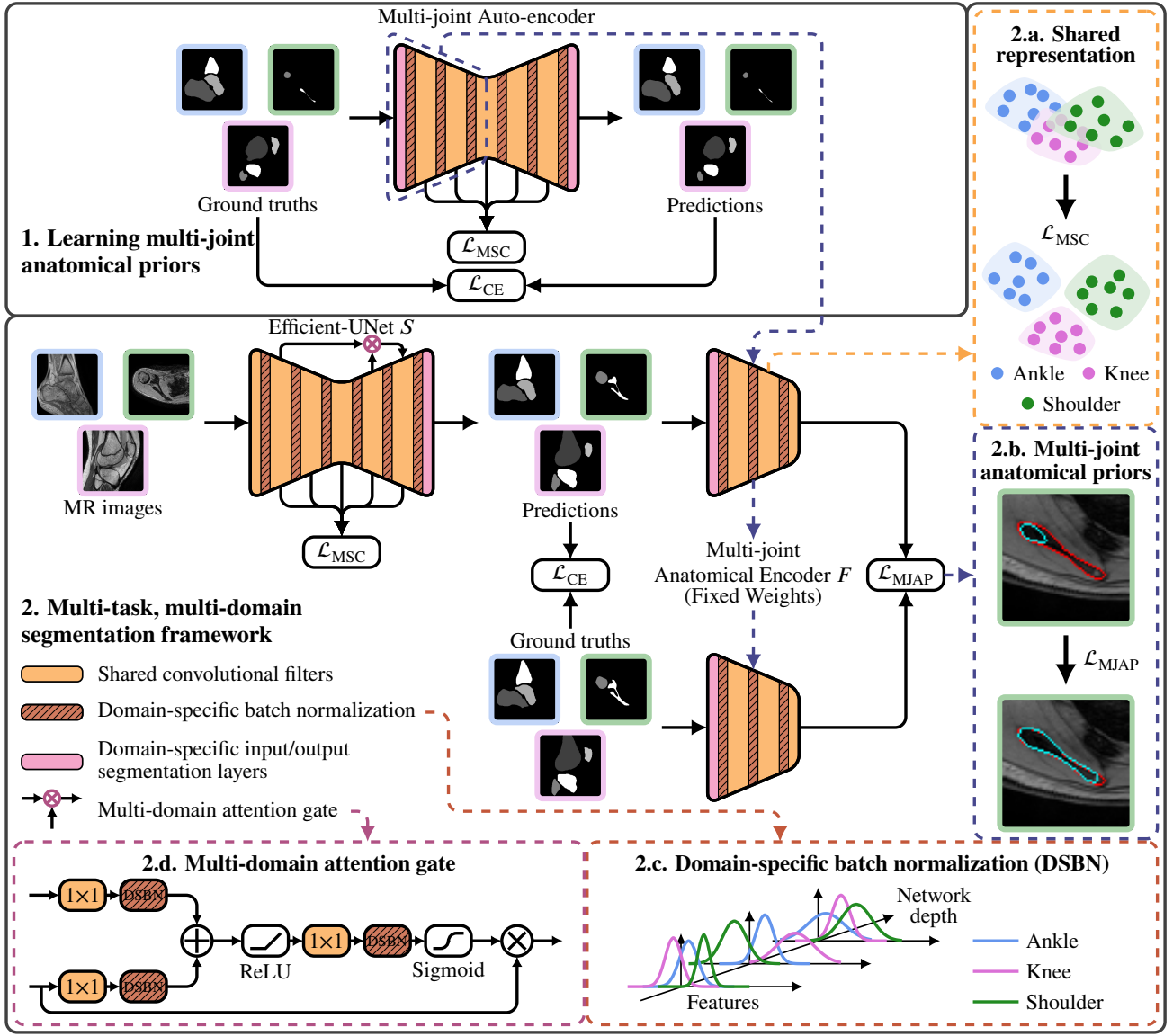

\begin{figure*}[ht!]
\centering
\begin{adjustbox}{width=\textwidth}
\tikzstyle{dashed}=[dash pattern=on 10pt off 10pt]
\begin{tikzpicture}[every node/.style={inner sep=0,outer sep=0}]

\draw[line width=0.01mm, fill=black] (4.75,-2) rectangle (4.85,2);
\draw[line width=0.01mm, fill=black] (5.85,-1) rectangle (6.75,1);
\draw[line width=0.01mm, fill=black] (7.75,-1) rectangle (8.25,1);

\draw[line width=0.01mm, fill=black] (9.25,-.5) rectangle (9.95,.5);

\draw[line width=0.01mm, fill=black] (10.95,-.25) rectangle (12.05,.25);

\draw[line width=0.01mm, fill=black] (13.05,-.125) rectangle (14.75,.125);
\draw[line width=0.01mm, fill=black] (15.75,-.125) rectangle (17.75,.125);

\draw[line width=0.01mm, fill=black] (18.75,-0.075) rectangle (21.15,0.075);
\draw[line width=0.01mm, fill=black] (22.15,-0.075) rectangle (24.85,0.075);

\draw[line width=0.01mm, fill=black] (25.85,-.125) rectangle (27.85,.125);
\draw[line width=0.01mm] (27.85,-.125) rectangle (29.85,.125);
\draw[line width=0.01mm, fill=black] (30.85,-.125) rectangle (32.85,.125);
\draw[line width=0.01mm, fill=black] (33.85,-.125) rectangle (35.85,.125);

\draw[line width=0.01mm, fill=black] (36.85,-.25) rectangle (37.95,.25);
\draw[line width=0.01mm] (37.95,-.25) rectangle (39.05,.25);
\draw[line width=0.01mm, fill=black] (40.05,-.25) rectangle (41.15,.25);

\draw[line width=0.01mm, fill=black] (42.15,-.5) rectangle (42.85,.5);
\draw[line width=0.01mm] (42.85,-.5) rectangle (43.55,.5);
\draw[line width=0.01mm, fill=black] (44.55,-.5) rectangle (45.25,.5);

\draw[line width=0.01mm, fill=black] (46.25,-1) rectangle (46.75,1);
\draw[line width=0.01mm] (46.75,-1) rectangle (47.25,1);
\draw[line width=0.01mm, fill=black] (48.25,-1) rectangle (48.75,1);

\draw[line width=0.01mm, fill=black] (49.75,-2) rectangle (50.65,2);
\draw[line width=0.01mm, fill=black] (51.65,-2) rectangle (52.55,2);
\draw[line width=0.01mm, fill=black] (53.65,-2) rectangle (53.75,2);

\draw[line width=1mm, -{Latex[length=15pt, width=15pt]}] (16.75,.125) -- (16.75, 1.325) -- (26.5, 1.325);
\draw[line width=1mm, -{Latex[length=15pt, width=15pt]}] (26.85,.075) -- (26.85,.975);
\draw[line width=1mm, -{Latex[length=15pt, width=15pt]}] (27.2,1.325) -- (28.85,1.325) -- (28.85,.125);

\draw[line width=1mm] (26.85, 1.325) circle (.35);
\draw[line width=1mm] (27.0975,1.0775) -- (26.6025,1.5725);
\draw[line width=1mm] (27.0975,1.5725) -- (26.6025,1.0775);

\draw[line width=1mm, -{Latex[length=15pt, width=15pt]}] (11.5,.25) -- (11.5,1.925) -- (37.05, 1.925);
\draw[line width=1mm, -{Latex[length=15pt, width=15pt]}] (37.4,.25) -- (37.4,1.575);
\draw[line width=1mm, -{Latex[length=15pt, width=15pt]}] (37.75,1.925) -- (38.5,1.925) -- (38.5,.25);

\draw[line width=1mm] (37.4, 1.925) circle (.35);
\draw[line width=1mm] (37.6475,1.6775) -- (37.1525,2.1725);
\draw[line width=1mm] (37.6475,2.1725) -- (37.1525,1.6775);

\draw[line width=1mm, -{Latex[length=15pt, width=15pt]}] (9.6,.5) -- (9.6,2.575) -- (42.15,2.475);
\draw[line width=1mm, -{Latex[length=15pt, width=15pt]}] (42.5,.5) -- (42.5,2.125);
\draw[line width=1mm, -{Latex[length=15pt, width=15pt]}] (42.85,2.475) -- (43.2,2.475) -- (43.2,.5);

\draw[line width=1mm] (42.5, 2.475) circle (.35);
\draw[line width=1mm] (42.7475,2.2275) -- (42.2525,2.7225);
\draw[line width=1mm] (42.7475,2.7225) -- (42.2525,2.2275);

\draw[line width=1mm, -{Latex[length=15pt, width=15pt]}] (8, 1) -- (8,3.035) -- (46.15,3.035);
\draw[line width=1mm, -{Latex[length=15pt, width=15pt]}] (46.5,1) -- (46.5,2.725);
\draw[line width=1mm, -{Latex[length=15pt, width=15pt]}] (46.85,3.035) -- (47,3.035) -- (47,1);

\draw[line width=1mm] (46.5, 3.035) circle (.35);
\draw[line width=1mm] (46.7475,2.7875) -- (46.2525,3.2825);
\draw[line width=1mm] (46.7475,3.2825) -- (46.2525,2.7875);

\draw[line width=0.01mm, fill=ForestGreen] (5.05,.4) -- (5.05,-.4) -- (5.65,0) -- cycle;
\draw[line width=0.01mm, fill=MidnightBlue] (6.95,.4) -- (6.95,-.4) -- (7.55,0) -- cycle;
\draw[line width=0.01mm, fill=Cyan] (8.45,.4) -- (8.45,-.4) -- (9.05,0) -- cycle;

\draw[line width=0.01mm, fill=Apricot] (10.15,.4) -- (10.15,-.4) -- (10.75,0) -- cycle;

\draw[line width=0.01mm, fill=Cyan] (12.25,.4) -- (12.25,-.4) -- (12.95,0) -- cycle;

\draw[line width=0.01mm, fill=Apricot] (14.95,.4) -- (14.95,-.4) -- (15.55,0) -- cycle;
\draw[line width=0.01mm, fill=Apricot] (17.95,.4) -- (17.95,-.4) -- (18.65,0) -- cycle;

\draw[line width=0.01mm, fill=Cyan] (21.35,.4) -- (21.35,-.4) -- (21.95,0) -- cycle;
\draw[line width=0.01mm, fill=Orchid] (25.05,.4) -- (25.005,-.4) -- (25.65,0) -- cycle;

\draw[line width=0.01mm, fill=Cyan] (30.05,.4) -- (30.05,-.4) -- (30.65,0) -- cycle;
\draw[line width=0.01mm, fill=Apricot] (33.05,.4) -- (33.05,-.4) -- (33.65,0) -- cycle;
\draw[line width=0.01mm, fill=Orchid] (36.05,.4) -- (36.05,-.4) -- (36.65,0) -- cycle;

\draw[line width=0.01mm, fill=Cyan] (39.25,.4) -- (39.25,-.4) -- (39.85,0) -- cycle;
\draw[line width=0.01mm, fill=Orchid] (41.35,.4) -- (41.35,-.4) -- (41.95,0) -- cycle;

\draw[line width=0.01mm, fill=Apricot] (43.75,.4) -- (43.75,-.4) -- (44.35,0) -- cycle;
\draw[line width=0.01mm, fill=Orchid] (45.45,.4) -- (45.45,-.4) -- (46.05,0) -- cycle;

\draw[line width=0.01mm, fill=MidnightBlue] (47.45,.4) -- (47.45,-.4) -- (48.05,0) -- cycle;
\draw[line width=0.01mm, fill=Orchid] (48.95,.4) -- (48.95,-.4) -- (49.55,0) -- cycle;

\draw[line width=0.01mm, fill=ForestGreen] (50.85,.4) -- (50.85,-.4) -- (51.45,0) -- cycle;
\draw[line width=0.01mm, fill=CarnationPink] (52.75,.4) -- (52.75,-.4) -- (53.35,0) -- cycle;

\draw[line width=0.01mm, fill=BurntOrange] (7.6,-2.3) -- (8.4,-2.3) -- (8,-2.9) -- cycle;
\draw[line width=0.01mm, fill=BurntOrange] (9.2,-1.8) -- (10,-1.8) -- (9.6,-2.4) -- cycle;
\draw[line width=0.01mm, fill=BurntOrange] (11.1,-1.55) -- (11.9,-1.55) -- (11.5,-2.15) -- cycle;
\draw[line width=0.01mm, fill=BurntOrange] (16.35,-1.425) -- (17.15,-1.425) -- (16.75,-2.025) -- cycle;
\draw[line width=0.01mm, fill=BurntOrange] (23.1,-1.375) -- (23.9,-1.375) -- (23.5,-1.975) -- cycle;
\draw[line width=0.01mm, fill=BurntOrange] (34.45,-1.425) -- (35.25,-1.425) -- (34.85,-2.025) -- cycle;
\draw[line width=0.01mm, fill=BurntOrange] (40.2,-1.55) -- (41,-1.5) -- (40.6,-2.15) -- cycle;
\draw[line width=0.01mm, fill=BurntOrange] (44.5,-1.8) -- (45.3,-1.8) -- (44.9,-2.4) -- cycle;
\draw[line width=0.01mm, fill=BurntOrange] (51.7,-3.3) -- (52.5,-3.3) -- (52.1,-3.9) -- cycle;

\node[rotate=90] at (4.15,0) {\scalebox{3.25}{256$\times$256}};
\node at (4.8,-2.6) {\scalebox{3.25}{3}};
\node at (6.3,-1.6) {\scalebox{3.25}{40}};
\node at (8,-1.6) {\scalebox{3.25}{24}};

\node at (9.6,-1.1) {\scalebox{3.25}{32}};

\node at (11.5,-.85) {\scalebox{3.25}{48}};

\node at (13.9,-.725) {\scalebox{3.25}{96}};
\node at (16.75,-.725) {\scalebox{3.25}{136}};

\node at (19.95,-.675) {\scalebox{3.25}{232}};
\node at (23.5,-.675) {\scalebox{3.25}{384}};

\node at (27.85,-.725) {\scalebox{3.25}{136}};
\node at (27.85,-1.525) {\scalebox{2}{$\times$2}};
\node at (31.85,-.725) {\scalebox{3.25}{136}};
\node at (34.85,-.725) {\scalebox{3.25}{136}};

\node at (37.95,-.85) {\scalebox{3.25}{48}};
\node at (37.95,-1.65) {\scalebox{2}{$\times$2}};
\node at (40.6,-.85) {\scalebox{3.25}{48}};

\node at (42.85,-1.1) {\scalebox{3.25}{32}};
\node at (42.85,-1.9) {\scalebox{2}{$\times$2}};
\node at (44.9,-1.1) {\scalebox{3.25}{32}};

\node at (46.75,-1.6) {\scalebox{3.25}{24}};
\node at (46.75,-2.4) {\scalebox{2}{$\times$2}};
\node at (48.5,-1.6) {\scalebox{3.25}{24}};

\node at (50.2,-2.6) {\scalebox{3.25}{40}};
\node at (52.1,-2.6) {\scalebox{3.25}{40}};
\node at (53.7,-2.6) {\scalebox{3.25}{$\vert\mathcal{C}_{k}\vert$}};

\node at (8,-3.5) {\scalebox{3.25}{$z_{s_1}$}};
\node at (9.6,-3) {\scalebox{3.25}{$z_{s_2}$}};
\node at (11.5,-2.75) {\scalebox{3.25}{$z_{s_3}$}};
\node at (16.75,-2.625) {\scalebox{3.25}{$z_{s_4}$}};
\node at (23.5,-2.575) {\scalebox{3.25}{$z_{s_5}$}};
\node at (34.85,-2.625) {\scalebox{3.25}{$z_{s_6}$}};
\node at (40.6,-2.75) {\scalebox{3.25}{$z_{s_7}$}};
\node at (44.9,-3) {\scalebox{3.25}{$z_{s_8}$}};
\node at (52.1,-4.5) {\scalebox{3.25}{$z_{s_9}$}};

\node[rotate=90] at (7.25,.8) {\scalebox{2}{$\times$2}};
\node[rotate=90] at (8.75,.8) {\scalebox{2}{$\times$3}};
\node[rotate=90] at (10.45,.8) {\scalebox{2}{$\times$3}};
\node[rotate=90] at (12.55,.8) {\scalebox{2}{$\times$5}};
\node[rotate=90] at (15.25,.8) {\scalebox{2}{$\times$5}};
\node[rotate=90] at (18.25,.8) {\scalebox{2}{$\times$6}};
\node[rotate=90] at (21.65,.8) {\scalebox{2}{$\times$2}};
\node[rotate=90] at (30.35,.8) {\scalebox{2}{$\times$5}};
\node[rotate=90] at (33.35,.8) {\scalebox{2}{$\times$5}};
\node[rotate=90] at (39.55,.8) {\scalebox{2}{$\times$3}};
\node[rotate=90] at (44.05,.8) {\scalebox{2}{$\times$3}};
\node[rotate=90] at (47.75,.8) {\scalebox{2}{$\times$2}};
\node[rotate=90] at (51.15,.8) {\scalebox{2}{$\times$2}};

\node at (29.25,3.9) {\scalebox{3.25}{\textbf{Multi-task, multi-domain segmentation network $S$ based on Efficient-UNet}}};

\draw[line width=0.01mm, fill=black] (10.15,-9.5) rectangle (10.25,-5.5);
\draw[line width=0.01mm, fill=black] (11.25,-9.5) rectangle (11.75,-5.5);
\draw[line width=0.01mm, fill=black] (12.75,-9.5) rectangle (13.25,-5.5);

\draw[line width=0.01mm, fill=black] (14.25,-8.5) rectangle (14.75,-6.5);
\draw[line width=0.01mm, fill=black] (15.75,-8.5) rectangle (16.45,-6.5);

\draw[line width=0.01mm, fill=black] (17.45,-8) rectangle (18.15,-7);
\draw[line width=0.01mm, fill=black] (19.15,-8) rectangle (20.25,-7);

\draw[line width=0.01mm, fill=black] (21.25,-7.75) rectangle (22.35,-7.25);
\draw[line width=0.01mm, fill=black] (23.35,-7.75) rectangle (25.05,-7.25);

\draw[line width=0.01mm, fill=black] (26.05,-7.625) rectangle (27.75,-7.375);
\draw[line width=0.01mm, fill=black] (28.75,-7.625) rectangle (31.25,-7.375);

\draw[line width=0.01mm, fill=black] (32.25,-7.75) rectangle (33.95,-7.25);
\draw[line width=0.01mm, fill=black] (34.95,-7.75) rectangle (36.65,-7.25);

\draw[line width=0.01mm, fill=black] (37.65,-8) rectangle (38.75,-7);
\draw[line width=0.01mm, fill=black] (39.75,-8) rectangle (40.85,-7);

\draw[line width=0.01mm, fill=black] (41.85,-8.5) rectangle (42.55,-6.5);
\draw[line width=0.01mm, fill=black] (43.55,-8.5) rectangle (44.25,-6.5);

\draw[line width=0.01mm, fill=black] (45.25,-9.5) rectangle (45.75,-5.5);
\draw[line width=0.01mm, fill=black] (46.75,-9.5) rectangle (47.25,-5.5);
\draw[line width=0.01mm, fill=black] (48.25,-9.5) rectangle (48.35,-5.5);

\draw[line width=0.01mm, fill=SeaGreen] (10.45,-7.1) -- (10.45,-7.9) -- (11.05,-7.5) -- cycle;
\draw[line width=0.01mm, fill=ForestGreen] (11.95,-7.1) -- (11.95,-7.9) -- (12.55,-7.5) -- cycle;
\draw[line width=0.01mm, fill=BrickRed] (13.45,-7.1) -- (13.45,-7.9) -- (14.05,-7.5) -- cycle; 

\draw[line width=0.01mm, fill=ForestGreen] (14.95,-7.1) -- (14.95,-7.9) -- (15.55,-7.5) -- cycle;
\draw[line width=0.01mm, fill=BrickRed] (16.65,-7.1) -- (16.65,-7.9) -- (17.25,-7.5) -- cycle;

\draw[line width=0.01mm, fill=ForestGreen] (18.35,-7.1) -- (18.35,-7.9) -- (18.95,-7.5) -- cycle;
\draw[line width=0.01mm, fill=BrickRed] (20.45,-7.1) -- (20.45,-7.9) -- (21.05,-7.5) -- cycle;

\draw[line width=0.01mm, fill=ForestGreen] (22.55,-7.1) -- (22.55,-7.9) -- (23.15,-7.5) -- cycle;
\draw[line width=0.01mm, fill=BrickRed] (25.25,-7.1) -- (25.25,-7.9) -- (25.85,-7.5) -- cycle;

\draw[line width=0.01mm, fill=ForestGreen] (27.95,-7.1) -- (27.95,-7.9) -- (28.55,-7.5) -- cycle;

\draw[line width=0.01mm, fill=Orchid] (31.45,-7.1) -- (31.45,-7.9) -- (32.05,-7.5) -- cycle;
\draw[line width=0.01mm, fill=ForestGreen] (34.15,-7.1) -- (34.15,-7.9) -- (34.75,-7.5) -- cycle;

\draw[line width=0.01mm, fill=Orchid] (36.85,-7.1) -- (36.85,-7.9) -- (37.45,-7.5) -- cycle;
\draw[line width=0.01mm, fill=ForestGreen] (38.95,-7.1) -- (38.95,-7.9) -- (39.55,-7.5) -- cycle;

\draw[line width=0.01mm, fill=Orchid] (41.05,-7.1) -- (41.05,-7.9) -- (41.65,-7.5) -- cycle;
\draw[line width=0.01mm, fill=ForestGreen] (42.75,-7.1) -- (42.75,-7.9) -- (43.35,-7.5) -- cycle;

\draw[line width=0.01mm, fill=Orchid] (44.45,-7.1) -- (44.45,-7.9) -- (45.05,-7.5) -- cycle;
\draw[line width=0.01mm, fill=ForestGreen] (45.95,-7.1) -- (45.95,-7.9) -- (46.55,-7.5) -- cycle;
\draw[line width=0.01mm, fill=CarnationPink] (47.45,-7.1) -- (47.45,-7.9) -- (48.05,-7.5) -- cycle;

\draw[line width=0.01mm, fill=BurntOrange] (12.6,-10.8) -- (13.4,-10.8) -- (13,-11.4) -- cycle;
\draw[line width=0.01mm, fill=BurntOrange] (15.7,-9.8) -- (16.5,-9.8) -- (16.1,-10.4) -- cycle;
\draw[line width=0.01mm, fill=BurntOrange] (19.3,-9.3) -- (20.1,-9.3) -- (19.7,-9.9) -- cycle;
\draw[line width=0.01mm, fill=BurntOrange] (23.8,-9.05) -- (24.6,-9.05) -- (24.2,-9.65) -- cycle;
\draw[line width=0.01mm, fill=BurntOrange] (29.6,-8.925) -- (30.4,-8.925) -- (30,-9.525) -- cycle;
\draw[line width=0.01mm, fill=BurntOrange] (35.4,-9.05) -- (36.2,-9.05) -- (35.8,-9.65) -- cycle;
\draw[line width=0.01mm, fill=BurntOrange] (39.9,-9.3) -- (40.7,-9.3) -- (40.3,-9.9) -- cycle;
\draw[line width=0.01mm, fill=BurntOrange] (43.6,-9.8) -- (44.4,-9.8) -- (44,-10.4) -- cycle;
\draw[line width=0.01mm, fill=BurntOrange] (46.6,-10.8) -- (47.4,-10.8) -- (47,-11.4) -- cycle;

\node[rotate=90] at (9.55,-7.5) {\scalebox{3.25}{256$\times$256}};
\node at (10.2,-10.1) {\scalebox{3.25}{$\vert\mathcal{C}_{k}\vert$}};
\node at (11.5,-10.1) {\scalebox{3.25}{32}};
\node at (13,-10.1) {\scalebox{3.25}{32}};

\node at (14.5,-9.1) {\scalebox{3.25}{32}};
\node at (16.1,-9.1) {\scalebox{3.25}{64}};

\node at (17.8,-8.6) {\scalebox{3.25}{64}};
\node at (19.7,-8.6) {\scalebox{3.25}{128}};

\node at (21.8,-8.35) {\scalebox{3.25}{128}};
\node at (24.2,-8.35) {\scalebox{3.25}{256}};

\node at (26.9,-8.225) {\scalebox{3.25}{256}};
\node at (30,-8.225) {\scalebox{3.25}{512}};

\node at (33.1,-8.35) {\scalebox{3.25}{256}};
\node at (35.8,-8.35) {\scalebox{3.25}{256}};

\node at (38.2,-8.6) {\scalebox{3.25}{128}};
\node at (40.3,-8.6) {\scalebox{3.25}{128}};

\node at (42.3,-9.1) {\scalebox{3.25}{64}};
\node at (44,-9.1) {\scalebox{3.25}{64}};

\node at (45.5,-10.1) {\scalebox{3.25}{32}};
\node at (47,-10.1) {\scalebox{3.25}{32}};
\node at (48.3,-10.1) {\scalebox{3.25}{$\vert\mathcal{C}_{k}\vert$}};

\node at (13,-12) {\scalebox{3.25}{$\Tilde{z}_{s_1}$}};
\node at (16.1,-11) {\scalebox{3.25}{$\Tilde{z}_{s_2}$}};
\node at (19.7,-10.5) {\scalebox{3.25}{$\Tilde{z}_{s_3}$}};
\node at (24.2,-10.25) {\scalebox{3.25}{$\Tilde{z}_{s_4}$}};
\node at (30,-10.125) {\scalebox{3.25}{$\Tilde{z}_{s_5}$}};
\node at (35.8,-10.25) {\scalebox{3.25}{$\Tilde{z}_{s_6}$}};
\node at (40.3,-10.5) {\scalebox{3.25}{$\Tilde{z}_{s_7}$}};
\node at (44,-11) {\scalebox{3.25}{$\Tilde{z}_{s_8}$}};
\node at (47,-12) {\scalebox{3.25}{$\Tilde{z}_{s_9}$}};

\node[rotate=90] at (15.25,-6.7) {\scalebox{2}{$\times$2}};
\node[rotate=90] at (18.65,-6.7) {\scalebox{2}{$\times$2}};
\node[rotate=90] at (22.85,-6.7) {\scalebox{2}{$\times$2}};
\node[rotate=90] at (28.25,-6.7) {\scalebox{2}{$\times$2}};
\node[rotate=90] at (34.45,-6.7) {\scalebox{2}{$\times$2}};
\node[rotate=90] at (39.25,-6.7) {\scalebox{2}{$\times$2}};
\node[rotate=90] at (43.05,-6.7) {\scalebox{2}{$\times$2}};
\node[rotate=90] at (46.25,-6.7) {\scalebox{2}{$\times$2}};

\node at (29.25,-4.5) {\scalebox{3.25}{\textbf{Multi-joint auto-encoder}}};
\draw[line width=.1mm] (31.35,-6.75) -- (31.35,-5.25);
\node[anchor=east] at (31.25,-5.75) {\scalebox{3.25}{Encoder $F$}};
\node[anchor=west] at (31.45,-5.75) {\scalebox{3.25}{Decoder $G$}};

\draw[dashed, line width=1mm, color=darkgray, rounded corners=10] (0,-13) rectangle (58.5,-19.25);

\draw[line width=0.01mm, fill=MidnightBlue] (1,-13.85) -- (1,-14.65) -- (1.6,-14.25) -- cycle;
\draw[line width=0.01mm, fill=Cyan] (1,-15.1) -- (1,-15.9) -- (1.6,-15.5) -- cycle;
\draw[line width=0.01mm, fill=Apricot] (1,-16.35) -- (1,-17.15) -- (1.6,-16.75) -- cycle;

\draw[line width=0.01mm, fill=ForestGreen] (16.5,-13.85) -- (16.5,-14.65) -- (17.1,-14.25) -- cycle;
\draw[line width=0.01mm, fill=SeaGreen] (16.5,-15.1) -- (16.5,-15.9) -- (17.1,-15.5) -- cycle;
\draw[line width=0.01mm, fill=CarnationPink] (16.5,-16.35) -- (16.5,-17.15) -- (17.1,-16.75) -- cycle;

\draw[line width=0.01mm, fill=Orchid] (33.5,-14.475) -- (33.5,-15.275) -- (34.1,-14.875) -- cycle;
\draw[line width=0.01mm, fill=BurntOrange] (33.5,-16.35) -- (33.5,-17.15) -- (34.1,-16.75) -- cycle;

\draw[line width=0.01mm, fill=BrickRed] (48.35,-13.85) -- (48.35,-14.65) -- (48.95,-14.25) -- cycle;
\draw[line width=1mm, -{Latex[length=15pt, width=15pt]}] (47.6,-15.5) -- (49.1,-15.5);
\draw[line width=1mm] (48.6, -17.375) circle (.35);
\draw[line width=1mm] (48.8475,-17.1275) -- (48.3525,-17.6225);
\draw[line width=1mm] (48.8475,-17.6225) -- (48.3525,-17.1275);

\node[anchor=west] at (2.1,-14.25) {\scalebox{3.25}{Multi-domain MBConv1 3$\times$3}};
\node[anchor=west] at (2.1,-15.5) {\scalebox{3.25}{Multi-domain MBConv6 3$\times$3}};
\node[anchor=west] at (2.1,-16.75) {\scalebox{3.25}{Multi-domain MBConv6 5$\times$5}};
\node[anchor=west] at (2.1,-18) {\scalebox{3.25}{\small{$\hookrightarrow$ (Shared Conv 1$\times$1, DSBN, Activ), Shared DW Conv $k\times k$, DSBN, Activ, SE, Shared Conv 1$\times$1, DSBN}}};

\node[anchor=west] at (17.6,-14.25) {\scalebox{3.25}{Shared Conv 3$\times$3, DSBN, Activ}};
\node[anchor=west] at (17.6,-15.5) {\scalebox{3.25}{DS Conv 3$\times$3, DSBN, Activ}};
\node[anchor=west] at (17.6,-16.75) {\scalebox{3.25}{DS Conv 1$\times$1, Softmax}};

\node[anchor=west] at (34.6,-14.25) {\scalebox{3.25}{Up-sampling 2$\times$2, Shared}};
\node[anchor=west] at (34.6,-15.5) {\scalebox{3.25}{Conv 3$\times$3, DSBN, Activ}};
\node[anchor=west] at (34.6,-16.75) {\scalebox{3.25}{Global Average Pooling}};

\node[anchor=west] at (49.6,-14.25) {\scalebox{3.25}{Max-pooling 2$\times$2}};
\node[anchor=west] at (49.6,-15.5) {\scalebox{3.25}{Skip Connection}};
\node[anchor=west] at (49.6,-16.75) {\scalebox{3.25}{Multi-domain}};
\node[anchor=west] at (49.6,-18) {\scalebox{3.25}{Attention Gate}};

\end{tikzpicture}
\end{adjustbox}
\caption{Proposed neural network architectures: multi-task, multi-domain segmentation network $S$ based on Efficient-UNet (top) \citep{ronneberger_u-net_2015, tan_efficientnet_2019} and multi-joint auto-encoder (bottom) comprising encoder $F$ and decoder $G$. The multi-scale embedding ($z_{s_1}, ..., z_{s_9}$) and ($\Tilde{z}_{s_1}, ..., \Tilde{z}_{s_9}$) are obtained via global average pooling. $\vert\mathcal{C}_{k}\vert$ denotes the number of classes in the $k^{\text{th}}$ segmentation task while activations (Activ) correspond to either \texttt{SiLU} (Efficient-UNet) or \texttt{ReLU} (auto-encoder) functions. The multi-domain MBConv block integrates shared point-wise (1$\times$1) and depth-wise (DW) convolutions, domain-specific batch normalization (DSBN) and squeeze-and-excite (SE) modules \citep{tan_efficientnet_2019}.}
\label{fig:architecture}
\end{figure*}

In this section, we first describe the proposed multi-task, multi-domain segmentation network (Section \ref{sec:multi-task_multi-domain_deep_segmentation}) built upon Efficient-UNet (Section \ref{sec:efficient_segmentation_network_with_pre-trained_encoder}) and domain-specific layers (Section \ref{sec:domain-specific_layers}). We then incorporate the multi-scale contrastive regularization (Section \ref{sec:multi-scale_contrastive_regularization}) along with the multi-joint anatomical priors (Section \ref{sec:multi-joint_anatomical_priors}) into our model.

\subsection{Multi-task, multi-domain deep segmentation}
\label{sec:multi-task_multi-domain_deep_segmentation}

Let $\mathcal{D}_{1}, ..., \mathcal{D}_{K}$ be $K$ different datasets organized such that the $k^{\text{th}}$ dataset $\mathcal{D}_k = \{ x_{i}^{k}, y_{i}^{k} \}_{i=1}^{n_{k}}$ contains $n_k$ pairs of greyscale images $x_{i}^{k}$ in intensity domain $\mathcal{I}_{k}$ and their corresponding class label images $y_{i}^{k}$ in label space $\mathcal{C}_{k}$. Each intensity domain $\mathcal{I}_{1}, ..., \mathcal{I}_{K}$ is characterized by its own intensity distribution, while the label spaces $\mathcal{C}_{1}, ..., \mathcal{C}_{K}$ represent separate segmentation tasks constituted of different anatomical structure of interest (plus background). Hence, the goal of multi-task, multi-domain deep segmentation is to learn a single mapping $S$ between each intensity domain and its corresponding label space, formally $\forall k \in [1, ..., K]$, $S: \mathcal{I}_{k} \to \mathcal{C}_{k}$. 

In what follows, the function $S$ is approximated by a segmentation network composed of a succession of layers whose parameters must be learnt during training. More specifically, $S: x_{i}^{k} \mapsto S(x_{i}^{k}; \Theta, \Gamma)$ is composed of shared parameters $\Theta$ and domain-specific weights $\Gamma = \{\Gamma_{k}\}_{k=1}^{K}$ selected based on the domain $k$ of the input image. During training, we used the stochastic gradient descent algorithm to optimize the cross-entropy loss defined in a multi-task and multi-domain setting:
\begin{equation}
    \mathcal{L}_{\text{CE}} = - \dfrac{1}{K} \sum_{k=1}^{K} \dfrac{1}{n_{k} \vert \mathcal{C}_{k} \vert} \sum_{i=1}^{n_{k}} \sum_{c \in \mathcal{C}_{k}} y_{i,c}^{k} \log(\hat{y}_{i,c}^{k})
\end{equation}
\noindent where $\hat{y}_{i}^{k} = S(x_{i}^{k}; \Theta, \Gamma)$ was the predicted segmentation and $\vert \mathcal{C}_{k} \vert$ denoted the cardinality of the $k^{\text{th}}$ label space. The shared parameters and domain-specific weights were simultaneously derived through this novel optimization scheme. In consequence, the network $S$ learnt to segment all structures of interest defined in label spaces $\mathcal{C}_{1}, ..., \mathcal{C}_{K}$ across all intensity domains $\mathcal{I}_{1}, ..., \mathcal{I}_{K}$.

\subsubsection{Efficient segmentation network with pre-trained encoder}
\label{sec:efficient_segmentation_network_with_pre-trained_encoder}

The architecture of the neural network $S$ was based on UNet \citep{ronneberger_u-net_2015} whose encoder branch was replaced by a classification network with weights previously trained on an image classification task (Fig. \ref{fig:architecture}). Following previous work on transfer learning and fine tuning from large datasets such as ImageNet \citep{russakovsky_imagenet_2015}, we assumed that leveraging a pre-trained encoder would lead to better segmentation outcomes compared to models with randomly initialized weights \citep{conze_healthy_2020}. Performance improvements have been particularly reported in low data regime \citep{raghu_transfusion_2019} similar to our scarce pediatric dataset setting. We further advanced this strategy by integrating an Efficient classification network from the Efficient-Net family as encoder \citep{tan_efficientnet_2019}. Specifically, we employed the \texttt{EfficientB3} encoder which incorporates mobile inverted bottlenecks convolutional blocks (MBConv) to simultaneously balance the network depth, width and resolution while improving predictive performance \citep{tan_efficientnet_2019}. 

First, to fit the \texttt{EfficientB3} image dimensions, we concatenated three copies of each MR slice to extend them from single greyscale channel to three channels. The encoder branch was then built on classical convolution, batch normalization, and \texttt{SiLU} non-linearity along with MBConv blocks (MBConv1-6, Fig. \ref{fig:architecture}) consisting of point-wise and depth-wise convolutions, as well as additional squeeze-and-excite modules \citep{tan_efficientnet_2019}. Specifically, combination of point-wise and depth-wise convolutions layers allows to reduce the number of parameter by leveraging the decoupling of cross-channel correlations and spatial correlations \citep{chollet_xception_2017}. For their parts, squeeze-and-excite modules aim at improving performance by adaptively recalibrating channel-wise features through explicit modeling of interdependencies between channels \citep{hu_squeeze-and-excitation_2020}. The overall architecture (i.e. depth, width, and resolution) of \texttt{EfficientB3} encoder is then defined in a principled way using a compound scaling coefficient \citep{tan_efficientnet_2019}. Ultimately, \texttt{EfficientB3} produced a $384$ dimensional output and the resulting feature-map corresponded to the central part between the contracting and expanding paths of $S$ (Fig. \ref{fig:architecture}). Next, we constructed a symmetrical decoder branch with up-sampling layers, classical convolutions and MBConv blocks (Fig. \ref{fig:architecture}). Contrary to encoder weights that are pre-trained on ImageNet \citep{russakovsky_imagenet_2015}, the decoder weights were randomly initialized. Finally, to improve both model interpretability and performance, we employed spatial attention gates to implicitly suppress irrelevant regions of the input image while highlighting salient features \citep{oktay_attention_2018}. These modules attached to the skip connections selected important features using contextual information from the decoding branch (Fig. \ref{fig:framework}).

\subsubsection{Domain-specific layers (DSL)}
\label{sec:domain-specific_layers}

Batch normalization is a ubiquitous transformation found in deep convolutional models which aims at improving convergence speed and generalization abilities of neural networks by normalizing their internal features \citep{ioffe_batch_2015}. However, in multi-domain learning, as the individual statistics of the intensity domains $\mathcal{I}_{1}, ..., \mathcal{I}_{K}$ can be very different from each other (Fig. \ref{fig:framework}), a domain-agnostic batch normalization layer could lead to defective features \citep{karani_lifelong_2018, chang_domain-specific_2019, dou_unpaired_2020, liu_ms-net_2020}. Specifically, if we consider the $m^{\text{th}}$ features at the $l^{\text{th}}$ layer, the mean activation over domains $K^{-1}\sum_{k=1}^{K}\mu_{l,m}^{k}$ could be null while the domain-specific means $\mu_{l,m}^{k}$ are non-zero, making a domain-agnostic normalization meaningless.

Thus, to more carefully calibrate the internal features of the model, we employed domain-specific batch normalization functions (\texttt{DSBN}) \citep{karani_lifelong_2018, chang_domain-specific_2019, dou_unpaired_2020, liu_ms-net_2020}:
\begin{equation}
    \texttt{DSBN}_{\beta_{l,m}^{k}, \gamma_{l,m}^{k}}(v_{i,l,m}^{k}) = \gamma_{l,m}^{k} \dfrac{v_{i,l,m}^{k} - \mu_{l,m}^{k}}{\sqrt{(\sigma_{l,m}^{k})^{2} + \epsilon}}  + \beta_{l,m}^{k}
\end{equation}
\noindent where $v_{i,l,m}^{k}$ denoted the $m^{\text{th}}$ feature-map at the $l^{\text{th}}$ layer produced by the $i^{\text{th}}$ image of the $k^{\text{th}}$ dataset, $\mu_{l,m}^{k}$ and $\sigma_{l,m}^{k}$ the domain-specific mini-batch mean and standard deviation respectively. $\epsilon = 1\text{e-}5$ was added for numerical stability. The \texttt{DSBN} weights $\Lambda_{k} = \{\beta_{l,m}^{k}, \gamma_{l,m}^{k}\}_{l,m}$ thus comprised the domain-specific trainable shift and scale of each feature, at each layer.

Following the definition of \texttt{DSBN}, we modified the elementary block of convolutional models (i.e. sequence of convolution, batch normalization, and activation) for multi-domain learning. This novel multi-domain block was based on shared convolution, \texttt{DSBN}, and an activation function: 
\begin{equation}
\label{eq:multi-domain_block}
    u_{i,l+1,m}^{k} = \rho(\texttt{DSBN}_{\beta_{l,m}^{k}, \gamma_{l,m}^{k}}(\Theta_{l,m} *  u_{i,l}^{k}))
\end{equation}
Here, $u_{i,l+1,m}^{k}$ was the $m^{\text{th}}$ output activations generated by the $(l+1)^{\text{th}}$ block with the $i^{\text{th}}$ image of the $k^{\text{th}}$ dataset as input, $\rho$ was a non-linearity (e.g. \texttt{ReLU}, \texttt{SiLU}, \texttt{Sigmoid}, etc.), and $u_{i,l}^{k}$ was the output of the $l^{\text{th}}$ layer. As a convention, the input image corresponded to the input of the first layer $u_{i,0}^{k} = x_{i}^{k}$. As indicated in \citep{ioffe_batch_2015}, the bias of the convolutional layer can be ignored, as its role is subsumed by the shift of the subsequent normalization transformation. Thus, the shared convolutional parameters $\Theta = \{ \Theta_{l,m}\}_{l,m}$ comprised solely the convolutional filters (classical, point-wise, or depth-wise). Based on this new multi-domain block, MBConv modules \citep{tan_efficientnet_2019} and attention gates \citep{oktay_attention_2018} (Section \ref{sec:efficient_segmentation_network_with_pre-trained_encoder}) were consequently adapted to the multi-domain setting (Fig. \ref{fig:architecture}). In practice, this corresponded to the modification of each batch normalization layer into its domain-specific equivalent. For instance, as attention gates select spatial regions based on feature activations (e.g. \texttt{Sigmoid} activation) \citep{oktay_attention_2018}, we hypothesized that their multi-domain counterpart could help highlight different areas in each domain thanks to domain-specific feature calibration (Fig. \ref{fig:framework}).

As intensity domains and segmentation tasks were similar in nature (i.e. pediatric bone in MR images), we assumed that low-level features (e.g. edges, gradients, etc.) as well as high-level features (e.g. bone texture, bone shape, etc.) were similar across tasks and domains. We therefore hypothesized that shared convolutional kernels would leverage features shared among tasks and domains to be more robust than their individual counterparts, while the \texttt{DSBN} would enable better generalization capabilities thanks to the domain-specific calibration of the internal features.

Furthermore, as the $K$ segmentation tasks were distinct (Fig. \ref{fig:introduction}), a domain-agnostic segmentation layer may predict classes from each label space $\mathcal{C}_{1}, ..., \mathcal{C}_{K}$, which is counterproductive \citep{fourure_semantic_2016} (e.g. predicting ankle bones from a shoulder image). Hence, it was essential to employ a dedicated output layer for each domain and task pair. Specifically, if $u_{i}^{k}$ denotes the output of the penultimate layer then:
\begin{equation}
    \hat{y}_i^k = \texttt{softmax}(W_{k} * u_{i}^{k} + b_{k})
\end{equation}
\noindent was a domain-specific segmentation layer which produced a segmentation mask $\hat{y}_i^k$ with $\vert \mathcal{C}_{k} \vert$ classes (Fig. \ref{fig:architecture}). Here, the weights of the domain-specific output segmentation layer $\Xi_{k} = \{W_{k}, b_{k}\}$ corresponded to the final 1$\times$1 (i.e. point-wise) convolutional filter and associated bias.

To recapitulate, the domain-specific layers (DSL) $\Gamma_k = \{\Lambda_k, \Xi_k\}$ comprised the \texttt{DSBN} weights $\Lambda_k$ and the weights $\Xi_k$ of the domain-specific output segmentation layers, whereas the shared parameters $\Theta$ corresponded to the classical, point-wise, and depth-wise convolutional filters. Most notably, the domain-specific weights represented a minimal supplementary parameterization with regards to the total number of shared convolutional kernels.

\subsection{Multi-scale contrastive regularization}
\label{sec:multi-scale_contrastive_regularization}
Each multi-domain block (Eq. \ref{eq:multi-domain_block}) mapped its input to a shared representation in which features were shifted and scaled according to their domain before applying a non-linear activation. Here, we hypothesized that learning shared representations with domain-specific clusters would enhance the generalization capabilities of the model and improve the accuracy of the segmentation predictions. More precisely, we assumed that a local variation in the output of each multi-domain block should preserve the category of the domain \citep{bengio_representation_2013}. Hence, we designed a novel regularization term  aimed at disentangling domain representations by conserving intra-domain cohesion and inter-domain separation in the shared latent space (Fig. \ref{fig:framework}). The proposed contrastive regularization was adapted from image classification \citep{khosla_supervised_2020} to multi-task, multi-domain segmentation using the known domains labels.

However, rather than applying the contrastive regularization after each multi-domain block (i.e. after each non-linearity), we imposed the clusterization constraints at each scale of the model to reduce computational complexity. To this end, we considered an ensemble of layers indices $\mathcal{S}$ corresponding to the different spatial scale of the segmentation network, which were symmetrically distributed between the encoder and the decoder (Fig. \ref{fig:architecture}). Hence, unlike the previously proposed single-scale contrastive loss that only regularized the bottleneck activations \citep{boutillon_multi-task_2021}, our multi-scale approach untangled the domain representations at each stage of the encoder and decoder modules in a deeply-supervised manner. Since the semantic information extracted and captured by the neural network differed at each scale as well as across scales, we hypothesized that it was necessary to enforce a multi-scale regularization to achieve better generalization capabilities compared to the single scale constraint.

Let $z^{k}_{i,s} = \texttt{GlobalAveragePooling}(u^{k}_{i,s})$ be the embedding of $x_{i}^{k}$ at scale $s \in \mathcal{S}$ to which we applied \texttt{GlobalAveragePooling} to project the data in a lower-dimensional space $\mathbb{R}^{d}$ invariant to spatial transformations (e.g. rotation, translation, flipping), allowing global comparison of image representations originating from different domains (Fig. \ref{fig:architecture}). The dimensionality $d$ of the representations were thus distinct at each scale and $z^{k}_{i,s}$ was then normalized to lie on the unit hyper-sphere, which enabled to measure distances by using an inner product \citep{khosla_supervised_2020}. 

We note $\mathcal{P}_{i}^{k} = \{j \in [1, ..., n_{k}]: j \neq i \}$ the set of indexes of all images from the same domain as $x_{i}^{k}$ (i.e. positive pairs) and $n = \sum_{k=1}^{K} n_{k}$ the total number of images across domains. The multi-scale contrastive loss was defined as follows:

\begin{equation}
    \hspace{-.8cm} \mathcal{L}_{\text{MSC}} = - \frac{1}{n\vert\mathcal{S}\vert} \sum_{s\in\mathcal{S}} \sum_{\substack{1\leq k\leq K\\1\leq i\leq n_{k}}} \frac{1}{\vert\mathcal{P}_{i}^{k}\vert} \sum_{j\in\mathcal{P}_{i}^{k}} \log \left( \frac{\exp(z^{k}_{i,s} \cdot z^{k}_{j,s} / \tau)}{\sum\limits_{\substack{(k^{\prime},i^{\prime})\\\neq(k,i)}} \exp(z^{k}_{i,s} \cdot z^{k^{\prime}}_{i^{\prime},s} / \tau)} \right)
\end{equation}
\noindent where $z_{i,s}^{k} \cdot z_{j,s}^{k}$ denoted the inner product between two $L^2$ normalized representations (i.e. cosine similarity) and $\tau$ was the temperature hyper-parameter which controlled the smoothness of the loss as well as imposed hard negative/positive predictions \citep{chen_simple_2020, khosla_supervised_2020}. As the cosine similarity was bounded in the interval $[-1, 1]$ regardless of the dimensionality of the representations, we assumed that the temperature $\tau$ should be constant over scales. Optimization of $\mathcal{L}_{\text{MSC}}$ encouraged the model to produce, at each scale, closely aligned representations for all pairs from the same domain and orthogonal representations for negative couples. Thus, the multi-scale contrastive regularization gathered the embedding from the same domain, while simultaneously separating clusters from different domains (Fig. \ref{fig:framework}).

\subsection{Multi-joint anatomical priors}
\label{sec:multi-joint_anatomical_priors}

Although incorporation of the multi-scale contrastive loss improved the generalization capabilities of the model by imposing clusterization in its internal representations, a supplementary constraint on output predictions could further increase performance. In this direction, recent works have proposed to integrate into the segmentation network a shape representation of the anatomy, which is learnt from ground truth segmentation masks by a deep auto-encoder \citep{ravishankar_learning_2017, dalca_anatomical_2018, oktay_anatomically_2018}. An auto-encoder is a neural network composed of an encoder $F$ which maps its input to a low-dimensional feature space that compactly encodes the characteristics of the anatomy and a decoder $G$ which reconstructs the original input from the compact representation (Fig. \ref{fig:architecture}) \citep{ravishankar_learning_2017, dalca_anatomical_2018, oktay_anatomically_2018}.

We extended the standard anatomical priors framework to the multi-task, multi-domain setting by designing a multi-joint auto-encoder $AE: y_{i}^{k} \mapsto G(F(y_{i}^{k}; \Theta_{F}, \Gamma_{F}); \Theta_{G}, \Gamma_{G})$ which simultaneously learns the anatomical representation of multiple joints (Fig. \ref{fig:framework}). The weights $\Theta_{F}$ and $\Theta_{G}$ corresponded to the shared convolutional kernels of $F$ and $G$, whereas $\Gamma_{F}$ and $\Gamma_{G}$ comprised the weights of the \texttt{DSBN} and domain-specific input and output segmentation layers of $F$ and $G$ respectively (Fig. \ref{fig:architecture}). Similar to the design of the segmentation network, the multi-joint auto-encoder integrated \texttt{DSBN} functions to efficiently normalize its internal feature distributions, while the input and output convolutional filters operated on the distinct anatomical structures of interest.

As all segmentation tasks solely comprised pediatric bo\-nes, we assumed that our multi-joint learning scheme would leverage anatomical features common between musculoskeletal joints to obtain a more robust representations of the anatomy. The multi-joint auto-encoder training procedure was based on the cross-entropy loss function which penalizes the reconstruction of each joint to be dissimilar from the original input \citep{ravishankar_learning_2017, oktay_anatomically_2018}. Moreover, based on our multi-scale contrastive regularization, we imposed a clusterization constraint on the shared anatomical representations ($\Tilde{z}_{s_{1}}, ..., \Tilde{z}_{s_{9}}$ as denoted in Fig. \ref{fig:architecture}) of the auto-encoder, to promote separated low-dimensional manifold for each anatomical joint. Hence, the loss of the auto-encoder becomes:
\begin{equation}
    \mathcal{L}_{\text{AE}} = \mathcal{L}_{\text{CE}} + \lambda_{1} \mathcal{L}_{\text{MSC}}
\end{equation}
\noindent with empirically set weighting factor $\lambda_1$.

After training the multi-joint auto-encoder, we integrated its encoder component $F$ into the segmentation framework by computing a multi-joint anatomical priors term (Fig. \ref{fig:framework}). To this end, both predictions and ground truth labels of each joint were projected onto the multi-joint latent anatomical space by $F$ with learnt weights $\Theta_F$ and $\Gamma_F$. The multi-joint anatomical priors loss computed the Euclidean distance between both latent anatomical representations \citep{oktay_anatomically_2018}, as follows: 
\begin{equation}
    \mathcal{L}_{\text{MJAP}} = \frac{1}{K} \sum_{k=1}^{K} \frac{1}{n_{k}} \sum_{i=1}^{n_{k}} \norm{F(y_{i}^{k}; \Theta_{F}, \Gamma_{F}) - F(\hat{y}_{i}^{k}; \Theta_{F}, \Gamma_{F})}^{2}
\end{equation}
\noindent Minimization of this loss enforced the predicted segmentation of each joint to be in the same low-dimensional manifold as the corresponding ground truth mask \citep{oktay_anatomically_2018} and thus encouraged anatomically consistent delineations (Fig. \ref{fig:framework}). More precisely, minimizing the Euclidean distance led to similar anatomical codes for each pair of segmentation masks. It should be emphasized that anatomical codes were represented as 2D feature maps (i.e. auto-encoder bottleneck, Fig. \ref{fig:architecture}) with each value encoding a distinct feature of the anatomy. As the weights of the anatomical encoder remained fixed during this step, the two feature maps were in correspondence, with each value encoding the same global anatomical feature for both ground truth and predicted segmentation masks. Anatomical features typically encompass position, orientation, size, and shape information of each structure of interest as well as their respective intra- and inter-structure correlations. However, due to the black-box nature of deep learning models, the interpretability of each anatomical feature remained limited in practice.

The segmentation network $S$ was ultimately trained using the proposed loss function based on a combination of cross-entropy, multi-scale contrastive regularization and multi-joint anatomical priors losses:
\begin{equation}
\mathcal{L} = \mathcal{L}_{\text{CE}} + \lambda_{2} \mathcal{L}_{\text{MSC}} + \lambda_{3} \mathcal{L}_{\text{MJAP}}
\end{equation}
\noindent where $\lambda_2$ and $\lambda_3$ were empirically set weighting factors.

\section{Experiments}
\label{sec:experiments}

In this section, we explain the experiments conducted with the proposed multi-task, multi-domain network on the pediatric musculoskeletal datasets. We first present the ankle, knee and shoulder joint MR imaging datasets (Section \ref{sec:imaging_datasets}) followed by the compared segmentation strategies (Section \ref{sec:experimental_setups}). We describe the implementation details (Section \ref{sec:implementation_details}) as well as the conducted predicted segmentation assessment (Section \ref{sec:assessment_of_predicted_segmentation}).

\subsection{Imaging datasets}
\label{sec:imaging_datasets}

Experiments were performed on pediatric MR imaging datasets of three musculoskeletal joints: ankle, knee, and shoulder. Ankle and shoulder imaging datasets were acquired at Centre Hospitalier Régional Universitaire (CHRU) La Cavale Blanche, Brest, France, using a 3.0T Achieva scanner (Philips Healthcare, Best, Netherlands) while knee imaging datasets were obtained retrospectively from the Children's Mercy Hospital, Kansas City, United States. The knee data was acquired using a 3.0T MRI scanner (MAGNETOM Skyra, Siemens Healthineers, Siemens AG). MRI data acquisition was performed in line with the principles of the Declaration of Helsinki. Ethical approvals were respectively granted by the Ethics Committee (Comit\'e Protection de Personnes Ouest VI) of CHRU Brest (2015-A01409-40) and by the research ethics committee of the Children's Mercy Hospital, Kansas City, United States. \\

\noindent \textbf{Ankle joint dataset.} The ankle joint dataset contained $20$ MR examinations ($\mathrm{A_{1}}$, ..., $\mathrm{A_{20}}$) acquired on pediatric individuals aged from 7 to 13 years (average age: $10.1 \pm 2.1$ years). We included three additional pediatric ankle examinations compared to our previous study \citep{boutillon_multi-task_2021}. A T1-weighted gradient echo sequence was employed during image acquisition (TR: 7.9 ms, TE: 2.8 ms, FOV: $140\times161$ mm$^{2}$), with resolutions varying from $0.25\times0.25\times0.50$ mm$^3$ to $0.28\times0.28\times0.80$ mm$^3$. All images were annotated by a medically trained expert (15 years of experience) to get ground truth delineations of calcaneus, talus and tibia (distal) bones, with specific label for each bone using the ITK-SNAP software (\url{http://www.itksnap.org/}). \\

\noindent \textbf{Knee joint dataset.} The knee imaging dataset consisted of $17$ MR examinations ($\mathrm{K_{1}}$, ..., $\mathrm{K_{17}}$) extracted from a pediatric cohort composed of patients aged from 13 to 18 years old (average age: $15.4 \pm 1.6$ years). Images were acquired using a 3D Gradient Recall Echo (GRE) sequence (TR: 13.0 ms, TE: 4.4 ms, FOV: $320X320$ mm$^{2}$), with resolutions ranging from $0.47\times0.47\times0.5$ mm$^3$ to $0.625\times0.625\times0.63$ mm$^3$. Segmentation masks of the femur (distal), fibula (proximal), patella and tibia (proximal) bones were manually derived using ITK-SNAP. \\

\noindent \textbf{Shoulder joint dataset.} MR images of $15$ shoulder joints ($\mathrm{S_{1}}$, ..., $\mathrm{S_{15}}$) were obtained from pediatric individuals aged from 5 to 17 years old (average age: $11.6 \pm 4.4$ years). Images were acquired using an eTHRIVE (enhanced T1-weighted High-Resolution Isotropic Volume Examination) sequence (TR: 8.4 ms, TE: 4.2 ms, FOV: $260\times210$ mm$^{2}$). Image resolution varied across subjects from $0.24\times0.24\times0.60$ mm$^3$ to $0.37\times0.37\times1.00$ mm$^3$. Ground truth delineations of the humerus and scapula bones were produced following the same protocol as for ankle and knee datasets. \\

For each dataset, all 2D slices were downsampled to $256 \times 256$ pixels and intensities were normalized to have zero-mean and unit variance.

\subsection{Experimental setups}
\label{sec:experimental_setups}

\subsubsection{Multi-task, multi-domain strategies}
\label{sec:multi-task_multi-domain_strategies}

\begin{figure*}[ht!]
\centering
\begin{adjustbox}{width=\textwidth}
\tikzstyle{dashed}=[dash pattern=on .85pt off .85pt]
\begin{tikzpicture}[every node/.style={inner sep=0,outer sep=0}]

\draw[line width=0.15mm, color=darkgray, rounded corners=1] (3.75,.35) -- (3.75,2.025) -- (1.35,2.025) -- (1.35,.35) -- cycle;
\draw[line width=0.15mm, color=darkgray, rounded corners=1] (6.15,.35) -- (6.15,2.025) -- (3.75,2.025) -- (3.75,.35) -- cycle;
\draw[line width=0.15mm, color=darkgray, rounded corners=1] (8.55,.35) -- (8.55,1.1875) -- (6.15,1.1875) -- (6.15,.35) -- cycle;
\draw[line width=0.15mm, color=darkgray, rounded corners=1] (8.55,1.1875) -- (8.55,2.025) -- (6.15,2.025) -- (6.15,1.1875) -- cycle;

\node[inner sep=0pt] (mri_a) at (1.8,1.675)
    {\includegraphics[width=.02\textwidth]{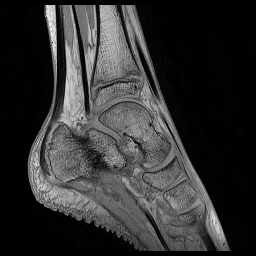}};
\draw [CornflowerBlue!40, rounded corners=1.25, line width=1] (mri_a.north west) -- (mri_a.north east) -- (mri_a.south east) -- (mri_a.south west) -- cycle;

\node[inner sep=0pt] (pred_a) at (1.8,0.575)
    {\includegraphics[width=.02\textwidth]{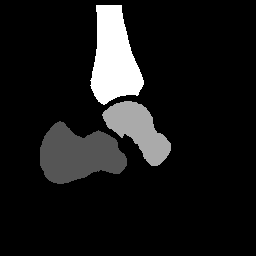}};
\draw [CornflowerBlue!40, rounded corners=1.25, line width=1] (pred_a.north west) -- (pred_a.north east) -- (pred_a.south east) -- (pred_a.south west) -- cycle;

\node[inner sep=0pt] (mri_k) at (2.55,1.675)
    {\includegraphics[width=.02\textwidth]{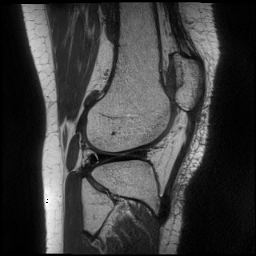}};
\draw [Orchid!40, rounded corners=1.25, line width=1] (mri_k.north west) -- (mri_k.north east) -- (mri_k.south east) -- (mri_k.south west) -- cycle;

\node[inner sep=0pt] (pred_k) at (2.55,0.575)
    {\includegraphics[width=.02\textwidth]{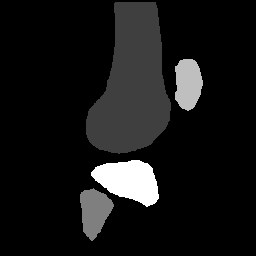}};
\draw [Orchid!40, rounded corners=1.25, line width=1] (pred_k.north west) -- (pred_k.north east) -- (pred_k.south east) -- (pred_k.south west) -- cycle;

\node[inner sep=0pt] (mri_s) at (3.3,1.675)
    {\includegraphics[width=.02\textwidth]{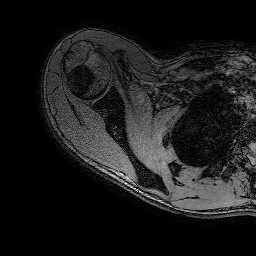}};
\draw [ForestGreen!40, rounded corners=1.25, line width=1] (mri_s.north west) -- (mri_s.north east) -- (mri_s.south east) -- (mri_s.south west) -- cycle;

\node[inner sep=0pt] (pred_s) at (3.3,0.575)
    {\includegraphics[width=.02\textwidth]{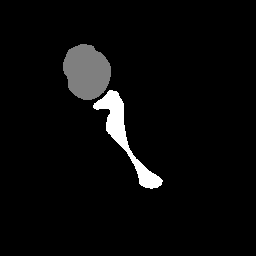}};   
\draw [ForestGreen!40, rounded corners=1.25, line width=1] (pred_s.north west) -- (pred_s.north east) -- (pred_s.south east) -- (pred_s.south west) -- cycle;

\node[inner sep=0pt] (mri_a) at (4.2,1.675)
    {\includegraphics[width=.02\textwidth]{ankle_mri.png}};
\draw [CornflowerBlue!40, rounded corners=1.25, line width=1] (mri_a.north west) -- (mri_a.north east) -- (mri_a.south east) -- (mri_a.south west) -- cycle;

\node[inner sep=0pt] (pred_a) at (4.2,0.575)
    {\includegraphics[width=.02\textwidth]{ankle_pred.png}};
\draw [CornflowerBlue!40, rounded corners=1.25, line width=1] (pred_a.north west) -- (pred_a.north east) -- (pred_a.south east) -- (pred_a.south west) -- cycle;

\node[inner sep=0pt] (mri_k) at (4.95,1.675)
    {\includegraphics[width=.02\textwidth]{knee_mri.png}};
\draw [Orchid!40, rounded corners=1.25, line width=1] (mri_k.north west) -- (mri_k.north east) -- (mri_k.south east) -- (mri_k.south west) -- cycle;

\node[inner sep=0pt] (pred_k) at (4.95,0.575)
    {\includegraphics[width=.02\textwidth]{knee_pred.png}};
\draw [Orchid!40, rounded corners=1.25, line width=1] (pred_k.north west) -- (pred_k.north east) -- (pred_k.south east) -- (pred_k.south west) -- cycle;

\node[inner sep=0pt] (mri_s) at (5.7,1.675)
    {\includegraphics[width=.02\textwidth]{shoulder_mri.png}};
\draw [ForestGreen!40, rounded corners=1.25, line width=1] (mri_s.north west) -- (mri_s.north east) -- (mri_s.south east) -- (mri_s.south west) -- cycle;

\node[inner sep=0pt] (pred_s) at (5.7,0.575)
    {\includegraphics[width=.02\textwidth]{shoulder_pred.png}};   
\draw [ForestGreen!40, rounded corners=1.25, line width=1] (pred_s.north west) -- (pred_s.north east) -- (pred_s.south east) -- (pred_s.south west) -- cycle;

\node[inner sep=0pt] (mri_a) at (6.4,1.80625)
    {\includegraphics[width=.02\textwidth]{ankle_mri.png}};
\draw [CornflowerBlue!40, rounded corners=1.25, line width=1] (mri_a.north west) -- (mri_a.north east) -- (mri_a.south east) -- (mri_a.south west) -- cycle;

\node[inner sep=0pt] (mri_k) at (6.6,1.40625)
    {\includegraphics[width=.02\textwidth]{knee_mri.png}};
\draw [Orchid!40, rounded corners=1.25, line width=1] (mri_k.north west) -- (mri_k.north east) -- (mri_k.south east) -- (mri_k.south west) -- cycle;
    
\node[inner sep=0pt] (mri_s) at (6.8,1.80625)
    {\includegraphics[width=.02\textwidth]{shoulder_mri.png}};
\draw [ForestGreen!40, rounded corners=1.25, line width=1] (mri_s.north west) -- (mri_s.north east) -- (mri_s.south east) -- (mri_s.south west) -- cycle;

\node[inner sep=0pt] (pred_a) at (7.9,1.80625)
    {\includegraphics[width=.02\textwidth]{ankle_pred.png}};
\draw [CornflowerBlue!40, rounded corners=1.25, line width=1] (pred_a.north west) -- (pred_a.north east) -- (pred_a.south east) -- (pred_a.south west) -- cycle;

\node[inner sep=0pt] (pred_k) at (8.1,1.40625)
    {\includegraphics[width=.02\textwidth]{knee_pred.png}};
\draw [Orchid!40, rounded corners=1.25, line width=1] (pred_k.north west) -- (pred_k.north east) -- (pred_k.south east) -- (pred_k.south west) -- cycle;

\node[inner sep=0pt] (pred_s) at (8.3,1.80625)
    {\includegraphics[width=.02\textwidth]{shoulder_pred.png}};
\draw [ForestGreen!40, rounded corners=1.25, line width=1] (pred_s.north west) -- (pred_s.north east) -- (pred_s.south east) -- (pred_s.south west) -- cycle;

\node[inner sep=0pt] (mri_a) at (6.4,0.96875)
    {\includegraphics[width=.02\textwidth]{ankle_mri.png}};
\draw [CornflowerBlue!40, rounded corners=1.25, line width=1] (mri_a.north west) -- (mri_a.north east) -- (mri_a.south east) -- (mri_a.south west) -- cycle;

\node[inner sep=0pt] (mri_k) at (6.6,0.56875)
    {\includegraphics[width=.02\textwidth]{knee_mri.png}};
\draw [Orchid!40, rounded corners=1.25, line width=1] (mri_k.north west) -- (mri_k.north east) -- (mri_k.south east) -- (mri_k.south west) -- cycle;
    
\node[inner sep=0pt] (mri_s) at (6.8,0.96875)
    {\includegraphics[width=.02\textwidth]{shoulder_mri.png}};
\draw [ForestGreen!40, rounded corners=1.25, line width=1] (mri_s.north west) -- (mri_s.north east) -- (mri_s.south east) -- (mri_s.south west) -- cycle;

\node[inner sep=0pt] (pred_a) at (7.9,0.96875)
    {\includegraphics[width=.02\textwidth]{ankle_pred.png}};
\draw [CornflowerBlue!40, rounded corners=1.25, line width=1] (pred_a.north west) -- (pred_a.north east) -- (pred_a.south east) -- (pred_a.south west) -- cycle;

\node[inner sep=0pt] (pred_k) at (8.1,0.56875)
    {\includegraphics[width=.02\textwidth]{knee_pred.png}};
\draw [Orchid!40, rounded corners=1.25, line width=1] (pred_k.north west) -- (pred_k.north east) -- (pred_k.south east) -- (pred_k.south west) -- cycle;

\node[inner sep=0pt] (pred_s) at (8.3,0.96875)
    {\includegraphics[width=.02\textwidth]{shoulder_pred.png}};
\draw [ForestGreen!40, rounded corners=1.25, line width=1] (pred_s.north west) -- (pred_s.north east) -- (pred_s.south east) -- (pred_s.south west) -- cycle;

\draw[line width=0.1mm, fill=CornflowerBlue!60, rounded corners = 1] (1.55,.825) -- (2.05,.825) -- (1.925,1.125) -- (2.05,1.425) -- (1.55,1.425) --  (1.675,1.125) -- cycle;

\draw[line width=0.1mm, fill=Orchid!60, rounded corners = 1] (2.3,.825) -- (2.8,.825) -- (2.675,1.125) -- (2.8,1.425) -- (2.3,1.425) --  (2.425,1.125) -- cycle;

\draw[line width=0.1mm, fill=ForestGreen!60, rounded corners = 1] (3.05,.825) -- (3.55,.825) -- (3.425,1.125) -- (3.55,1.425) -- (3.05,1.425) --  (3.175,1.125) -- cycle;

\draw[line width=0.1mm, fill=CornflowerBlue!60, rounded corners = 1] (3.95,.825) -- (4.45,.825) -- (4.325,1.125) -- (4.45,1.425) -- (3.95,1.425) --  (4.075,1.125) -- cycle;

\draw[line width=0.1mm, fill=Orchid!60, rounded corners = 1] (4.7,.825) -- (5.2,.825) -- (5.075,1.125) -- (5.2,1.425) -- (4.7,1.425) -- (4.825,1.125) -- cycle;

\draw[line width=0.1mm, fill=ForestGreen!60, rounded corners = 1] (5.45,.825) -- (5.95,.825) -- (5.825,1.125) -- (5.95,1.425) -- (5.45,1.425) --  (5.575,1.125) -- cycle;

\draw[line width=0.1mm, fill=BurntOrange!60, rounded corners = 1] (7.65,1.85625) -- (7.35,1.73125) -- (7.05,1.85625) -- (7.05,1.35625) -- (7.35,1.48125) --  (7.65,1.35625) -- cycle;

\draw[line width=0.1mm, fill=BurntOrange!60, rounded corners = 1] (7.585,0.9854) -- (7.35,0.8875) -- (7.05,1.0125) -- (7.05,0.5125) -- (7.35,0.6375) -- (7.585,0.5396);
\draw[line width=0.1mm, fill=CarnationPink!70, rounded corners = 1] (7.585,0.9854) -- (7.65,1.0125) -- (7.65,0.5125) --  (7.585,0.5396);
\draw[line width=0.01mm, fill=BrickRed!60] (7.085,0.9979) -- (7.115,0.9854) -- (7.115,0.5396) -- (7.085,0.5271) -- cycle;
\draw[line width=0.01mm, pattern={Lines[angle=45,distance={3pt/sqrt(2)}]}, pattern color=black, solid] (7.085,0.9979) -- (7.115,0.9854) -- (7.115,0.5396) -- (7.085,0.5271) -- cycle;
\draw[line width=0.01mm, fill=BrickRed!60] (7.185,0.9562) -- (7.215,0.9437) -- (7.215,0.5813) -- (7.185,0.5688) -- cycle;
\draw[line width=0.01mm, pattern={Lines[angle=45,distance={3pt/sqrt(2)}]}, pattern color=black] (7.185,0.9562) -- (7.215,0.9437) -- (7.215,0.5813) -- (7.185,0.5688) -- cycle;
\draw[line width=0.01mm, fill=BrickRed!60] (7.285,0.9145) -- (7.315,0.902) -- (7.315,0.623) -- (7.285,0.6105) -- cycle;
\draw[line width=0.01mm, pattern={Lines[angle=45,distance={3pt/sqrt(2)}]}, pattern color=black] (7.285,0.9145) -- (7.315,0.902) -- (7.315,0.623) -- (7.285,0.6105) -- cycle;
\draw[line width=0.01mm, fill=BrickRed!60] (7.385,0.902) -- (7.415,0.9145) -- (7.415,0.6105) -- (7.385,0.623) -- cycle;
\draw[line width=0.01mm,  pattern={Lines[angle=45,distance={3pt/sqrt(2)}]}, pattern color=black] (7.385,0.902) -- (7.415,0.9145) -- (7.415,0.6105) -- (7.385,0.623) -- cycle;
\draw[line width=0.01mm, fill=BrickRed!60] (7.485,0.9437) -- (7.515,0.9562) -- (7.515,0.5688) -- (7.485,0.5813) -- cycle;
\draw[line width=0.01mm, pattern={Lines[angle=45,distance={3pt/sqrt(2)}]}, pattern color=black] (7.485,0.9437) -- (7.515,0.9562) -- (7.515,0.5688) -- (7.485,0.5813) -- cycle;
\draw[line width=0.01mm, fill=BrickRed!60] (7.585,0.9854) -- (7.615,0.9979) -- (7.615,0.5271) -- (7.585,0.5396) -- cycle;
\draw[line width=0.01mm, pattern={Lines[angle=45,distance={3pt/sqrt(2)}]}, pattern color=black]  (7.585,0.9854) -- (7.615,0.9979) -- (7.615,0.5271) -- (7.585,0.5396) -- cycle;
\draw[line width=0.1mm, rounded corners = 1] (7.05,1.0125) -- (7.35,0.8875) -- (7.65,1.0125) -- (7.65,0.5125) -- (7.35,0.6375) -- (7.05,0.5125) -- cycle;

\node at (2.55,1.925) {\scalebox{.35}{\textbf{(a) Individual}}};
\node[rotate=90] at (1.8,1.125) {\scalebox{.35}{Ankle}};
\node[rotate=90]  at (2.55,1.125) {\scalebox{.35}{Knee}};
\node[rotate=90]  at (3.3,1.125) {\scalebox{.35}{Shoulder}};
\node at (4.95,1.925) {\scalebox{.35}{\textbf{(b) Transfer}}};
\node[rotate=90] at (4.2,1.125) {\scalebox{.35}{Ankle}};
\node[rotate=90]  at (4.95,1.125) {\scalebox{.35}{Knee}};
\node[rotate=90]  at (5.7,1.125) {\scalebox{.35}{Shoulder}};
\node at (7.35,1.925) {\scalebox{.35}{\textbf{(c) Shared}}};
\node at (7.35,1.67625) {\scalebox{.35}{Shared}};
\node at (7.35,1.53625) {\scalebox{.35}{Weights}};
\node at (7.35,1.08125) {\scalebox{.35}{\textbf{(d) DSL}}};

\draw[line width=0.2mm, color=darkgray, dashed, -{Latex[length=2.5pt, width=2.5pt]}] (4.325,1.125) to[out=25, in=155] (4.825,1.125);  
\draw[line width=0.2mm, color=darkgray, dashed, -{Latex[length=2.5pt, width=2.5pt]}] (4.325,1.125) to[out=35, in=145] (5.575,1.125); 
\node at (4.575,1.075) {\scalebox{.275}{Weights}};
\node at (4.575,.975) {\scalebox{.275}{Transfer}};

\end{tikzpicture}
\end{adjustbox}
  \caption{Proposed multi-task, multi-domain segmentation strategies: (a) individual strategy constituted of domain-specific networks, (b) transfer strategy in which weights learnt on one domain were transferred to other domains for initialization, (c) shared strategy comprising a single network with all parameters shared between domains, and (d) domain-specific layers (DSL) strategy based on a model with shared convolutional filters along with domain-specific batch normalization and segmentation layers. The transfer strategy encompassed all possible combinations of transfer learning between the three domains including transfer$_{\text{Ankle}}$ (as depicted here), transfer$_{\text{Knee}}$, and transfer$_{\text{Shoulder}}$ (both omitted for brevity).}
  \label{fig:experiments}
\end{figure*}

As a first experiment, we investigated various multi-task, multi-domain segmentation strategies with Att-UNet \citep{oktay_attention_2018} as backbone architecture to assess which one would provide the best segmentation results. The compared methods built upon Att-UNet comprised four approaches (Fig. \ref{fig:experiments}): individual (trained on individual domains), transfer (pre-trained on one domain and fine-tuned on the others), shared (trained on all domains at once, with all parameters shared between domains) and DSL (trained on all domains at once, with shared and domain-specific parameters). The shared approach differed from the DSL scheme by its domain-agnostic batch normalization and shared segmentation layer which predicted bones of interest from all domains with distinct labels (plus background). In this sense, the shared approach was analogous to that developed by \citep{moeskops_deep_2016}, although their network architecture differed from Att-UNet and lacked the efficiency to provide dense segmentation predictions. In addition, all networks were trained from scratch with randomly distributed weights except in the transfer scheme in which weights learnt on one domain were transferred to other domains for initialization (Fig. \ref{fig:experiments}). In the transfer scheme, models were not tested on their domain of origin because re-training on the same dataset would not have corresponded to a transfer of knowledge between domains. Hence, transfer$_{\text{Ankle}}$ denoted models pre-trained on ankle images and fine-tuned on either knee or shoulder domains. We investigated all possible combinations of transfer learning between the three datasets, and defined transfer$_{\text{Knee}}$ and transfer$_{\text{Shoulder}}$ schemes in a similar manner.

Furthermore, to evaluate the contributions of the multi-scale contrastive regularization and multi-joint anatomical priors, we performed an ablation study by setting the hyper-parameters weighting factors $\lambda_1$, $\lambda_2$, and $\lambda_3$ to zero respectively. Specifically, as the intensity domains $\mathcal{I}_{1}, ..., \mathcal{I}_{K}$ were not differentiated in the shared approach, the multi-scale contrastive regularization $\mathcal{L}_{\text{MSC}}$ could only be integrated in the DSL scheme. Meanwhile, the multi-joint anatomical priors $\mathcal{L}_{\text{MJAP}}$ were incorporated in both shared (using a multi-joint auto-encoder with all parameters shared) and DSL (using a multi-joint auto-encoder with shared and domain-specific parameters) approaches. Finally, we assessed the advantages of the multi-scale contrastive over the previously proposed single-scale contrastive ($\mathcal{L}_{\text{SSC}}$) method \citep{boutillon_multi-task_2021} which only constrain the network bottleneck (i.e. encoder output).

\subsubsection{Pre-trained architectures}
\label{sec:pre-trained_architecture}

As a second experiment, we evaluated the performance of our method based on Efficient-UNet with pre-trained \texttt{Effi\-cientB3} encoder \citep{tan_efficientnet_2019}, DSL, multi-scale contrastive regularization and multi-joint anatomical priors against Inception-Net and Dense-Net backbone architectures \citep{szegedy_rethinking_2016, huang_densely_2017} similarly pre-trained on large natural image database \citep{russakovsky_imagenet_2015}. Specifically, the pre-trained models Inception-UNet \citep{szegedy_rethinking_2016}, Dense-UNet \citep{huang_densely_2017} and Efficient-UNet \citep{tan_efficientnet_2019} were compared using individual, shared with multi-joint anatomical priors ($\text{shared}+\mathcal{L}_{\text{MJAP}}$), and DSL with multi-scale contrastive regularization and multi-joint anatomical priors ($\text{DSL}+\mathcal{L}_{\text{MSC}}+\mathcal{L}_{\text{MJAP}}$) schemes. For the shared and DSL strategies, we only retained the best approach observed within each during Att-UNet experiments (i.e. $\text{shared}+\mathcal{L}_{\text{MJAP}}$ and $\text{DSL}+\mathcal{L}_{\text{MSC}}+\mathcal{L}_{\text{MJAP}}$, Section \ref{sec:quantitative_assessment}). Finally, the transfer scheme was discarded in this experimental setup as networks were already partially pre-trained on the ImageNet database \citep{russakovsky_imagenet_2015}.

The compared Inception and Dense-UNet architectures referred to UNet models with encoder respectively replaced by either an \texttt{InceptionV3} \citep{szegedy_rethinking_2016} or a \texttt{DenseNet121} \citep{huang_densely_2017} classifier network pre-trained on a natural image classification task (Table \ref{tab:experiments}). Similarly to Efficient-UNet, the decoder components of both Inception-UNet and Dense-UNet were designed to be symmetrical from their respective encoder branches. Consequently, their decoders were extended from original UNet design by adding convolutional filters and more features, as well as Inception modules \citep{szegedy_rethinking_2016} and dense blocks \citep{huang_densely_2017} respectively. In addition, as for Efficient-UNet, spatial attention gates were incorporated to the skip connections of both Incep\-tion-UNet and Dense-UNet pre-trained architectures (Table \ref{tab:experiments}).

\subsection{Implementation details}
\label{sec:implementation_details}

\begin{table*}[ht!]
\caption{Summary of the networks employed during experiments and their corresponding architecture design, including: pre-trained encoder \citep{szegedy_rethinking_2016, huang_densely_2017, tan_efficientnet_2019}, attention gate (AG) \citep{oktay_attention_2018} and number of trainable parameters in individual, shared and DSL learning schemes; along with their corresponding training hyper-parameter values: batch size, number of epochs and learning rate. }
\centering
    \begin{tabular}{|M{2.5cm}||M{2.2cm}|M{.65cm}|M{.9cm}|M{1.25cm}|M{1.25cm}|M{1.45cm}|M{1.45cm}|M{1.45cm}|} 
    \hline
    \multirow{2}{*}{Network} & \multirow{2}{*}{\shortstack{Pre-trained\\Encoder}} & \multirow{2}{*}{AG} & \multirow{2}{*}{\shortstack{Batch\\Size}} & \multirow{2}{*}{$\#$Epochs} & \multirow{2}{*}{\shortstack{Learning\\Rate}} & \multicolumn{3}{c|}{$\#$Parameters} \\\cline{7-9}
    & & & & & & Individual & Shared & DSL \\\hline\hline
    Auto-encoder & \--- & \--- & 24 & 8 & $1\mathrm{e}{-4}$ & \--- & 7.9M & 7.9M \\
    \hline\hline
    Att-UNet & \--- & \checkmark &  18 & 6 & $5\mathrm{e}{-4}$ & $3\times$8.7M & 8.7M & 8.7M \\\hline
    Inception-UNet & \texttt{InceptionV3} & \checkmark & 18 & 6 & $5\mathrm{e}{-4}$ & $3\times$48.1M & 48.1M & 48.3M\\\hline
    Dense-UNet & \texttt{DenseNet121} & \checkmark & 12 & 4 & $1\mathrm{e}{-4}$ & $3\times$23.3M & 23.3M & 23.6M\\\hline
    Efficient-UNet & \texttt{EfficientNetB3} & \checkmark & 12 & 4 & $5\mathrm{e}{-4}$ & $3\times$14.6M & 14.6M & 14.8M\\\hline
    \end{tabular}
\label{tab:experiments}
\end{table*}

Each of the networks employed through the experiments was characterized by specific implementation and architecture designs (Table \ref{tab:experiments}). As previously indicated, all networks integrated attention gates with the exception of the auto-encoder due to the lack of skip connections. Moreover, \texttt{ReLU} and \texttt{SiLU} non-linear activation functions were implemented in Eq. \ref{eq:multi-domain_block}, in accordance with the original design of the employed models \citep{szegedy_rethinking_2016, huang_densely_2017, oktay_attention_2018, tan_efficientnet_2019}. During training, all networks were optimized using the Adam optimizer with distinct batch size, number of epochs and learning rate for each (Table \ref{tab:experiments}), and these hyper-parameters values remained fixed across all multi-task, multi-domain segmentation strategies (individual, transfer, shared, and DSL). In shared and DSL schemes, the image batch was equally split between each dataset to prevent domain-bias during optimization. The number of scales employed in the multi-scale contrastive regularization $\mathcal{L}_{\text{MSC}}$ remained fixed across the networks with $\vert\mathcal{S}\vert = 9$. Meanwhile, the single scale contrastive constraint $\mathcal{L}_{\text{SSC}}$ only involved the $5^{th}$ spatial scale corresponding to the network bottleneck (i.e. encoder output) \citep{boutillon_multi-task_2021}. Additionally, model complexity (i.e. number of trainable parameters) varied across architectures and learning schemes with a maximum of $3\times48.1$M (millions) parameters for individual Inception-UNet (Table \ref{tab:experiments}). Most notably, domain-specific weights represented at maximum $3.0\%$ of the total of trainable parameters and DSL frameworks were highly compact compared to individual schemes which required $K = 3$ times as many parameters. Finally, we explored various values for the hyper-parameters of the multi-scale contrastive regularization $\mathcal{L}_{\text{MSC}}$ and multi-joint anatomical priors $\mathcal{L}_{\text{MJAP}}$, and found $\tau = 0.1$, $\lambda_1 = 0.05$, and $\lambda_3 = 0.1$ to be optimal. The optimal value of $\lambda_2$ varied between architecture with $\lambda_2 = 0.5$ for Att-UNet and Inception-UNet whereas $\lambda_2 = 0.05$ for Dense-UNet and Efficient-UNet. For its part, the single scale contrastive regularization $\mathcal{L}_{\text{SSC}}$ was weighted by an hyper-parameter set to $0.1$ \citep{boutillon_multi-task_2021}.

Implementation of the deep learning architectures was carried out in PyTorch. Training and inference were performed using an Nvidia RTX 2080 Ti GPU with 12 GB of RAM. All the models were trained on 2D slices with extensive on-the-fly data augmentation due to limited available training data. Data augmentation comprised random rotation ($\pm22.5^{\circ}$), shifting ($\pm10\%$), and flipping in both directions to teach the networks the desired invariance, covariance and robustness properties. Furthermore, the same post-processing was employed after each method: first, the obtained 2D segmentation masks were stacked together to form a 3D volume, then we selected the largest connected set of each anatomical structure as final 3D predicted mask, and we finally applied morphological closing (5$\times$5$\times$5 spherical kernel) to smooth the resulting boundaries.

\subsection{Assessment of predicted segmentation}
\label{sec:assessment_of_predicted_segmentation}

Assessment of the 3D delineations generated by the different methods relied on a comparison against manually annotated ground truths. For each dataset, Dice coefficient, sensitivity, specificity, maximum symmetric surface distance (MSSD) (i.e. symmetric Hausdorff distance), average symmetric surface distance (ASSD), and relative absolute volume difference (RAVD) metrics were computed for each bone and the average score was reported in Tables \ref{tab:leave-one-out_quantitative_assessment_of_att-unet} and \ref{tab:leave-one-out_quantitative_assessment_of_pre-trained_architectures}. Dice coefficient assesses the similarity between the two voxels sets while sensitivity and specificity measure the true positive and true negative rates respectively. Surface distances (MSSD and ASSD) determine the models' ability to generate the same 3D contours as those produced manually. Finally, RAVD computes the volumetric difference between volumes. Moreover, an expert visually validated the global anatomical consistency and plausibility of each predicted segmentation. Please refer to the supplementary material for mathematical definition of the employed metrics. 

Due to the scarce amount of pediatric examinations, experiments were performed in a leave-one-out manner such that, for each dataset, one examination was retained for validation, one for test, and the remaining data were used to train the model. We iterated through the datasets simultaneously to compute the mean and standard deviation of each metric, and used each examination at maximum once for test. We did not test all combinations between datasets, as this would have introduced redundant observations in the results and drastically increased computation time (i.e. $20\times17\times15 = 5100$ possible combinations). Consequently, as the shoulder joint dataset contained the fewest number of MR image volumes, $5$ ankle and $2$ knee joint examinations were never included in the test sets since all $15$ shoulder samples were already tested. Specifically, the imaging dataset with the fewest samples defined the total number of steps in the leave-one-out evaluation, as we refrained from testing examinations from this dataset multiple times to avoid redundant results and associated bias. All experiments followed the same protocol and imaging examinations with the same index (i.e. $\mathrm{A_{i}}$, $\mathrm{K_{i}}$, and $\mathrm{S_{i}}$) indicated 3D samples tested in the same $i^{\text{th}}$ fold of the leave-one-out evaluation. Following standard machine learning practice, the hyper-parameters values ($\tau$, $\lambda_1$, $\lambda_2$, $\lambda_3$, batch size, epochs, learning rate, etc.) were selected based on the performance of the model on the validation set (Table \ref{tab:experiments}).

The limited amount of 3D examinations also forced us to perform the statistical analysis between methods on the 2D MR images. To compare the multi-task, multi-domain strategies, we concatenated the 2D scores obtained on each dataset to create a unique distribution per metric. Specifically, we employed the Kolmogorov-Smirnov non-parametric test using Dice, sensitivity, and specificity scores obtained from the $2649$ ankle, $3041$ knee, and $3682$ shoulder 2D slices which corresponded to the $45$ MR image volumes in the test sets. Nevertheless, to avoid distorting the scores distributions, we retained only the scores obtained from the $1294$ ankle, $2283$ knee, and $3357$ shoulder 2D images with at least one anatomical structure of interest. The non-normality of the 2D results distributions was preliminary verified using D’Agostino and Pearson normality test. Moreover, due to the skew of the non-normal distributions, we reported their mean and the distances from the mean to the upper and lower bound of the $68\%$ confidence interval, which correspond to the 16 and 84 percentiles, as in \citep{schnider_3d_2020}. Since transfer models (transfer$_{\text{Ankle}}$, transfer$_{\text{Knee}}$, and transfer$_{\text{Shoulder}}$) were not tested on their original domain, we used the 2D scores obtained in the individual scheme as substitute. For each backbone architecture (Att-UNet, Inception-UNet, Dense-UNet, and Efficient-UNet) we evaluated the statistical significance of the performance obtained by our methodology based on DSL with multi-scale contrastive regularization and multi-joint anatomical priors ($\text{DSL}+\mathcal{L}_{\text{MSC}}+\mathcal{L}_{\text{MJAP}}$) compared to other multi-task, multi-domain strategies and reported the results in Table \ref{tab:statistical_analysis}.

Finally, we performed visual comparison of predicted segmentation masks at two levels. First, we evaluated the benefits in segmentation quality of the proposed multi-scale contrastive regularization ($\mathcal{L}_{\text{MSC}}$) along with multi-joint anat\-omical priors ($\mathcal{L}_{\text{MJAP}}$) using Att-UNet as backbone architecture in shared and DSL schemes. Second, we compared the segmentation obtained by the proposed Efficient-UNet pre-trained architecture in individual, $\text{shared}+\mathcal{L}_{\text{MJAP}}$, and $\text{DSL}+\mathcal{L}_{\text{MSC}}+\mathcal{L}_{\text{MJAP}}$ optimization schemes. We also provide attention maps computed by multi-domain attention gates to assess the interpretability of the proposed multi-task, multi-domain deep learning architectures (Att-UNet, Inception-U\-Net, Dense-UNet, and Efficient-UNet in $\text{DSL}+\mathcal{L}_{\text{MSC}}+\mathcal{L}_{\text{MJAP}}$ learning scheme). Specifically, we visualized the attention maps extracted by the spatial attention gate with highest resolution, which were up-sampled to original image resolution (i.e. $256 \times 256$) for Inception-UNet, Dense-UNet, and Efficient-UNet models.

\subsection{Assessment of learnt shared representations}
\label{sec:assessment_of_learnt_shared_representations}

To assess the benefits of the proposed multi-scale contrastive regularization on the internal features of multi-domain neural networks, we compared the shared representations learnt by Att-UNet and the multi-joint auto-encoder in shared, DSL, and $\text{DSL}+\mathcal{L}_{\text{MSC}}$ schemes. First, we computed the multi-scale embeddings $z_{s_1}, ..., z_{s_9}$ of Att-UNet (respectively $\Tilde{z}_{s_1}, ..., \Tilde{z}_{s_9}$ of the multi-joint auto-encoder, Fig. \ref{fig:architecture}) using ankle, knee, and shoulder 2D MR images (respectively 2D segmentation masks) originating from the training and validation sets. The 2D segmentation masks consisting of solely background were discarded during the process. Then, we applied the dimensionality reduction procedure recommended in \citep{maaten_visualizing_2008}, to visualize the high dimensional feature vectors belonging to $\mathbb{R}^d$ with $d$ ranging from 32 to 512 (Fig. \ref{fig:architecture}). For vector space dimension $d > 50$, we first employed principal component analysis to reduce the representations to 50 dimensional feature vectors. We ultimately used the t-SNE algorithm with perplexity and learning rate respectively set to 30 and 200, to embed the data into a 2D space. 

Finally, to provide a quantitative validation of the multi-scale contrastive regularization, we computed and compared the mean inter- and intra-domain cosine similarity of Att-UNet representations learnt in shared, $\text{DSL}+\mathcal{L}_{\text{SSC}}$, and $\text{DSL}+\mathcal{L}_{\text{MSC}}$ schemes. As evaluating the similarity measure of each possible data points pairs was computationally expensive, we randomly selected $10^{5}$ pairs within and between each domain, and reported their respective mean cosine similarity and standard deviation in Table \ref{tab:quantitative_assessment_of_shared_representations}.

\section{Results}
\label{sec:results}

The proposed method based on Efficient-UNet with pre-trained encoder, DSL, multi-scale contrastive regularization, and multi-joint anatomical priors was evaluated on three pediatric imaging domains and segmentation tasks. In this section, we report the quantitative results (Section \ref{sec:quantitative_assessment}) and qualitative comparisons (Section \ref{sec:qualitative_assessment}) of the multi-task, multi-domain strategies with different backbone architectures.

\subsection{Quantitative assessment}
\label{sec:quantitative_assessment}

\begin{table*}[ht!]
\caption{Leave-one-out quantitative assessment of Att-UNet \citep{oktay_attention_2018} using individual, transfer, shared, and DSL strategies employed with single-scale contrastive regularization $\mathcal{L}_{\text{SSC}}$, multi-scale contrastive regularization $\mathcal{L}_{\text{MSC}}$, and multi-joint anatomical priors $\mathcal{L}_{\text{MJAP}}$ on ankle, knee, and shoulder datasets. Metrics include Dice (\%), sensitivity (\%), specificity (\%), MSSD (mm), ASSD (mm), and RAVD (\%). Mean scores and standard deviations reported in bold and underlined respectively correspond to the first and second best results obtained for each dataset.}
\centering
    \begin{tabular}{|P{.25cm}|P{.25cm}|P{3.05cm}||P{1.6cm}|P{1.6cm}|P{1.6cm}|P{1.6cm}|P{1.6cm}|P{1.6cm}|} 
    \hline
    \multicolumn{3}{|c||}{Method} & Dice $\uparrow$ & Sens. $\uparrow$ & Spec. $\uparrow$ & MSSD $\downarrow$ & ASSD $\downarrow$ & RAVD $\downarrow$\\ 
    \hline\hline
     
    \multirow{30}{*}{\rotatebox[origin=c]{90}{Att-UNet}} & \multirow{10}{*}{\rotatebox[origin=c]{90}{Ankle}} & Individual & $88.2\pm1.9$ & $88.1\pm5.4$ & \underline{$99.8\pm0.1$} & $17.9\pm10.8$ & $1.9\pm1.1$ & $14.1\pm4.6$ \\\cline{3-9}
    & & Transfer$_{\text{Knee}}$ & $89.5\pm5.7$ & $88.3\pm6.0$ & $\mathbf{99.9\pm0.1}$ & $12.6\pm10.2$ & $1.6\pm1.7$ & $14.0\pm10.9$ \\\cline{3-9}
    & & Transfer$_{\text{Shoulder}}$ & $89.3\pm4.2$ & $87.5\pm6.5$ & $\mathbf{99.9\pm0.1}$ & $11.6\pm5.0$ & $1.3\pm0.6$ & $12.9\pm8.6$ \\\cline{3-9}
    & & Shared & $88.8\pm2.5$ & $87.6\pm6.3$ & $\mathbf{99.9\pm0.1}$ & $13.4\pm8.1$ & $1.5\pm0.8$ & $12.5\pm7.0$ \\\cline{3-9}
    & & $\text{Shared}+\mathcal{L}_{\text{MJAP}}$ & $89.6\pm1.6$ & \underline{$90.6\pm5.3$} & \underline{$99.8\pm0.1$} & $13.4\pm4.2$ & $1.3\pm0.3$ & $13.1\pm4.9$ \\\cline{3-9}
    & & DSL & $90.6\pm2.3$ & $88.5\pm4.6$ & $\mathbf{99.9\pm0.1}$ & $11.0\pm7.4$ & $1.2\pm0.8$ & $10.9\pm5.6$ \\\cline{3-9}
    & & $\text{DSL}+\mathcal{L}_{\text{MJAP}}$ & $90.9\pm1.9$ & $89.1\pm4.6$ & $\mathbf{99.9\pm0.1}$ & $12.7\pm9.2$ & $1.3\pm1.2$ & $10.5\pm4.4$ \\\cline{3-9}
    & & $\text{DSL}+\mathcal{L}_{\text{SSC}}$ & $90.6\pm2.1$ & $87.7\pm4.9$ & $\mathbf{99.9\pm0.1}$ & $\mathbf{9.0\pm3.0}$ & \underline{$1.0\pm0.3$} & $11.3\pm4.7$ \\\cline{3-9}
    & & $\text{DSL}+\mathcal{L}_{\text{MSC}}$ & \underline{$91.5\pm2.0$} & $\mathbf{90.7\pm4.7}$ & $\mathbf{99.9\pm0.1}$ & \underline{$9.7\pm3.7$} & \underline{$1.0\pm0.3$} & \underline{$9.6\pm4.2$} \\\cline{3-9}
    & & $\text{DSL}+\mathcal{L}_{\text{MSC}}+\mathcal{L}_{\text{MJAP}}$ & $\mathbf{91.8\pm1.8}$ & $\mathbf{90.7\pm4.8}$ & $\mathbf{99.9\pm0.1}$ & $\mathbf{9.0\pm2.9}$ & $\mathbf{0.9\pm0.3}$ & $\mathbf{8.8\pm4.7}$ \\
    \hhline{|~|========}
    
    & \multirow{10}{*}{\rotatebox[origin=c]{90}{Knee}} & Individual & $91.1\pm3.6$ & $88.9\pm5.5$ & $\mathbf{99.9\pm0.1}$ & $16.5\pm12.1$ & $1.6\pm1.5$ & $10.7\pm6.1$ \\\cline{3-9}
    & & Transfer$_{\text{Ankle}}$ & $92.8\pm2.9$ & $91.1\pm3.3$ & $\mathbf{99.9\pm0.1}$ & $12.4\pm10.3$ & $1.0\pm0.9$ & $7.6\pm5.4$ \\\cline{3-9}
    & & Transfer$_{\text{Shoulder}}$ & $92.5\pm2.4$ & $90.7\pm4.0$ & $\mathbf{99.9\pm0.1}$ & $13.1\pm11.3$ & $1.0\pm0.8$ & $7.8\pm4.5$ \\\cline{3-9}
    & & Shared & $91.7\pm3.2$ & $88.5\pm4.8$ & $\mathbf{99.9\pm0.1}$ & $12.5\pm9.0$ & $1.4\pm1.4$ & $9.5\pm6.0$ \\\cline{3-9}
    & & $\text{Shared}+\mathcal{L}_{\text{MJAP}}$ & $93.6\pm1.8$ & $91.9\pm3.1$ & $\mathbf{99.9\pm0.1}$ & \underline{$7.9\pm8.4$} & $0.8\pm0.9$ & $6.4\pm3.1$ \\\cline{3-9}
    & & DSL & $93.3\pm2.5$ & $92.8\pm3.5$ & $\mathbf{99.9\pm0.1}$ & $12.8\pm12.1$ & $1.1\pm1.3$ & $6.0\pm4.0$ \\\cline{3-9}
    & & $\text{DSL}+\mathcal{L}_{\text{MJAP}}$ & \underline{$93.8\pm2.5$} & \underline{$93.0\pm4.2$} & $\mathbf{99.9\pm0.1}$ & $9.4\pm5.9$ & \underline{$0.7\pm0.4$} & $6.8\pm4.1$ \\\cline{3-9}
    & & $\text{DSL}+\mathcal{L}_{\text{SSC}}$ & $93.7\pm1.6$ & $92.8\pm2.3$ & $\mathbf{99.9\pm0.1}$ & $11.3\pm8.9$ & $1.1\pm1.2$ & $6.4\pm3.7$ \\\cline{3-9}
    & & $\text{DSL}+\mathcal{L}_{\text{MSC}}$ & $\mathbf{94.3\pm2.0}$ & $\mathbf{93.2\pm3.9}$ & $\mathbf{99.9\pm0.1}$ & $8.9\pm9.7$ & $0.8\pm0.7$ & $\mathbf{5.5\pm4.0}$ \\\cline{3-9}
    & & $\text{DSL}+\mathcal{L}_{\text{MSC}}+\mathcal{L}_{\text{MJAP}}$ & $\mathbf{94.3\pm1.4}$ & $92.5\pm2.5$ & $\mathbf{99.9\pm0.1}$ & $\mathbf{5.6\pm2.1}$ & $\mathbf{0.5\pm0.2}$ & \underline{$5.9\pm3.0$} \\
    \hhline{|~|========}
    
    & \multirow{10}{*}{\rotatebox[origin=c]{90}{Shoulder}} & Individual & $80.9\pm10.1$ & $77.7\pm14.9$ & $\mathbf{99.9\pm0.1}$ & $26.9\pm14.1$ & $2.4\pm1.8$ & $15.2\pm16.7$ \\\cline{3-9}
    & & Transfer$_{\text{Ankle}}$ & $82.6\pm8.8$ & $79.8\pm12.5$ & $\mathbf{99.9\pm0.1}$ & $26.9\pm17.2$ & $2.2\pm1.8$ & $17.0\pm10.1$ \\\cline{3-9}
    & & Transfer$_{\text{Knee}}$ & \underline{$83.3\pm10.1$} & \underline{$80.5\pm12.5$} & $\mathbf{99.9\pm0.1}$ & \underline{$24.0\pm14.3$} & \underline{$2.1\pm2.4$} & \underline{$13.7\pm13.2$} \\\cline{3-9}
    & & Shared & $80.1\pm9.6$ & $76.6\pm12.9$ & $\mathbf{99.9\pm0.1}$ & $28.1\pm12.2$ & $2.7\pm1.8$ & $18.3\pm12.4$ \\\cline{3-9}
    & & $\text{Shared}+\mathcal{L}_{\text{MJAP}}$ & $80.7\pm9.0$ & $79.2\pm12.3$ & $\mathbf{99.9\pm0.1}$ & $25.0\pm15.6$ & $2.3\pm1.8$ & $19.4\pm12.0$ \\\cline{3-9}
    & & DSL & $80.9\pm7.3$ & $77.6\pm11.6$ & $\mathbf{99.9\pm0.1}$ & $34.4\pm19.1$ & $3.3\pm2.3$ & $19.4\pm12.2$ \\\cline{3-9}
    & & $\text{DSL}+\mathcal{L}_{\text{MJAP}}$ & $81.4\pm9.0$ & $79.2\pm14.3$ & $\mathbf{99.9\pm0.1}$ & $31.0\pm18.5$ & $2.5\pm2.2$ & $15.7\pm12.2$ \\\cline{3-9}
    & & $\text{DSL}+\mathcal{L}_{\text{SSC}}$ & $81.3\pm9.1$ & $78.1\pm12.2$ & $\mathbf{99.9\pm0.1}$ & $25.4\pm13.7$ & $2.7\pm2.7$ & $17.1\pm13.3$ \\\cline{3-9}
    & & $\text{DSL}+\mathcal{L}_{\text{MSC}}$ & $82.1\pm8.0$ & $79.8\pm9.1$ & $\mathbf{99.9\pm0.1}$ & $27.5\pm11.6$ & $2.7\pm1.8$ & $15.6\pm13.1$ \\\cline{3-9}
    & & $\text{DSL}+\mathcal{L}_{\text{MSC}}+\mathcal{L}_{\text{MJAP}}$ & $\mathbf{84.9\pm6.3}$ & $\mathbf{82.9\pm9.2}$ & $\mathbf{99.9\pm0.1}$ & $\mathbf{17.6\pm8.0}$ & $\mathbf{1.5\pm1.1}$ & $\mathbf{13.5\pm10.5}$ \\
    \hline
    
    \end{tabular}
\label{tab:leave-one-out_quantitative_assessment_of_att-unet}
\end{table*}

\begin{table*}[ht!]
\caption{Leave-one-out quantitative assessment of the pre-trained architectures: Inception-UNet \citep{szegedy_rethinking_2016}, Dense-UNet \citep{huang_densely_2017}, and Efficient-UNet \citep{tan_efficientnet_2019} on ankle, knee, and shoulder datasets. Individual, shared, and DSL strategies are employed with multi-scale contrastive regularization $\mathcal{L}_{\text{MSC}}$ and multi-joint anatomical priors $\mathcal{L}_{\text{MJAP}}$. Metrics include Dice (\%), sensitivity (\%), specificity (\%), MSSD (mm), ASSD (mm), and RAVD (\%). Mean scores and standard deviations reported in bold and underlined respectively correspond to the first and second best results obtained for each dataset.}
\centering
    \begin{tabular}{|P{.25cm}|P{.25cm}|P{3.05cm}||P{1.6cm}|P{1.6cm}|P{1.6cm}|P{1.6cm}|P{1.6cm}|P{1.6cm}|} 
    \hline
    \multicolumn{3}{|c||}{Method} & Dice $\uparrow$ & Sens. $\uparrow$ & Spec. $\uparrow$ & MSSD $\downarrow$ & ASSD $\downarrow$ & RAVD $\downarrow$\\
    \hline\hline
    
    \multirow{9}{*}{\rotatebox[origin=c]{90}{Inception-UNet}} & \multirow{3}{*}{\rotatebox[origin=c]{90}{Ankle}} & Individual & \underline{$91.4\pm2.6$} & $\mathbf{92.2\pm3.3}$ & $\mathbf{99.9\pm0.1}$ & \underline{$8.5\pm4.2$} & \underline{$0.9\pm0.4$} & \underline{$9.7\pm5.9$} \\\cline{3-9}
    & & $\text{Shared}+\mathcal{L}_{\text{MJAP}}$ & $91.3\pm2.2$ & \underline{$91.6\pm5.1$} & $\mathbf{99.9\pm0.1}$ & $9.8\pm4.1$ & $1.0\pm0.3$ & $10.0\pm5.0$ \\\cline{3-9}
    & & $\text{DSL}+\mathcal{L}_{\text{MSC}}+\mathcal{L}_{\text{MJAP}}$ & $\mathbf{93.2\pm1.5}$ & $\mathbf{92.2\pm3.8}$ & $\mathbf{99.9\pm0.1}$ & $\mathbf{6.5\pm2.0}$ & $\mathbf{0.7\pm0.2}$ & $\mathbf{7.4\pm3.4}$ \\\hhline{|~|========}
    
    & \multirow{3}{*}{\rotatebox[origin=c]{90}{Knee}} & Individual & $93.9\pm2.2$ & $92.1\pm3.7$ & $\mathbf{99.9\pm0.1}$ & $\mathbf{5.5\pm2.6}$ & $\mathbf{0.5\pm0.2}$ & $6.9\pm4.5$ \\\cline{3-9}
    & & $\text{Shared}+\mathcal{L}_{\text{MJAP}}$ & \underline{$94.2\pm2.1$} & \underline{$93.0\pm3.7$} & $\mathbf{99.9\pm0.1}$ & $6.4\pm3.0$ & $\mathbf{0.5\pm0.2}$ & \underline{$6.0\pm3.4$} \\\cline{3-9}
    & & $\text{DSL}+\mathcal{L}_{\text{MSC}}+\mathcal{L}_{\text{MJAP}}$ & $\mathbf{94.5\pm1.1}$ & $\mathbf{93.6\pm2.0}$ & $\mathbf{99.9\pm0.1}$ & \underline{$6.3\pm2.2$} & $\mathbf{0.5\pm0.2}$ & $\mathbf{5.2\pm2.7}$ \\
    \hhline{|~|========}
    
    & \multirow{3}{*}{\rotatebox[origin=c]{90}{\small{Shoulder}}} & Individual & $82.8\pm7.3$ & $79.6\pm9.3$ & $\mathbf{99.9\pm0.1}$ & $21.8\pm10.9$ & $2.1\pm1.5$ & \underline{$15.9\pm10.7$} \\\cline{3-9}
    & & $\text{Shared}+\mathcal{L}_{\text{MJAP}}$ & \underline{$83.1\pm5.8$} & $\mathbf{81.2\pm10.6}$ & $\mathbf{99.9\pm0.1}$ & $\mathbf{20.0\pm11.2}$ & \underline{$1.7\pm1.5$} & $16.4\pm10.5$ \\\cline{3-9}
    & & $\text{DSL}+\mathcal{L}_{\text{MSC}}+\mathcal{L}_{\text{MJAP}}$ & $\mathbf{84.5\pm5.8}$ & \underline{$80.5\pm9.2$} & $\mathbf{99.9\pm0.1}$ & \underline{$20.2\pm6.5$} & $\mathbf{1.6\pm0.7}$ & $\mathbf{14.7\pm9.7}$ \\
    \hline\hline
    
    \multirow{9}{*}{\rotatebox[origin=c]{90}{Dense-UNet}} & \multirow{3}{*}{\rotatebox[origin=c]{90}{Ankle}} & Individual & \underline{$92.4\pm1.7$} & $91.4\pm4.7$ & $\mathbf{99.9\pm0.1}$ & $7.4\pm2.4$ & \underline{$0.8\pm0.2$} & \underline{$7.9\pm4.5$} \\\cline{3-9}
    & & $\text{Shared}+\mathcal{L}_{\text{MJAP}}$ & $\mathbf{93.4\pm1.5}$ & $\mathbf{92.8\pm4.4}$ & $\mathbf{99.9\pm0.1}$ & \underline{$6.9\pm2.1$} & $\mathbf{0.7\pm0.2}$ & $\mathbf{6.6\pm4.6}$ \\\cline{3-9}
    & & $\text{DSL}+\mathcal{L}_{\text{MSC}}+\mathcal{L}_{\text{MJAP}}$ & $\mathbf{93.4\pm1.3}$ & \underline{$92.5\pm4.0$} & $\mathbf{99.9\pm0.1}$ & $\mathbf{6.4\pm1.7}$ & $\mathbf{0.7\pm0.2}$ & $\mathbf{6.6\pm4.1}$ \\
    \hhline{|~|========}
    
    & \multirow{3}{*}{\rotatebox[origin=c]{90}{Knee}} & Individual & \underline{$94.3\pm1.3$} & $92.6\pm2.5$ & $\mathbf{99.9\pm0.1}$ & $5.2\pm2.2$ & $\mathbf{0.5\pm0.1}$ & $5.6\pm2.9$ \\\cline{3-9}
    & & $\text{Shared}+\mathcal{L}_{\text{MJAP}}$ & $\mathbf{95.1\pm1.6}$ & $\mathbf{94.9\pm2.4}$ & $\mathbf{99.9\pm0.1}$ & \underline{$4.6\pm1.8$} & $\mathbf{0.5\pm0.2}$ & $\mathbf{4.5\pm3.2}$ \\\cline{3-9}
    & & $\text{DSL}+\mathcal{L}_{\text{MSC}}+\mathcal{L}_{\text{MJAP}}$ & $\mathbf{95.1\pm1.2}$ & \underline{$93.7\pm2.5$} & $\mathbf{99.9\pm0.1}$ & $\mathbf{4.5\pm1.4}$ & $\mathbf{0.5\pm0.1}$ & \underline{$4.7\pm2.2$} \\
    \hhline{|~|========}
    
    & \multirow{3}{*}{\rotatebox[origin=c]{90}{\small{Shoulder}}} & Individual & $82.5\pm9.2$ & $79.5\pm12.4$ & $\mathbf{99.9\pm0.1}$ & $22.1\pm11.6$ & $1.9\pm1.5$ & $16.2\pm14.4$ \\\cline{3-9}
    & & $\text{Shared}+\mathcal{L}_{\text{MJAP}}$ & \underline{$84.9\pm4.8$} & \underline{$85.2\pm8.1$} & $\mathbf{99.9\pm0.1}$ & \underline{$20.2\pm13.0$} & \underline{$1.4\pm1.0$} & \underline{$14.9\pm8.2$} \\\cline{3-9}
    & & $\text{DSL}+\mathcal{L}_{\text{MSC}}+\mathcal{L}_{\text{MJAP}}$ & $\mathbf{86.6\pm4.3}$ & $\mathbf{87.0\pm5.3}$ & $\mathbf{99.9\pm0.1}$ & $\mathbf{16.0\pm6.5}$ & $\mathbf{1.2\pm0.6}$ & $\mathbf{11.5\pm6.5}$ \\
    \hline\hline
    
    \multirow{9}{*}{\rotatebox[origin=c]{90}{Efficient-UNet}} & \multirow{3}{*}{\rotatebox[origin=c]{90}{Ankle}} & Individual & \underline{$92.3\pm1.5$} & \underline{$92.0\pm3.8$} & $\mathbf{99.9\pm0.1}$ & $7.0\pm2.1$ & \underline{$0.8\pm0.2$} & $8.2\pm4.1$ \\\cline{3-9}
    & & $\text{Shared}+\mathcal{L}_{\text{MJAP}}$ & $\mathbf{93.8\pm0.9}$ & $\mathbf{93.5\pm2.8}$ & $\mathbf{99.9\pm0.1}$ & \underline{$6.5\pm1.6$} & $\mathbf{0.6\pm0.1}$ & $\mathbf{5.9\pm2.2}$ \\\cline{3-9}
    & & $\text{DSL}+\mathcal{L}_{\text{MSC}}+\mathcal{L}_{\text{MJAP}}$ & $\mathbf{93.8\pm1.3}$ & $\mathbf{93.5\pm4.0}$ & $\mathbf{99.9\pm0.1}$ & $\mathbf{5.6\pm1.8}$ & $\mathbf{0.6\pm0.2}$ & \underline{$6.9\pm3.7$} \\
    \hhline{|~|========}
    
    & \multirow{3}{*}{\rotatebox[origin=c]{90}{Knee}} & Individual & $94.1\pm1.3$ & $93.0\pm2.9$ & $\mathbf{99.9\pm0.1}$ & \underline{$4.7\pm1.2$} & \underline{$0.5\pm0.1$} & $5.7\pm2.5$ \\\cline{3-9}
    & & $\text{Shared}+\mathcal{L}_{\text{MJAP}}$ & \underline{$95.0\pm1.2$} & \underline{$94.3\pm2.4$} & $\mathbf{99.9\pm0.1}$ & $4.8\pm1.7$ & \underline{$0.5\pm0.2$} & \underline{$4.1\pm2.3$} \\\cline{3-9}
    & & $\text{DSL}+\mathcal{L}_{\text{MSC}}+\mathcal{L}_{\text{MJAP}}$ & $\mathbf{95.4\pm1.1}$ & $\mathbf{95.0\pm2.0}$ & $\mathbf{99.9\pm0.1}$ & $\mathbf{4.2\pm1.3}$ & $\mathbf{0.4\pm0.1}$ & $\mathbf{3.8\pm1.6}$ \\
    \hhline{|~|========}
    
    & \multirow{3}{*}{\rotatebox[origin=c]{90}{\small{Shoulder}}} & Individual & \underline{$87.7\pm4.0$} & $86.8\pm5.7$ & $\mathbf{99.9\pm0.1}$ & $16.0\pm5.4$ & \underline{$1.0\pm0.5$} & \underline{$8.4\pm6.1$} \\\cline{3-9}
    & & $\text{Shared}+\mathcal{L}_{\text{MJAP}}$ & $86.9\pm4.1$ & $\mathbf{89.0\pm4.8}$ & $\mathbf{99.9\pm0.1}$ & $\mathbf{14.3\pm5.1}$ & $\mathbf{0.9\pm0.3}$ & $10.7\pm8.2$ \\\cline{3-9}
    & & $\text{DSL}+\mathcal{L}_{\text{MSC}}+\mathcal{L}_{\text{MJAP}}$ & $\mathbf{87.9\pm3.8}$ & \underline{$87.4\pm4.8$} & $\mathbf{99.9\pm0.1}$ & \underline{$15.6\pm5.5$} & \underline{$1.0\pm0.5$} & $\mathbf{7.3\pm5.0}$ \\
    \hline
  
    \end{tabular}
\label{tab:leave-one-out_quantitative_assessment_of_pre-trained_architectures}
\end{table*}

\begin{table*}[ht!]
\renewcommand\arraystretch{1.15}
\caption{Statistical analysis between the proposed methods using the four backbone architectures: Att-UNet \citep{oktay_attention_2018}, Inception-UNet \citep{szegedy_rethinking_2016}, Dense-UNet \citep{huang_densely_2017}, and Efficient-UNet \citep{tan_efficientnet_2019}. Multi-task, multi-domain strategies include: individual, transfer, shared, and DSL employed with single-scale contrastive regularization $\mathcal{L}_{\text{SSC}}$, multi-scale contrastive regularization $\mathcal{L}_{\text{MSC}}$, and multi-joint anatomical priors $\mathcal{L}_{\text{MJAP}}$. Statistical analysis performed through Kolmogorov-Smirnov non-parametric test using Dice ($\%$), sensitivity ($\%$), and specificity ($\%$) computed on 2D slices from ankle, knee and shoulder datasets. Bold \textit{p}-values ($<0.01$) highlight statistically significant results for each metric, while first and second best 2D results are reported in bold and underlined respectively. Mean 2D scores and the distances from the mean to the upper and lower bound of the $68\%$ confidence interval are reported.}
\centering
    \begin{tabular}{|P{.35cm}|P{3.4cm}||P{1.65cm}|P{1.65cm}|P{1.65cm}|P{1.65cm}|P{1.65cm}|P{1.65cm}|} 
    \hline
    \multicolumn{2}{|c||}{Method} & Dice 2D $\uparrow$ & \textit{p}-value & Sens. 2D $\uparrow$ & \textit{p}-value & Spec. 2D $\uparrow$ & \textit{p}-value\\ 
    \hline\hline
     
    \multirow{11}{*}{\rotatebox[origin=c]{90}{Att-UNet}} & Individual & $84.1^{+13.9}_{-14.6}$ & $\mathbf{5.8\times10^{-37}}$ & $84.8^{+13.7}_{-14.2}$ & $\mathbf{4.5\times10^{-31}}$ & \underline{$99.8^{+0.2}_{-0.1}$} & $\mathbf{1.9\times10^{-35}}$ \\\cline{2-8}
    & Transfer$_{\text{Ankle}}$ & $85.1^{+13.1}_{-12.5}$ & $\mathbf{1.6\times10^{-20}}$ & $85.8^{+12.5}_{-13.5}$ & $\mathbf{3.0\times10^{-17}}$ & \underline{$99.8^{+0.2}_{-0.1}$} & $\mathbf{3.7\times10^{-28}}$ \\\cline{2-8}
    & Transfer$_{\text{Knee}}$ & $84.7^{+13.3}_{-13.8}$ & $\mathbf{7.5\times10^{-23}}$ & $85.4^{+13.3}_{-13.7}$ & $\mathbf{2.5\times10^{-22}}$ & $\mathbf{99.9^{+0.1}_{-0.1}}$ & $\mathbf{1.4\times10^{-13}}$ \\\cline{2-8}
    & Transfer$_{\text{Shoulder}}$ & $84.8^{+13.4}_{-15.4}$ & $\mathbf{9.0\times10^{-23}}$ & $85.0^{+13.6}_{-15.7}$ & $\mathbf{3.7\times10^{-25}}$ & $\mathbf{99.9^{+0.1}_{-0.1}}$ & $\mathbf{1.1\times10^{-19}}$ \\\cline{2-8}
    & Shared & $83.6^{+14.3}_{-15.0}$ & $\mathbf{3.2\times10^{-55}}$ & $83.3^{+15.1}_{-15.9}$ & $\mathbf{1.8\times10^{-71}}$ & $\mathbf{99.9^{+0.1}_{-0.1}}$ & $\mathbf{1.9\times10^{-14}}$ \\\cline{2-8}
    & $\text{Shared}+\mathcal{L}_{\text{MJAP}}$ & $85.4^{+12.8}_{-12.7}$ & $\mathbf{8.3\times10^{-15}}$ & \underline{$87.5^{+11.4}_{-13.8}$} & $\mathbf{2.3\times10^{-6}}$ & $\mathbf{99.9^{+0.1}_{-0.1}}$ & $\mathbf{1.3\times10^{-34}}$ \\\cline{2-8}
    & DSL & $84.2^{+13.9}_{-16.8}$ & $\mathbf{9.0\times10^{-30}}$ & $84.3^{+14.4}_{-16.6}$ & $\mathbf{3.2\times10^{-38}}$ & $\mathbf{99.9^{+0.1}_{-0.1}}$ & $\mathbf{1.0\times10^{-15}}$ \\\cline{2-8} 
    & $\text{DSL}+\mathcal{L}_{\text{MJAP}}$ & $84.8^{+13.6}_{-16.2}$ & $\mathbf{1.7\times10^{-17}}$ & $86.6^{+12.3}_{-14.5}$ & $\mathbf{1.9\times10^{-6}}$ & $\mathbf{99.9^{+0.1}_{-0.1}}$ & $\mathbf{3.5\times10^{-19}}$ \\\cline{2-8}
    & $\text{DSL}+\mathcal{L}_{\text{SSC}}$ & $85.1^{+13.1}_{-15.2}$ & $\mathbf{2.3\times10^{-18}}$ & $84.9^{+13.6}_{-15.9}$ & $\mathbf{7.9\times10^{-28}}$ & $\mathbf{99.9^{+0.1}_{-0.1}}$ & $\mathbf{7.7\times10^{-6}}$ \\\cline{2-8}
    & $\text{DSL}+\mathcal{L}_{\text{MSC}}$ & \underline{$86.1^{+12.2}_{-12.8}$} & $\mathbf{1.4\times10^{-9}}$ & $86.2^{+12.3}_{-13.4}$ & $\mathbf{1.3\times10^{-16}}$ & $\mathbf{99.9^{+0.1}_{-0.1}}$ & $\mathbf{2.8\times10^{-13}}$ \\\cline{2-8}
    & $\text{DSL}+\mathcal{L}_{\text{MSC}}+\mathcal{L}_{\text{MJAP}}$ & $\mathbf{88.2^{+10.2}_{-10.9}}$ & \--- & $\mathbf{88.3^{+10.4}_{-12.9}}$ & \--- & $\mathbf{99.9^{+0.1}_{-0.1}}$ & \--- \\\hline\hline

    \multirow{3}{*}{\rotatebox[origin=c]{90}{Inception}} & Individual & $86.4^{+11.8}_{-12.1}$ & $\mathbf{1.5\times10^{-7}}$ & $86.4^{+12.1}_{-12.2}$ & $\mathbf{7.3\times10^{-9}}$ & $\mathbf{99.9^{+0.1}_{-0.1}}$ & $\mathbf{5.2\times10^{-12}}$ \\\cline{2-8}
    & $\text{Shared}+\mathcal{L}_{\text{MJAP}}$ & \underline{$86.5^{+11.8}_{-11.8}$} & $\mathbf{4.0\times10^{-7}}$ & $\mathbf{87.8^{+10.9}_{-13.2}}$ & $\mathbf{2.8\times10^{-13}}$ & $\mathbf{99.9^{+0.1}_{-0.1}}$ & $\mathbf{5.7\times10^{-16}}$ \\\cline{2-8}
    & $\text{DSL}+\mathcal{L}_{\text{MSC}}+\mathcal{L}_{\text{MJAP}}$ & $\mathbf{88.2^{+10.3}_{-11.0}}$ & \--- & \underline{$87.7^{+11.0}_{-12.6}$} & \---  & $\mathbf{99.9^{+0.1}_{-0.1}}$ & \--- \\\hline\hline
    
    \multirow{3}{*}{\rotatebox[origin=c]{90}{Dense}} & Individual & $87.7^{+10.6}_{-9.8}$ & $\mathbf{3.2\times10^{-11}}$ & $87.4^{+11.1}_{-12.2}$ & $\mathbf{2.8\times10^{-13}}$ & $\mathbf{99.9^{+0.1}_{-0.1}}$ & $\mathbf{3.5\times10^{-8}}$ \\\cline{2-8}
    & $\text{Shared}+\mathcal{L}_{\text{MJAP}}$ & \underline{$89.0^{+9.4}_{-8.9}$} & $\mathbf{2.3\times10^{-4}}$ & $\mathbf{90.7^{+8.3}_{-8.6}}$ & $\mathbf{5.9\times10^{-12}}$ & $\mathbf{99.9^{+0.1}_{-0.1}}$ & $\mathbf{2.8\times10^{-12}}$ \\\cline{2-8}
    & $\text{DSL}+\mathcal{L}_{\text{MSC}}+\mathcal{L}_{\text{MJAP}}$ & $\mathbf{89.5^{+8.8}_{-6.9}}$ & \--- & \underline{$89.8^{+8.9}_{-8.4}$} & \--- & $\mathbf{99.9^{+0.1}_{-0.1}}$ &  \--- \\\hline\hline
    
    \multirow{3}{*}{\rotatebox[origin=c]{90}{Efficient}} & Individual & $89.6^{+8.7}_{-6.0}$ & $\mathbf{4.7\times10^{-10}}$ & $90.0^{+8.5}_{-7.9}$ & $\mathbf{1.8\times10^{-15}}$ & $\mathbf{99.9^{+0.1}_{-0.1}}$ & $1.2\times10^{-1}$ \\\cline{2-8}
    & $\text{Shared}+\mathcal{L}_{\text{MJAP}}$ & \underline{$90.0^{+8.3}_{-6.5}$} & $\mathbf{3.1\times10^{-5}}$ & $\mathbf{91.7^{+6.9}_{-5.3}}$ & $\mathbf{1.5\times10^{-15}}$ & $\mathbf{99.9^{+0.1}_{-0.1}}$ & $\mathbf{5.9\times10^{-22}}$ \\\cline{2-8}
    & $\text{DSL}+\mathcal{L}_{\text{MSC}}+\mathcal{L}_{\text{MJAP}}$ & $\mathbf{90.1^{+8.4}_{-6.9}}$ & \--- & \underline{$90.6^{+8.2}_{-8.7}$} & \--- & $\mathbf{99.9^{+0.1}_{-0.1}}$ & \--- \\\hline
    
    \end{tabular}
    
\label{tab:statistical_analysis}
\end{table*}

Assessment of the multi-task, multi-domain segmentation strategies using Att-UNet architecture as backbone demonstrated that the segmentation method based on DSL with multi-scale contrastive regularization $\mathcal{L}_{\text{MSC}}$ and multi-joint anatomical priors $\mathcal{L}_{\text{MJAP}}$ achieved the best results on all metrics, except for sensitivity ($0.7\%$ lower than the best) and RAVD ($0.4\%$ higher than the best) on the knee dataset (Table \ref{tab:leave-one-out_quantitative_assessment_of_att-unet}). For ankle examinations, the method outperformed other approaches in Dice ($+0.3\%$), MSSD ($-0.7$ mm), ASSD ($-0.1$ mm) and RAVD ($-0.8\%$), while reaching sensitivity performance ($90.7\%$) comparable to $\text{DSL}+\mathcal{L}_{\text{MSC}}$ strategy. With respect to the scores obtained for knee bone segmentation, our approach improved MSSD ($-2.3$ mm) and ASSD ($-0.2$ mm), while achieving same Dice results ($94.3\%$) as $\text{DSL}+\mathcal{L}_{\text{MSC}}$ scheme. Additionally, for the shoulder dataset, our method outperformed other approaches in Dice ($+1.6\%$), sensitivity ($+2.4\%$), MSSD ($-6.4$ mm), ASSD ($-0.6$ mm), and RAVD ($-0.2\%$). All methods achieved excellent specificity scores on all datasets ($>99.8\%$, Table \ref{tab:leave-one-out_quantitative_assessment_of_att-unet}). Moreover, the statistical analysis performed on 2D slices using Dice, sensitivity and specificity metrics  indicated that the proposed method ($\text{DSL}+\mathcal{L}_{\text{MSC}}+\mathcal{L}_{\text{MJAP}}$) produced significant improvements in segmentation performance (\textit{p}-values $<0.01$, Table \ref{tab:statistical_analysis}). The 2D results also confirmed the overall performance improvements produced by our approach on Dice ($+2.1\%$) and sensitivity ($+0.8\%$) scores.

We then evaluated the performance of the backbone architectures with an encoder pre-trained on ImageNet using individual, $\text{shared}+\mathcal{L}_{\text{MJAP}}$, and $\text{DSL}+\mathcal{L}_{\text{MSC}}+\mathcal{L}_{\text{MJAP}}$ learning schemes (Table \ref{tab:leave-one-out_quantitative_assessment_of_pre-trained_architectures}). Results obtained with Inception-UNet, Dense-UNet, and Efficient-UNet models further illustrated the benefits of the proposed learning scheme based on DSL, multi-scale contrastive regularization $\mathcal{L}_{\text{MSC}}$, and multi-joint anatomical priors $\mathcal{L}_{\text{MJAP}}$. In Inception-UNet experiments, the $\text{DSL}+\mathcal{L}_{\text{MSC}}+\mathcal{L}_{\text{MJAP}}$ scheme ranked best in all metrics and in all datasets except for knee MSSD ($0.8$ mm higher than the best), shoulder sensitivity ($0.7\%$ lower than the best), and shoulder MSSD ($0.2$ mm higher than the best). Similarly, Dense-UNet backbone with $\text{DSL}+\mathcal{L}_{\text{MSC}}+\mathcal{L}_{\text{MJAP}}$ approach ranked best in all metrics and in all datasets except for ankle sensitivity ($1.3\%$ lower than the best), knee sensitivity ($1.2\%$ lower than the best), and knee RAVD ($0.2\%$ higher than the best). For its part, the proposed Efficient-UNet with $\text{DSL}+\mathcal{L}_{\text{MSC}}+\mathcal{L}_{\text{MJAP}}$ achieved the best performance in all metrics and in all datasets except for ankle RAVD ($1.0\%$ higher than the best), shoulder sensitivity ($1.6\%$ lower than the best), and shoulder MSSD ($1.3$ mm higher than the best). Moreover, with respect to the 2D results, the proposed $\text{DSL}+\mathcal{L}_{\text{MSC}}+\mathcal{L}_{\text{MJAP}}$ scheme consistently reached the best Dice performance while $\text{Shared}+\mathcal{L}_{\text{MJAP}}$ achieved the best sensitivity within each backbone (Table \ref{tab:statistical_analysis}). The obtained $p$-values indicated that proposed $\text{DSL}+\mathcal{L}_{\text{MSC}}+\mathcal{L}_{\text{MJAP}}$ produced statistically significant different results ($p$-values $< 0.01$), except compared with the individual scheme using the Efficient backbone on the sensitivity metric. In this particular case, the difference between the 2D scores distributions was not statistically significant. However, as $\text{DSL}+\mathcal{L}_{\text{MSC}}+\mathcal{L}_{\text{MJAP}}$ produced statistically significant improvements on the remaining 2D metrics, we considered the overall improvements to be statistically significant.

Finally, when comparing the four backbone architectures (Att-UNet, Inception-UNet, Dense-UNet, and Efficient-UNet) with fixed $\text{DSL}+\mathcal{L}_{\text{MSC}}+\mathcal{L}_{\text{MJAP}}$ learning scheme (Tables \ref{tab:leave-one-out_quantitative_assessment_of_att-unet} and \ref{tab:leave-one-out_quantitative_assessment_of_pre-trained_architectures}), we observed that the proposed Efficient-UNet $\text{DSL}+\mathcal{L}_{\text{MSC}}+\mathcal{L}_{\text{MJAP}}$ reached the best performance in all metrics and in all datasets except for ankle RAVD ($0.3\%$ higher than Dense-UNet $\text{DSL}+\mathcal{L}_{\text{MSC}}+\mathcal{L}_{\text{MJAP}}$).

\subsection{Qualitative assessment}
\label{sec:qualitative_assessment}

\begin {figure*}[ht!]
\centering
\begin{adjustbox}{width=\textwidth}
\begin{tikzpicture}
\begin{scope}[spy using outlines=
      {circle, magnification=3, size=.3cm, connect spies, rounded corners}]

\node[inner sep=0pt]  at (0,0)
    {\includegraphics[width=.055\textwidth]{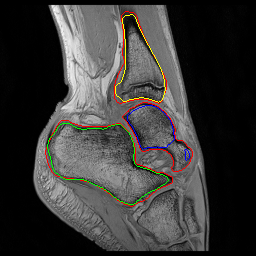}};
\node[inner sep=0pt]  at (0,-1)
    {\includegraphics[width=.055\textwidth]{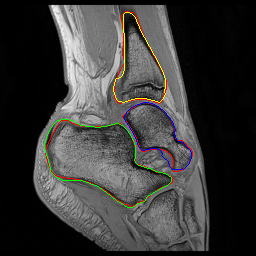}};
\node[inner sep=0pt]  at (0,-2)
    {\includegraphics[width=.055\textwidth]{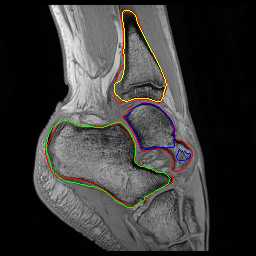}};
\node[inner sep=0pt]  at (0,-3)
    {\includegraphics[width=.055\textwidth]{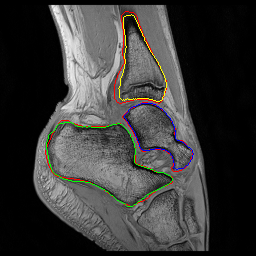}};
\node[inner sep=0pt]  at (0,-4)
    {\includegraphics[width=.055\textwidth]{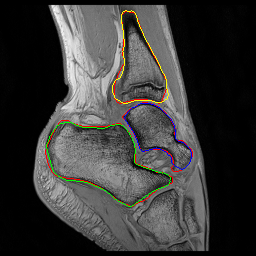}};
\node[inner sep=0pt]  at (0,-5)
    {\includegraphics[width=.055\textwidth]{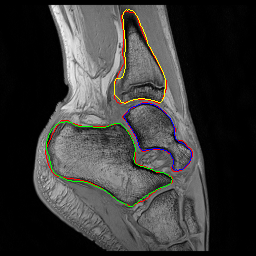}};
    
\node[inner sep=0pt]  at (1,0)
    {\includegraphics[width=.055\textwidth]{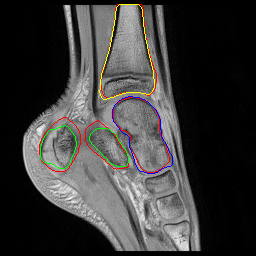}};
\node[inner sep=0pt]  at (1,-1)
    {\includegraphics[width=.055\textwidth]{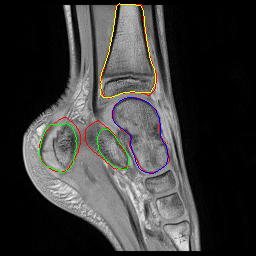}};
\node[inner sep=0pt]  at (1,-2)
    {\includegraphics[width=.055\textwidth]{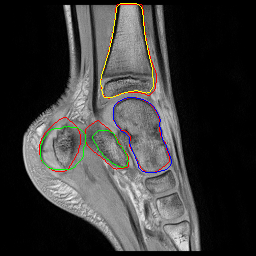}};
\node[inner sep=0pt]  at (1,-3)
    {\includegraphics[width=.055\textwidth]{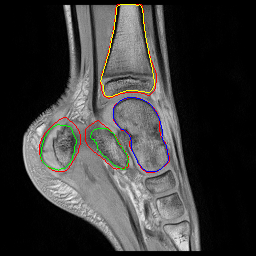}};
\node[inner sep=0pt]  at (1,-4)
    {\includegraphics[width=.055\textwidth]{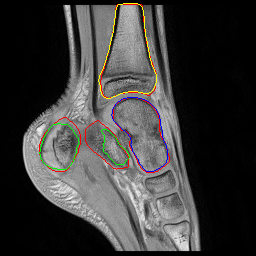}};
\node[inner sep=0pt]  at (1,-5)
    {\includegraphics[width=.055\textwidth]{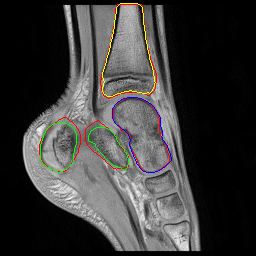}};
    
\node[inner sep=0pt]  at (2.1,0)
    {\includegraphics[width=.055\textwidth]{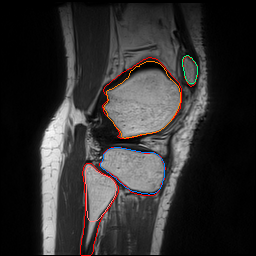}};
\node[inner sep=0pt]  at (2.1,-1)
    {\includegraphics[width=.055\textwidth]{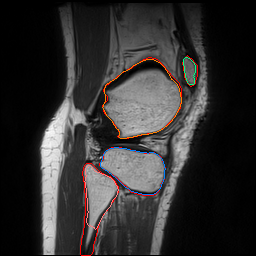}};
\node[inner sep=0pt]  at (2.1,-2)
    {\includegraphics[width=.055\textwidth]{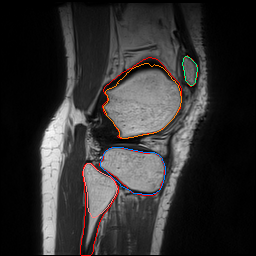}};
\node[inner sep=0pt]  at (2.1,-3)
    {\includegraphics[width=.055\textwidth]{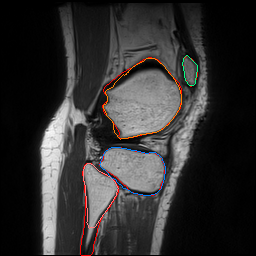}};
\node[inner sep=0pt]  at (2.1,-4)
    {\includegraphics[width=.055\textwidth]{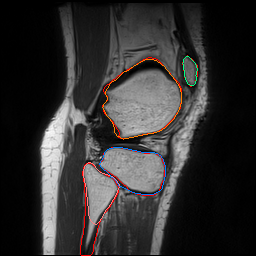}};
\node[inner sep=0pt]  at (2.1,-5)
    {\includegraphics[width=.055\textwidth]{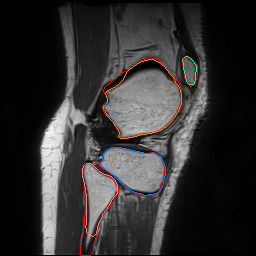}};
    
\node[inner sep=0pt]  at (3.1,0)
    {\includegraphics[width=.055\textwidth]{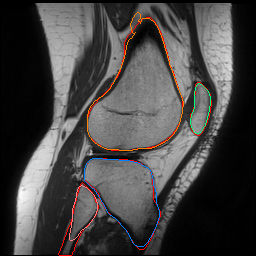}};
\node[inner sep=0pt]  at (3.1,-1)
    {\includegraphics[width=.055\textwidth]{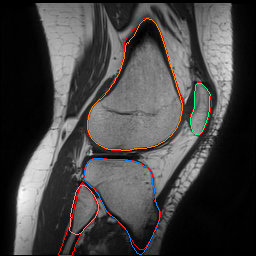}};
\node[inner sep=0pt]  at (3.1,-2)
    {\includegraphics[width=.055\textwidth]{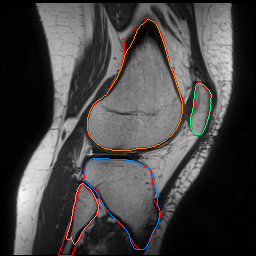}};
\node[inner sep=0pt]  at (3.1,-3)
    {\includegraphics[width=.055\textwidth]{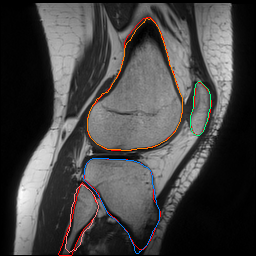}};
\node[inner sep=0pt]  at (3.1,-4)
    {\includegraphics[width=.055\textwidth]{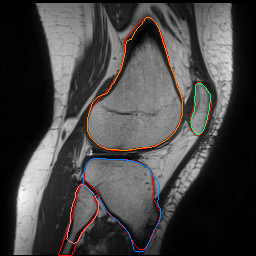}};
\node[inner sep=0pt]  at (3.1,-5)
    {\includegraphics[width=.055\textwidth]{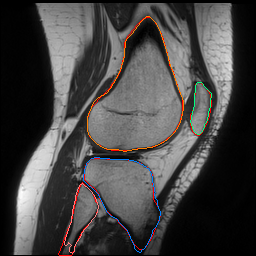}};
    
\node[inner sep=0pt]  at (4.2,0)
    {\includegraphics[width=.055\textwidth]{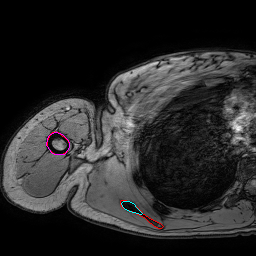}};
\node[inner sep=0pt]  at (4.2,-1)
    {\includegraphics[width=.055\textwidth]{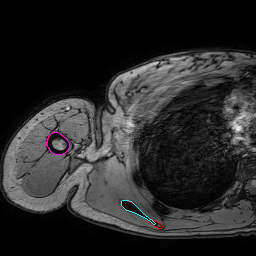}};
\node[inner sep=0pt]  at (4.2,-2)
    {\includegraphics[width=.055\textwidth]{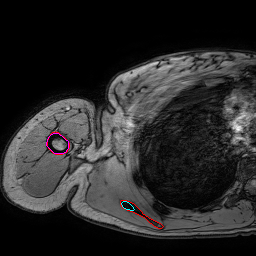}};
\node[inner sep=0pt]  at (4.2,-3)
    {\includegraphics[width=.055\textwidth]{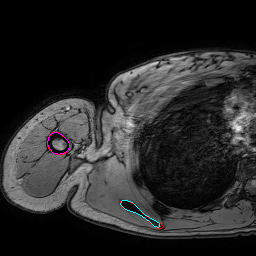}};
\node[inner sep=0pt]  at (4.2,-4)
    {\includegraphics[width=.055\textwidth]{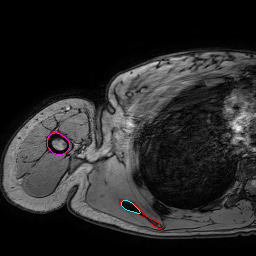}};
\node[inner sep=0pt]  at (4.2,-5)
    {\includegraphics[width=.055\textwidth]{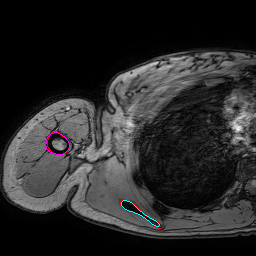}};
    
\node[inner sep=0pt]  at (5.2,0)
    {\includegraphics[width=.055\textwidth]{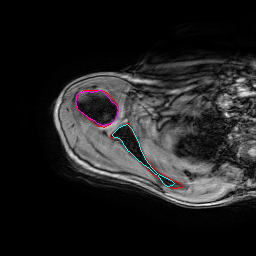}};
\node[inner sep=0pt]  at (5.2,-1)
    {\includegraphics[width=.055\textwidth]{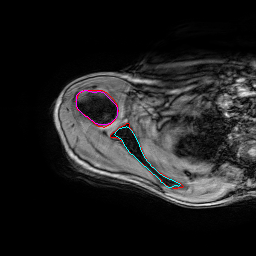}};
\node[inner sep=0pt]  at (5.2,-2)
    {\includegraphics[width=.055\textwidth]{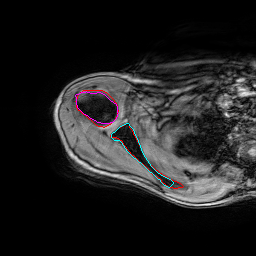}};
\node[inner sep=0pt]  at (5.2,-3)
    {\includegraphics[width=.055\textwidth]{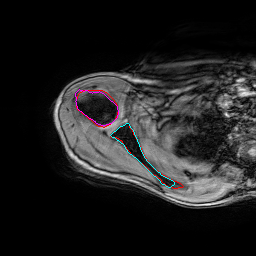}};
\node[inner sep=0pt]  at (5.2,-4)
    {\includegraphics[width=.055\textwidth]{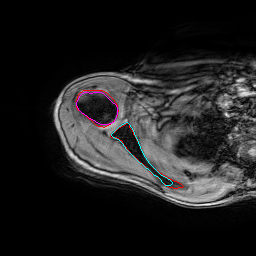}};
\node[inner sep=0pt]  at (5.2,-5)
    {\includegraphics[width=.055\textwidth]{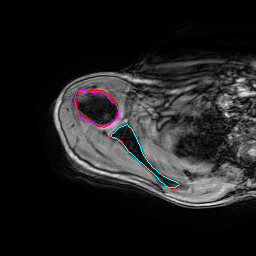}};
  
\spy [Dandelion] on (-.03,.12) in node [left] at (-.15,.3);
\spy [Dandelion] on (.17,-.18) in node [left] at (.45,.3);

\spy [Dandelion] on (-.03,-.88) in node [left] at (-.15,-.7);
\spy [Dandelion] on (.17,-1.18) in node [left] at (.45,-.7);

\spy [Dandelion] on (-.03,-1.88) in node [left] at (-.15,-1.7);
\spy [Dandelion] on (.17,-2.18) in node [left] at (.45,-1.7);

\spy [Dandelion] on (-.03,-2.88) in node [left] at (-.15,-2.7);
\spy [Dandelion] on (.17,-3.18) in node [left] at (.45,-2.7);

\spy [Dandelion] on (-.03,-3.88) in node [left] at (-.15,-3.7);
\spy [Dandelion] on (.17,-4.18) in node [left] at (.45,-3.7);

\spy [Dandelion] on (-.03,-4.88) in node [left] at (-.15,-4.7);
\spy [Dandelion] on (.17,-5.18) in node [left] at (.45,-4.7);

\spy [Dandelion] on (.76,.01) in node [left] at (.85,.3);
\spy [Dandelion] on (1.15,-.12) in node [left] at (1.45,.3);

\spy [Dandelion] on (.76,-.99) in node [left] at (.85,-.7);
\spy [Dandelion] on (1.15,-1.12) in node [left] at (1.45,-.7);

\spy [Dandelion] on (.76,-1.99) in node [left] at (.85,-1.7);
\spy [Dandelion] on (1.15,-2.12) in node [left] at (1.45,-1.7);

\spy [Dandelion] on (.76,-2.99) in node [left] at (.85,-2.7);
\spy [Dandelion] on (1.15,-3.12) in node [left] at (1.45,-2.7);

\spy [Dandelion] on (.76,-3.99) in node [left] at (.85,-3.7);
\spy [Dandelion] on (1.15,-4.12) in node [left] at (1.45,-3.7);

\spy [Dandelion] on (.76,-4.99) in node [left] at (.85,-4.7);
\spy [Dandelion] on (1.15,-5.12) in node [left] at (1.45,-4.7);

\spy [Dandelion] on (1.95,-.16) in node [left] at (1.95,.3);
\spy [Dandelion] on (2.18,.25) in node [left] at (2.55,-.3);

\spy [Dandelion] on (1.95,-1.16) in node [left] at (1.95,-.7);
\spy [Dandelion] on (2.18,-.75) in node [left] at (2.55,-1.3);

\spy [Dandelion] on (1.95,-2.16) in node [left] at (1.95,-1.7);
\spy [Dandelion] on (2.18,-1.75) in node [left] at (2.55,-2.3);

\spy [Dandelion] on (1.95,-3.16) in node [left] at (1.95,-2.7);
\spy [Dandelion] on (2.18,-2.75) in node [left] at (2.55,-3.3);

\spy [Dandelion] on (1.95,-4.16) in node [left] at (1.95,-3.7);
\spy [Dandelion] on (2.18,-3.75) in node [left] at (2.55,-4.3);

\spy [Dandelion] on (1.95,-5.16) in node [left] at (1.95,-4.7);
\spy [Dandelion] on (2.18,-4.75) in node [left] at (2.55,-5.3);

\spy [Dandelion] on (3.14,.41) in node [left] at (2.95,.3);
\spy [Dandelion] on (2.88,-.43) in node [left] at (3.55,-.3);

\spy [Dandelion] on (3.14,-.59) in node [left] at (2.95,-.7);
\spy [Dandelion] on (2.88,-1.43) in node [left] at (3.55,-1.3);

\spy [Dandelion] on (3.14,-1.59) in node [left] at (2.95,-1.7);
\spy [Dandelion] on (2.88,-2.43) in node [left] at (3.55,-2.3);

\spy [Dandelion] on (3.14,-2.59) in node [left] at (2.95,-2.7);
\spy [Dandelion] on (2.88,-3.43) in node [left] at (3.55,-3.3);

\spy [Dandelion] on (3.14,-3.59) in node [left] at (2.95,-3.7);
\spy [Dandelion] on (2.88,-4.43) in node [left] at (3.55,-4.3);

\spy [Dandelion] on (3.14,-4.59) in node [left] at (2.95,-4.7);
\spy [Dandelion] on (2.88,-5.43) in node [left] at (3.55,-5.3);

\spy [Dandelion] on (3.93,-.03) in node [left] at (4.05,.3);
\spy [Dandelion] on (4.31,-.37) in node [left] at (4.65,.3);

\spy [Dandelion] on (3.93,-1.03) in node [left] at (4.05,-.7);
\spy [Dandelion] on (4.31,-1.37) in node [left] at (4.65,-.7);

\spy [Dandelion] on (3.93,-2.03) in node [left] at (4.05,-1.7);
\spy [Dandelion] on (4.31,-2.37) in node [left] at (4.65,-1.7);

\spy [Dandelion] on (3.93,-3.03) in node [left] at (4.05,-2.7);
\spy [Dandelion] on (4.31,-3.37) in node [left] at (4.65,-2.7);

\spy [Dandelion] on (3.93,-4.03) in node [left] at (4.05,-3.7);
\spy [Dandelion] on (4.31,-4.37) in node [left] at (4.65,-3.7);

\spy [Dandelion] on (3.93,-5.03) in node [left] at (4.05,-4.7);
\spy [Dandelion] on (4.31,-5.37) in node [left] at (4.65,-4.7);

\spy [Dandelion] on (5.1,.02) in node [left] at (5.05,-.3);
\spy [Dandelion] on (5.39,-.22) in node [left] at (5.65,.3);

\spy [Dandelion] on (5.1,-.98) in node [left] at (5.05,-1.3);
\spy [Dandelion] on (5.39,-1.22) in node [left] at (5.65,-.7);

\spy [Dandelion] on (5.1,-1.98) in node [left] at (5.05,-2.3);
\spy [Dandelion] on (5.39,-2.22) in node [left] at (5.65,-1.7);

\spy [Dandelion] on (5.1,-2.98) in node [left] at (5.05,-3.3);
\spy [Dandelion] on (5.39,-3.22) in node [left] at (5.65,-2.7);

\spy [Dandelion] on (5.1,-3.98) in node [left] at (5.05,-4.3);
\spy [Dandelion] on (5.39,-4.22) in node [left] at (5.65,-3.7);

\spy [Dandelion] on (5.1,-4.98) in node [left] at (5.05,-5.3);
\spy [Dandelion] on (5.39,-5.22) in node [left] at (5.65,-4.7);

\end{scope}

\node[inner sep=0pt] at (.5,.575) {\scalebox{.325}{Ankle}};
\node[inner sep=0pt] at (2.6,.575) {\scalebox{.325}{Knee}};
\node[inner sep=0pt] at (4.7,.575) {\scalebox{.325}{Shoulder}};

\node[inner sep=0pt] at (0,-5.575) {\scalebox{.325}{$\mathrm{A_{14}}$}};
\node[inner sep=0pt] at (1,-5.575) {\scalebox{.325}{$\mathrm{A_{12}}$}};
\node[inner sep=0pt] at (2.1,-5.575) {\scalebox{.325}{$\mathrm{K_{3}}$}};
\node[inner sep=0pt] at (3.1,-5.575) {\scalebox{.325}{$\mathrm{K_{5}}$}};
\node[inner sep=0pt] at (4.2,-5.575) {\scalebox{.325}{$\mathrm{S_{11}}$}};
\node[inner sep=0pt] at (5.2,-5.575) {\scalebox{.325}{$\mathrm{S_{12}}$}};

\node[inner sep=0pt,rotate=90] at (-.575, 0) {\scalebox{.325}{Att-UNet Shared}};
\node[inner sep=0pt,rotate=90] at (-.7, -1) {\scalebox{.325}{Att-UNet}};
\node[inner sep=0pt,rotate=90] at (-.575, -1) {\scalebox{.325}{$\text{Shared}+\mathcal{L}_{\text{MJAP}}$}};
\node[inner sep=0pt,rotate=90] at (-.575, -2) {\scalebox{.325}{Att-UNet DSL}};
\node[inner sep=0pt,rotate=90] at (-.7, -3) {\scalebox{.325}{Att-UNet}};
\node[inner sep=0pt,rotate=90] at (-.575, -3) {\scalebox{.325}{$\text{DSL}+\mathcal{L}_{\text{MJAP}}$}};
\node[inner sep=0pt,rotate=90] at (-.7, -4) {\scalebox{.325}{Att-UNet}};
\node[inner sep=0pt,rotate=90] at (-.575, -4) {\scalebox{.325}{$\text{DSL}+\mathcal{L}_{\text{MSC}}$}};
\node[inner sep=0pt,rotate=90] at (-.7, -5) {\scalebox{.325}{Att-UNet}};
\node[inner sep=0pt,rotate=90] at (-.575, -5) {\scalebox{.325}{$\text{DSL}+\mathcal{L}_{\text{MSC}}+\mathcal{L}_{\text{MJAP}}$}};

\end{tikzpicture}
\end{adjustbox}
\caption{\textbf{Visual comparison of the multi-scale contrastive regularization $\mathcal{L}_{\text{MSC}}$ and multi-joint anatomical} priors $\mathcal{L}_{\text{MJAP}}$ using Att-UNet architecture. Automatic segmentation of ankle, knee, and shoulder bones based on Att-UNet \citep{oktay_attention_2018} employed in shared and DSL strategies. Ground truth delineations are in red (\textcolor{red}{\raisebox{1.8pt}{\rule{5pt}{1pt}}}) while predicted bones appear in green (\textcolor{green}{\raisebox{1.8pt}{\rule{5pt}{1pt}}}) for calcaneus, blue (\textcolor{blue}{\raisebox{1.8pt}{\rule{5pt}{1pt}}}) for talus, yellow (\textcolor{yellow}{\raisebox{1.8pt}{\rule{5pt}{1pt}}}) for tibia (distal), orange (\textcolor{orange}{\raisebox{1.8pt}{\rule{5pt}{1pt}}}) for femur (distal), pink (\textcolor{pink}{\raisebox{1.8pt}{\rule{5pt}{1pt}}}) for fibula (proximal), light green (\textcolor{LimeGreen}{\raisebox{1.8pt}{\rule{5pt}{1pt}}}) for patella, light blue (\textcolor{CornflowerBlue}{\raisebox{1.8pt}{\rule{5pt}{1pt}}}) for tibia (proximal), magenta (\textcolor{magenta}{\raisebox{1.8pt}{\rule{5pt}{1pt}}}) for humerus, and cyan (\textcolor{cyan}{\raisebox{1.8pt}{\rule{5pt}{1pt}}}) for scapula.}
\label{fig:comparison_att_unet}
\end{figure*}

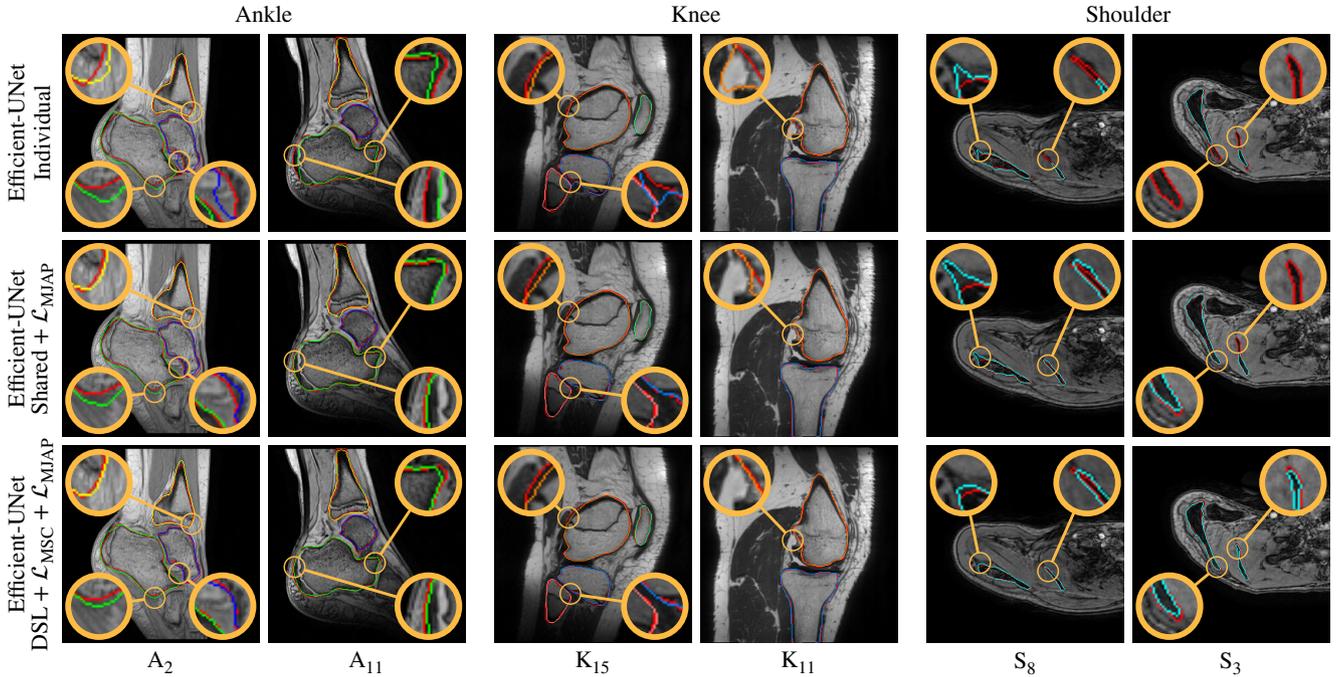
\begin {figure*}[ht!]
\centering
\begin{adjustbox}{width=\textwidth}
\begin{tikzpicture}
\begin{scope}[spy using outlines=
      {circle, magnification=3, size=.3cm, connect spies, rounded corners}]

\node[inner sep=0pt]  at (0,0)
    {\includegraphics[width=.055\textwidth]{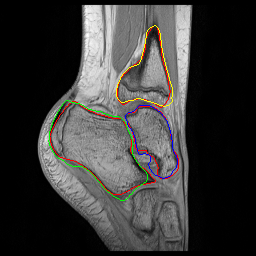}};
\node[inner sep=0pt]  at (0,-1)
    {\includegraphics[width=.055\textwidth]{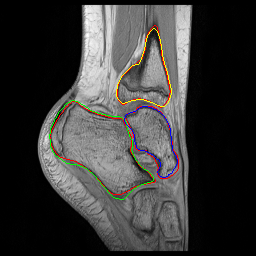}};
\node[inner sep=0pt]  at (0,-2)
    {\includegraphics[width=.055\textwidth]{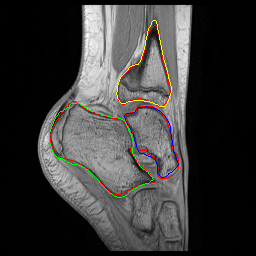}};
    
\node[inner sep=0pt]  at (1,0)
    {\includegraphics[width=.055\textwidth]{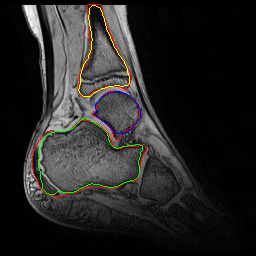}};
\node[inner sep=0pt]  at (1,-1)
    {\includegraphics[width=.055\textwidth]{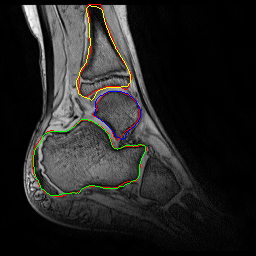}};
\node[inner sep=0pt]  at (1,-2)
    {\includegraphics[width=.055\textwidth]{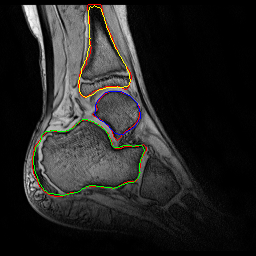}};
    
\node[inner sep=0pt]  at (2.1,0)
    {\includegraphics[width=.055\textwidth]{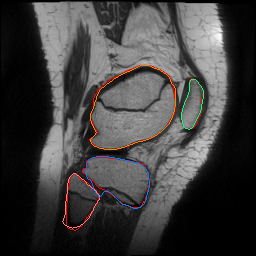}};
\node[inner sep=0pt]  at (2.1,-1)
    {\includegraphics[width=.055\textwidth]{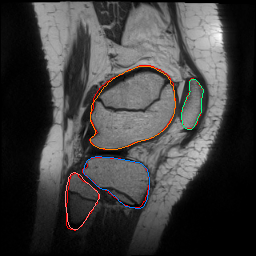}};
\node[inner sep=0pt]  at (2.1,-2)
    {\includegraphics[width=.055\textwidth]{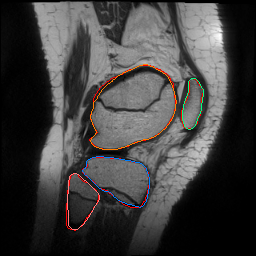}};
    
\node[inner sep=0pt]  at (3.1,0)
    {\includegraphics[width=.055\textwidth]{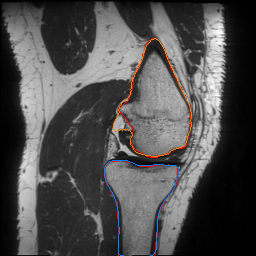}};
\node[inner sep=0pt]  at (3.1,-1)
    {\includegraphics[width=.055\textwidth]{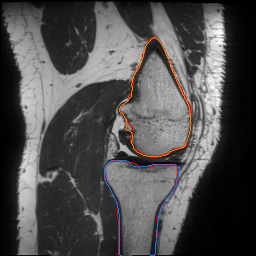}};
\node[inner sep=0pt]  at (3.1,-2)
    {\includegraphics[width=.055\textwidth]{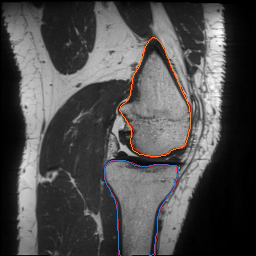}};
    
\node[inner sep=0pt]  at (4.2,0)
    {\includegraphics[width=.055\textwidth]{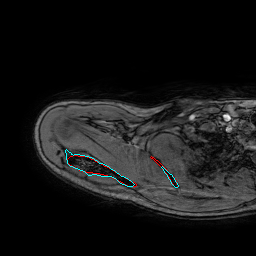}};
\node[inner sep=0pt]  at (4.2,-1)
    {\includegraphics[width=.055\textwidth]{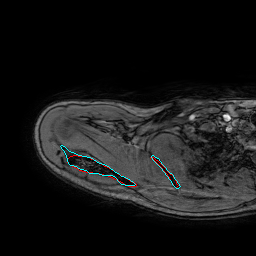}};
\node[inner sep=0pt]  at (4.2,-2)
    {\includegraphics[width=.055\textwidth]{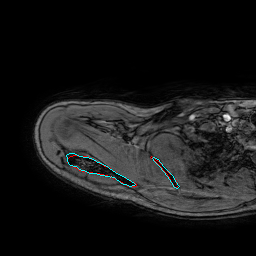}};
    
\node[inner sep=0pt]  at (5.2,0) {\includegraphics[width=.055\textwidth]{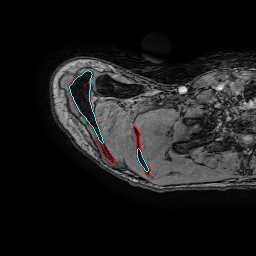}};
\node[inner sep=0pt]  at (5.2,-1)
    {\includegraphics[width=.055\textwidth]{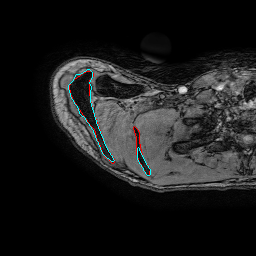}};
\node[inner sep=0pt]  at (5.2,-2)
    {\includegraphics[width=.055\textwidth]{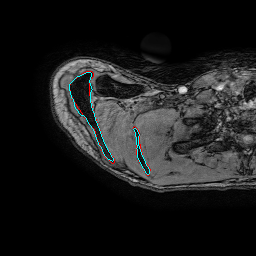}};
  
\spy [Dandelion] on (.15,.10) in node [left] at (-.15,.3);
\spy [Dandelion] on (-.03,-.26) in node [left] at (-.15,-.3);
\spy [Dandelion] on (.09,-.14) in node [left] at (.45,-.3);

\spy [Dandelion] on (.15,-.9) in node [left] at (-.15,-.7);
\spy [Dandelion] on (-.03,-1.26) in node [left] at (-.15,-1.3);
\spy [Dandelion] on (.09,-1.14) in node [left] at (.45,-1.3);

\spy [Dandelion] on (.15,-1.90) in node [left] at (-.15,-1.7);
\spy [Dandelion] on (-.03,-2.26) in node [left] at (-.15,-2.3);
\spy [Dandelion] on (.09,-2.14) in node [left] at (.45,-2.3);

\spy [Dandelion] on (1.04,-.09) in node [left] at (1.45,.3);
\spy [Dandelion] on (.65,-.11) in node [left] at (1.45,-.3);

\spy [Dandelion] on (1.04,-1.09) in node [left] at (1.45,-.7);
\spy [Dandelion] on (.65,-1.11) in node [left] at (1.45,-1.3);

\spy [Dandelion] on (1.04,-2.09) in node [left] at (1.45,-1.7);
\spy [Dandelion] on (.65,-2.11) in node [left] at (1.45,-2.3);

\spy [Dandelion] on (1.98,.13) in node [left] at (1.95,.3);
\spy [Dandelion] on (1.99,-.24) in node [left] at (2.55,-.3);

\spy [Dandelion] on (1.98,-.87) in node [left] at (1.95,-.7);
\spy [Dandelion] on (1.99,-1.24) in node [left] at (2.55,-1.3);

\spy [Dandelion] on (1.98,-1.87) in node [left] at (1.95,-1.7);
\spy [Dandelion] on (1.99,-2.24) in node [left] at (2.55,-2.3);

\spy [Dandelion] on (3.07,0.02) in node [left] at (2.95,.3);

\spy [Dandelion] on (3.07,-.98) in node [left] at (2.95,-.7);

\spy [Dandelion] on (3.07,-1.98) in node [left] at (2.95,-1.7);

\spy [Dandelion] on (3.98,-.09) in node [left] at (4.05,.3);
\spy [Dandelion] on (4.31,-.13) in node [left] at (4.65,.3);

\spy [Dandelion] on (3.98,-1.09) in node [left] at (4.05,-.7);
\spy [Dandelion] on (4.31,-1.13) in node [left] at (4.65,-.7);

\spy [Dandelion] on (3.98,-2.09) in node [left] at (4.05,-1.7);
\spy [Dandelion] on (4.31,-2.13) in node [left] at (4.65,-1.7);

\spy [Dandelion] on (5.13,-.11) in node [left] at (5.05,-.3);
\spy [Dandelion] on (5.23,-.02) in node [left] at (5.65,.3);

\spy [Dandelion] on (5.13,-1.11) in node [left] at (5.05,-1.3);
\spy [Dandelion] on (5.23,-1.02) in node [left] at (5.65,-.7);

\spy [Dandelion] on (5.13,-2.11) in node [left] at (5.05,-2.3);
\spy [Dandelion] on (5.23,-2.02) in node [left] at (5.65,-1.7);

\end{scope}

\node[inner sep=0pt] at (.5,.575) {\scalebox{.325}{Ankle}};
\node[inner sep=0pt] at (2.6,.575) {\scalebox{.325}{Knee}};
\node[inner sep=0pt] at (4.7,.575) {\scalebox{.325}{Shoulder}};

\node[inner sep=0pt] at (0,-2.575) {\scalebox{.325}{$\mathrm{A_{2}}$}};
\node[inner sep=0pt] at (1,-2.575) {\scalebox{.325}{$\mathrm{A_{11}}$}};
\node[inner sep=0pt] at (2.1,-2.575) {\scalebox{.325}{$\mathrm{K_{15}}$}};
\node[inner sep=0pt] at (3.1,-2.575) {\scalebox{.325}{$\mathrm{K_{11}}$}};
\node[inner sep=0pt] at (4.2,-2.575) {\scalebox{.325}{$\mathrm{S_{8}}$}};
\node[inner sep=0pt] at (5.2,-2.575) {\scalebox{.325}{$\mathrm{S_{3}}$}};

\node[inner sep=0pt,rotate=90] at (-.7, 0) {\scalebox{.325}{Efficient-UNet}};
\node[inner sep=0pt,rotate=90] at (-.575, 0) {\scalebox{.325}{Individual}};
\node[inner sep=0pt,rotate=90] at (-.7, -1) {\scalebox{.325}{Efficient-UNet}};
\node[inner sep=0pt,rotate=90] at (-.575, -1) {\scalebox{.325}{$\text{Shared}+\mathcal{L}_{\text{MJAP}}$}};
\node[inner sep=0pt,rotate=90] at (-.7, -2) {\scalebox{.325}{Efficient-UNet}};
\node[inner sep=0pt,rotate=90] at (-.575, -2) {\scalebox{.325}{$\text{DSL}+\mathcal{L}_{\text{MSC}}+\mathcal{L}_{\text{MJAP}}$}};

\end{tikzpicture}
\end{adjustbox}
\caption{\textbf{Visual comparison of the pre-trained Efficient-UNet models.} Automatic segmentation of ankle, knee, and shoulder bones based on Efficient-UNet \citep{tan_efficientnet_2019} employed in individual, $\text{shared}+\mathcal{L}_{\text{MJAP}}$, and $\text{DSL}+\mathcal{L}_{\text{MSC}}+\mathcal{L}_{\text{MJAP}}$ strategies. Ground truth delineations are in red (\textcolor{red}{\raisebox{1.8pt}{\rule{5pt}{1pt}}}) while predicted bones appear in green (\textcolor{green}{\raisebox{1.8pt}{\rule{5pt}{1pt}}}) for calcaneus, blue (\textcolor{blue}{\raisebox{1.8pt}{\rule{5pt}{1pt}}}) for talus, yellow (\textcolor{yellow}{\raisebox{1.8pt}{\rule{5pt}{1pt}}}) for tibia (distal), orange (\textcolor{orange}{\raisebox{1.8pt}{\rule{5pt}{1pt}}}) for femur (distal), pink (\textcolor{pink}{\raisebox{1.8pt}{\rule{5pt}{1pt}}}) for fibula (proximal), light green (\textcolor{LimeGreen}{\raisebox{1.8pt}{\rule{5pt}{1pt}}}) for patella, light blue (\textcolor{CornflowerBlue}{\raisebox{1.8pt}{\rule{5pt}{1pt}}}) for tibia (proximal), magenta (\textcolor{magenta}{\raisebox{1.8pt}{\rule{5pt}{1pt}}}) for humerus, and cyan (\textcolor{cyan}{\---}) for scapula.}
\label{fig:comparison_efficient_unet}
\end{figure*}

\begin {figure*}[ht!]
\centering
\begin{adjustbox}{width=\textwidth}
\begin{tikzpicture}

\node[inner sep=0pt]  at (0,0)
    {\includegraphics[width=.055\textwidth]{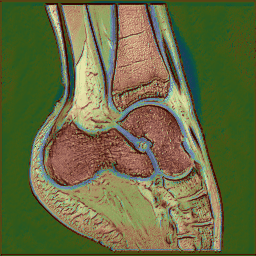}};
\node[inner sep=0pt]  at (0,-1.15)
    {\includegraphics[width=.055\textwidth]{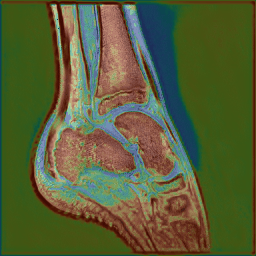}};
\node[inner sep=0pt]  at (0,-2.3)
    {\includegraphics[width=.055\textwidth]{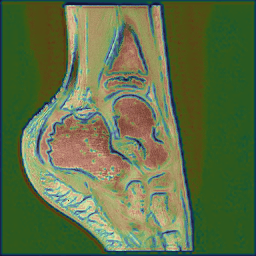}};
\node[inner sep=0pt]  at (0,-3.45)
    {\includegraphics[width=.055\textwidth]{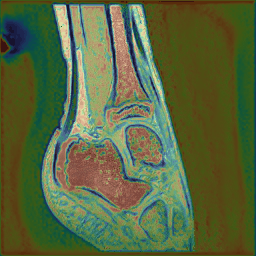}};
    
\node[inner sep=0pt]  at (1,0)
    {\includegraphics[width=.055\textwidth]{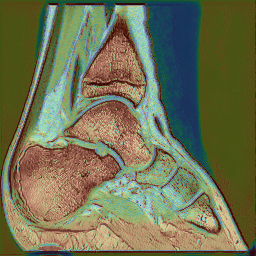}};
\node[inner sep=0pt]  at (1,-1.15)
    {\includegraphics[width=.055\textwidth]{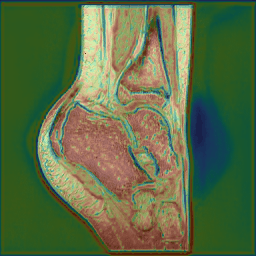}};
\node[inner sep=0pt]  at (1,-2.3)
    {\includegraphics[width=.055\textwidth]{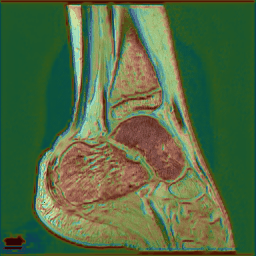}};
\node[inner sep=0pt]  at (1,-3.45)
    {\includegraphics[width=.055\textwidth]{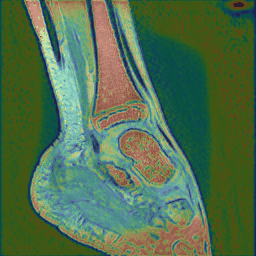}};

\node[inner sep=0pt]  at (2.1,0)
    {\includegraphics[width=.055\textwidth]{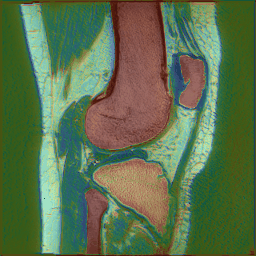}};
\node[inner sep=0pt]  at (2.1,-1.15)
    {\includegraphics[width=.055\textwidth]{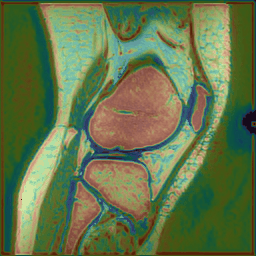}};
\node[inner sep=0pt]  at (2.1,-2.3)
    {\includegraphics[width=.055\textwidth]{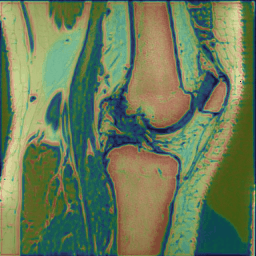}};
\node[inner sep=0pt]  at (2.1,-3.45)
    {\includegraphics[width=.055\textwidth]{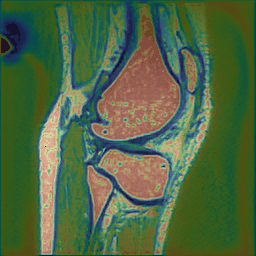}};
    
\node[inner sep=0pt]  at (3.1,0)
    {\includegraphics[width=.055\textwidth]{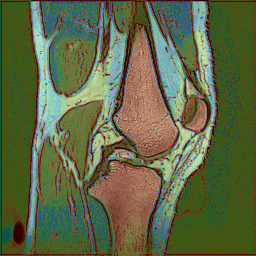}};
\node[inner sep=0pt]  at (3.1,-1.15)
    {\includegraphics[width=.055\textwidth]{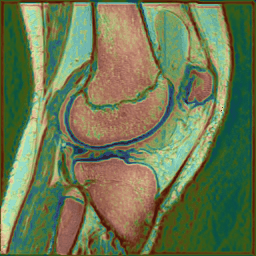}};
\node[inner sep=0pt]  at (3.1,-2.3)
    {\includegraphics[width=.055\textwidth]{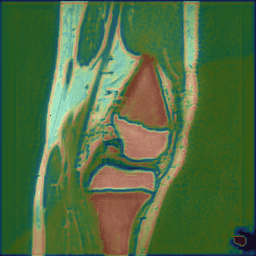}};
\node[inner sep=0pt]  at (3.1,-3.45)
    {\includegraphics[width=.055\textwidth]{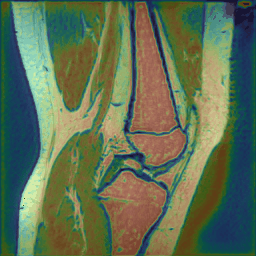}};
    
\node[inner sep=0pt]  at (4.2,0)
    {\includegraphics[width=.055\textwidth]{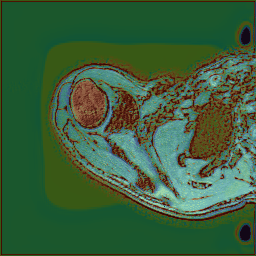}};
\node[inner sep=0pt]  at (4.2,-1.15)
    {\includegraphics[width=.055\textwidth]{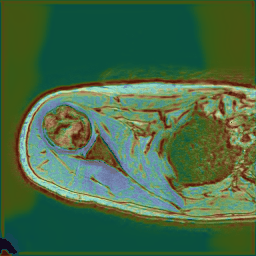}};
\node[inner sep=0pt]  at (4.2,-2.3)
    {\includegraphics[width=.055\textwidth]{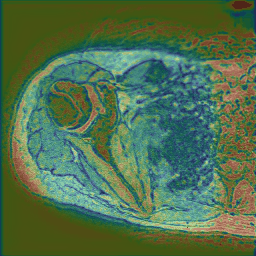}};
\node[inner sep=0pt]  at (4.2,-3.45)
    {\includegraphics[width=.055\textwidth]{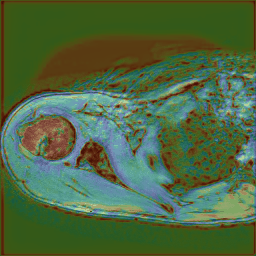}};
    
\node[inner sep=0pt]  at (5.2,0)
    {\includegraphics[width=.055\textwidth]{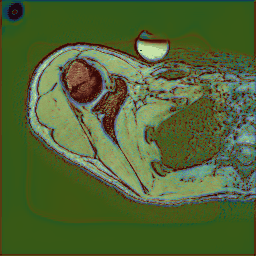}};
\node[inner sep=0pt]  at (5.2,-1.15)
    {\includegraphics[width=.055\textwidth]{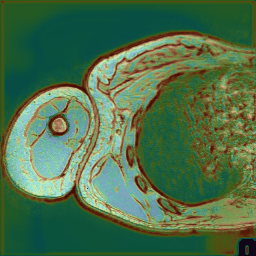}};
\node[inner sep=0pt]  at (5.2,-2.3)
    {\includegraphics[width=.055\textwidth]{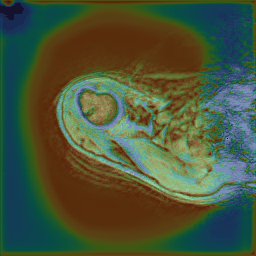}};
\node[inner sep=0pt]  at (5.2,-3.45)
    {\includegraphics[width=.055\textwidth]{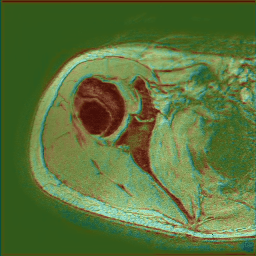}};
    
\node[inner sep=0pt, rotate=180]  at (5.85,-1.725)
    {\includegraphics[width=.0105\textwidth]{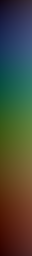}};

\node[inner sep=0pt] at (.5,.575) {\scalebox{.325}{Ankle}};
\node[inner sep=0pt] at (2.6,.575) {\scalebox{.325}{Knee}};
\node[inner sep=0pt] at (4.7,.575) {\scalebox{.325}{Shoulder}};

\node[inner sep=0pt] at (0,-.575) {\scalebox{.325}{$\mathrm{A_{14}}$}};
\node[inner sep=0pt] at (1,-.575) {\scalebox{.325}{$\mathrm{A_{10}}$}};
\node[inner sep=0pt] at (2.1,-.575) {\scalebox{.325}{$\mathrm{K_{10}}$}};
\node[inner sep=0pt] at (3.1,-.575) {\scalebox{.325}{$\mathrm{K_{6}}$}};
\node[inner sep=0pt] at (4.2,-.575) {\scalebox{.325}{$\mathrm{S_{3}}$}};
\node[inner sep=0pt] at (5.2,-.575) {\scalebox{.325}{$\mathrm{S_{6}}$}};

\node[inner sep=0pt] at (0,-1.725) {\scalebox{.325}{$\mathrm{A_{9}}$}};
\node[inner sep=0pt] at (1,-1.725) {\scalebox{.325}{$\mathrm{A_{2}}$}};
\node[inner sep=0pt] at (2.1,-1.725) {\scalebox{.325}{$\mathrm{K_{5}}$}};
\node[inner sep=0pt] at (3.1,-1.725) {\scalebox{.325}{$\mathrm{K_{4}}$}};
\node[inner sep=0pt] at (4.2,-1.725) {\scalebox{.325}{$\mathrm{S_{8}}$}};
\node[inner sep=0pt] at (5.2,-1.725) {\scalebox{.325}{$\mathrm{S_{4}}$}};

\node[inner sep=0pt] at (0,-2.875) {\scalebox{.325}{$\mathrm{A_{12}}$}};
\node[inner sep=0pt] at (1,-2.875) {\scalebox{.325}{$\mathrm{A_{1}}$}};
\node[inner sep=0pt] at (2.1,-2.875) {\scalebox{.325}{$\mathrm{K_{13}}$}};
\node[inner sep=0pt] at (3.1,-2.875) {\scalebox{.325}{$\mathrm{K_{8}}$}};
\node[inner sep=0pt] at (4.2,-2.875) {\scalebox{.325}{$\mathrm{S_{15}}$}};
\node[inner sep=0pt] at (5.2,-2.875) {\scalebox{.325}{$\mathrm{S_{12}}$}};

\node[inner sep=0pt] at (0,-4.025) {\scalebox{.325}{$\mathrm{A_{3}}$}};
\node[inner sep=0pt] at (1,-4.025) {\scalebox{.325}{$\mathrm{A_{15}}$}};
\node[inner sep=0pt] at (2.1,-4.025) {\scalebox{.325}{$\mathrm{K_{3}}$}};
\node[inner sep=0pt] at (3.1,-4.025) {\scalebox{.325}{$\mathrm{K_{9}}$}};
\node[inner sep=0pt] at (4.2,-4.025) {\scalebox{.325}{$\mathrm{S_{5}}$}};
\node[inner sep=0pt] at (5.2,-4.025) {\scalebox{.325}{$\mathrm{S_{1}}$}};

\node[inner sep=0pt,rotate=90] at (-.7, 0) {\scalebox{.325}{Att-UNet}};
\node[inner sep=0pt,rotate=90] at (-.575, 0) {\scalebox{.325}{$\text{DSL}+\mathcal{L}_{\text{MSC}}+\mathcal{L}_{\text{MJAP}}$}};
\node[inner sep=0pt,rotate=90] at (-.7, -1.15) {\scalebox{.325}{Inception-UNet}};
\node[inner sep=0pt,rotate=90] at (-.575, -1.15) {\scalebox{.325}{$\text{DSL}+\mathcal{L}_{\text{MSC}}+\mathcal{L}_{\text{MJAP}}$}};
\node[inner sep=0pt,rotate=90] at (-.7, -2.3) {\scalebox{.325}{Dense-UNet}};
\node[inner sep=0pt,rotate=90] at (-.575, -2.3) {\scalebox{.325}{$\text{DSL}+\mathcal{L}_{\text{MSC}}+\mathcal{L}_{\text{MJAP}}$}};
\node[inner sep=0pt,rotate=90] at (-.7, -3.45) {\scalebox{.325}{Efficient-UNet}};
\node[inner sep=0pt,rotate=90] at (-.575, -3.45) {\scalebox{.325}{$\text{DSL}+\mathcal{L}_{\text{MSC}}+\mathcal{L}_{\text{MJAP}}$}};

\node[inner sep=0pt] at (5.85, -.925) {\scalebox{.325}{1}};
\node[inner sep=0pt] at (5.85, -2.525) {\scalebox{.325}{0}};

\end{tikzpicture}
\end{adjustbox}
\caption{\textbf{Visualization of the attention maps computed by the multi-domain attention gates using $\text{DSL}+\mathcal{L}_{\text{MSC}}+\mathcal{L}_{\text{MJAP}}$ learning scheme.} Architectures encompassed Att-UNet \citep{oktay_attention_2018}, Inception-UNet \citep{szegedy_rethinking_2016}, Dense-UNet \citep{huang_densely_2017}, and Efficient-UNet \citep{tan_efficientnet_2019} employed on ankle, knee, and shoulder joint images. Pixel-wise coefficients ranging from 0 in blue to 1 in red indicated low to high attention.}
\label{fig:att_maps}
\end{figure*}

Visual comparison of the multi-scale contrastive regularization $\mathcal{L}_{\text{MSC}}$ and multi-joint anatomical priors $\mathcal{L}_{\text{MJAP}}$ provided visual evidence of gradual improvements in segmentation quality for both shared and DSL Att-UNet models (Fig. \ref{fig:comparison_att_unet}). Anatomical priors were clearly observed to promote globally more consistent and smoother contours for all anatomical joints by forcing the model to follow the learnt non-linear multi-joint anatomical representation. More specifically, incorporation of anatomical priors allowed the segmentation of the complete talus ($\mathrm{A_{14}}$), fibular ($\mathrm{K_{5}}$), and scapular shapes ($\mathrm{S_{11}}$ and $\mathrm{S_{12}}$), which were previously partially detected by both shared and DSL Att-UNet models. Additionally, the contrastive regularization encouraged more precise bone extraction in all domains ($\mathrm{A_{12}}$, $\mathrm{K_{3}}$, and $\mathrm{S_{11}}$) through more robust shared representations with domain-specific clusters. Meanwhile, the proposed $\text{DSL}+\mathcal{L}_{\text{MSC}}+\mathcal{L}_{\text{MJAP}}$ approach fostered the benefits of both previous terms and generated smoother and more realistic bone delineations ($\mathrm{A_{14}}$, $\mathrm{K_{5}}$ and $\mathrm{S_{11}}$).

We then visually compared the pre-trained Efficient-UNet models employed in individual, $\text{shared}+\mathcal{L}_{\text{MJAP}}$, and $\text{DSL}+\mathcal{L}_{\text{MSC}}+\mathcal{L}_{\text{MJAP}}$ learning strategies (Fig. \ref{fig:comparison_efficient_unet}). First, the qualitative comparison demonstrated that models with pre-trained encoder benefited from transfer learning to achieve robust feature extraction and produce highly accurate delineations in the three considered anatomical regions ($\mathrm{A_{2}}$, $\mathrm{K_{11}}$, and $\mathrm{S_{8}}$). However, we observed that individual models produced segmentation errors in several imaging examinations, for instance, by over-segmenting the femoral shape in knee joint ($\mathrm{K_{11}}$) or under-segmenting the scapular bone in shoulder joint ($\mathrm{S_{3}}$). Specifically, because the boundary between bone and ligament was not detected by the individual model, ligamentous tissues were erroneously classified as femur bone ($\mathrm{K_{11}}$). Furthermore, the thin structure of scapular bone led to its partial misclassification as background ($\mathrm{S_{3}}$). Additionally, the calcaneus shape was also under-segmented due to intensity difference within the bone ($\mathrm{A_{11}}$). While the $\text{shared}+\mathcal{L}_{\text{MJAP}}$ model produced segmentation improvements over its individual counterparts, it was essential to employ the $\text{DSL}+\mathcal{L}_{\text{MSC}}+\mathcal{L}_{\text{MJAP}}$ model incorporating layer specialization along with multiple regularizers to learn robust shared representations and achieve precise bone shape predictions on unseen images ($\mathrm{A_{11}}$, $\mathrm{K_{15}}$, and $\mathrm{S_{8}}$). 

Finally, we provide visualization of the attention maps computed by the multi-domain attention gates of the Att-UNet, Inception-UNet, Dense-UNet, and Efficient-UNet architectures employed in $\text{DSL}+\mathcal{L}_{\text{MSC}}+\mathcal{L}_{\text{MJAP}}$ learning scheme (Fig. \ref{fig:att_maps}). These attention maps were crucial in interpreting the inference process of deep neural networks. This visualization confirmed that the segmentation models exploited the spatial and contextual information from the encoder branch to focus on the bone of interest in each anatomical joint. Indeed, knee attention maps clearly equally highlighted each bone of interest (femur, fibula, patella, and tibia), and suppressed most of the irrelevant regions. In some cases, background elements were also included (e.g. $\mathrm{A_{9}}$ with Inception-UNet and $\mathrm{S_{12}}$ with Dense-UNet) and may help the inference process which remains difficult to interpret. We can note that attention maps computed on shoulder joint images highlighted the scapula less than the humerus bone. Meanwhile, ankle joint attention maps focused on the calcaneus, talus, and tibia bones, with some background structures also being highlighted. Finally, for each bone of interest, we observed a discontinuity in the attention coefficients at the bone borders (e.g. $\mathrm{K_{3}}$ with Efficient-UNet), that allowed the network to effectively distinguish and extract their shape from the rest of the image.

\section{Discussion}
\label{sec:discussion}

In this study, we developed and evaluated a novel multi-task, multi-domain deep segmentation framework with multi-scale contrastive regularization and multi-joint anatomical priors. To the best of our knowledge, the proposed multi-task, multi-domain segmentation method is the first illustration to optimize a single neural network over multiple pediatric musculoskeletal joints. Experiments performed on the ankle, knee, and shoulder joint imaging datasets demonstrated improved bone segmentation performance compared to individual, transfer, and shared learning schemes. The statistical analysis validated the significance of the results, while visual comparison of the predicted delineations further confirmed the enhancements in segmentation quality of the proposed framework. The proposed methodology could provide significant benefits to the management of pediatric imaging resources and have a major impact for any deep learning based medical image analysis framework.

\subsection{Segmentation performance}
\label{sec:segmentation_performance}

Regarding the performance of the multi-task, multi-domain strategies employed with Att-UNet architecture, we observed that all transfer learning schemes (Transfer$_\text{Ankle}$, Transfer$_\text{Knee}$, and Transfer$_\text{Shoulder}$) provided performance improvements compared to individual models on all datasets (Table \ref{tab:leave-one-out_quantitative_assessment_of_att-unet}), indicating better initialization than randomly set weights by exploiting features correlation and knowledge transfer between each task and domain pair. Compared to individual and transfer approaches, the results of shared and DSL schemes on both ankle and knee datasets indicated noticeable improvements while the results on shoulder examinations were less evident (Table \ref{tab:leave-one-out_quantitative_assessment_of_att-unet}). Hence, it was essential to employ both $\mathcal{L}_{\text{MJAP}}$ and $\mathcal{L}_{\text{MSC}}$ terms to benefit from the shared representation and layer specialization, and reach performance improvements over independent models on all datasets. This outcome was also supported by the results obtained on Incep\-tion-UNet, Dense-UNet, and Efficient-UNet models (Table \ref{tab:leave-one-out_quantitative_assessment_of_pre-trained_architectures}). It is also worth emphasizing that the multi-scale contrastive $\mathcal{L}_{\text{MSC}}$ regularization outperformed its single-scale $\mathcal{L}_{\text{SSC}}$ counterpart \citep{boutillon_multi-task_2021} on all datasets (Table \ref{tab:leave-one-out_quantitative_assessment_of_att-unet}), indicating that disentangling representations at each scale provided better generalization performance than focusing only on the features within the network's bottleneck. For instance, the ankle Dice score increased from $90.6\%$ to $91.5\%$, while knee and shoulder Dice metrics improved by $0.7\%$ and $0.8\%$ respectively.

Furthermore, both shared and DSL schemes offer an additional advantage compared to transfer approach by learning all task and domain pairs simultaneously rather than in a sequential manner that is prone to catastrophic forgetting. It should also be noted that in the shared segmentation scheme, predicted segmentation output that did not belong to the image task were considered as background in order to obtain a fair comparison against individual, transfer, and DSL strategies. In practice, confusion between tasks was very low, with the mean percentage of voxels per 3D examination labelled with a class foreign to the target segmentation classes (e.g. humerus identified in ankle MR images) being less than $0.001\%$ for all tasks. A low confusion between tasks was also reported by \citep{moeskops_deep_2016} in their multi-tasks segmentation framework. Finally, we observed that, within Att-UNet models, the proposed $\text{DSL}+\mathcal{L}_{\text{MSC}}+\mathcal{L}_{\text{MJAP}}$ scheme achieved important improvements in MSSD and ASSD metrics (Table \ref{tab:leave-one-out_quantitative_assessment_of_att-unet}), indicating lower surface errors. Qualitative assessment (Fig. \ref{fig:comparison_att_unet}) further confirmed this observation as compared methods were reported to partially segment the talus, fibular, and scapular shapes. In contrast, our method provided complete bone segmentation resulting in substantial surface metrics (i.e. MSSD and ASSD) improvements.

When comparing the performance of the four employed backbone architectures in the individual learning scheme (Tables \ref{tab:leave-one-out_quantitative_assessment_of_att-unet} and \ref{tab:leave-one-out_quantitative_assessment_of_pre-trained_architectures}), we observed that Inception-UNet, Dense-UNet, and Efficient-UNet outperformed Att-UNet in all metrics and in all datasets. As also highlighted in \citep{conze_healthy_2020} for shoulder muscle MR segmentation, this clearly indicated that designing a segmentation model with a pre-trained encoder resulted in better initialization through features learnt on ImageNet and better segmentation performance through a more complex and deeper CNN architecture. Indeed, compared to the complexity of the Att-UNet model, the number of trainable parameters in Inception-UNet, Dense-UNet and Efficient-UNet corresponded to an increase by a factor of five, three, and two respectively (Table \ref{tab:experiments}). However, to avoid overfitting it is also crucial to limit the number of trainable parameters, as models with too much capacity may learn the dataset and task too well. In practice, the optimal model capacity depends on the considered task and available imaging resources which are limited in the context of sparse pediatric datasets. In this sense, we observed step-wise performance improvements from Inception-UNet (48.3M parameters) to Dense-UNet (23.6M parameters) and ultimately Efficient-UNet (14.8M parameters) networks (Table \ref{tab:leave-one-out_quantitative_assessment_of_pre-trained_architectures}). For instance, shoulder Dice score increased from $84.5\%$ for Inception-UNet to $86.6\%$ for Dense-UNet and ultimately to $87.9\%$ using Effi\-cient-UNet. Finally, the proposed multi-task, multi-domain approach also allowed us to reduce the number of learnable parameters by a factor of $K=3$, and to consequently minimize overfitting and improve generalizability. Meanwhile, the supplementary parameterization introduced by the domain-specific layers was considered marginal (i.e. less than $3.0\%$).

As demonstrated through our experiments, the proposed $\text{DSL}+\mathcal{L}_{\text{MSC}}+\mathcal{L}_{\text{MJAP}}$ learning scheme is architecture-inde\-pendent, and thus can be effortlessly integrated into various existing CNN models and can improve the overall performance in all datasets. Indeed, although the obtained Dice and ASSD performance gains can be considered limited, these improvements are consistent and robust (i.e. lower standard deviation). Most importantly, experiments performed on In\-ception-UNet, Dense-UNet, and Efficient-UNet illustrated that the multi-scale contrastive regularization can be computed from the internal representations of networks composed of distinct building blocks (i.e. Inception, dense, or MBConv blocks) and diverse feature transformation operation (i.e. classical, point-wise, depth-wise, or asymmetrical convolutions and \texttt{ReLU} or \texttt{SiLU} non-linearity functions).

As indicated by the high variance in shoulder results (Tables \ref{tab:leave-one-out_quantitative_assessment_of_att-unet} and \ref{tab:leave-one-out_quantitative_assessment_of_pre-trained_architectures}), the shoulder dataset was more challenging to segment than the ankle and knee joints, due to more complex bone shapes (i.e thin scapular blade), higher variability among pediatric patients (i.e. different age groups), and the presence of examinations with a higher level of noise due to patient movements during acquisition. Interestingly, the attention maps (Fig. \ref{fig:att_maps}) could explain the lower performance for segmenting the scapular shape which appeared more challenging to detect than the humerus bone. Finally, compared to our previous experiments performed in \citep{boutillon_multi-task_2021}, we incorporated three additional ankle pediatric examinations with a higher level of noise in the test sets which led to a marginal drop in performance for ankle bone segmentation. Nevertheless, we still observed that for the ankle joint segmentation, the DSL scheme outperformed the shared approach, which in turn outranked the individual scheme.

\subsection{Assessment of learnt shared representations}
\label{sec:assessment_of_leanrt_shared_representations_discussion}

\begin{figure*}[ht!]
\centering
\begin{adjustbox}{width=\textwidth}
\begin{tikzpicture}


\node[inner sep=0pt] at (0,4)
    {\includegraphics[width=.11\textwidth]{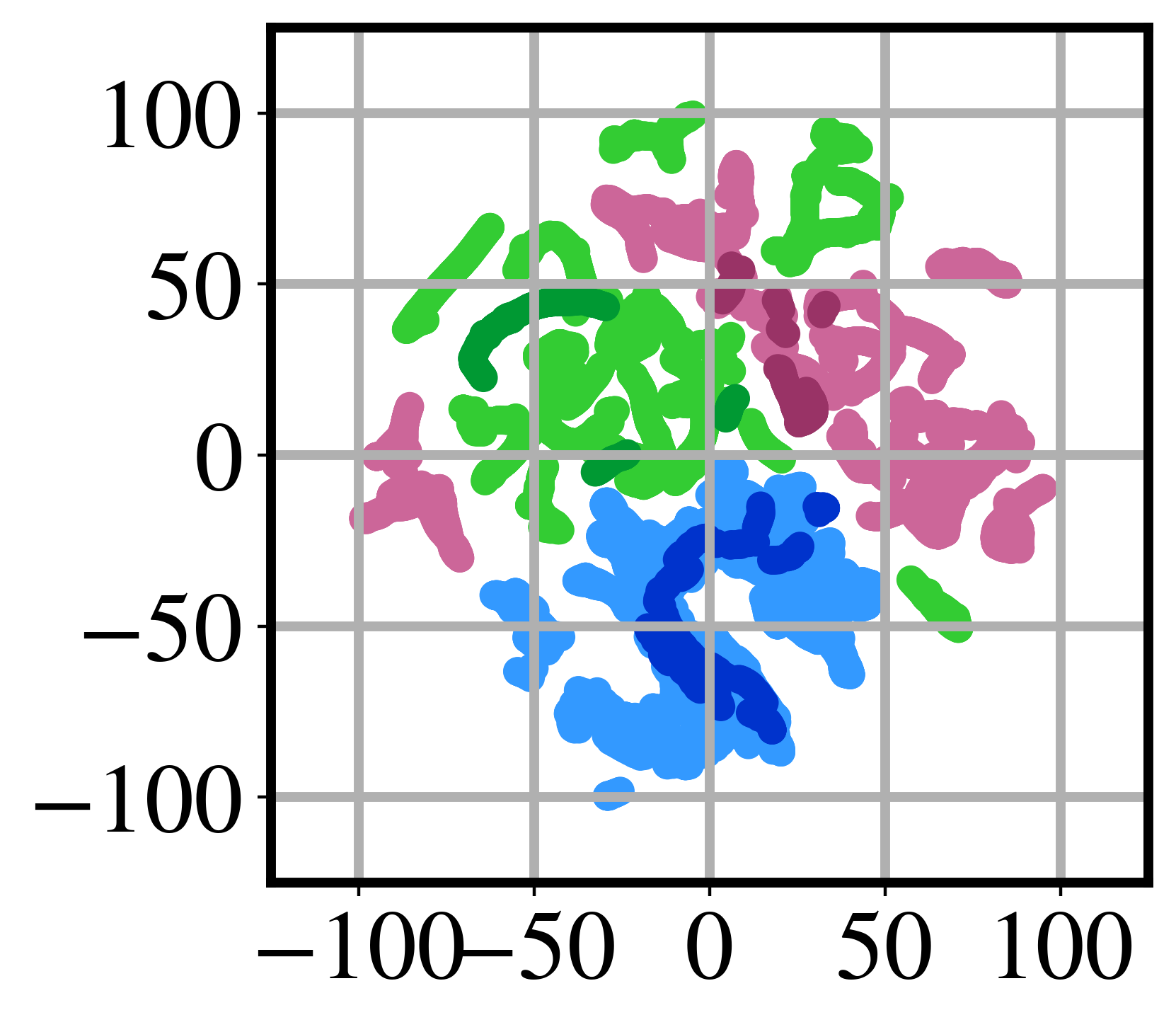}};
\node[inner sep=0pt] at (2,4)
    {\includegraphics[width=.11\textwidth]{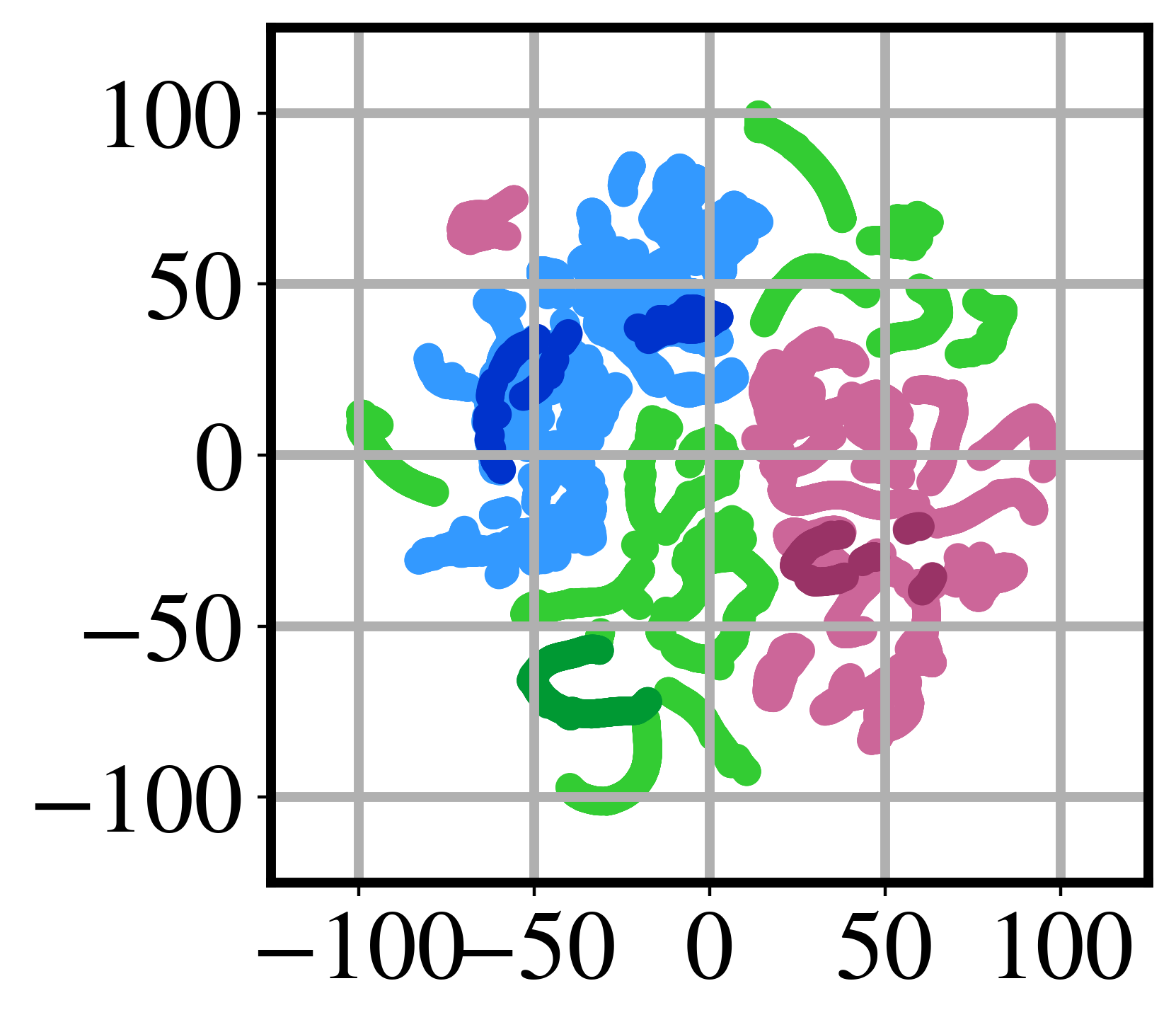}};
\node[inner sep=0pt] at (4,4)
    {\includegraphics[width=.11\textwidth]{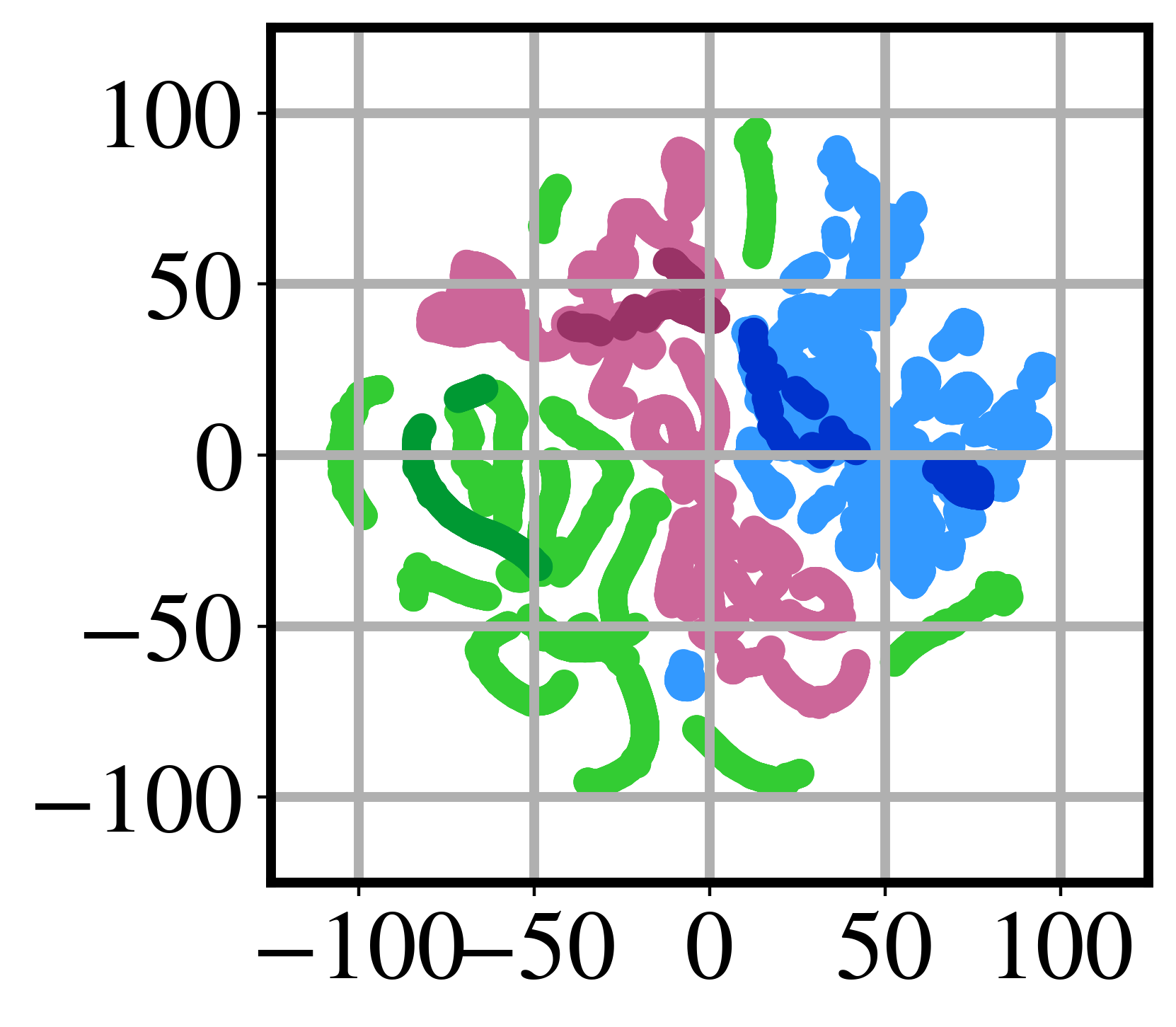}};
\node[inner sep=0pt] at (6,4)
    {\includegraphics[width=.11\textwidth]{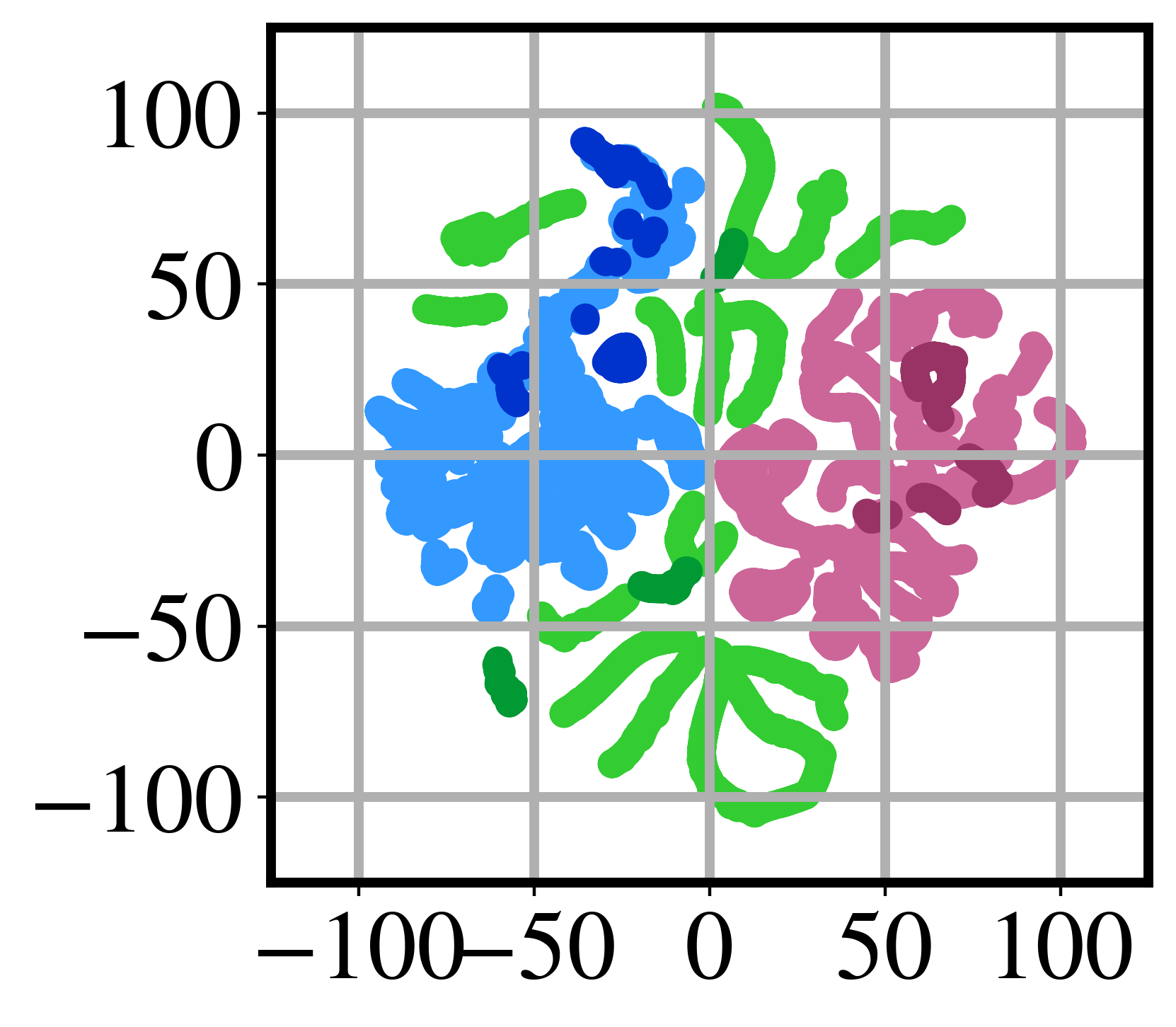}};
\node[inner sep=0pt] at (8,4)
    {\includegraphics[width=.11\textwidth]{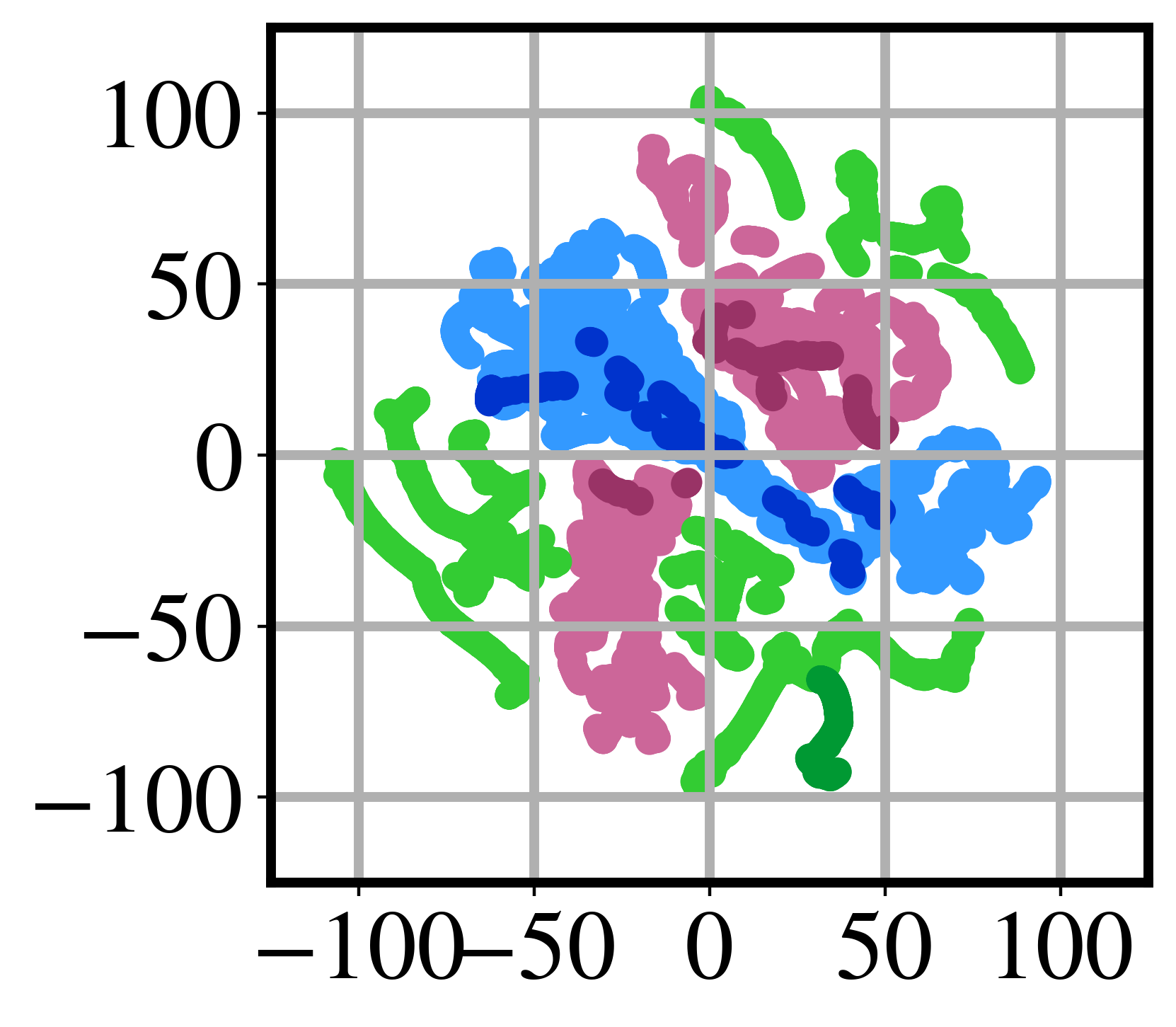}};
\node[inner sep=0pt] at (10,4)
    {\includegraphics[width=.11\textwidth]{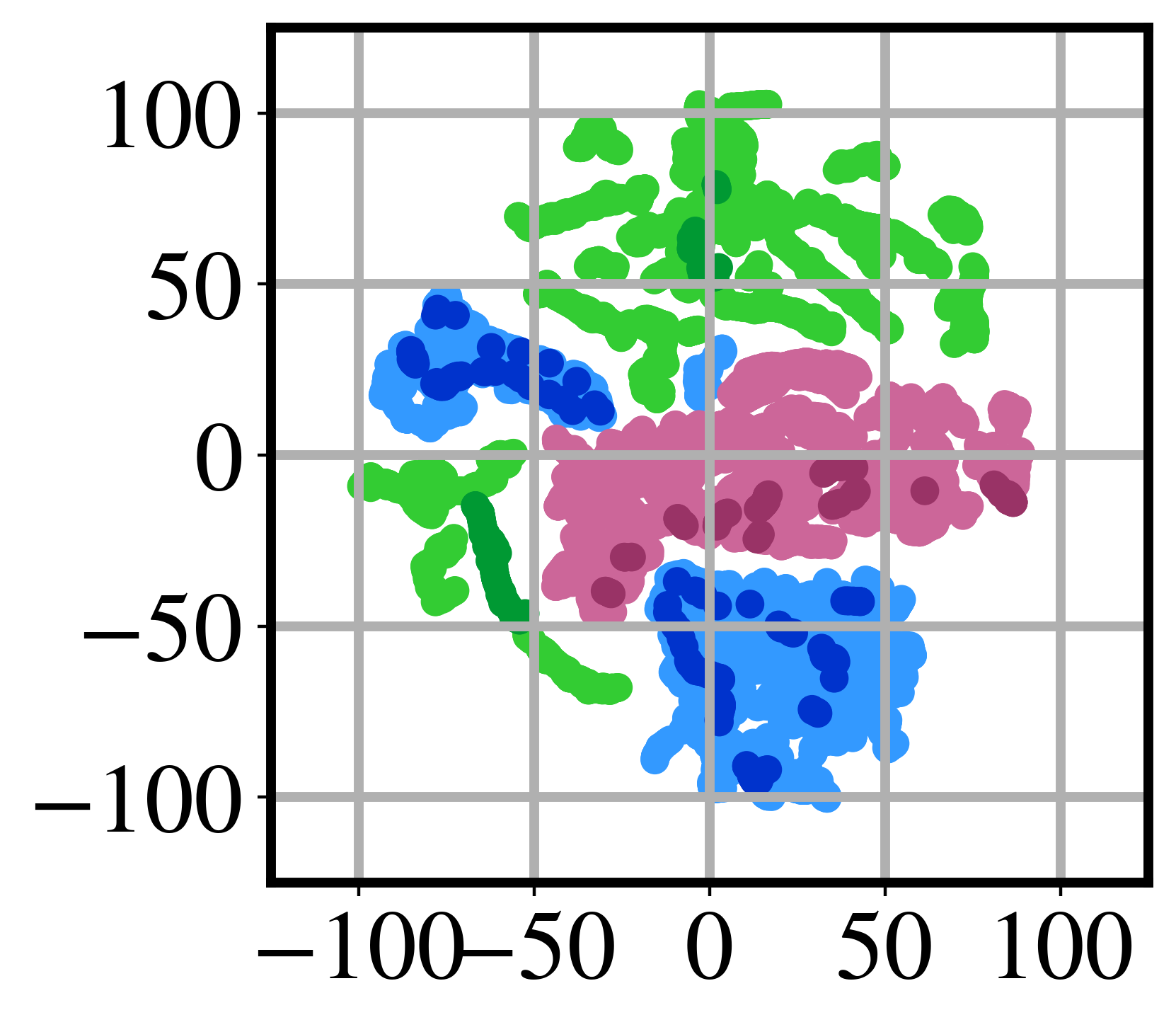}};
\node[inner sep=0pt] at (12,4)
    {\includegraphics[width=.11\textwidth]{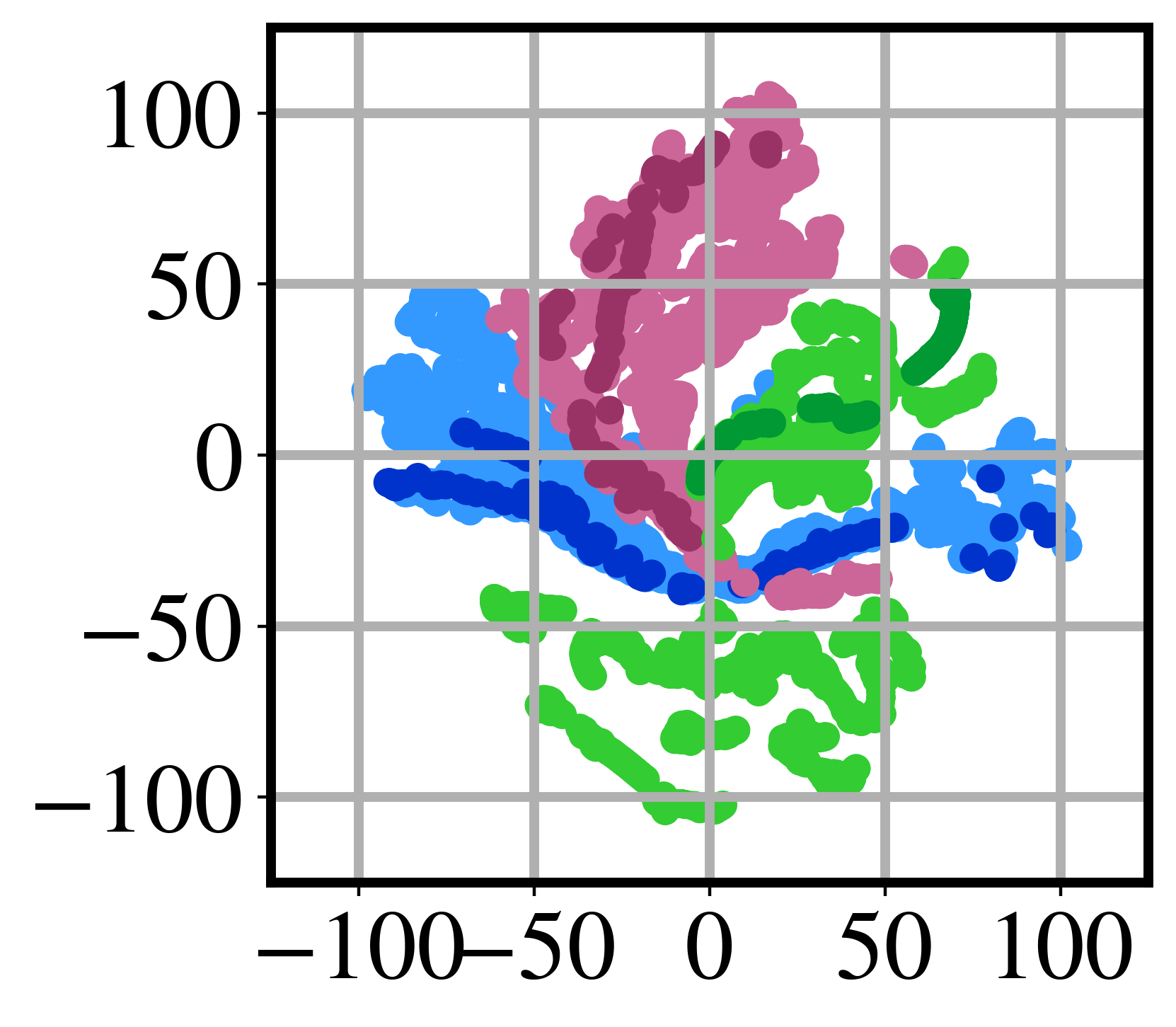}};
\node[inner sep=0pt] at (14,4)
    {\includegraphics[width=.11\textwidth]{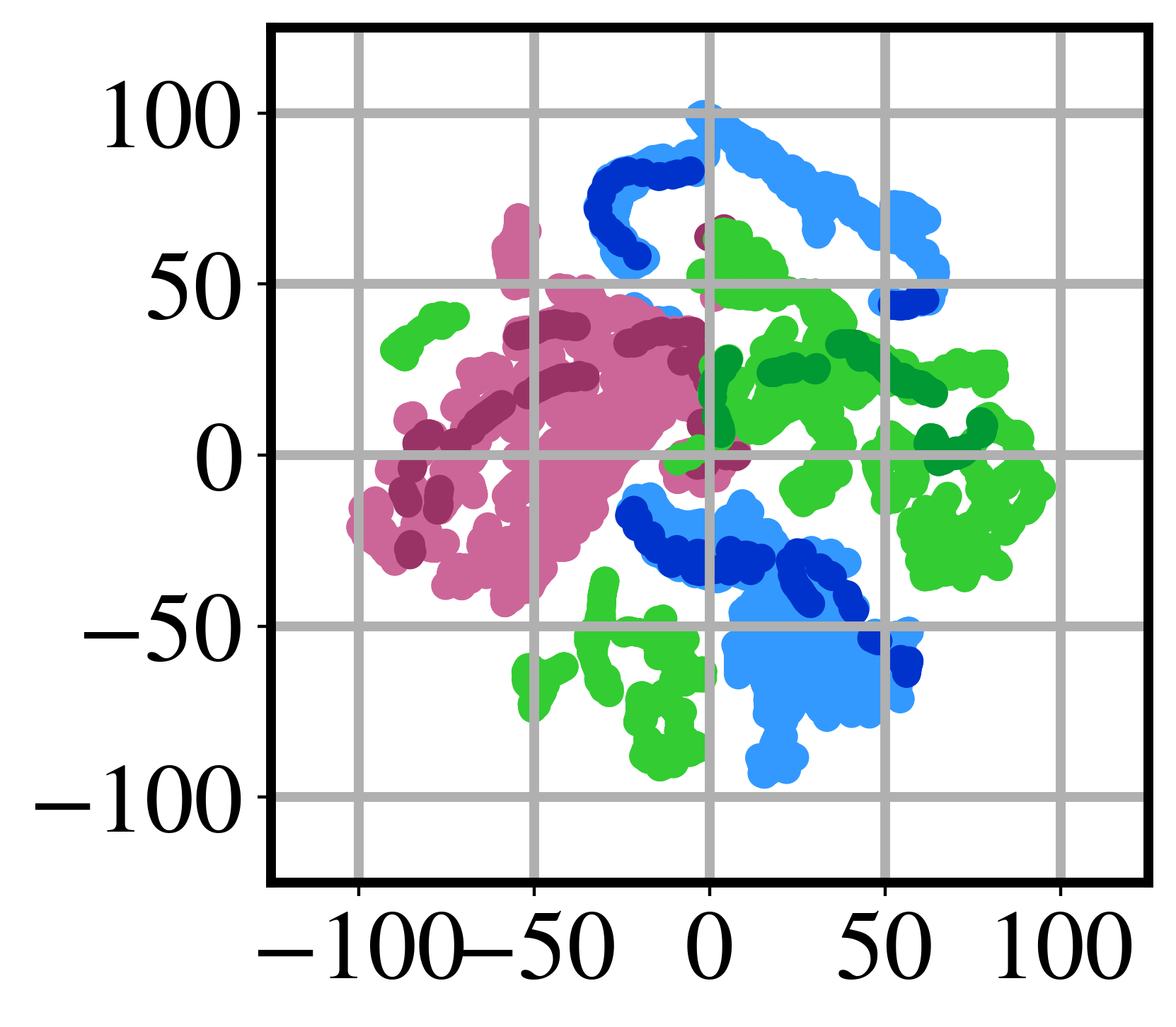}};
\node[inner sep=0pt] at (16,4)
    {\includegraphics[width=.11\textwidth]{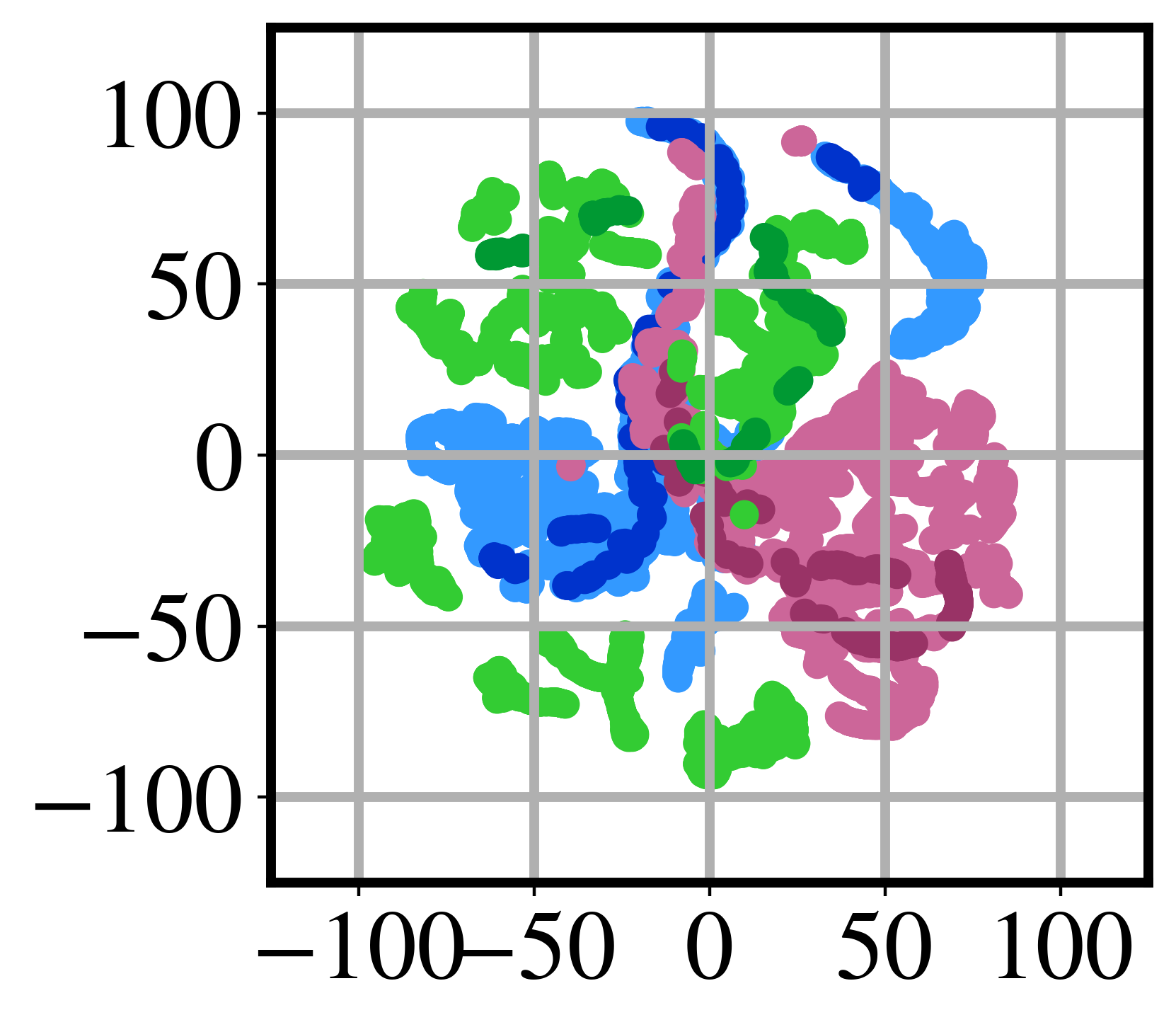}};
    
\node[inner sep=0pt] at (0,2)
    {\includegraphics[width=.11\textwidth]{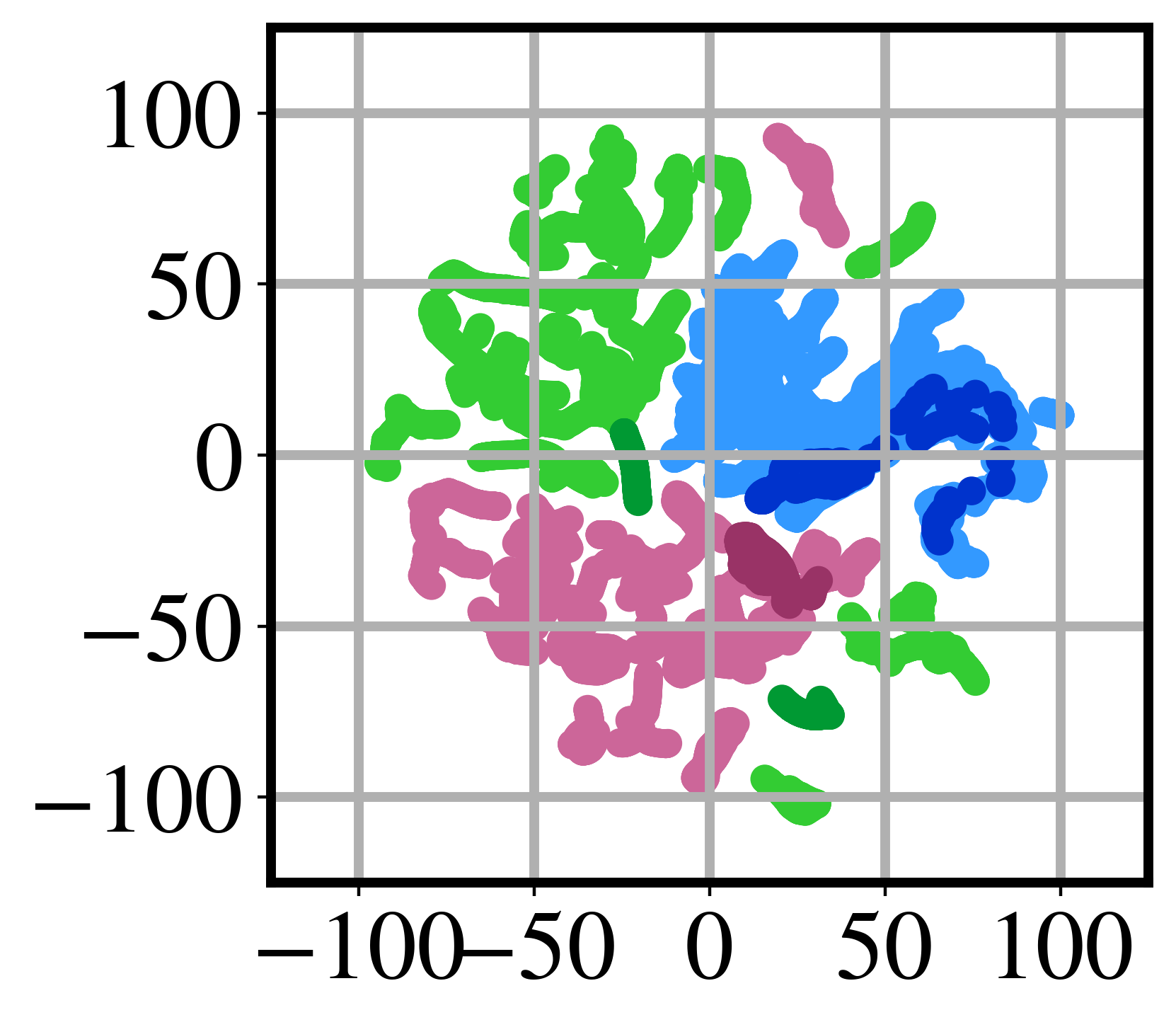}};
\node[inner sep=0pt] at (2,2)
    {\includegraphics[width=.11\textwidth]{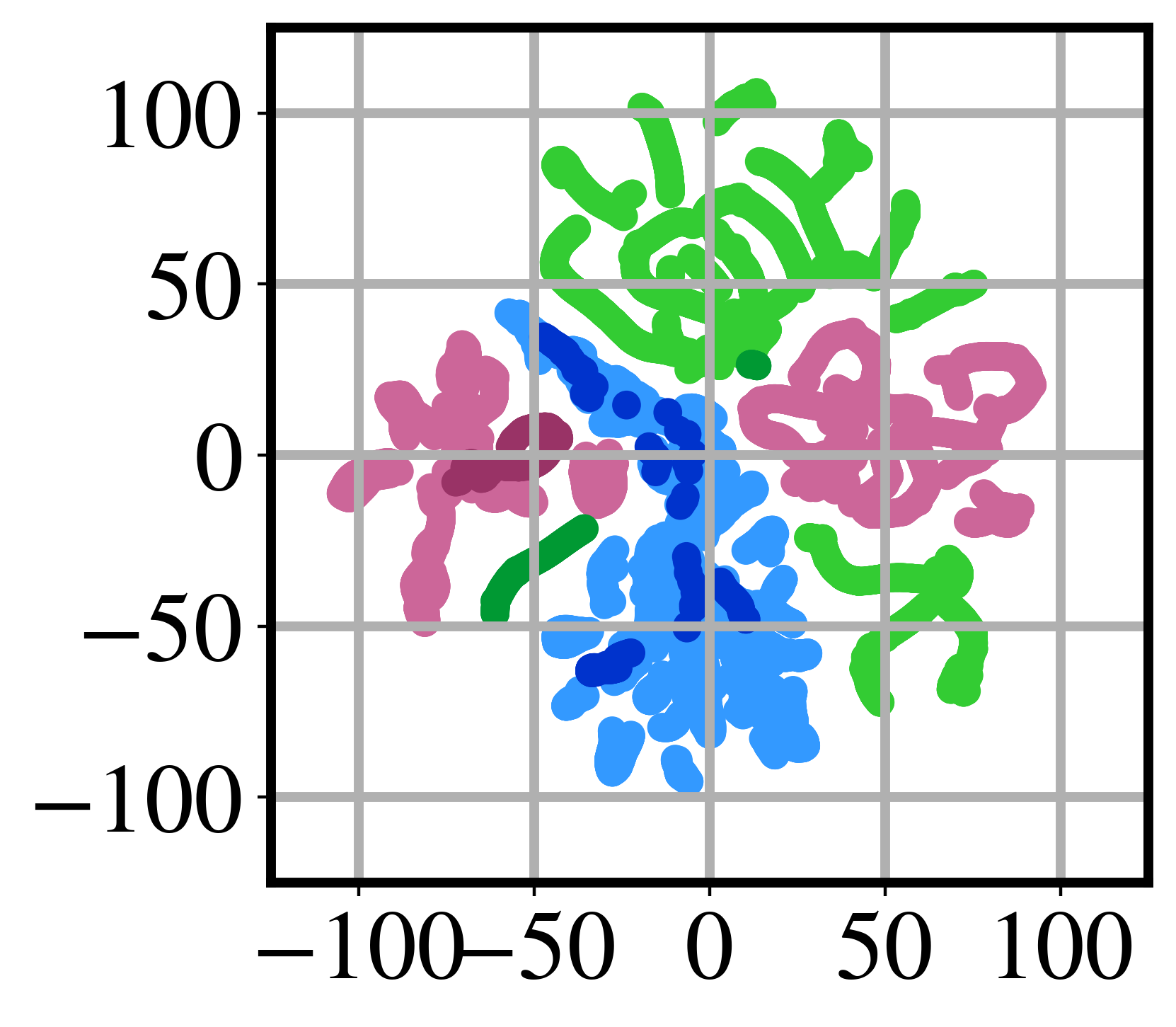}};
\node[inner sep=0pt] at (4,2)
    {\includegraphics[width=.11\textwidth]{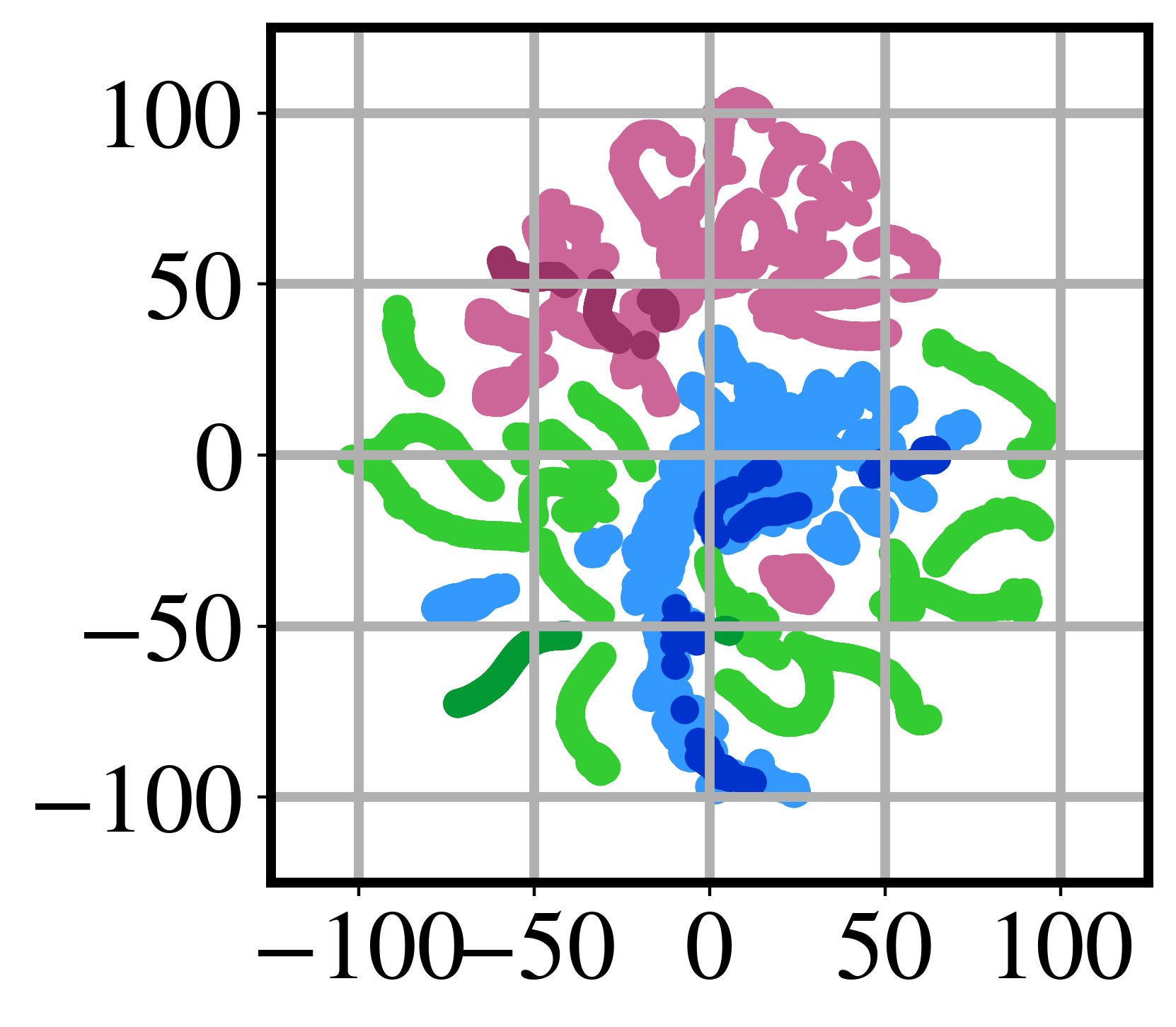}};
\node[inner sep=0pt] at (6,2)
    {\includegraphics[width=.11\textwidth]{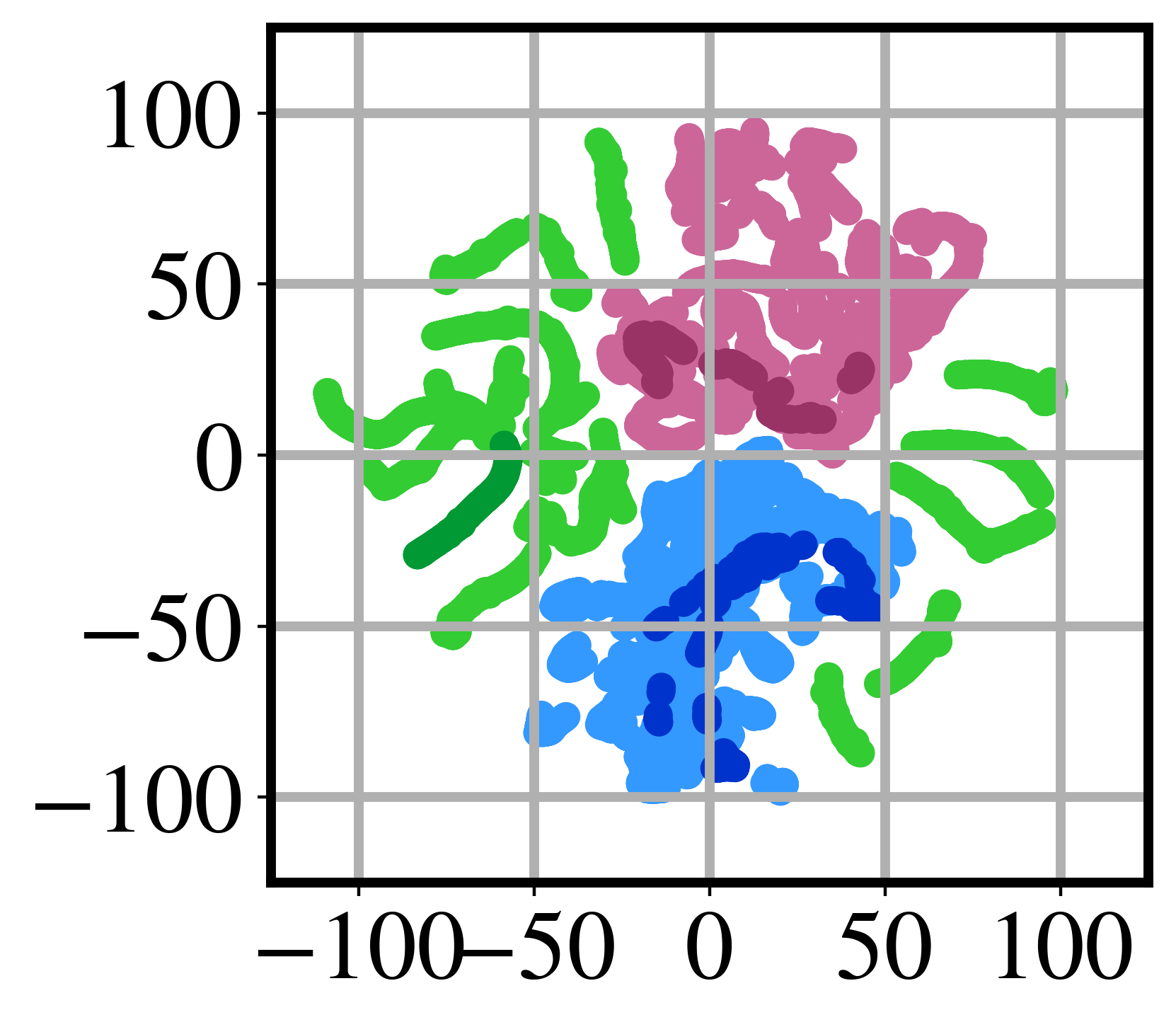}};
\node[inner sep=0pt] at (8,2)
    {\includegraphics[width=.11\textwidth]{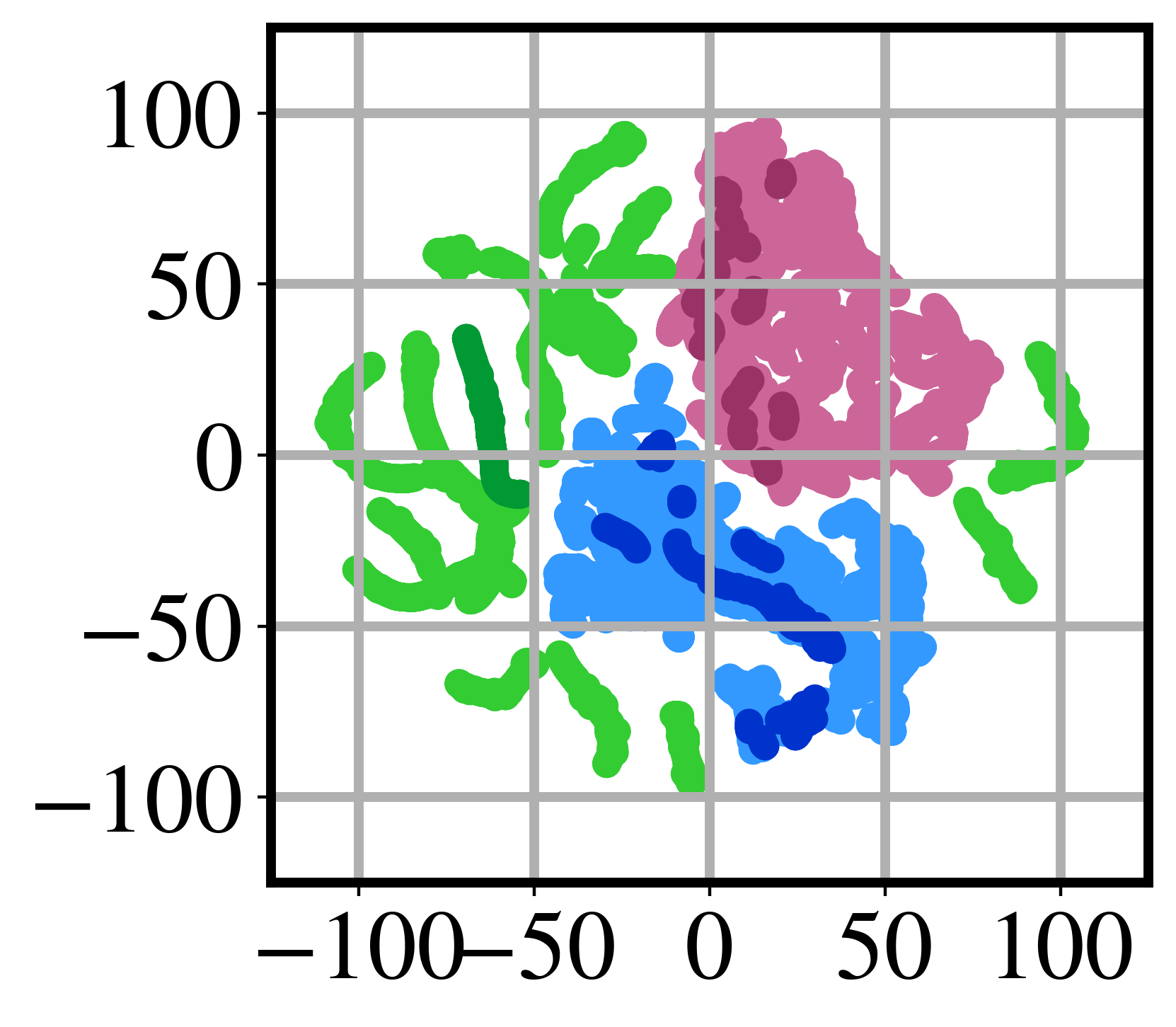}};
\node[inner sep=0pt] at (10,2)
    {\includegraphics[width=.11\textwidth]{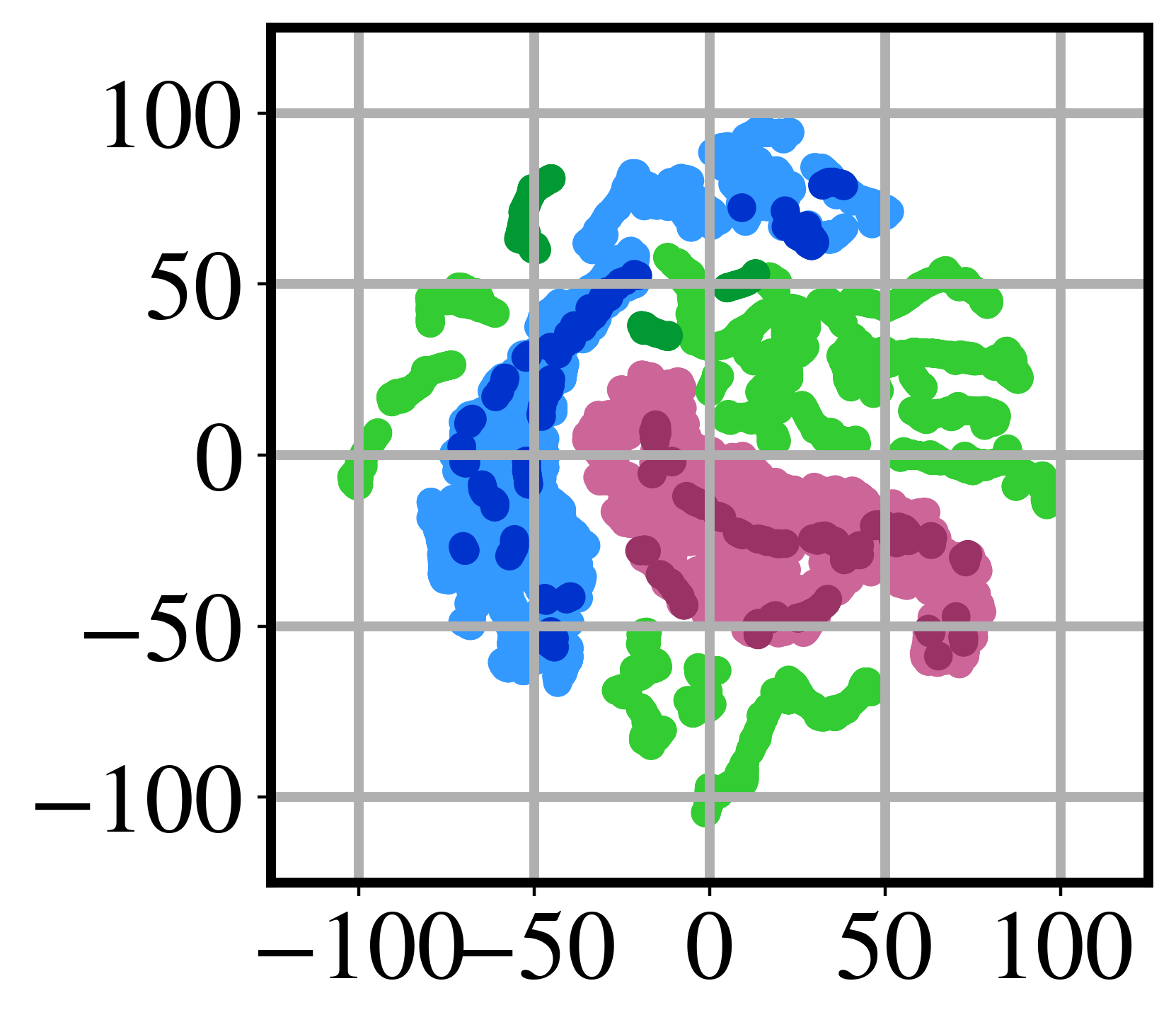}};
\node[inner sep=0pt] at (12,2)
    {\includegraphics[width=.11\textwidth]{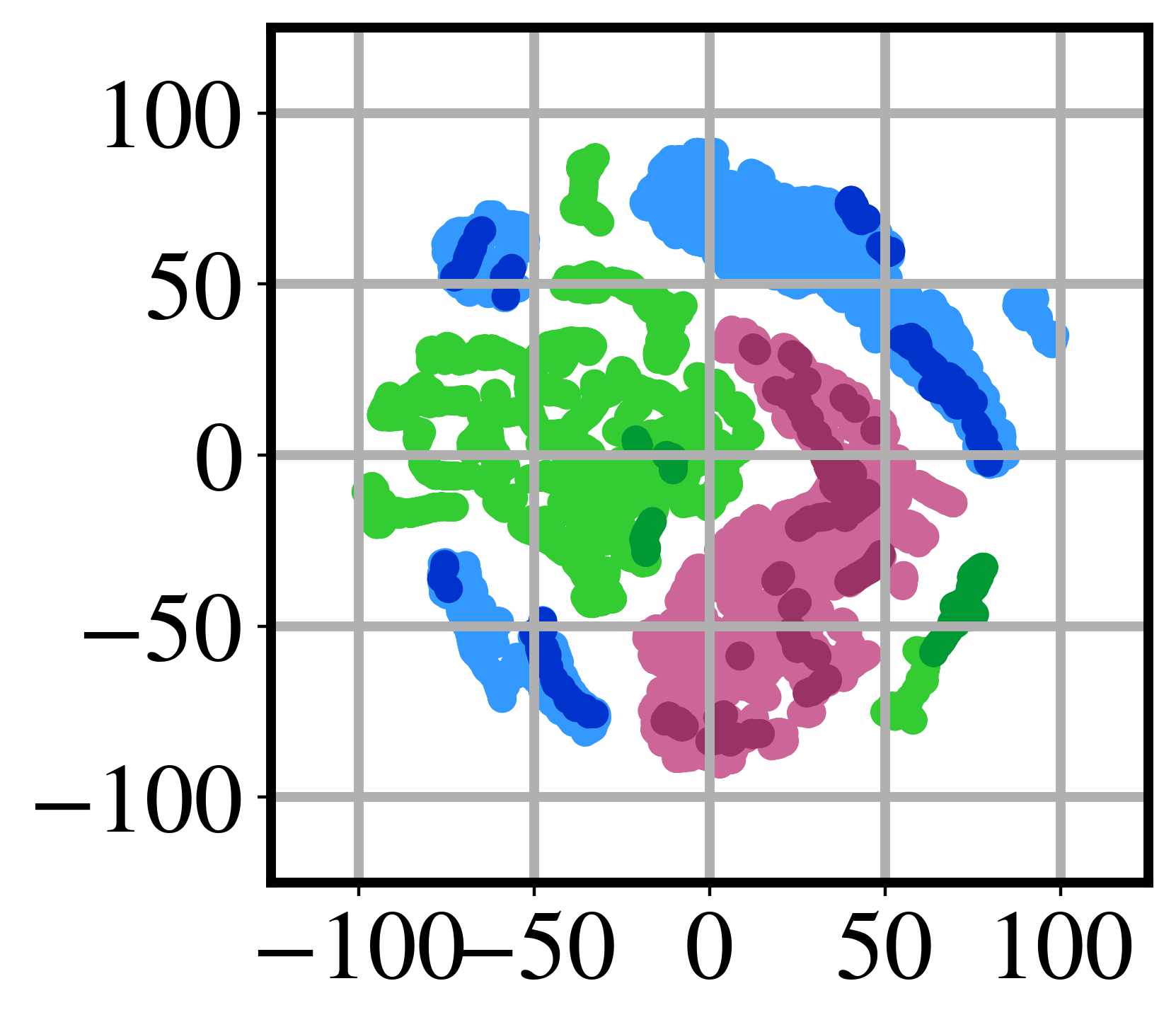}};
\node[inner sep=0pt] at (14,2)
    {\includegraphics[width=.11\textwidth]{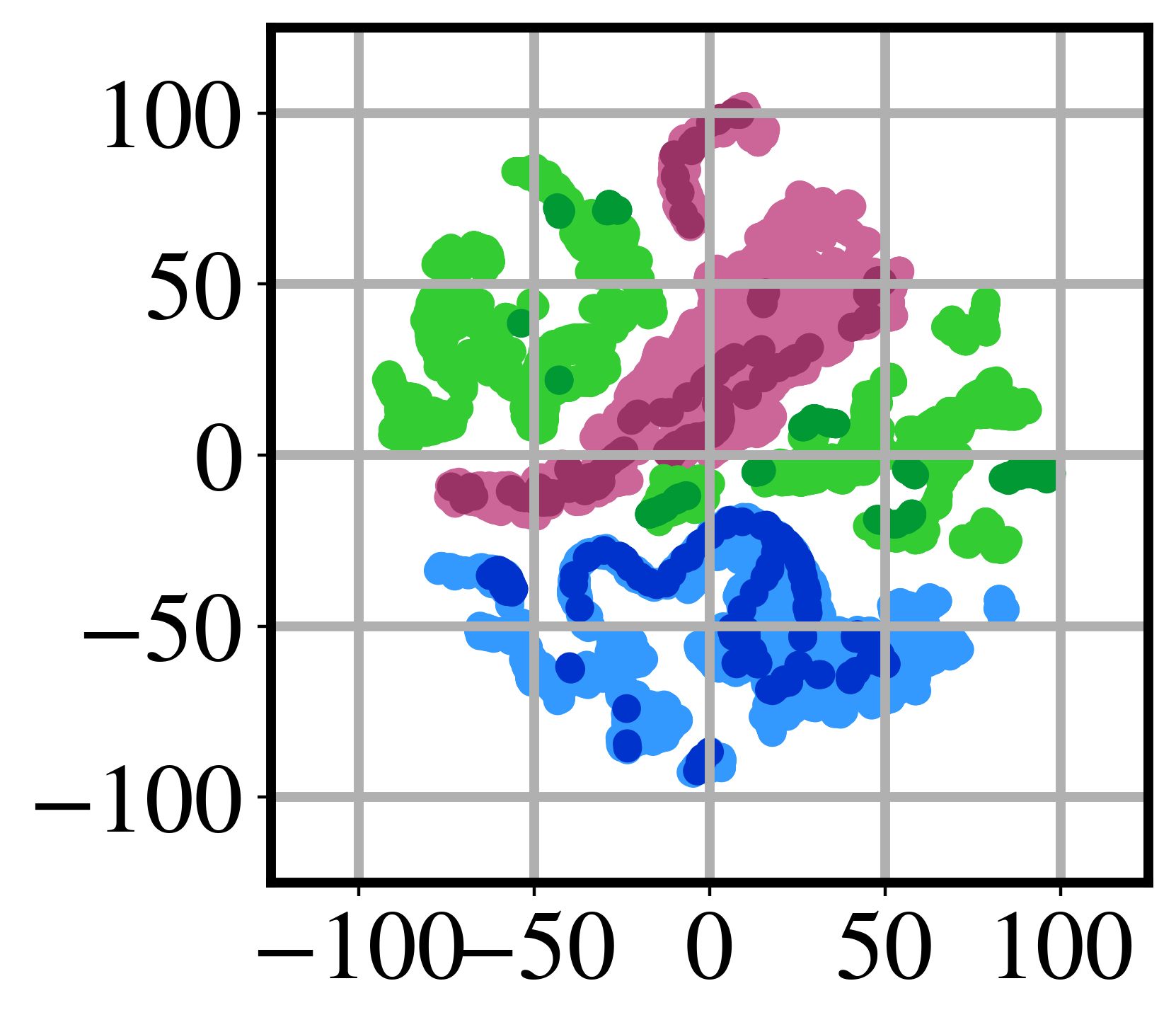}};
\node[inner sep=0pt] at (16,2)
    {\includegraphics[width=.11\textwidth]{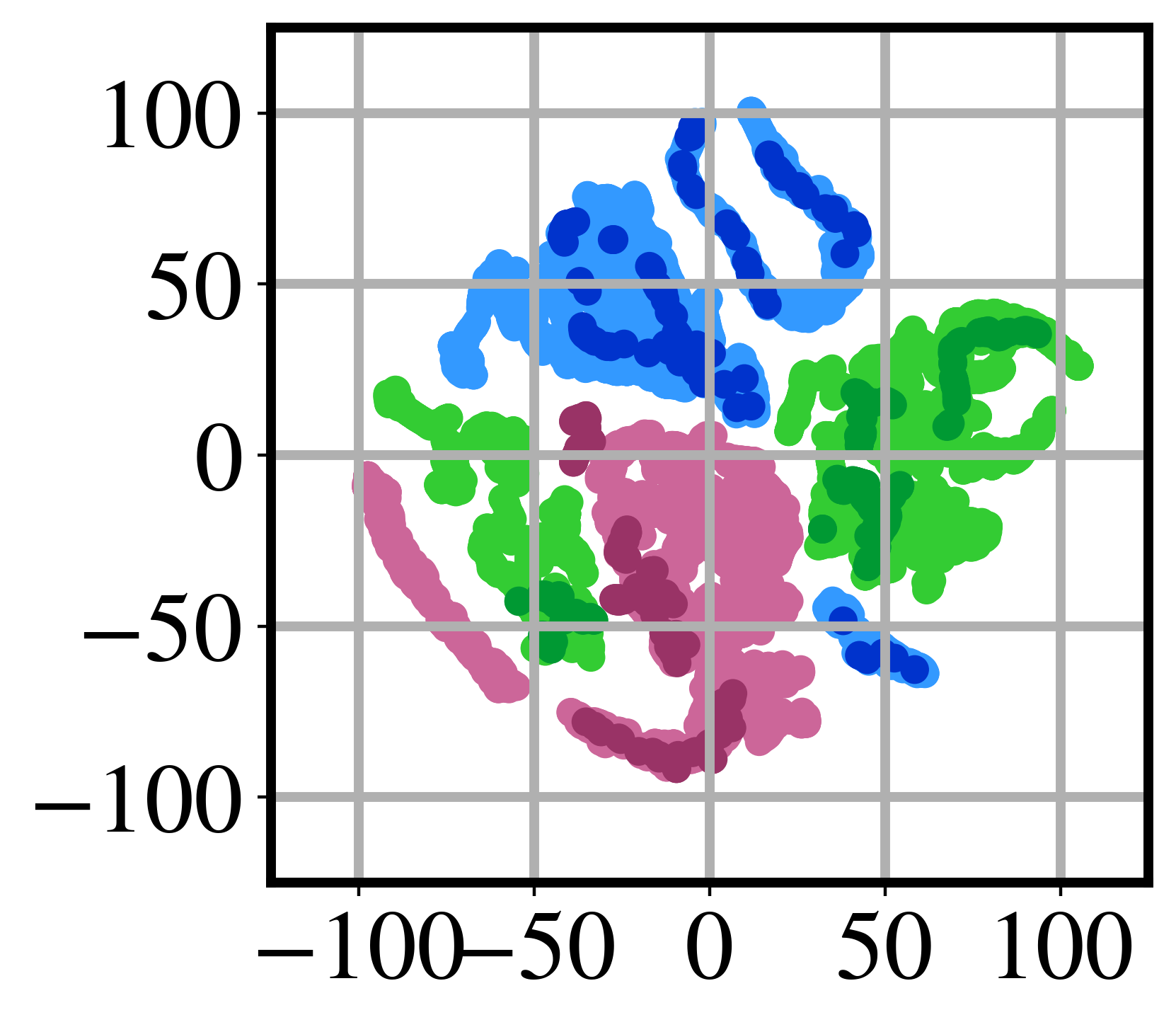}};
    
\node[inner sep=0pt] at (0,0)
    {\includegraphics[width=.11\textwidth]{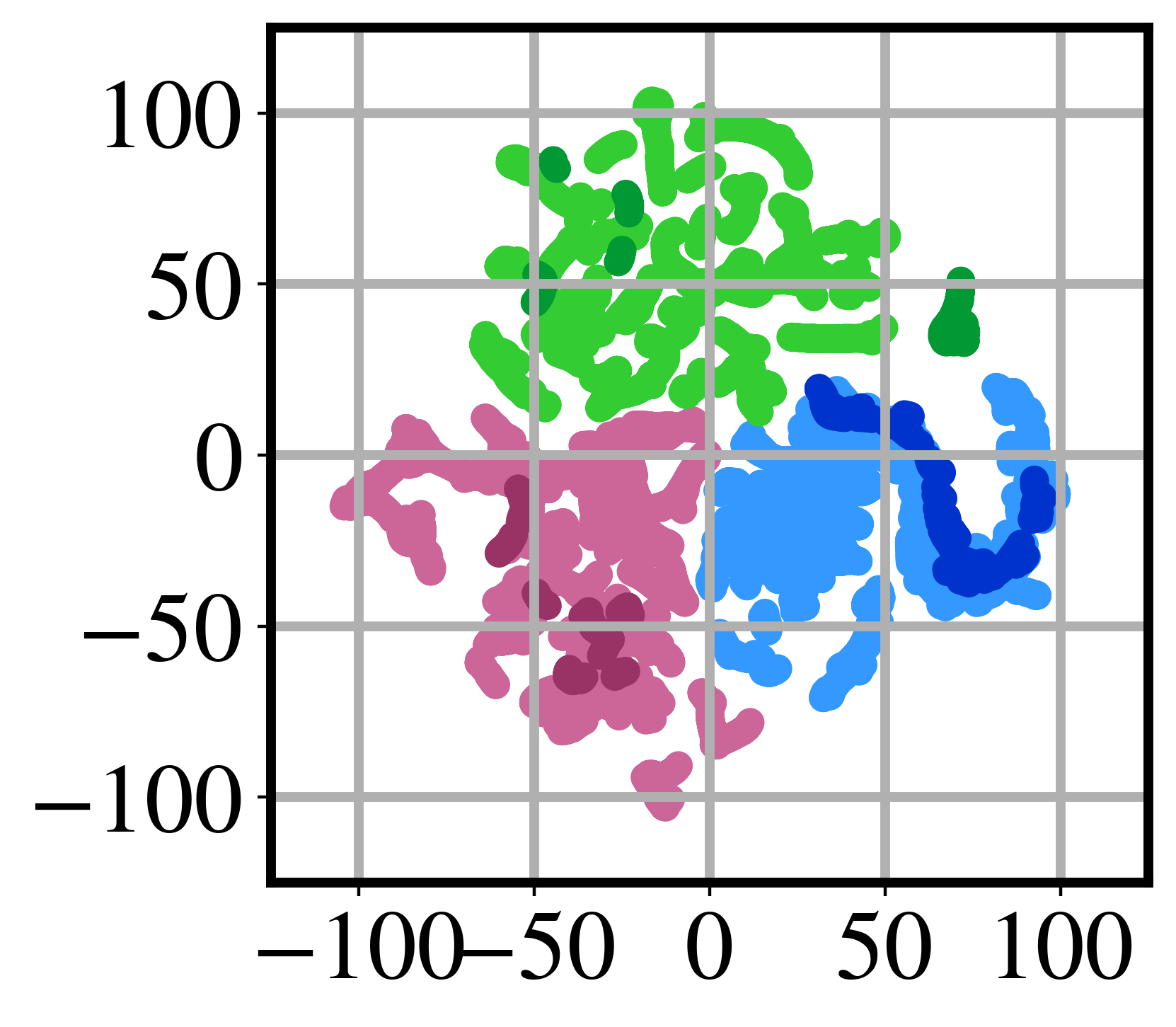}};
\node[inner sep=0pt] at (2,0)
    {\includegraphics[width=.11\textwidth]{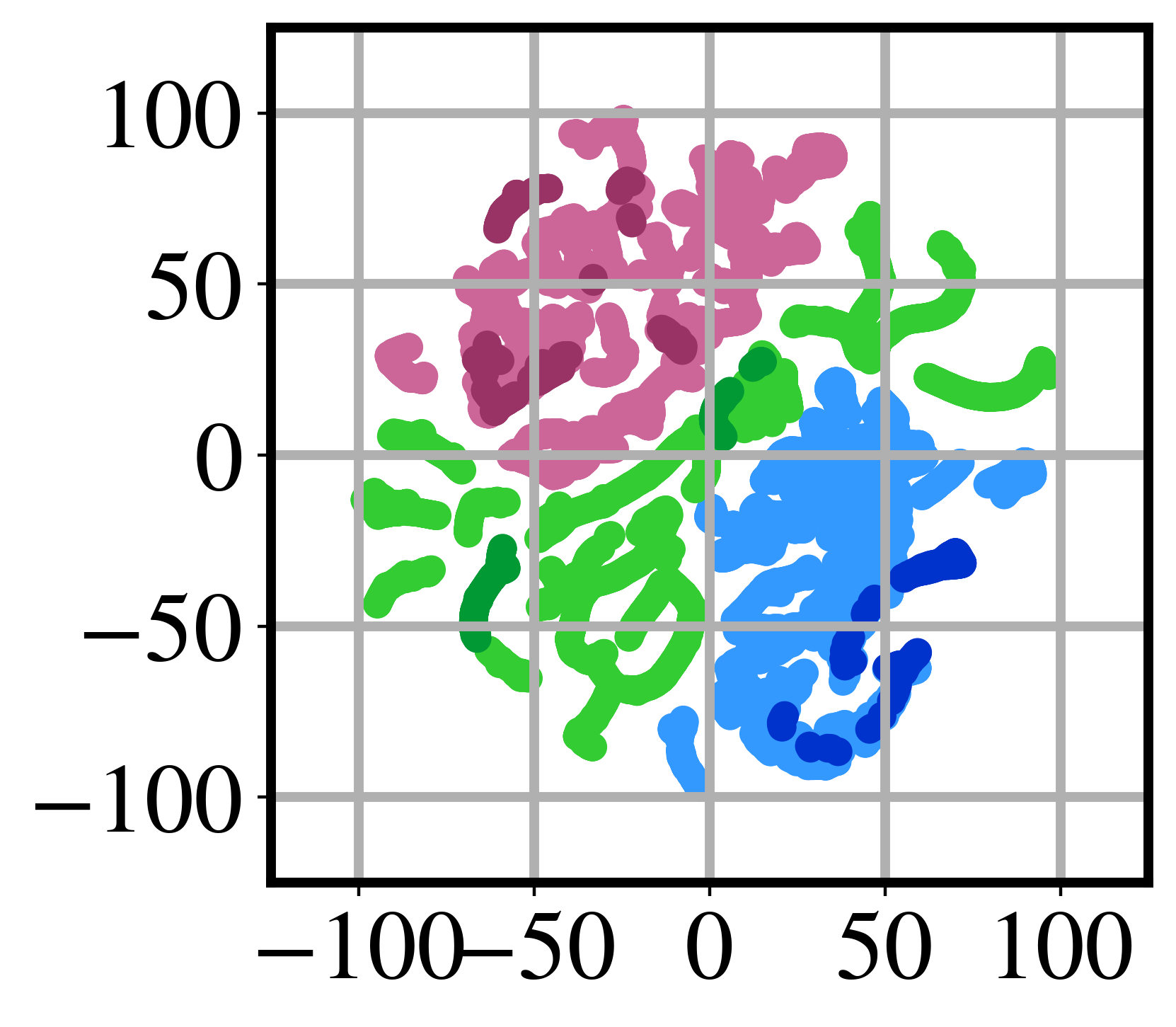}};
\node[inner sep=0pt] at (4,0)
    {\includegraphics[width=.11\textwidth]{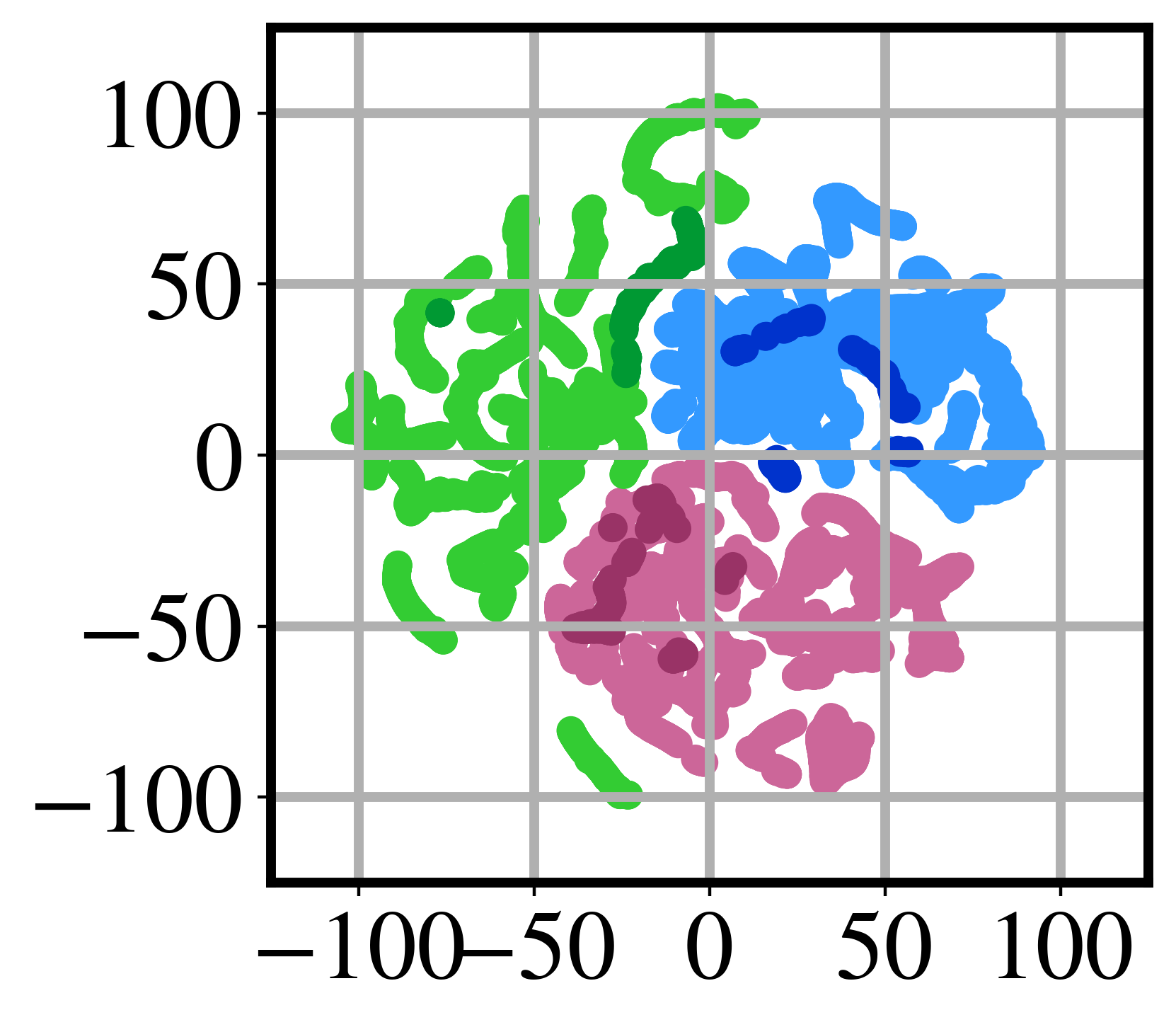}};
\node[inner sep=0pt] at (6,0)
    {\includegraphics[width=.11\textwidth]{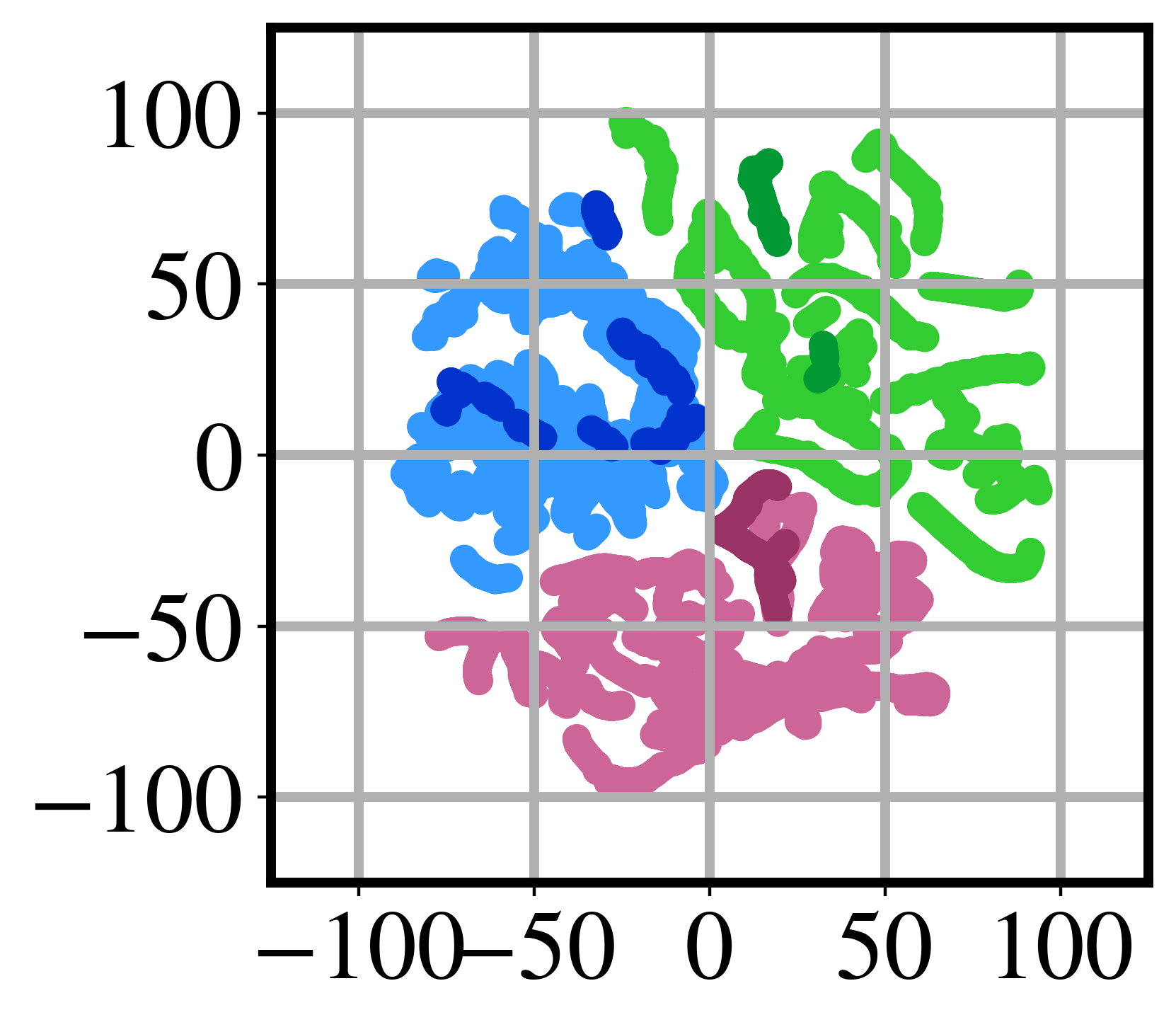}};
\node[inner sep=0pt] at (8,0)
    {\includegraphics[width=.11\textwidth]{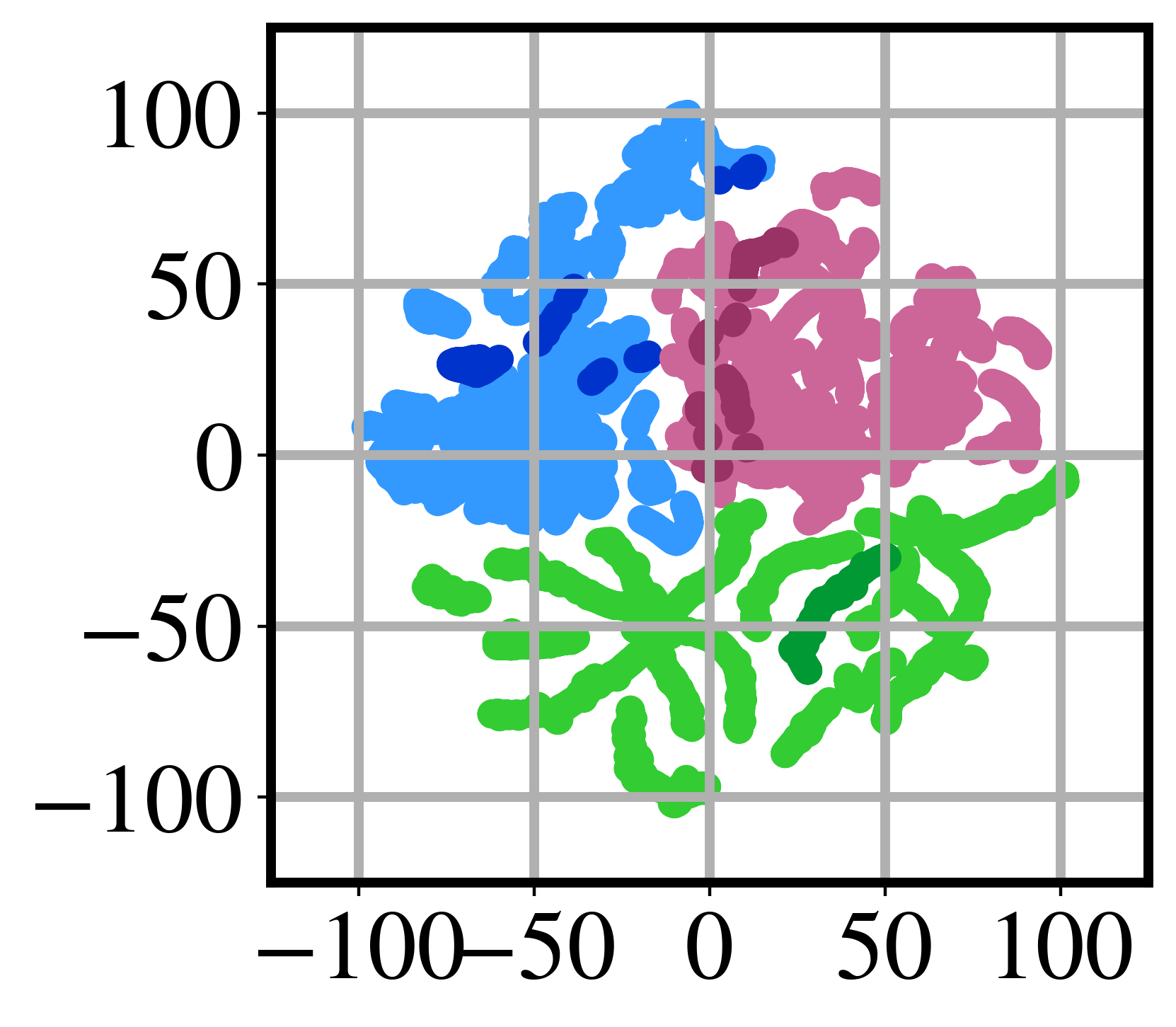}};
\node[inner sep=0pt] at (10,0)
    {\includegraphics[width=.11\textwidth]{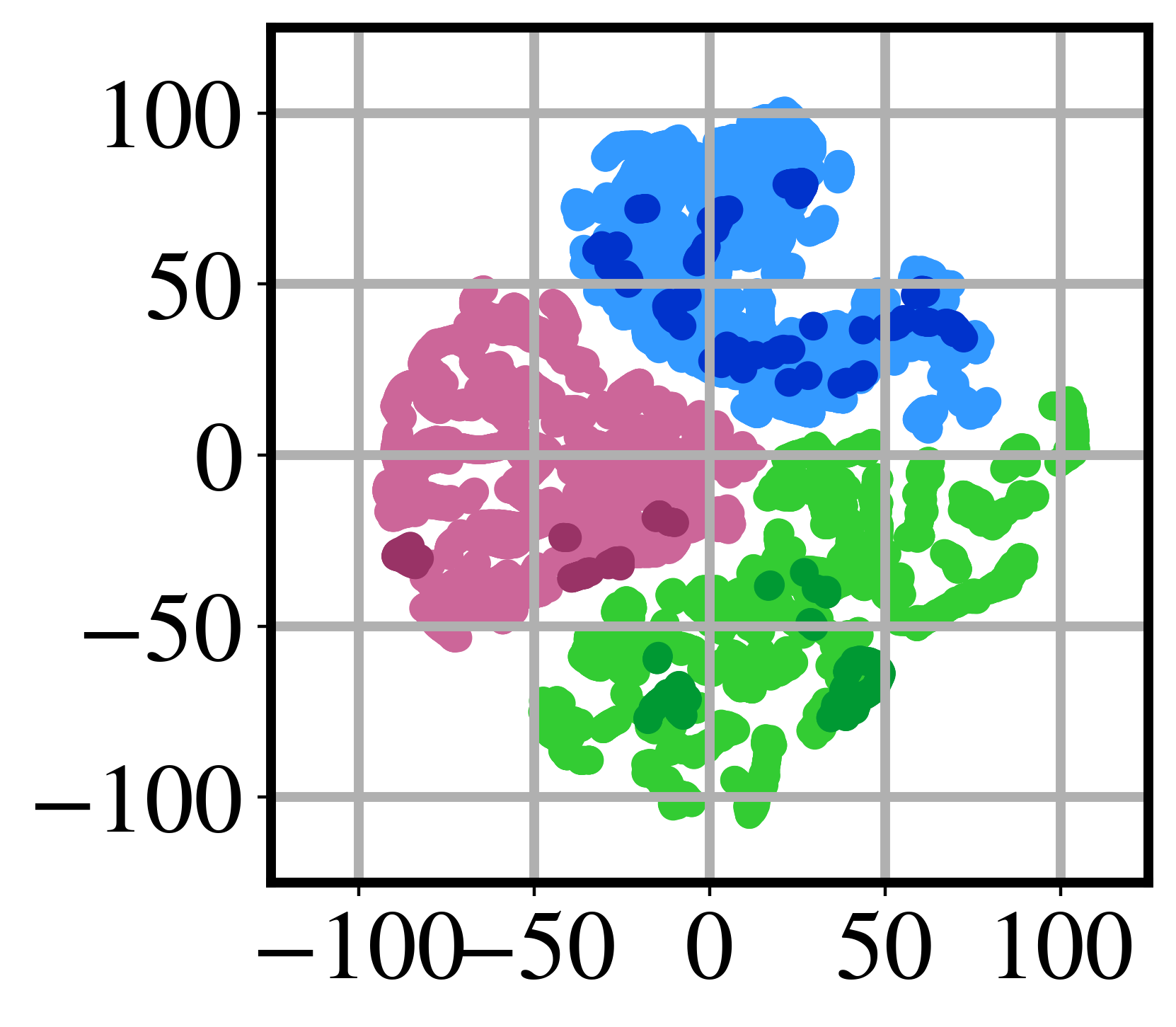}};
\node[inner sep=0pt] at (12,0)
    {\includegraphics[width=.11\textwidth]{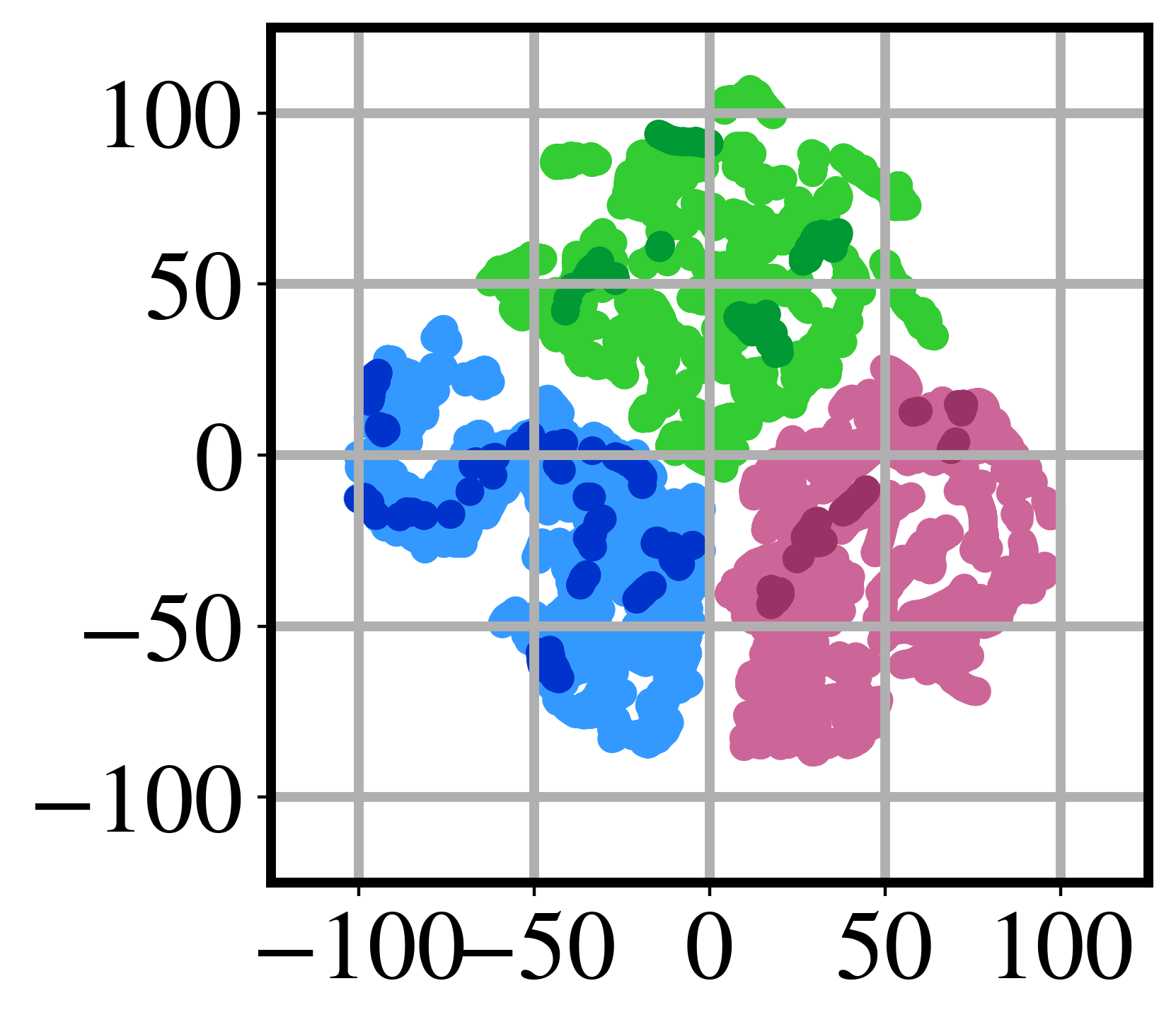}};
\node[inner sep=0pt] at (14,0)
    {\includegraphics[width=.11\textwidth]{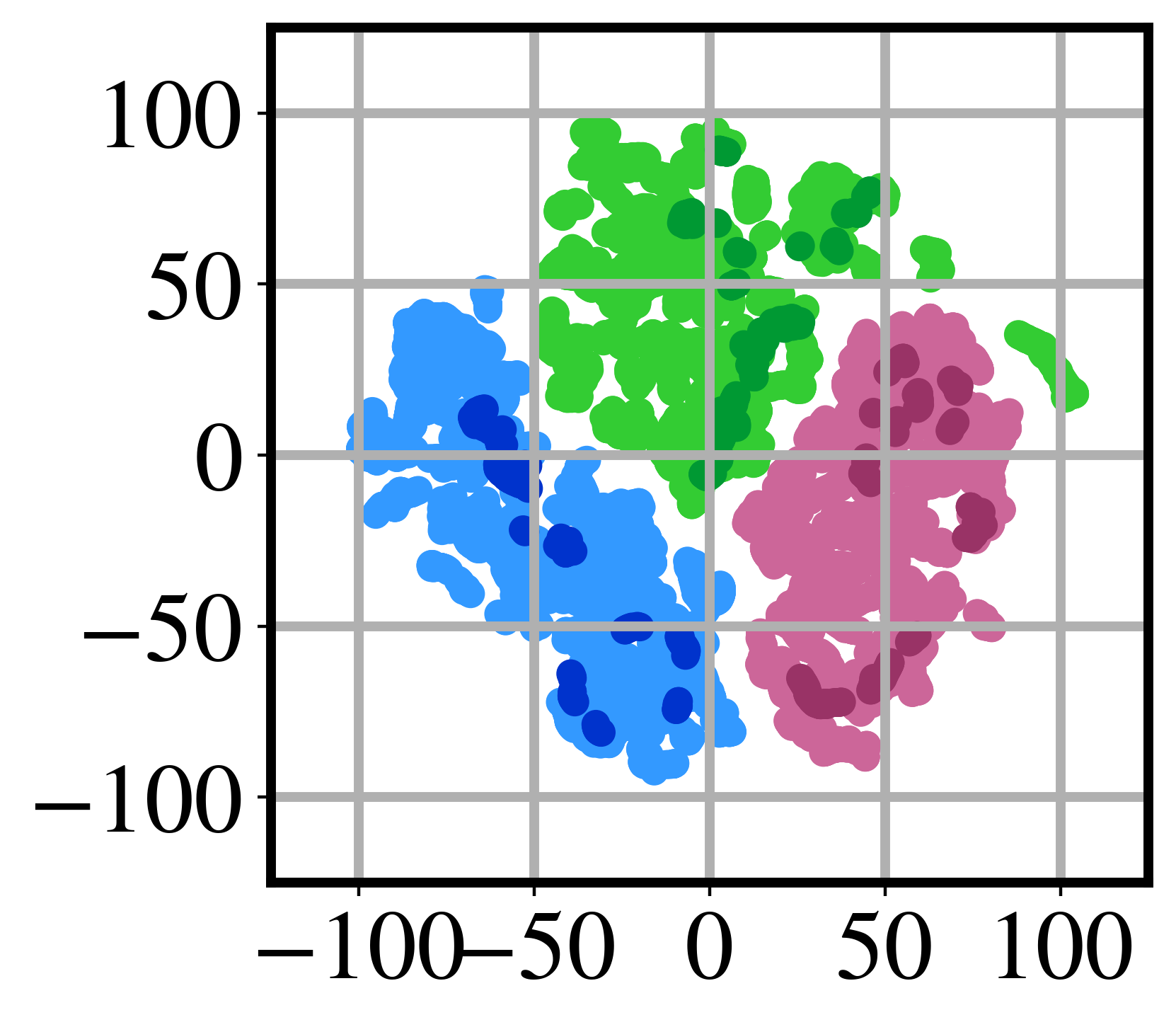}};
\node[inner sep=0pt] at (16,0)
    {\includegraphics[width=.11\textwidth]{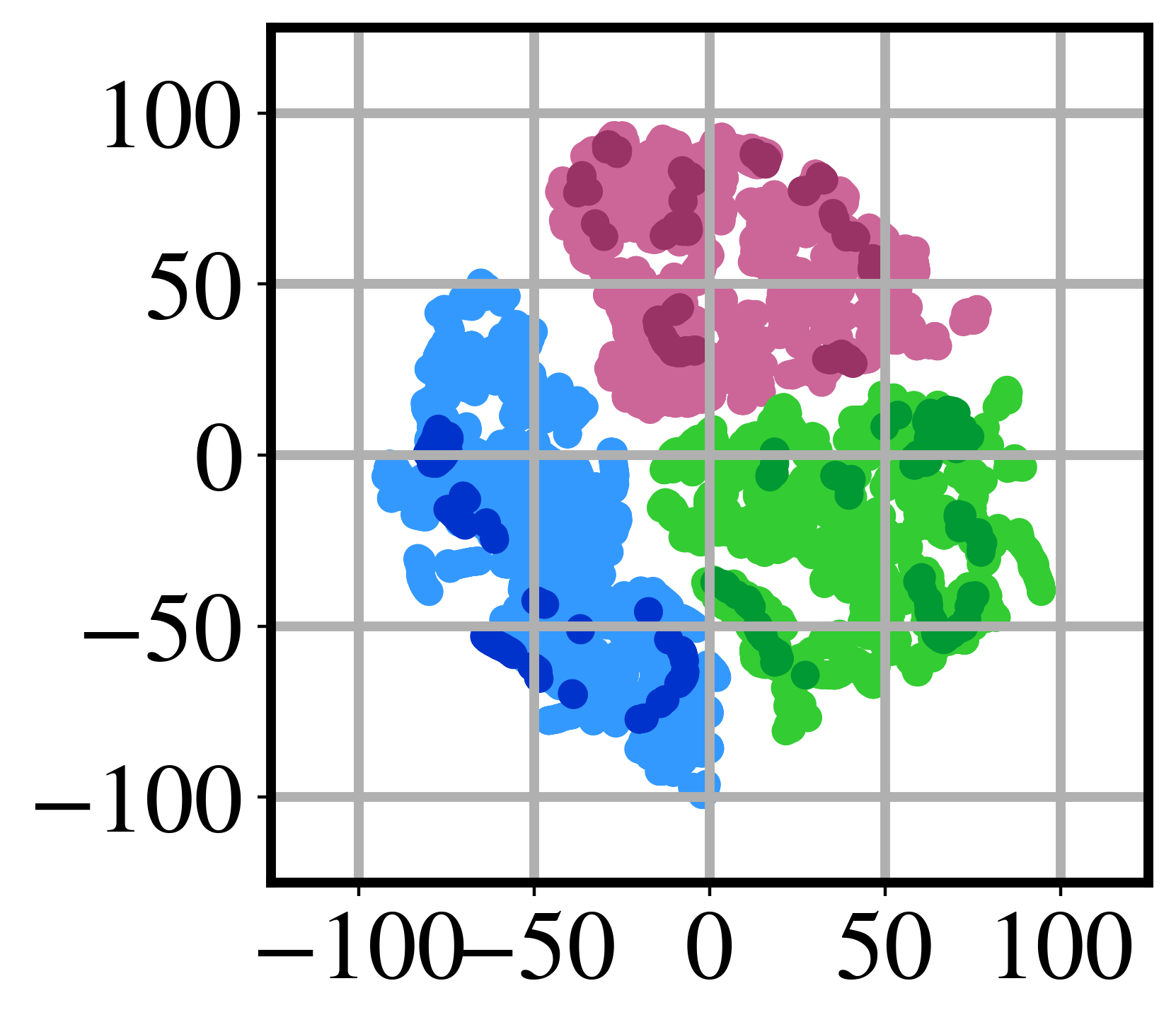}};
    
\node[inner sep=0pt] at (0,-3)
    {\includegraphics[width=.11\textwidth]{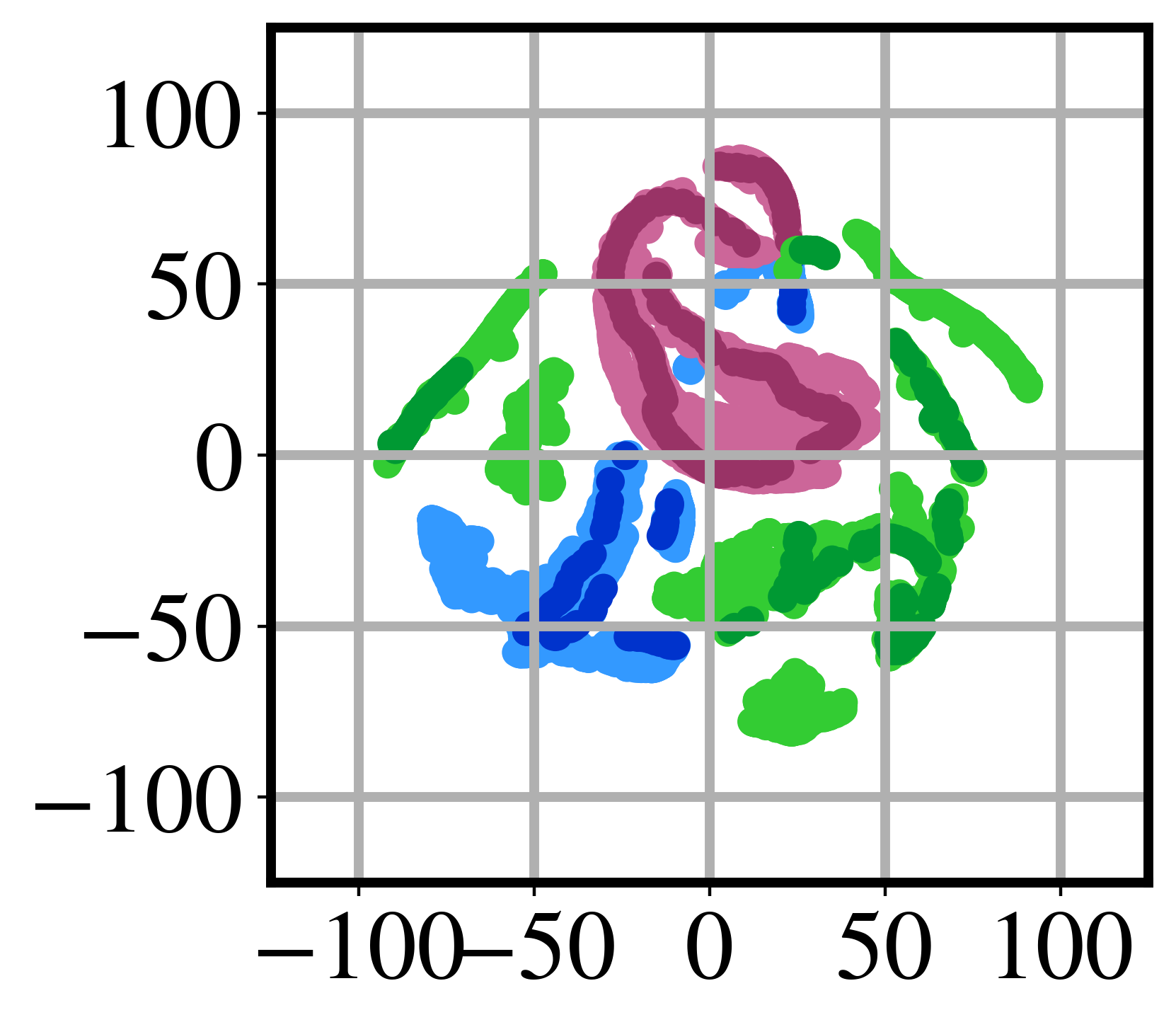}};
\node[inner sep=0pt] at (2,-3)
    {\includegraphics[width=.11\textwidth]{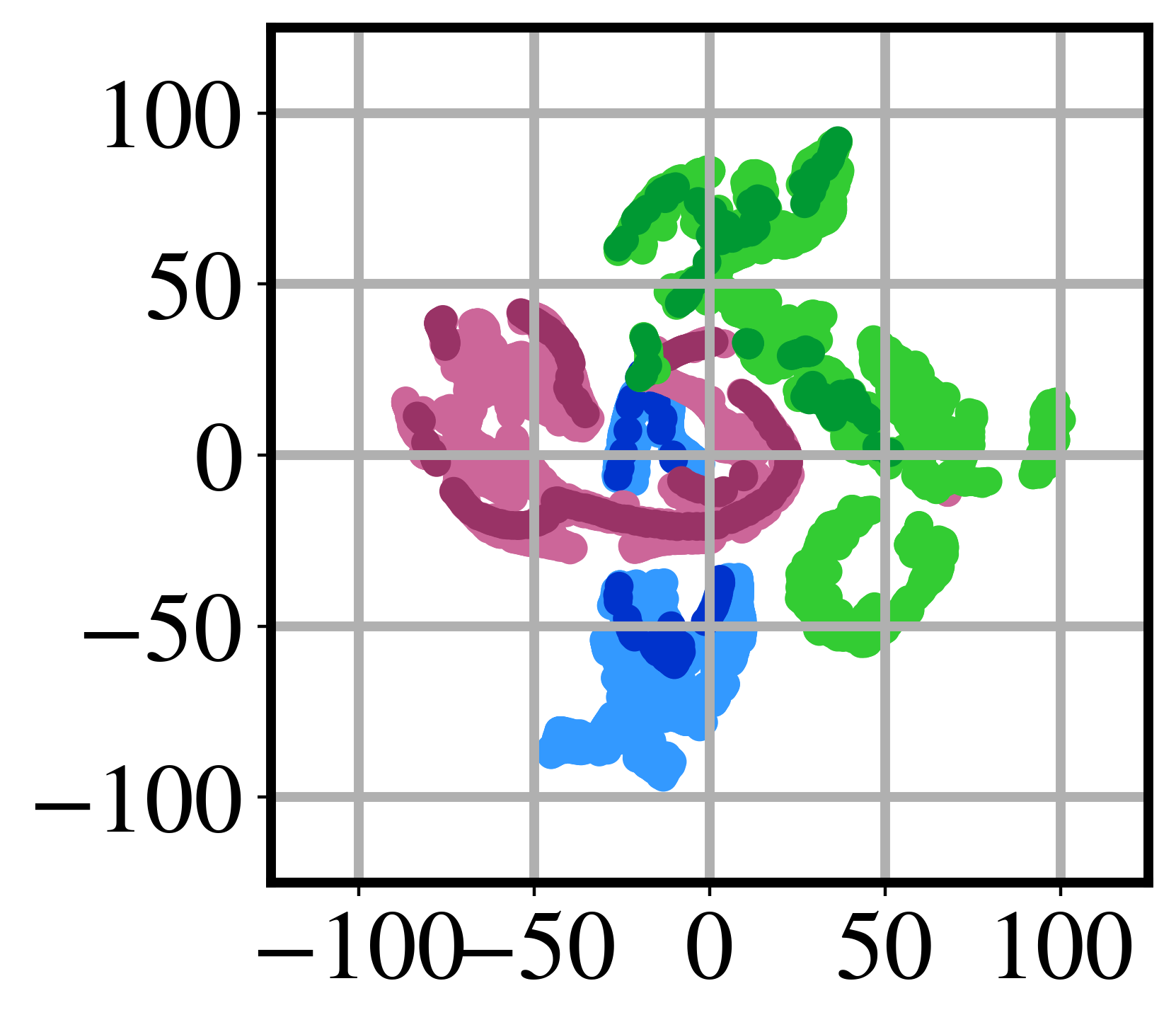}};
\node[inner sep=0pt] at (4,-3)
    {\includegraphics[width=.11\textwidth]{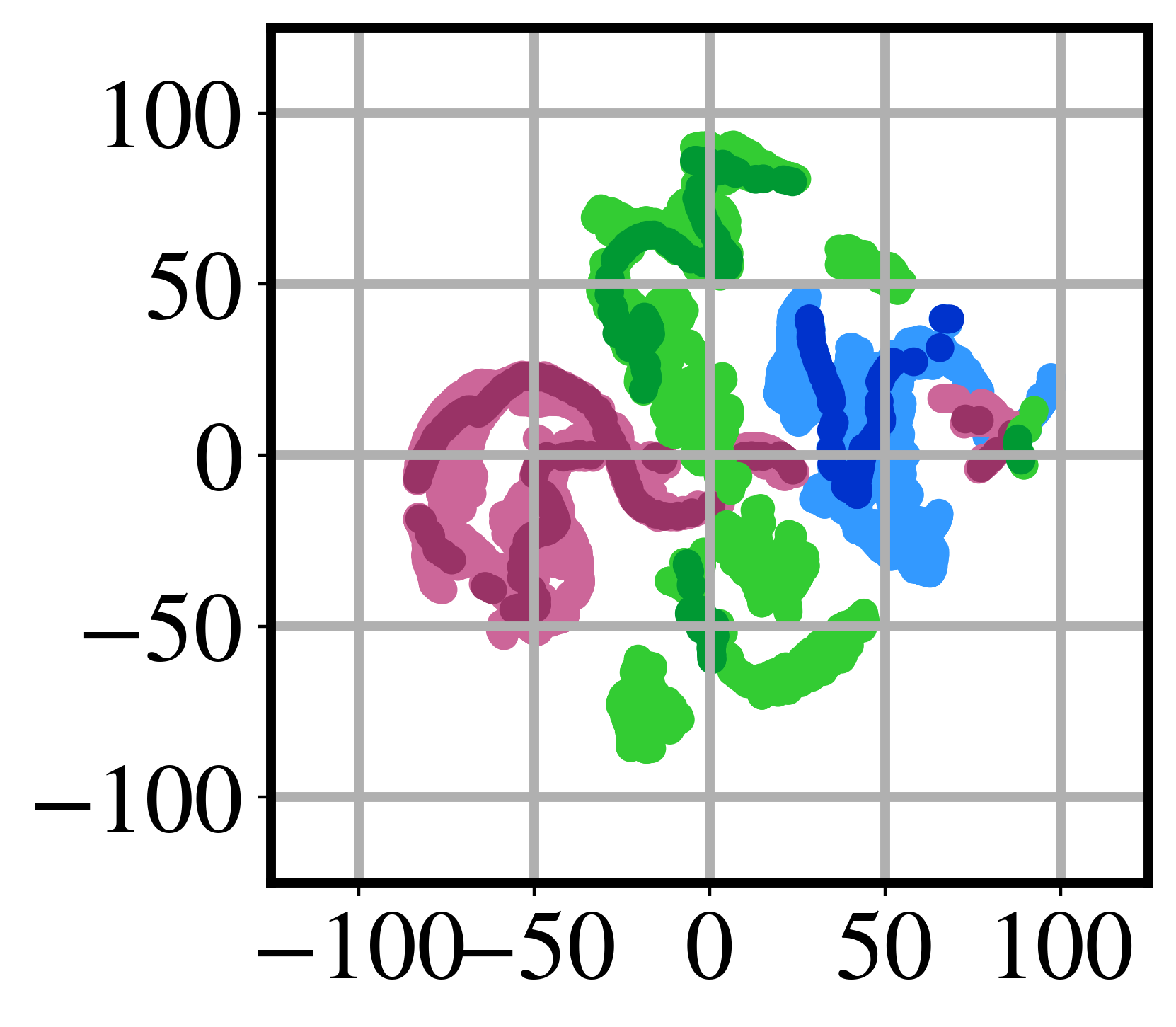}};
\node[inner sep=0pt] at (6,-3)
    {\includegraphics[width=.11\textwidth]{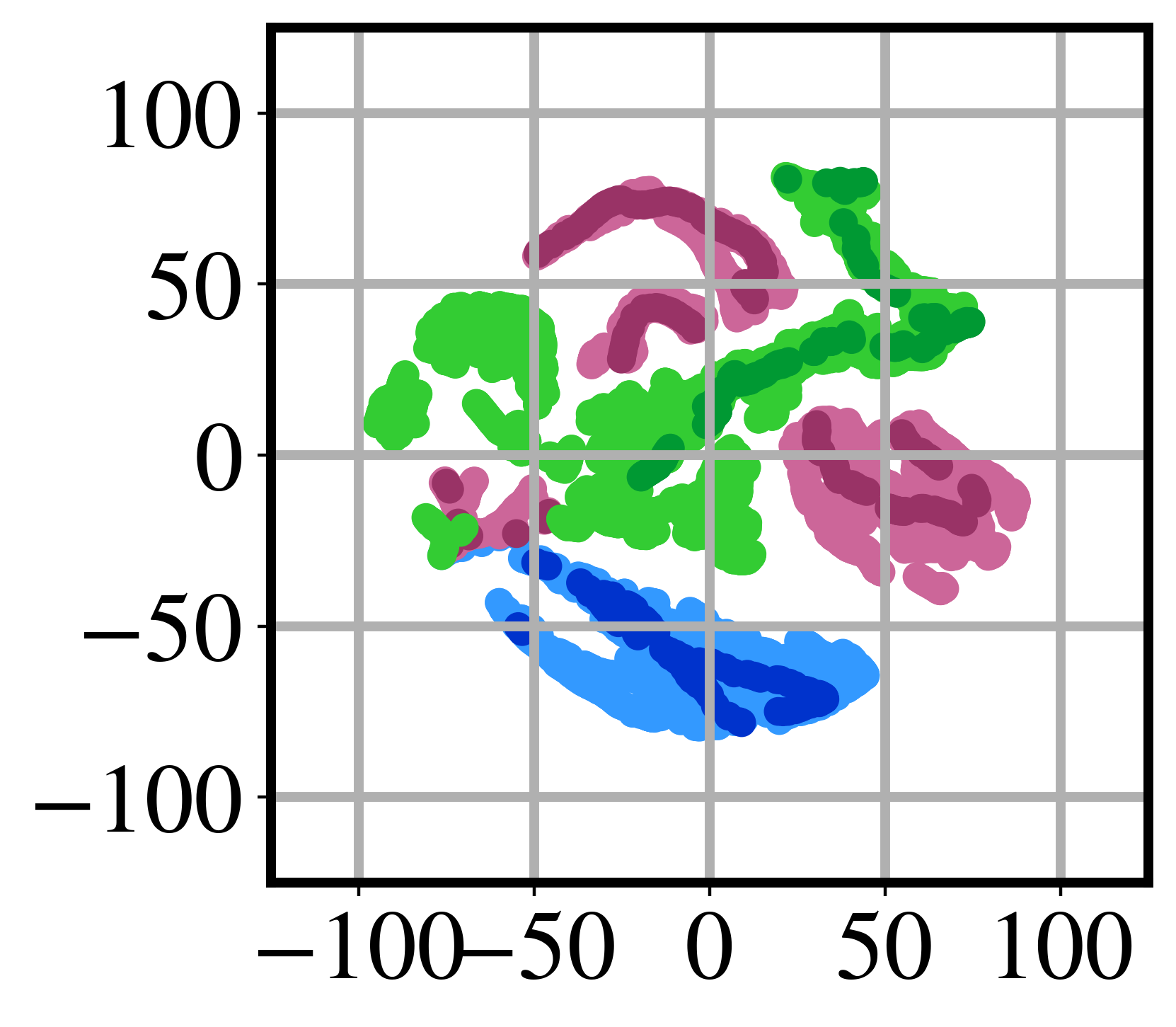}};
\node[inner sep=0pt] at (8,-3)
    {\includegraphics[width=.11\textwidth]{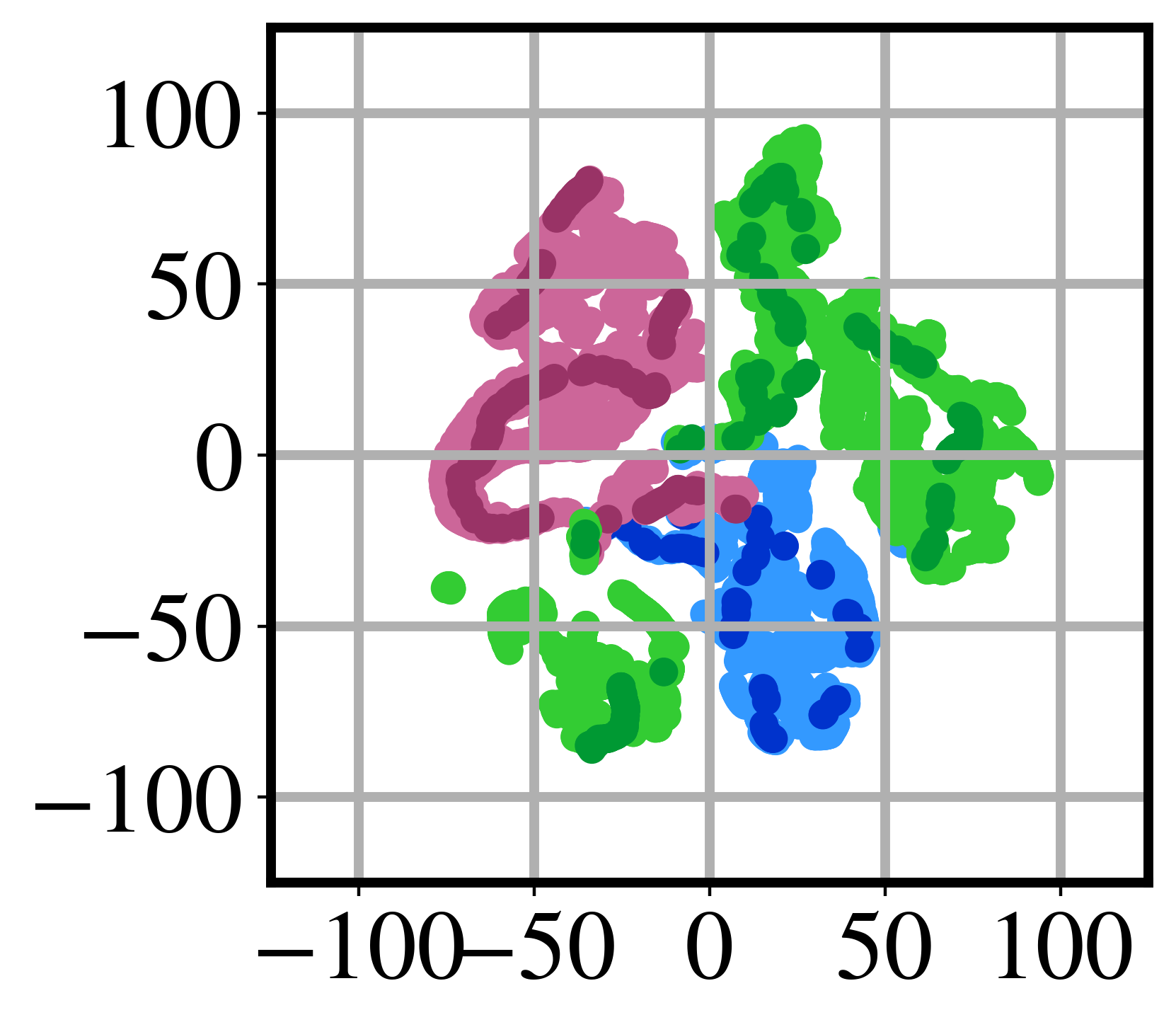}};
\node[inner sep=0pt] at (10,-3)
    {\includegraphics[width=.11\textwidth]{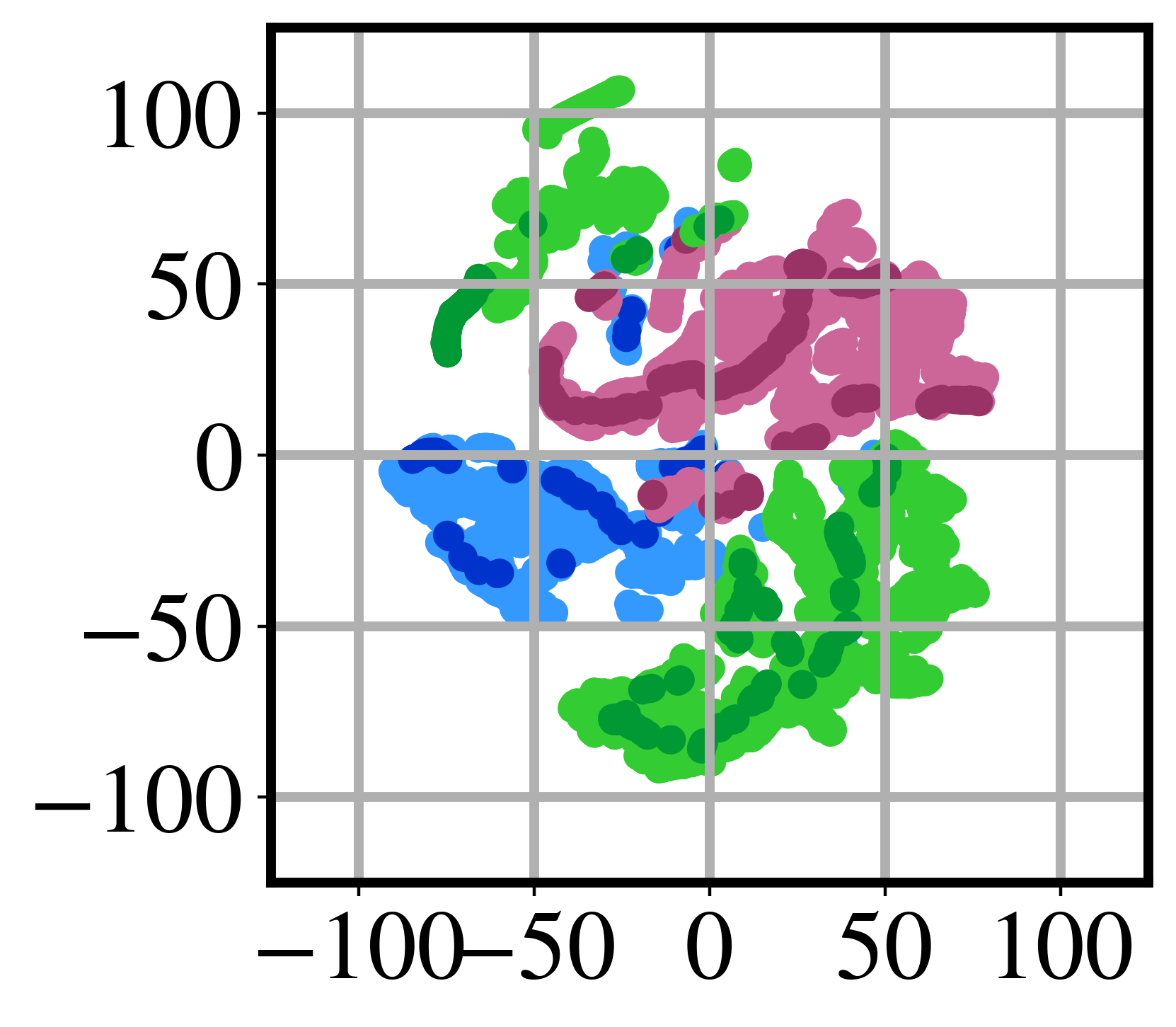}};
\node[inner sep=0pt] at (12,-3)
    {\includegraphics[width=.11\textwidth]{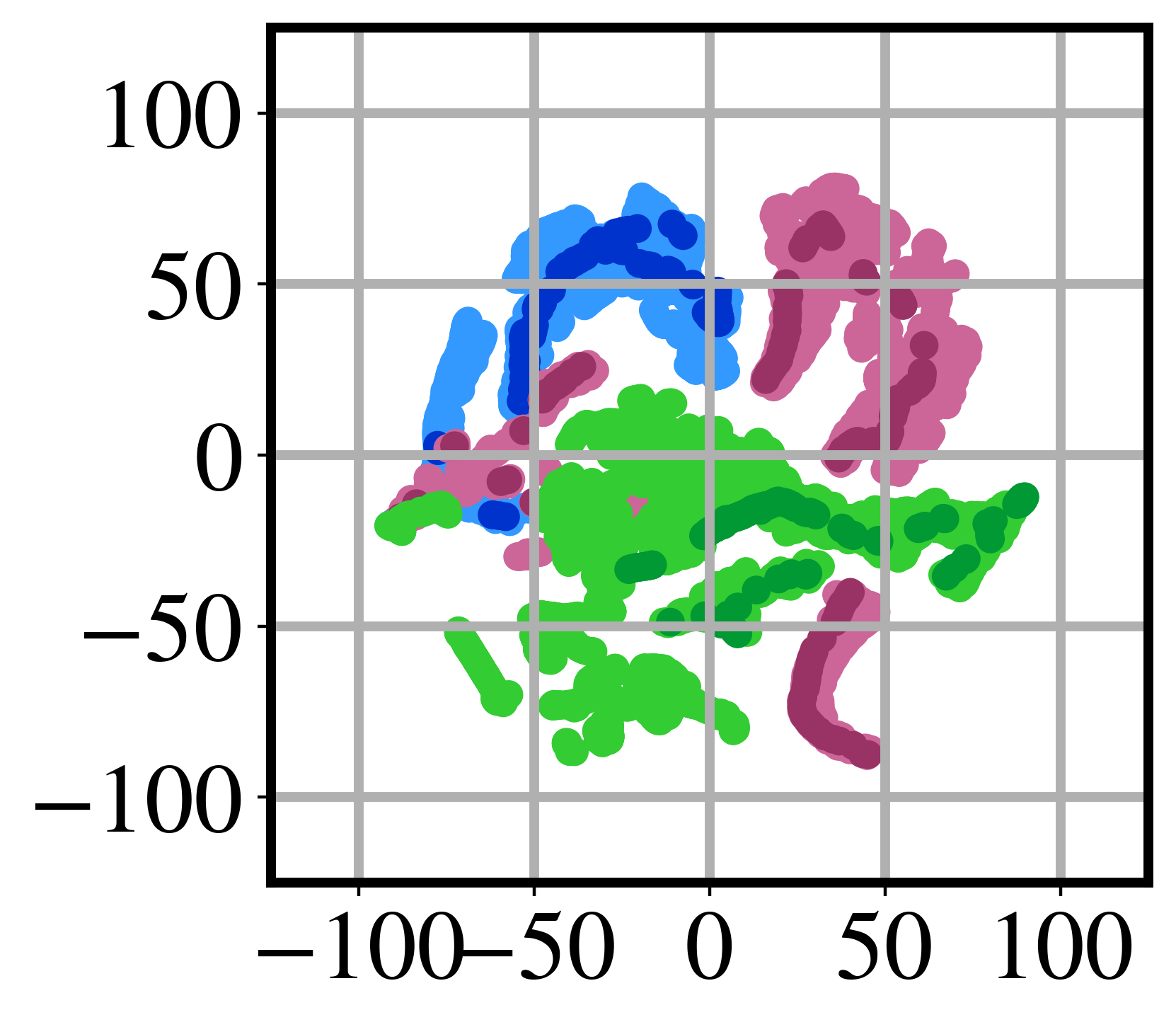}};
\node[inner sep=0pt] at (14,-3)
    {\includegraphics[width=.11\textwidth]{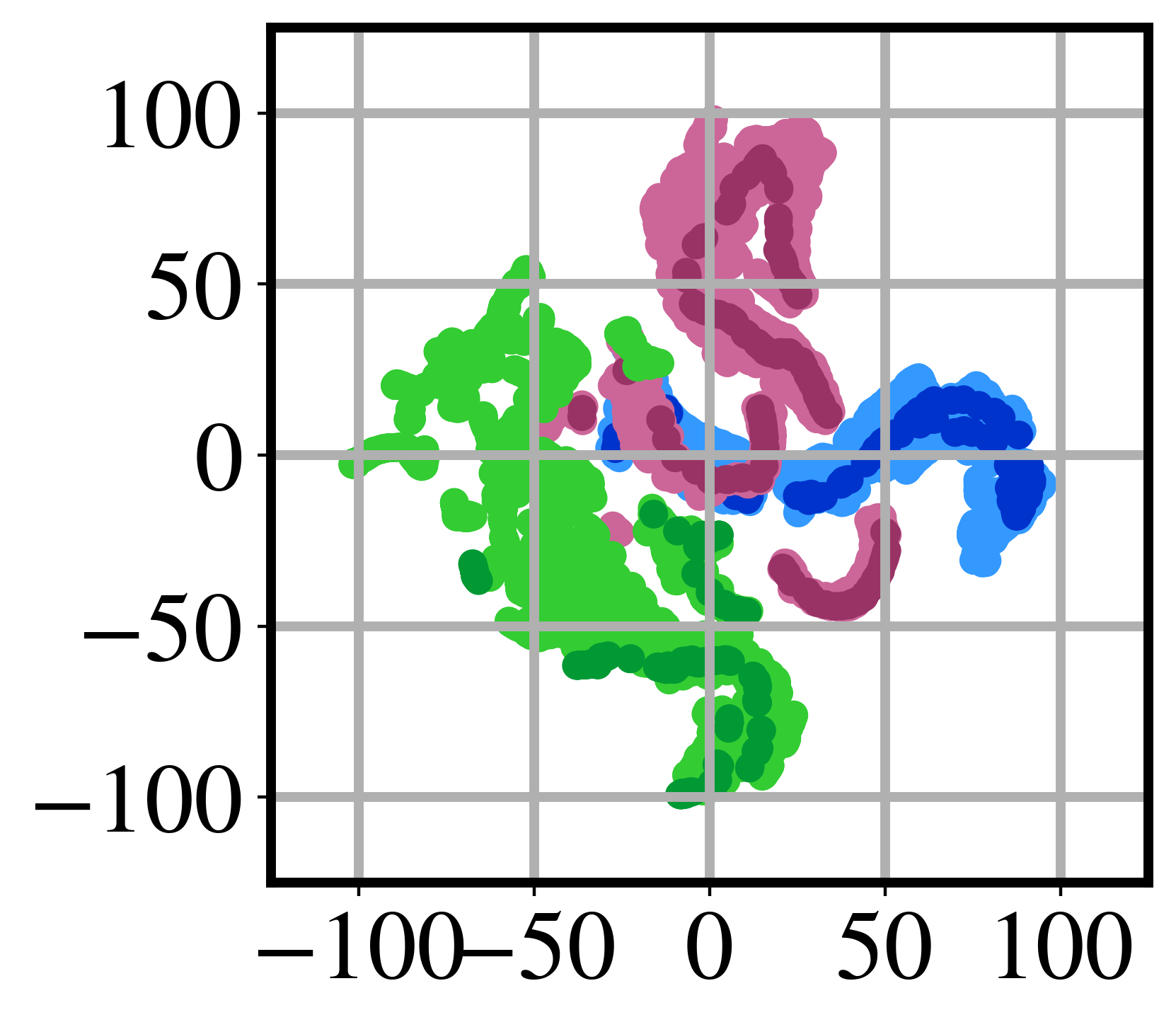}};
\node[inner sep=0pt] at (16,-3)
    {\includegraphics[width=.11\textwidth]{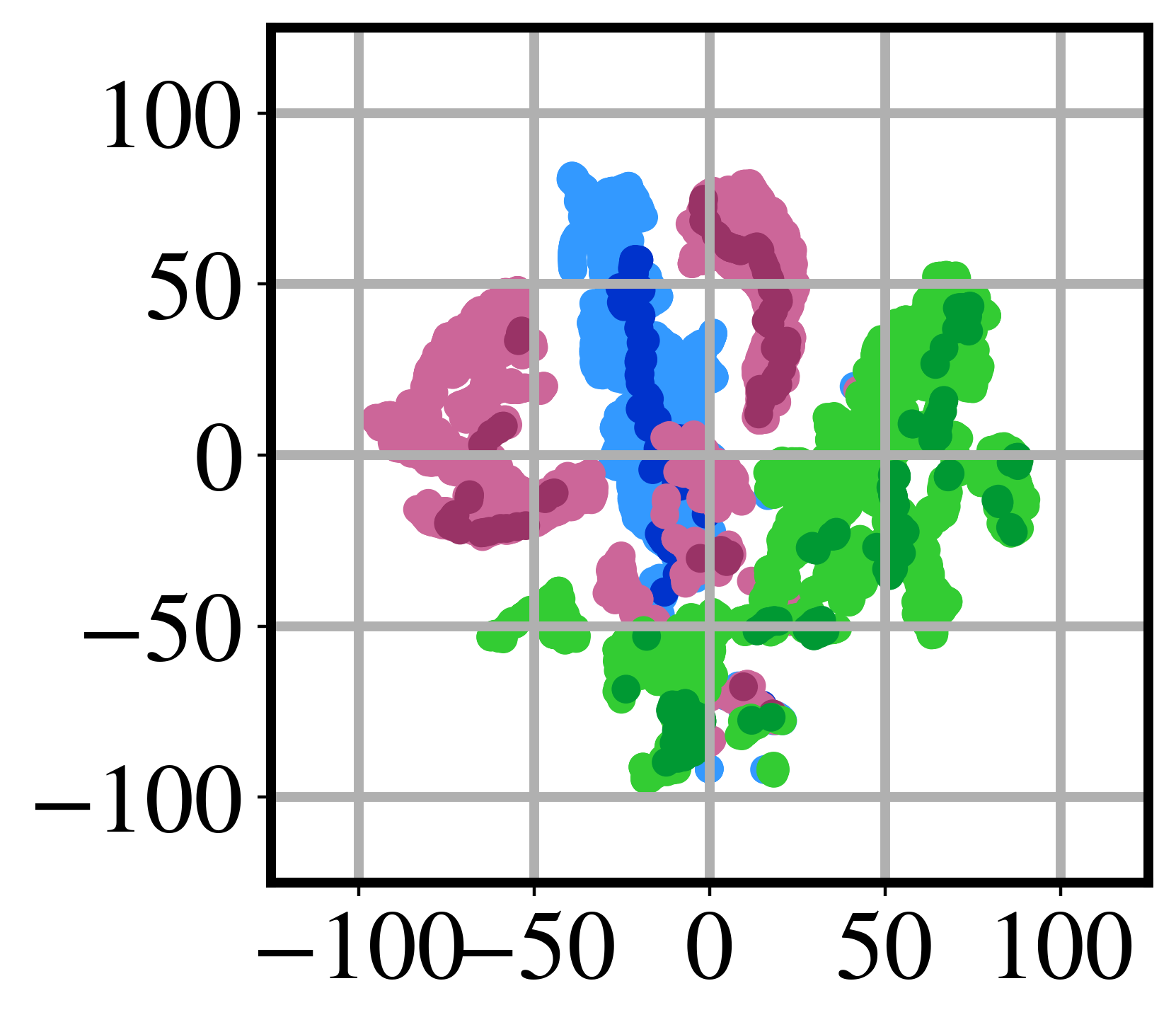}};
    
\node[inner sep=0pt] at (0,-5)
    {\includegraphics[width=.11\textwidth]{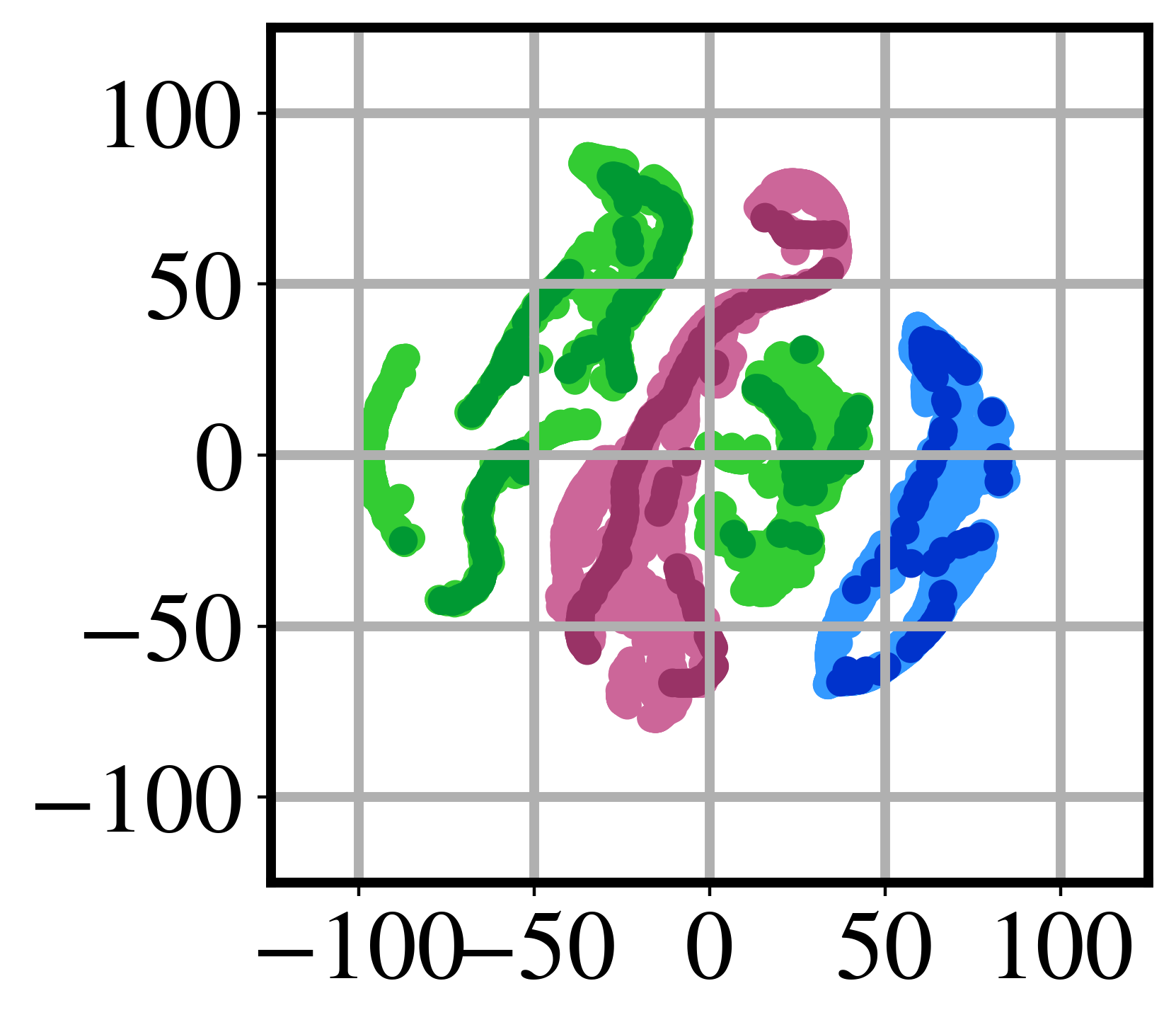}};
\node[inner sep=0pt] at (2,-5)
    {\includegraphics[width=.11\textwidth]{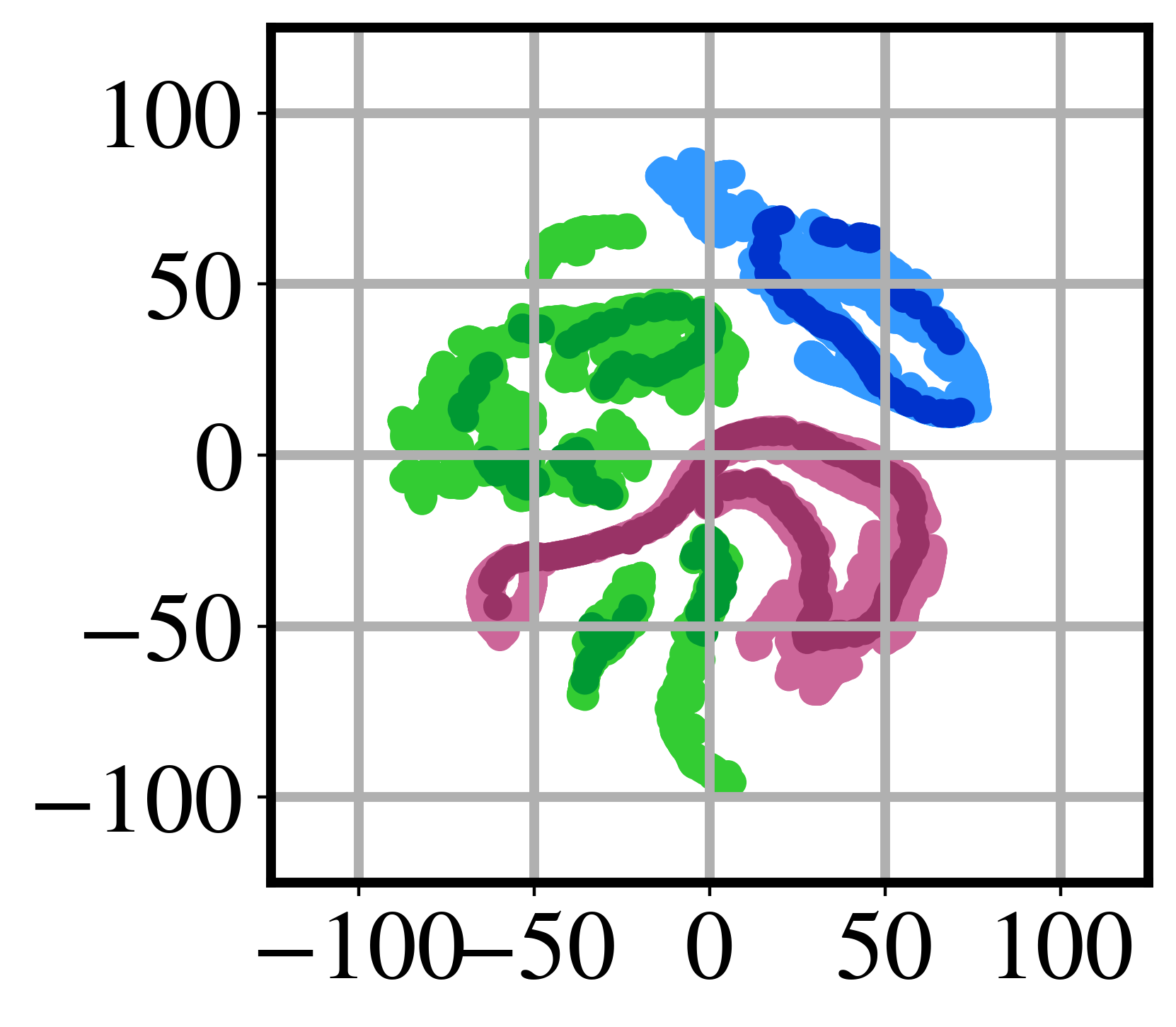}};
\node[inner sep=0pt] at (4,-5)
    {\includegraphics[width=.11\textwidth]{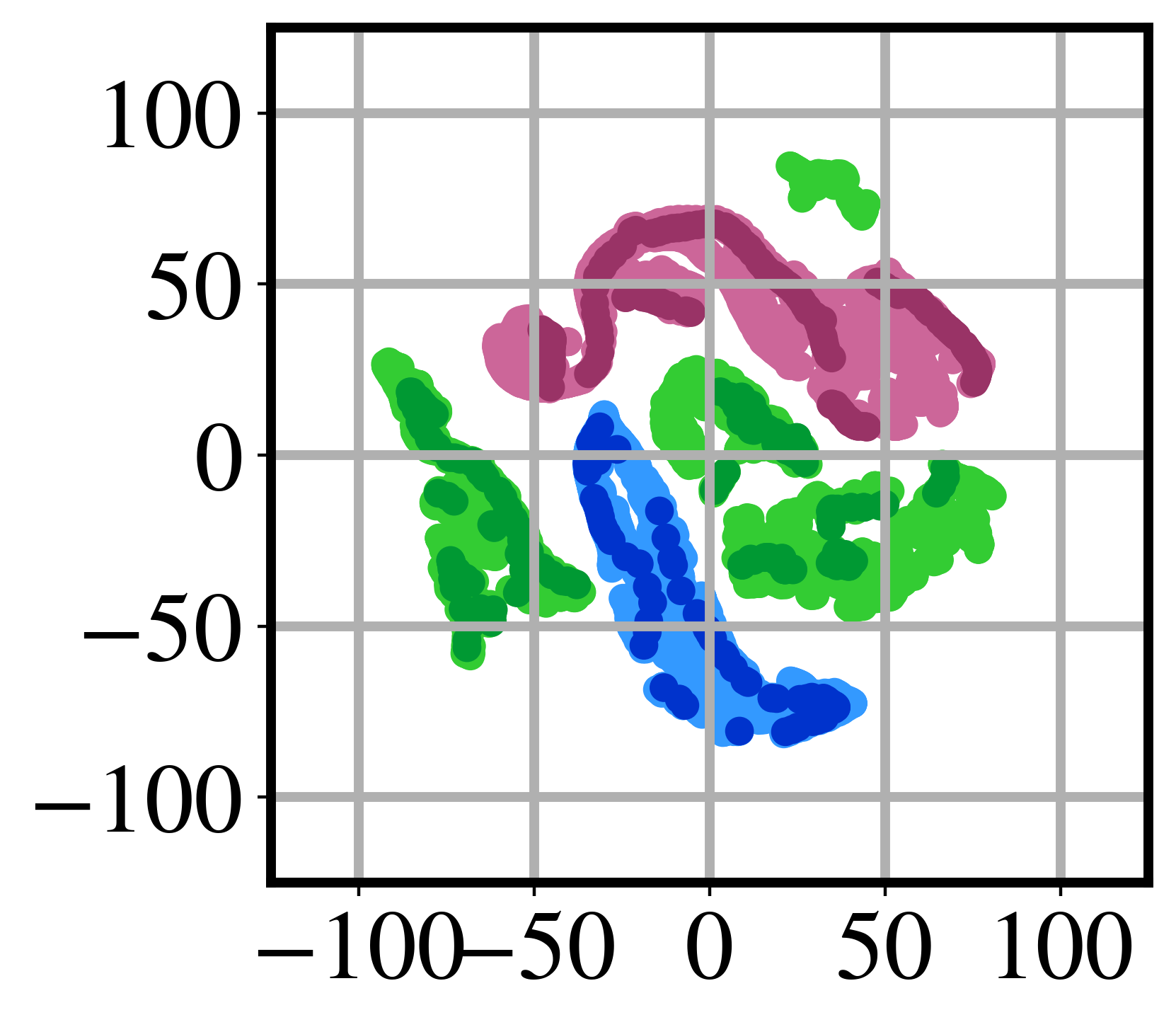}};
\node[inner sep=0pt] at (6,-5)
    {\includegraphics[width=.11\textwidth]{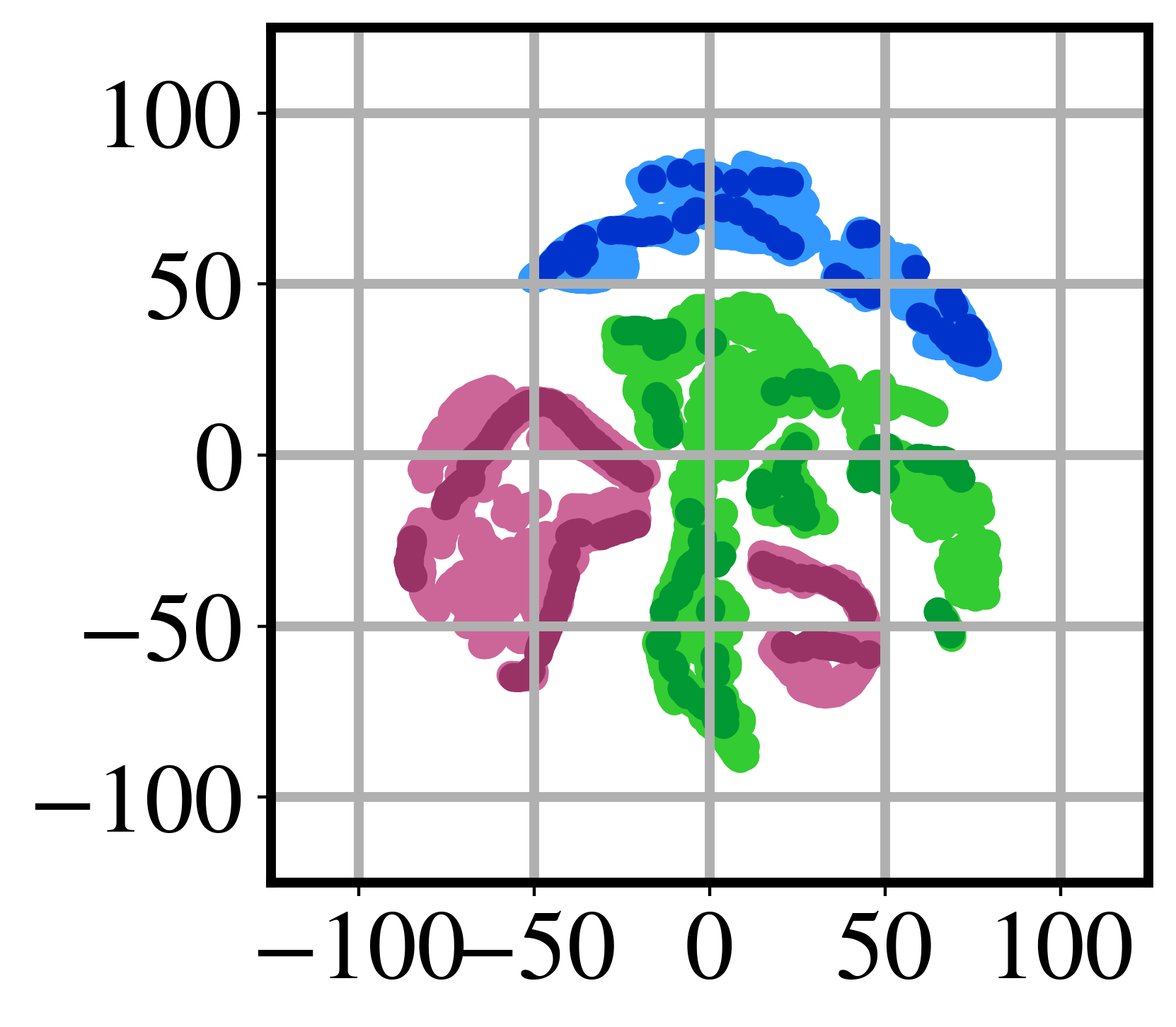}};
\node[inner sep=0pt] at (8,-5)
    {\includegraphics[width=.11\textwidth]{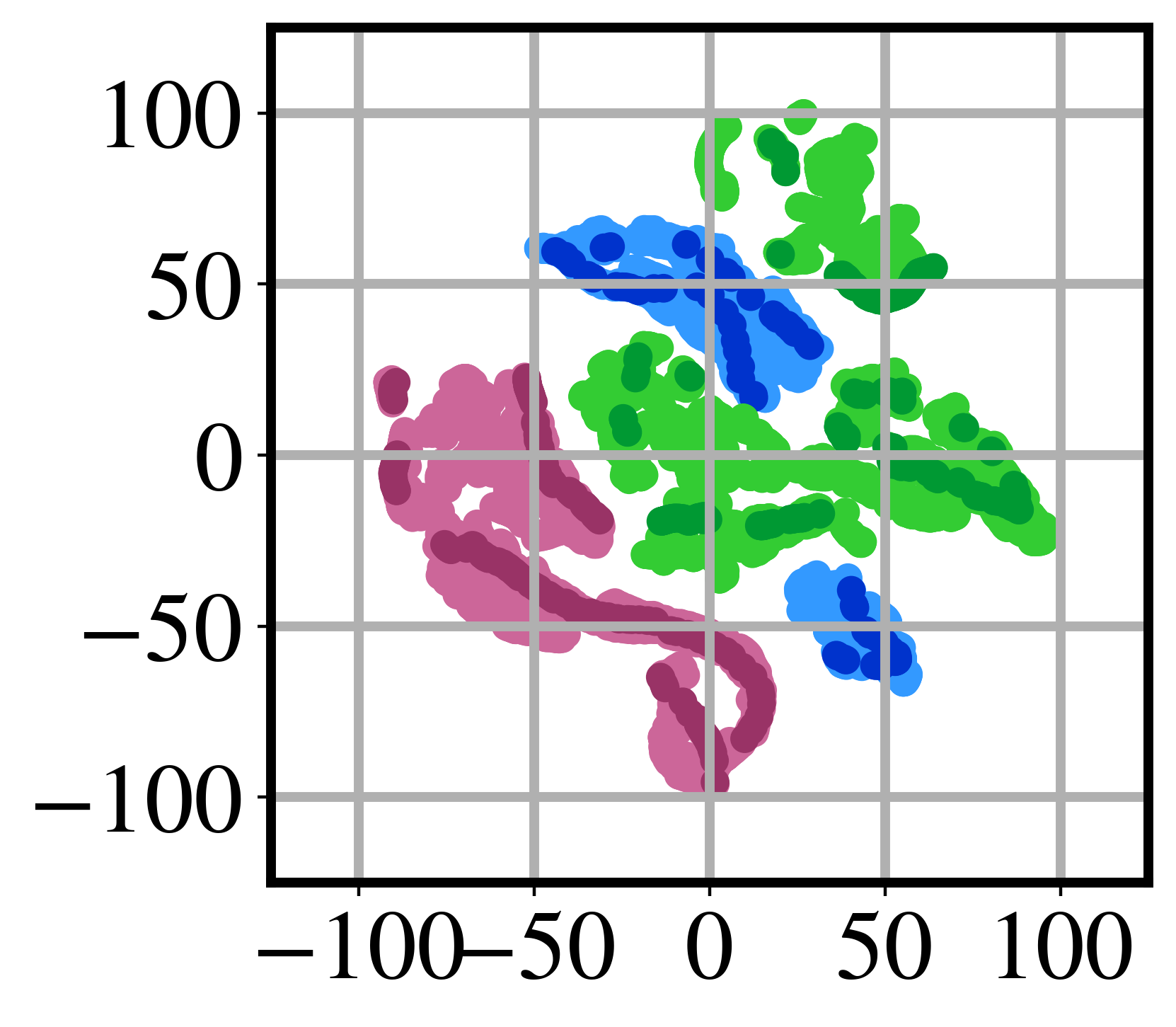}};
\node[inner sep=0pt] at (10,-5)
    {\includegraphics[width=.11\textwidth]{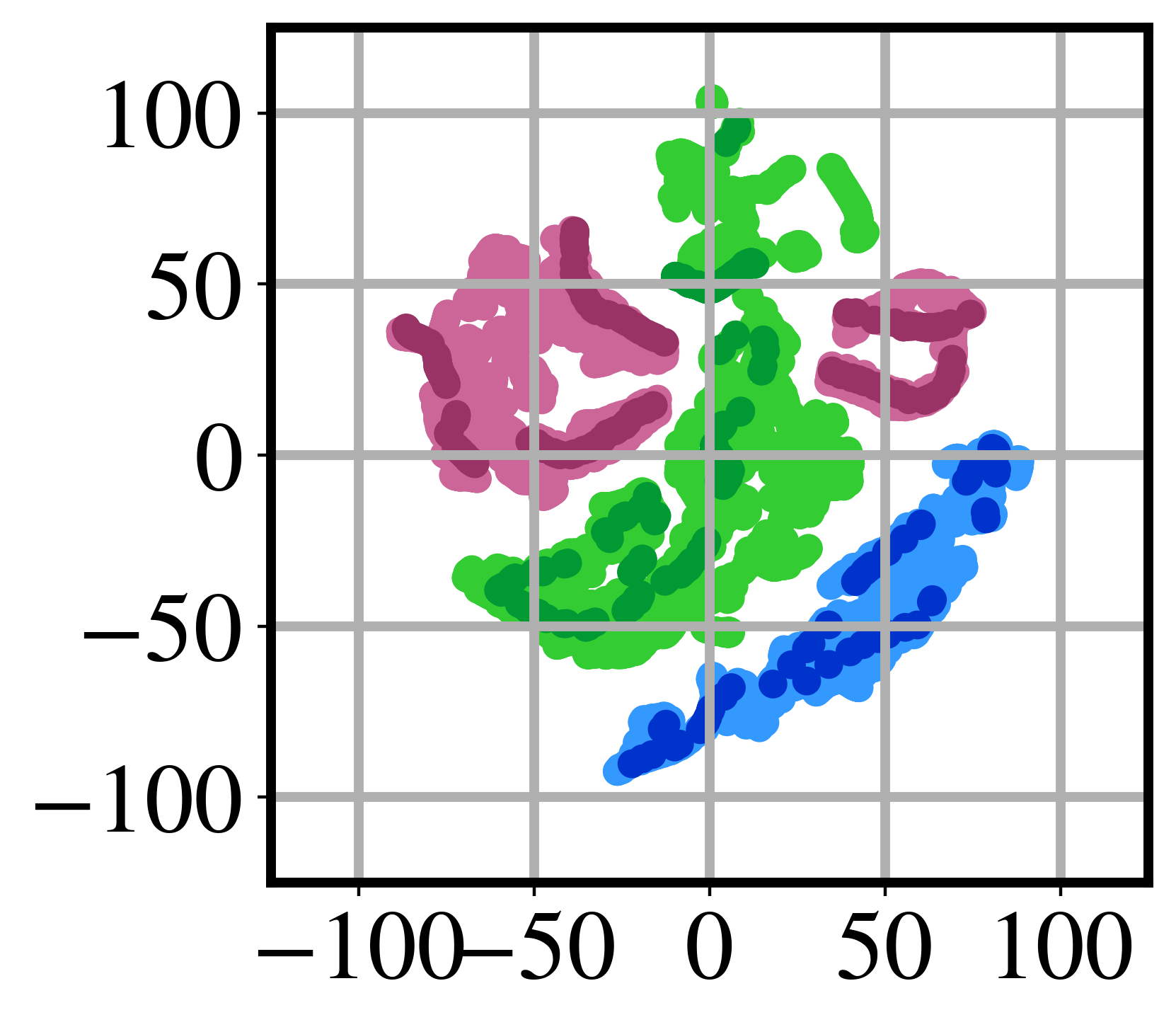}};
\node[inner sep=0pt] at (12,-5)
    {\includegraphics[width=.11\textwidth]{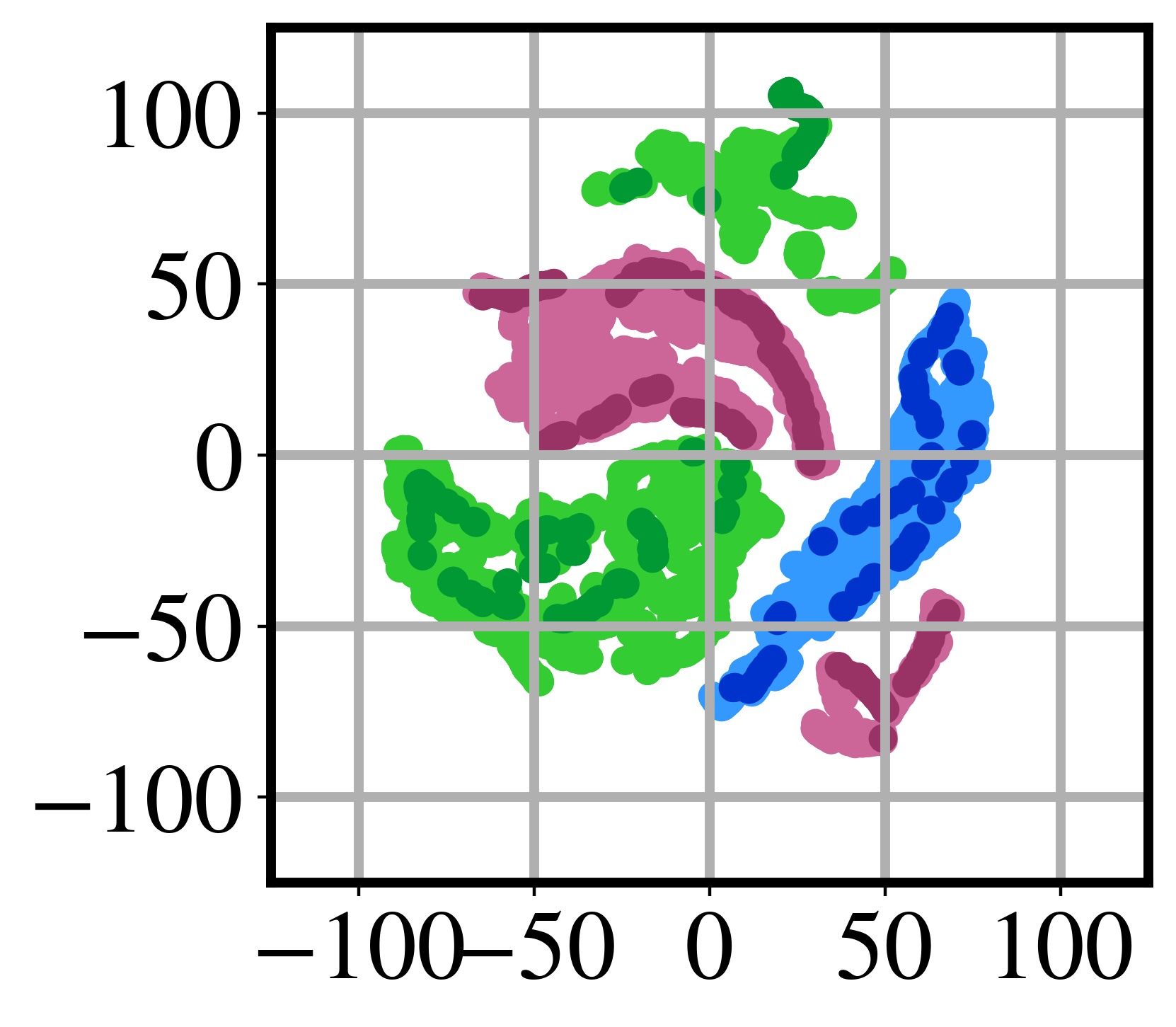}};
\node[inner sep=0pt] at (14,-5)
    {\includegraphics[width=.11\textwidth]{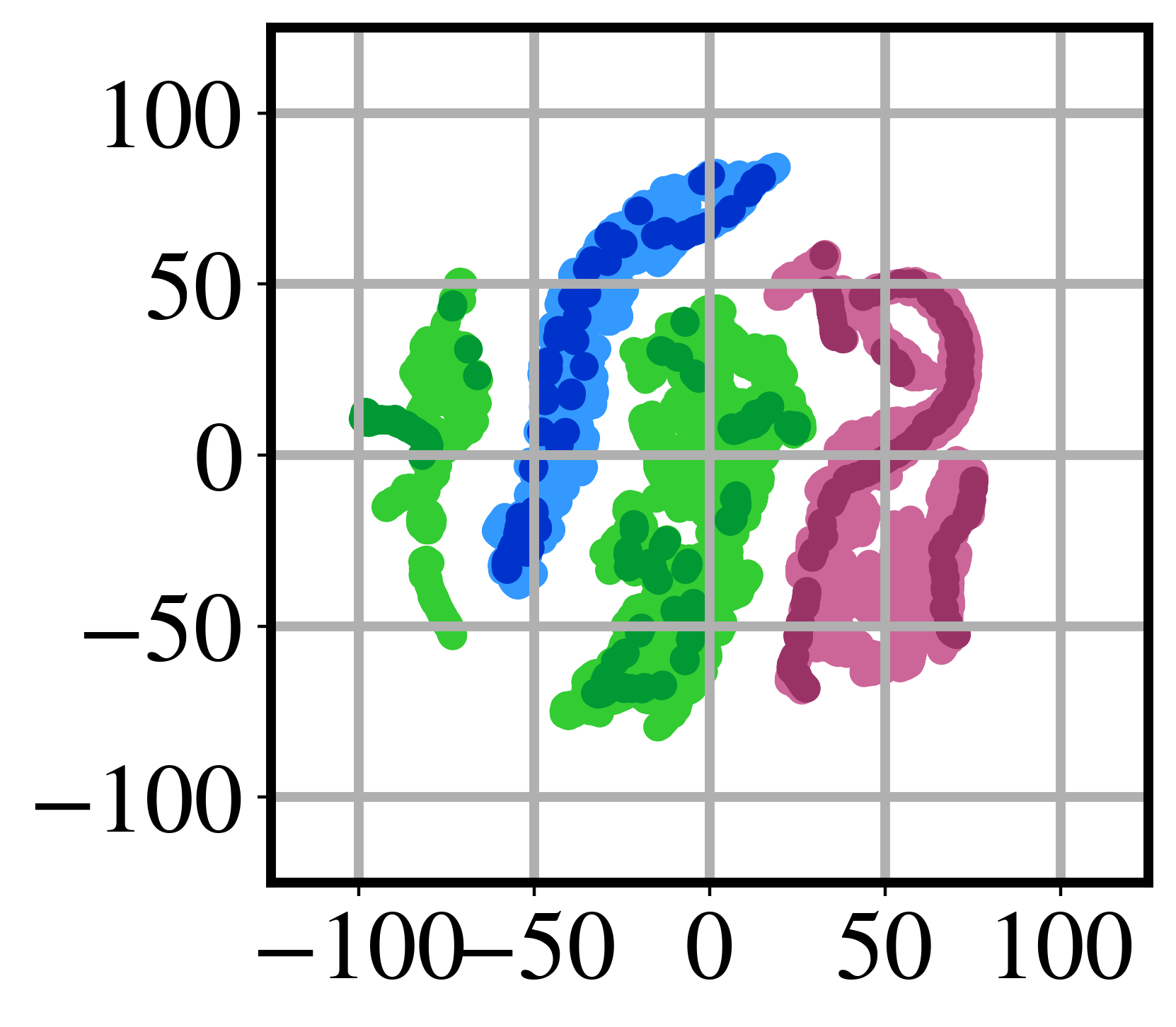}};
\node[inner sep=0pt] at (16,-5)
    {\includegraphics[width=.11\textwidth]{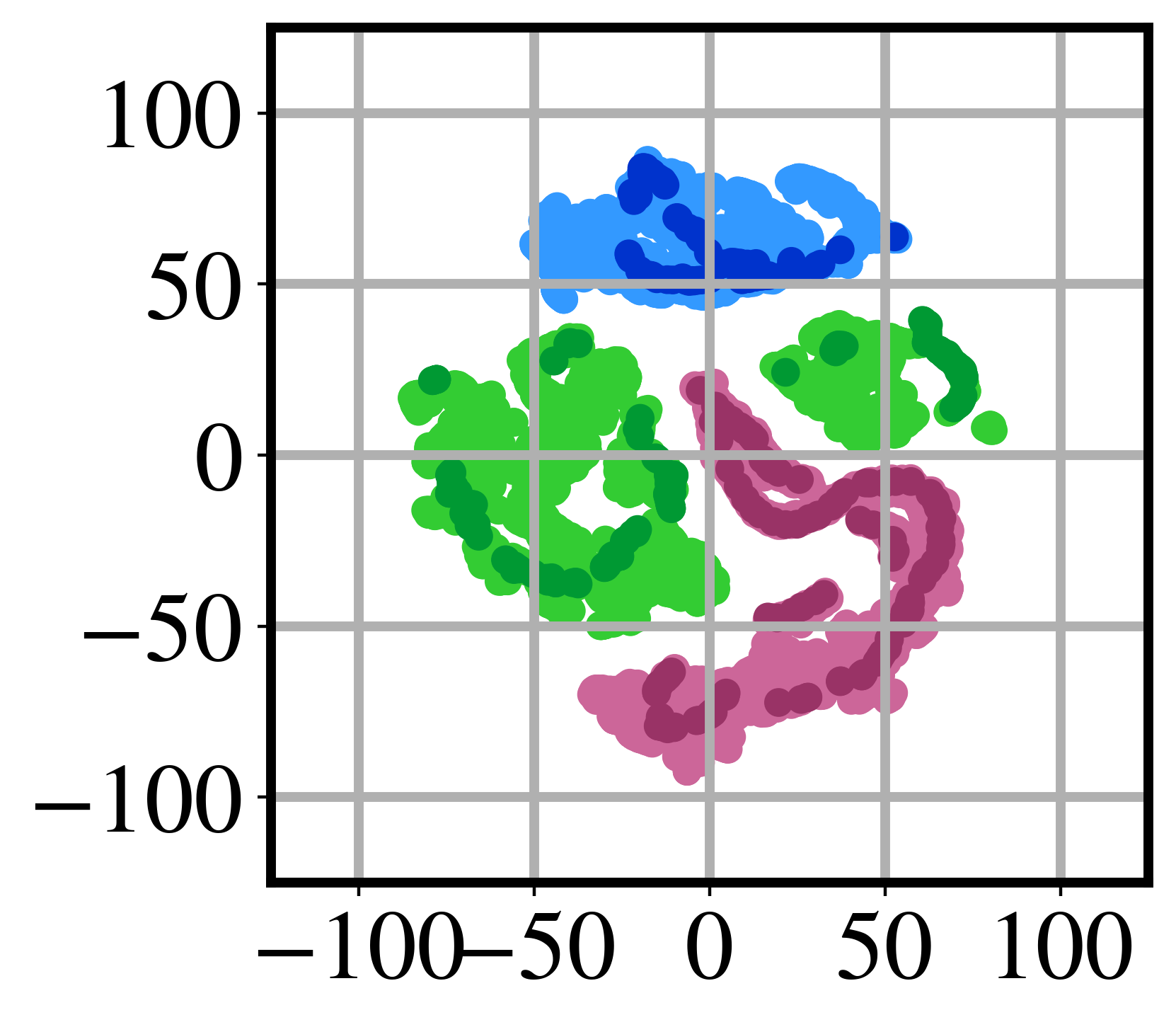}};
    
\node[inner sep=0pt] at (0,-7)
    {\includegraphics[width=.11\textwidth]{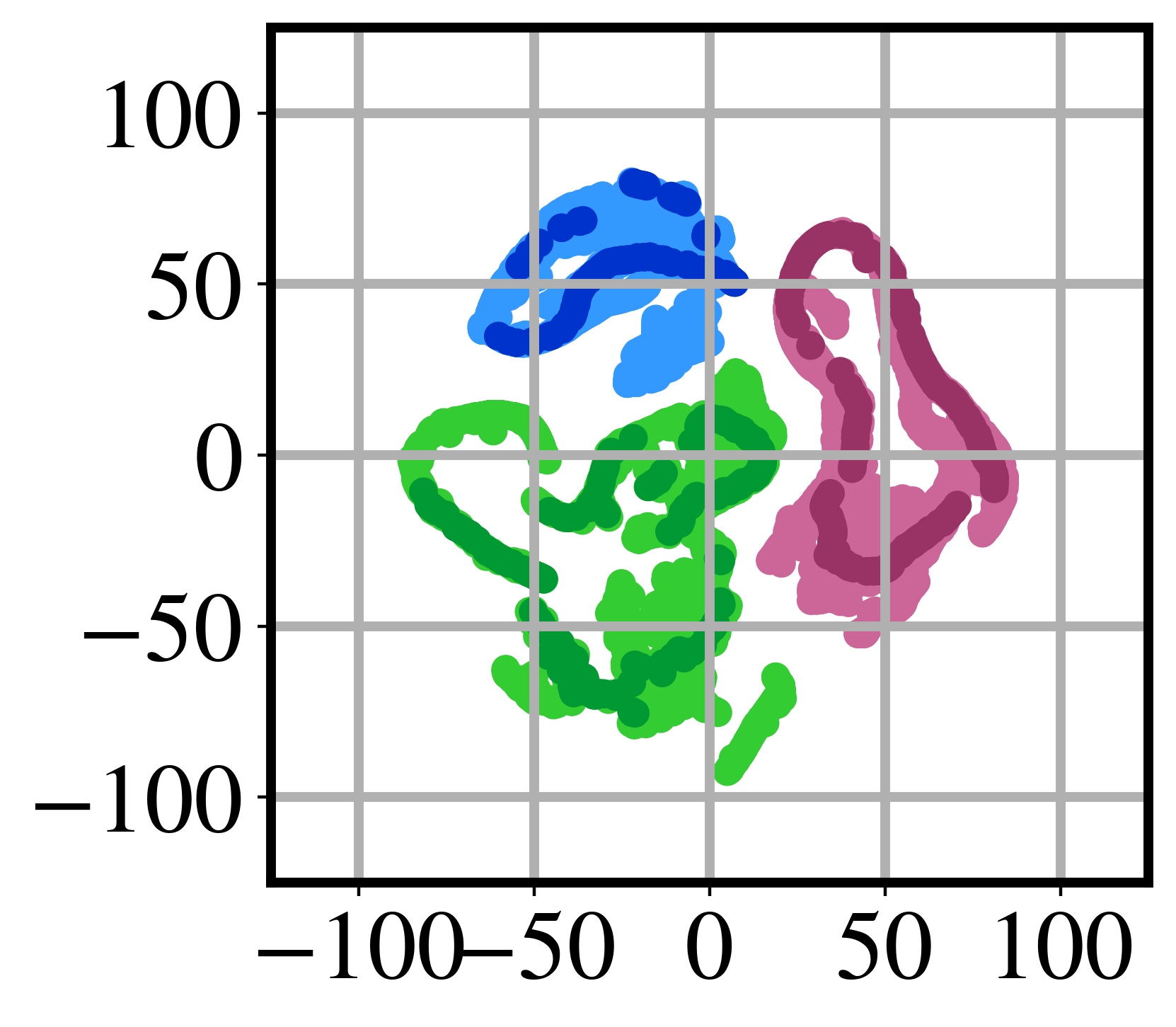}};
\node[inner sep=0pt] at (2,-7)
    {\includegraphics[width=.11\textwidth]{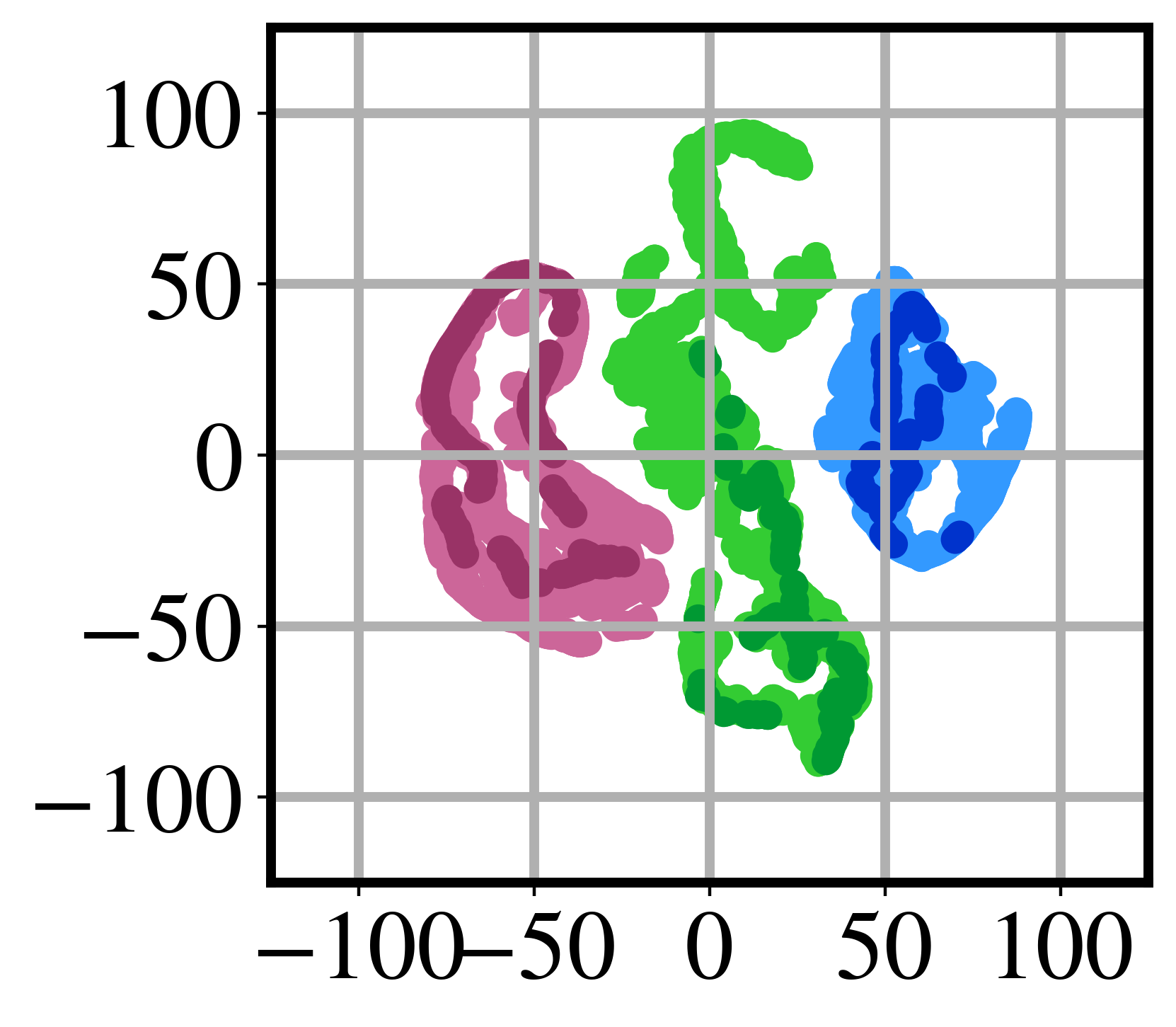}};
\node[inner sep=0pt] at (4,-7)
    {\includegraphics[width=.11\textwidth]{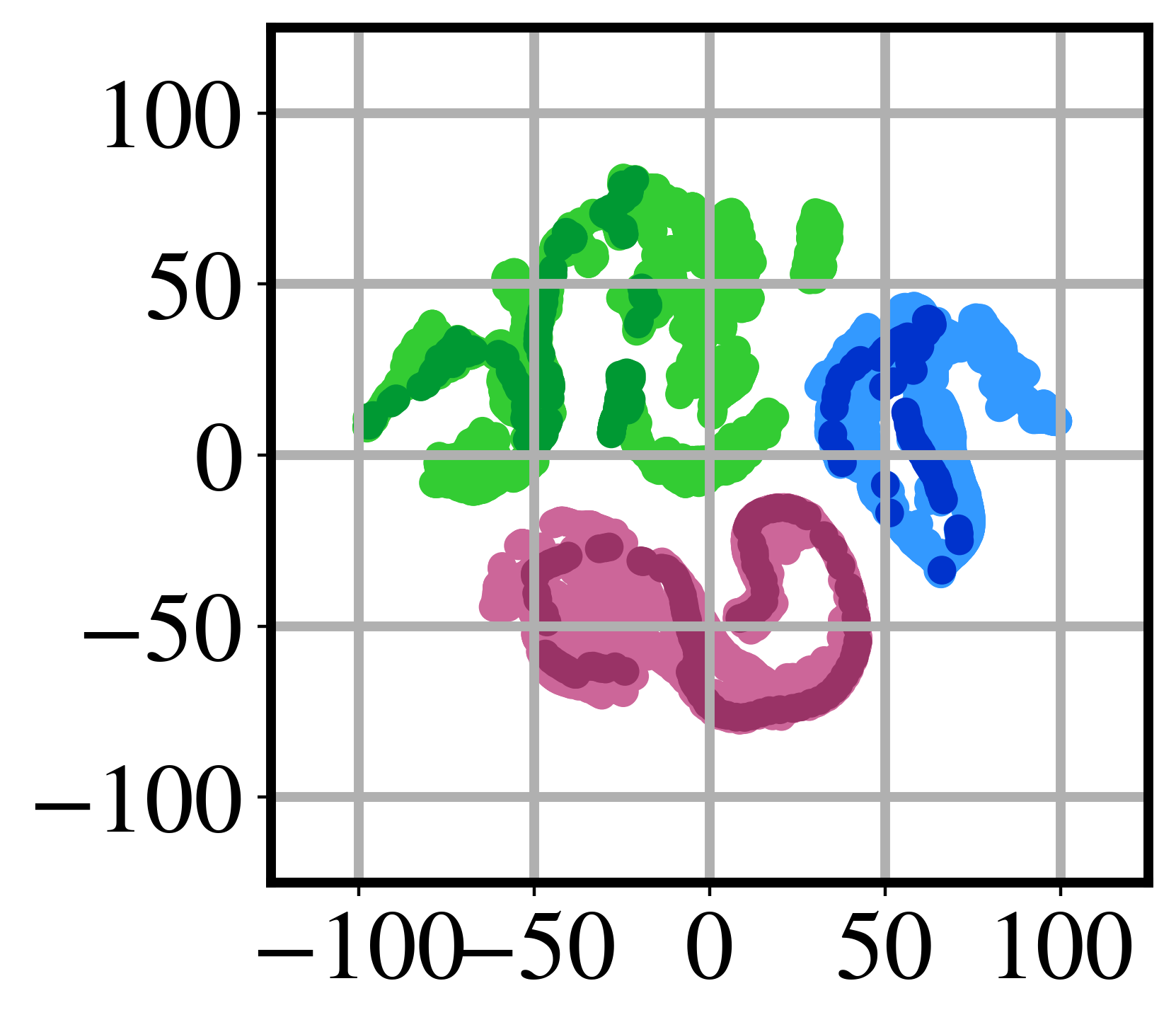}};
\node[inner sep=0pt] at (6,-7)
    {\includegraphics[width=.11\textwidth]{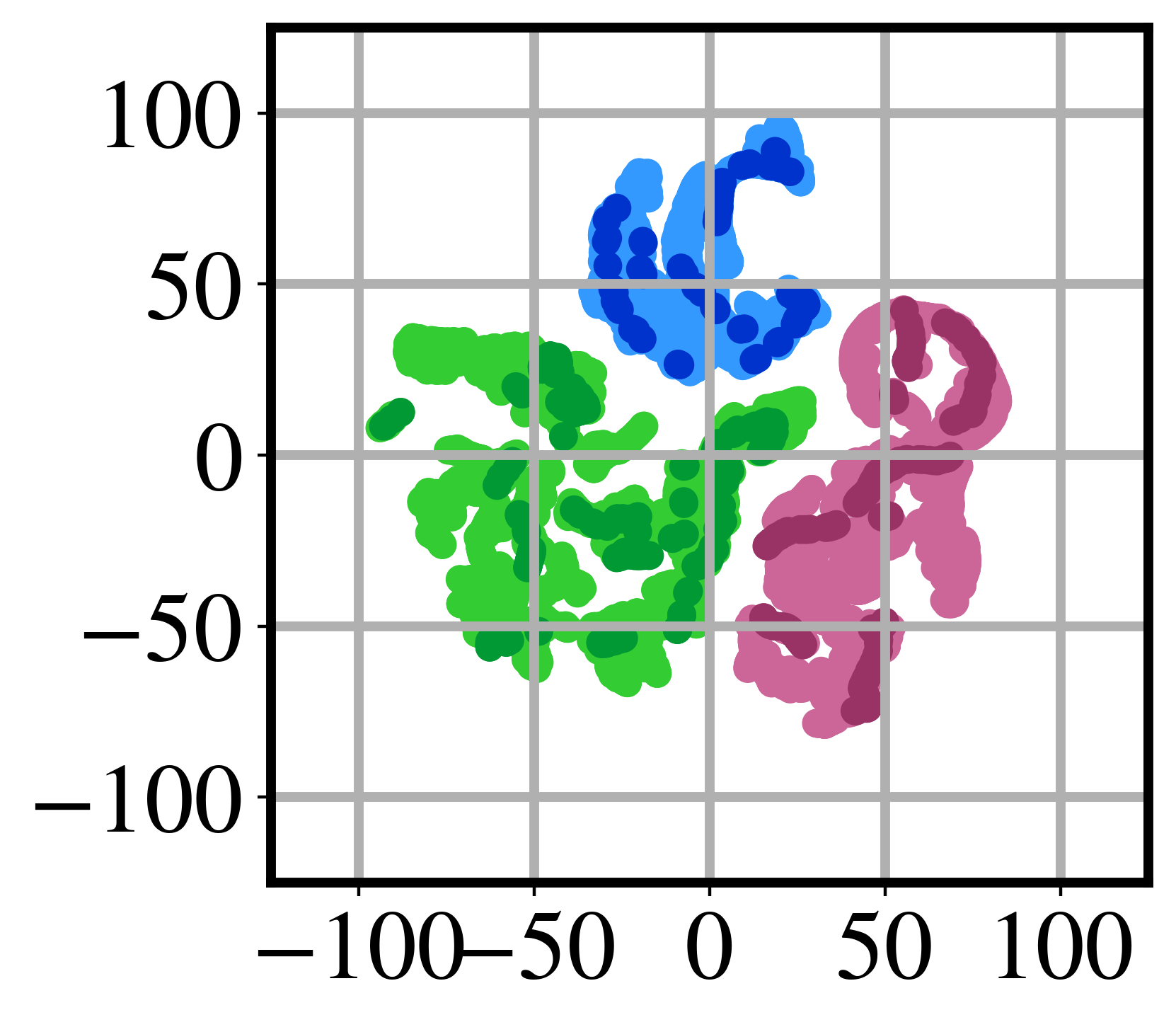}};
\node[inner sep=0pt] at (8,-7)
    {\includegraphics[width=.11\textwidth]{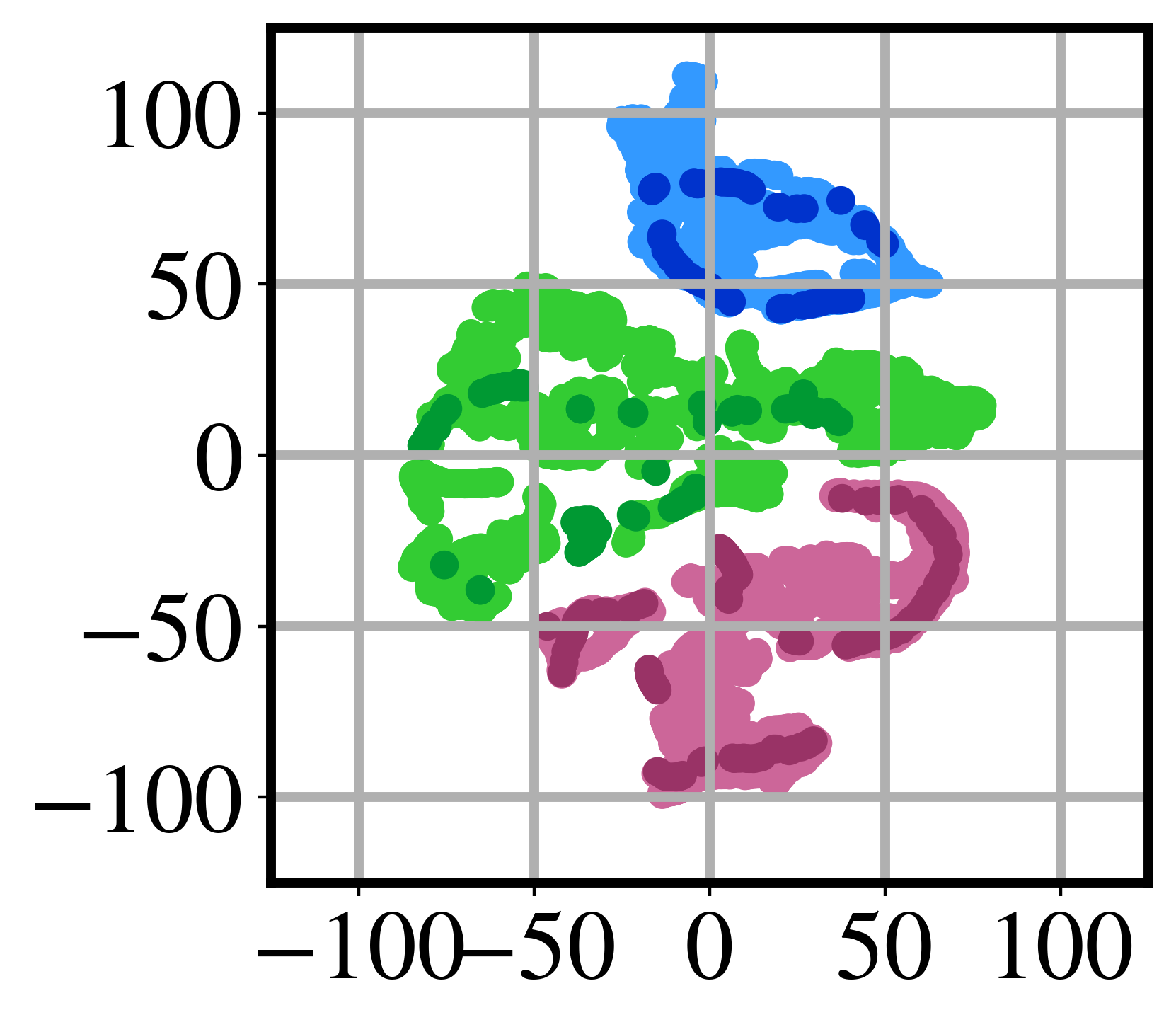}};
\node[inner sep=0pt] at (10,-7)
    {\includegraphics[width=.11\textwidth]{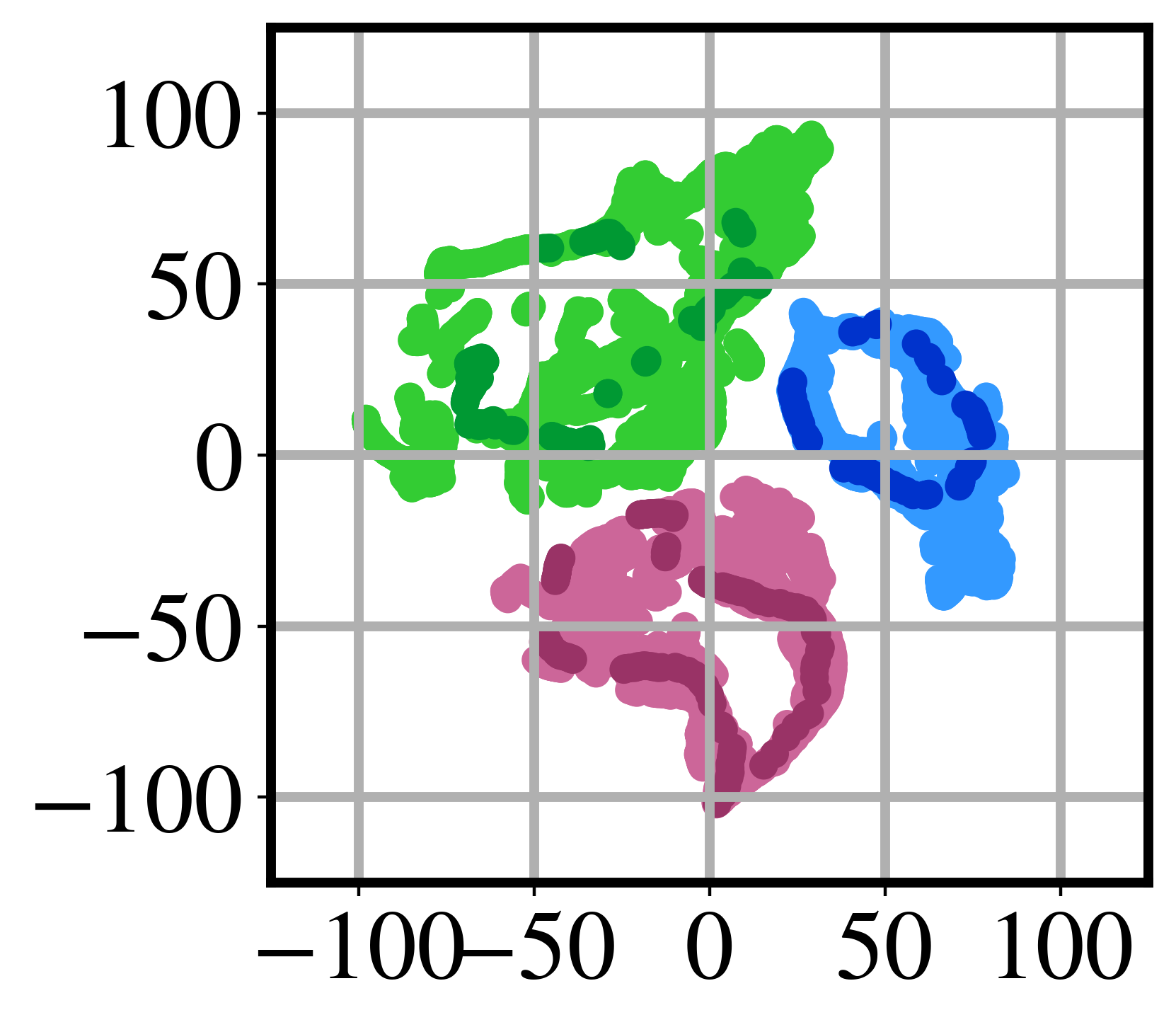}};
\node[inner sep=0pt] at (12,-7)
    {\includegraphics[width=.11\textwidth]{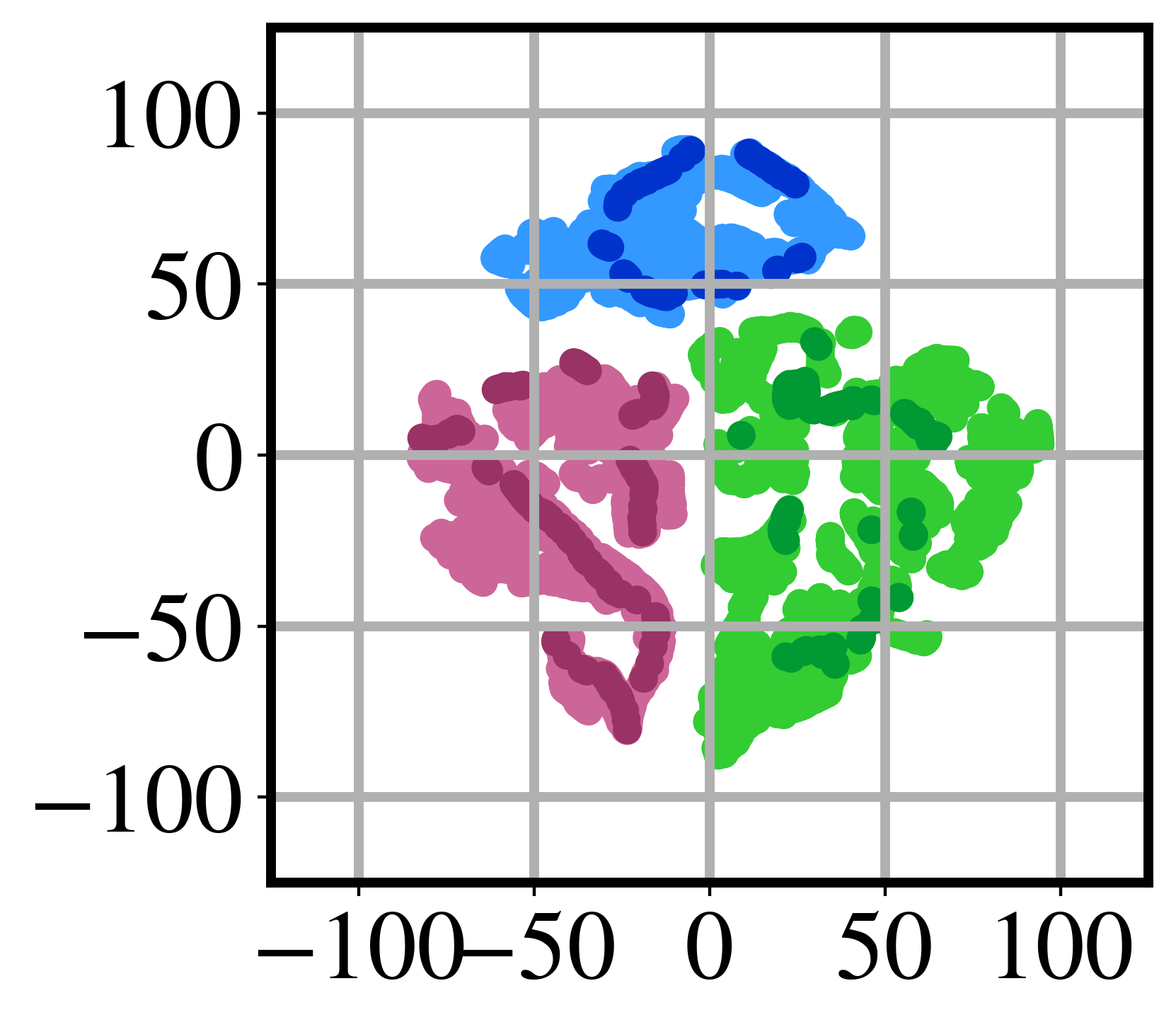}};
\node[inner sep=0pt] at (14,-7)
    {\includegraphics[width=.11\textwidth]{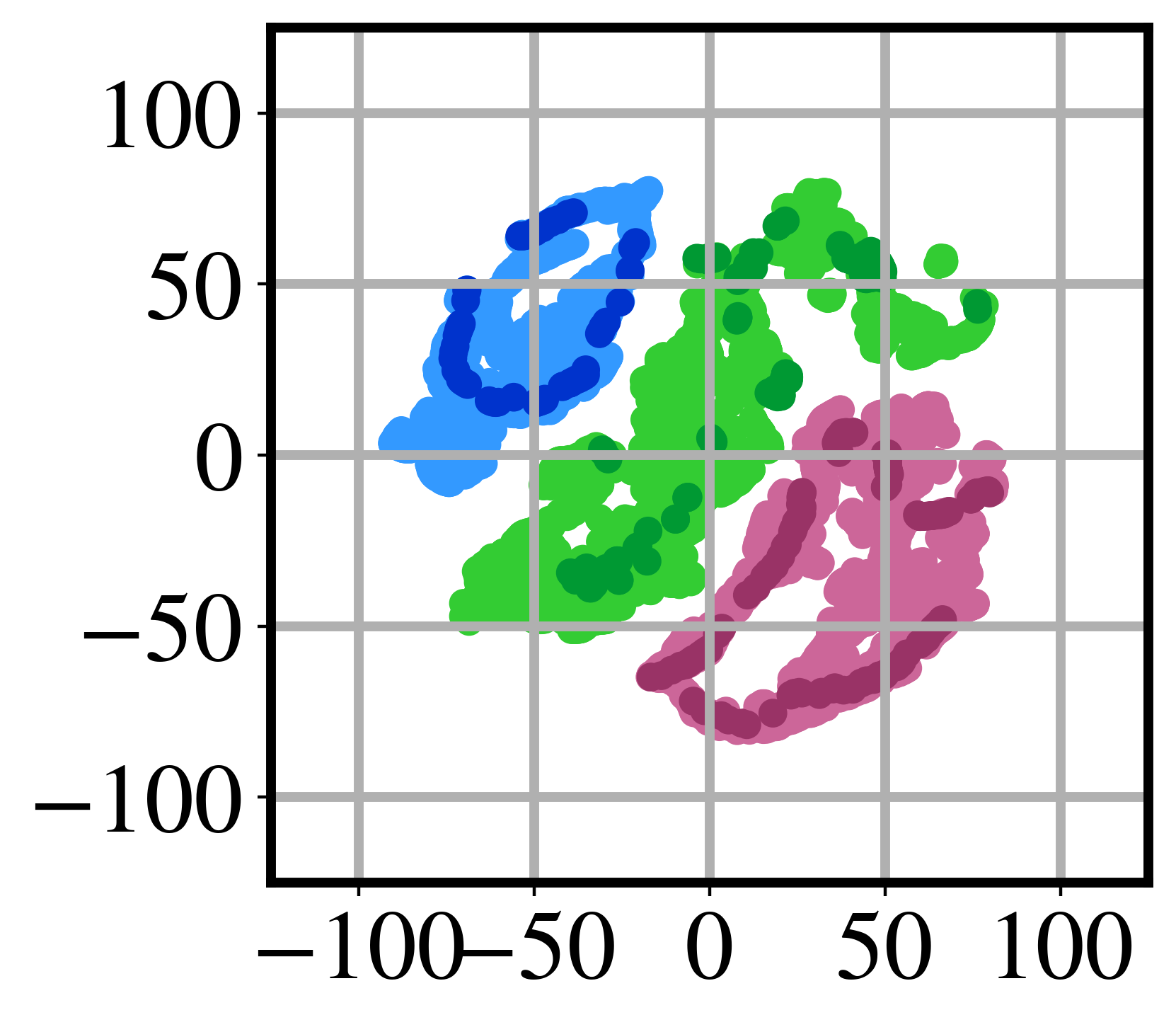}};
\node[inner sep=0pt] at (16,-7)
    {\includegraphics[width=.11\textwidth]{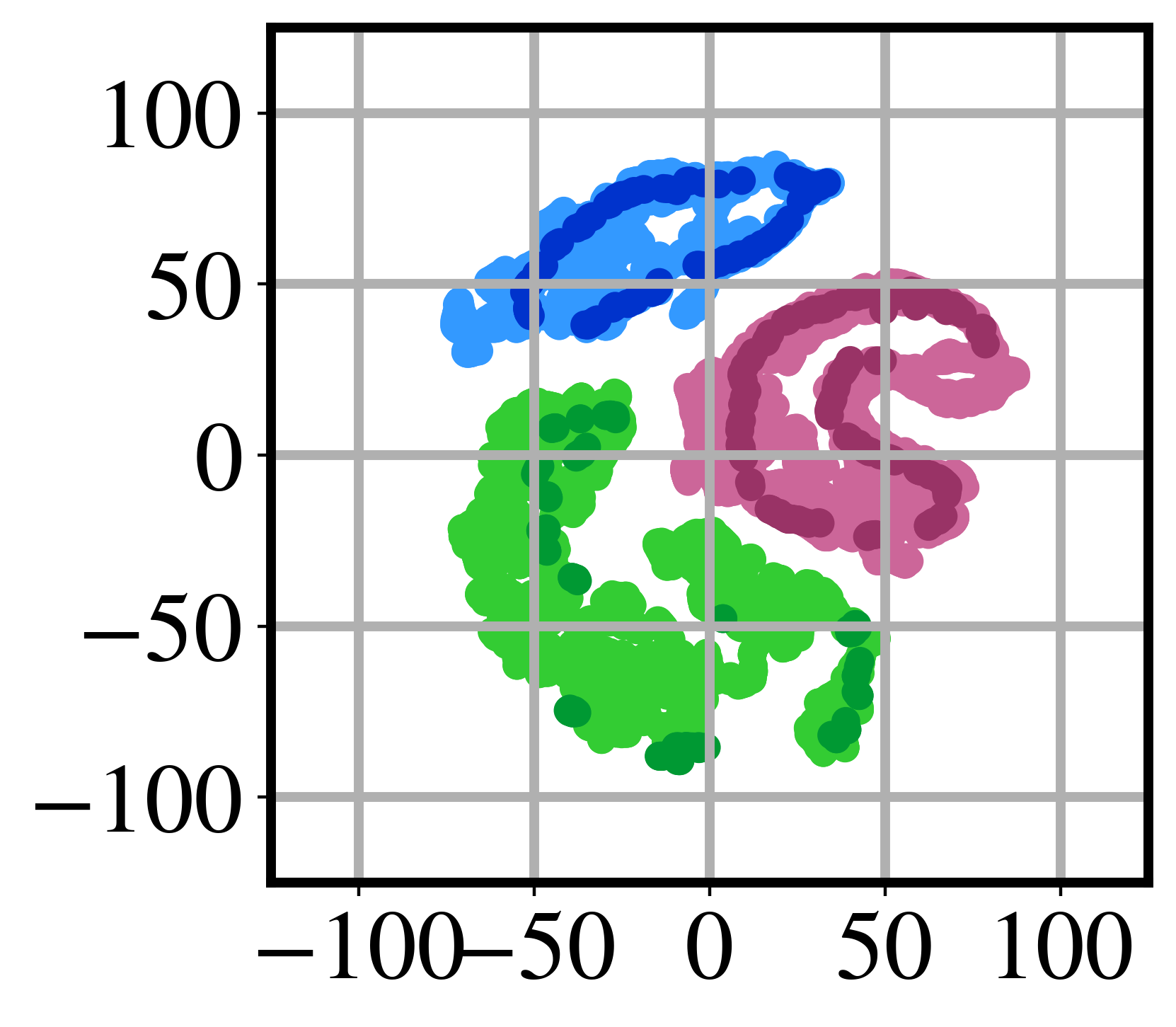}};

\node at (.175,5.05) {\scalebox{.85}{$z_{{s}_{1}} (d = 32)$}};
\node at (2.175,5.05) {\scalebox{.85}{$z_{{s}_{2}} (d = 64)$}};
\node at (4.175,5.05) {\scalebox{.85}{$z_{{s}_{3}} (d = 128)$}};
\node at (6.175,5.05) {\scalebox{.85}{$z_{{s}_{4}} (d = 256)$}};
\node at (8.175,5.05) {\scalebox{.85}{$z_{{s}_{5}} (d = 512)$}};
\node at (10.175,5.05) {\scalebox{.85}{$z_{{s}_{6}} (d = 256)$}};
\node at (12.175,5.05) {\scalebox{.85}{$z_{{s}_{7}} (d = 128)$}};
\node at (14.175,5.05) {\scalebox{.85}{$z_{{s}_{8}} (d = 64)$}};
\node at (16.175,5.05) {\scalebox{.85}{$z_{{s}_{9}} (d = 32)$}};

\node at (.175,-1.95 ) {\scalebox{.85}{$\Tilde{z}_{{s}_{1}} (d = 32)$}};
\node at (2.175,-1.95 ) {\scalebox{.85}{$\Tilde{z}_{{s}_{2}} (d = 64)$}};
\node at (4.175,-1.95 ) {\scalebox{.85}{$\Tilde{z}_{{s}_{3}} (d = 128)$}};
\node at (6.175,-1.95 ) {\scalebox{.85}{$\Tilde{z}_{{s}_{4}} (d = 256)$}};
\node at (8.175,-1.95 ) {\scalebox{.85}{$\Tilde{z}_{{s}_{5}} (d = 512)$}};
\node at (10.175,-1.95 ) {\scalebox{.85}{$\Tilde{z}_{{s}_{6}} (d = 256)$}};
\node at (12.175,-1.95 ) {\scalebox{.85}{$\Tilde{z}_{{s}_{7}} (d = 128)$}};
\node at (14.175,-1.95 ) {\scalebox{.85}{$\Tilde{z}_{{s}_{8}} (d = 64)$}};
\node at (16.175,-1.95 ) {\scalebox{.85}{$\Tilde{z}_{{s}_{9}} (d = 32)$}};

\node[rotate=90] at (-1.4,4.075) {\scalebox{.85}{Att-UNet}};
\node[rotate=90] at (-1.1,4.075) {\scalebox{.85}{Shared}};
\node[rotate=90] at (-1.4,2.075) {\scalebox{.85}{Att-UNet}};
\node[rotate=90] at (-1.1,2.075) {\scalebox{.85}{DSL}};
\node[rotate=90] at (-1.4,.075) {\scalebox{.85}{Att-UNet}};
\node[rotate=90] at (-1.1,.075) {\scalebox{.85}{$\text{DSL}+\mathcal{L}_{\text{MSC}}$}};

\node[rotate=90] at (-1.4,-2.925) {\scalebox{.85}{Auto-encoder}};
\node[rotate=90] at (-1.1,-2.925) {\scalebox{.85}{Shared}};
\node[rotate=90] at (-1.4,-4.925) {\scalebox{.85}{Auto-encoder}};
\node[rotate=90] at (-1.1,-4.925) {\scalebox{.85}{DSL}};
\node[rotate=90] at (-1.4,-6.925) {\scalebox{.85}{Auto-encoder}};
\node[rotate=90] at (-1.1,.-6.925) {\scalebox{.85}{$\text{DSL}+\mathcal{L}_{\text{MSC}}$}};

\node[inner sep=0pt] at (8.175,-8.4)
    {\includegraphics[width=.6\textwidth]{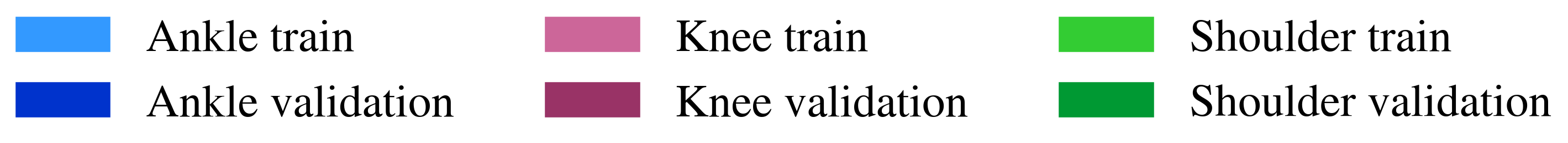}};

\draw[line width=.15mm] (2.95,-7.95) rectangle (13.4,-8.85);
\draw[line width=.15mm] (-.75,-1.25) rectangle (16.75,-1.25);

\end{tikzpicture}
\end{adjustbox}
  \caption{\textbf{Visual comparison of the shared representations learnt in shared, DSL and $\text{DSL}+\mathcal{L}_{\text{MSC}}$ learning schemes}. Architectures encompassed Att-UNet \citep{oktay_attention_2018} and the multi-joint auto-encoder. The multi-scale contrastive regularization $\mathcal{L}_{\text{MSC}}$ promoted intra-domain cohesion and inter-domain margins in embedded spaces at each scale. This visualization was obtained using the t-SNE algorithm \citep{maaten_visualizing_2008} in which each colored dot represented a 2D MR slice or segmentation mask from the training or validation set of the ankle, knee, or shoulder datasets.}
  \label{fig:visualization_contrastive_regularization}
\end{figure*}
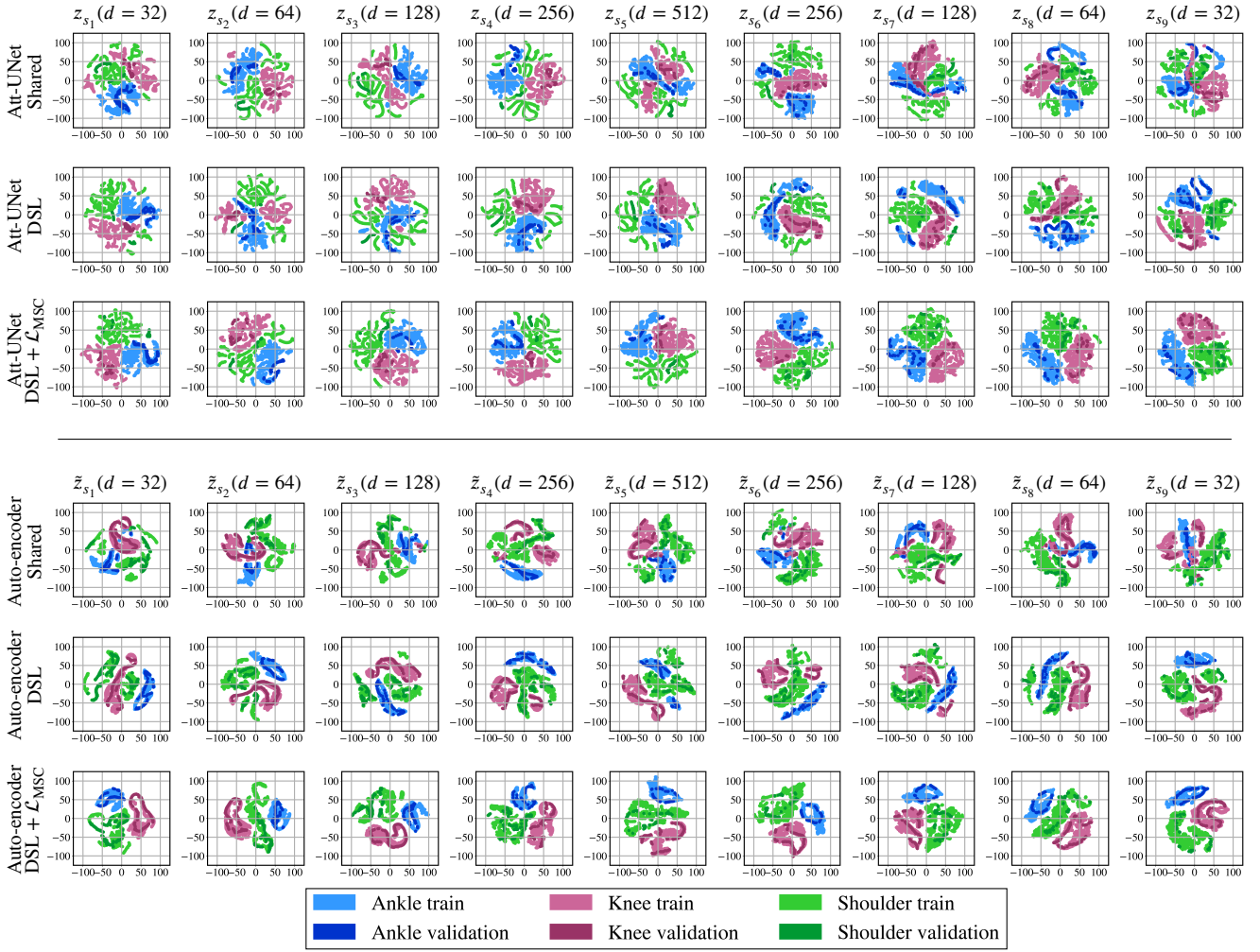

The visualization of the shared representation provided an indirect analysis of the inference process of deep neural networks and a qualitative validation of the benefits of the additional multi-scale contrastive regularization on both intra-domain cohesion and inter-domain separation (Fig. \ref{fig:visualization_contrastive_regularization}). In both Att-UNet and auto-encoder networks, the shared representation learnt using shared and DSL schemes did not present margins between domains. More specifically, shared models presented mixed features with most discriminative domain disentanglement in the network bottleneck ($s_{5}$) which corresponded to the higher dimensional vector space ($d = 512$) allowing more robust differentiation between domains. On the contrary, the addition of the contrastive regularization led to distinctive domain-specific clusters at each scale of both networks. Hence, the shared representations of our proposed neural networks were invariant to local variations and preserved the category of the input domain through the different scales of the models. Moreover, the generalization capabilities of the networks were visually attested as validation data points were located inside their respective domain clusters. 

\begin{table*}[ht!]
\caption{Quantitative analysis based on cosine similarity of the shared representations learnt by Att-UNet in shared, $\text{DSL}+\mathcal{L}_{\text{SSC}}$, and $\text{DSL}+\mathcal{L}_{\text{MSC}}$ strategies using ankle, knee, and shoulder datasets. The included scales correspond to the encoder first layer ($s_1$), network bottleneck ($s_5$), and decoder last layer ($s_9$). Mean and standard deviation similarity measures are reported.}
\centering
    \begin{tabular}{|P{.35cm}|P{.65cm}|P{1.1cm}||P{1.1cm}|P{1.1cm}|P{1.1cm}||P{1.1cm}|P{1.1cm}|P{1.1cm}||P{1.1cm}|P{1.1cm}|P{1.1cm}|} 
    \hline
    \multicolumn{3}{|c||}{\multirow{2}{*}{\shortstack{Cosine\\similarity}}} & \multicolumn{3}{c||}{$z_{{s}_{1}} \: (d = 32)$} & \multicolumn{3}{c||}{$z_{{s}_{5}} \: (d = 512)$} & \multicolumn{3}{c|}{$z_{{s}_{9}} \: (d = 32)$} \\\cline{4-12}
    \multicolumn{3}{|c||}{} & Ankle & Knee & Shoulder & Ankle & Knee & Shoulder & Ankle & Knee & Shoulder \\
    \hline\hline
    
    \multirow{9}{*}{\rotatebox[origin=c]{90}{Att-UNet}} & \multirow{3}{*}{\rotatebox[origin=c]{90}{Shared}} & Ankle & \cellcolor{Dandelion!84} 0.84\scriptsize{(0.16)} & \cellcolor{Dandelion!84} 0.84\scriptsize{(0.11)} & \cellcolor{Dandelion!83} 0.83\scriptsize{(0.10)} & \cellcolor{Dandelion!82} 0.82\scriptsize{(0.16)} & \cellcolor{Dandelion!58} 0.58\scriptsize{(0.20)} & \cellcolor{Dandelion!69} 0.69\scriptsize{(0.12)} & \cellcolor{Dandelion!86} 0.86\scriptsize{(0.14)} & \cellcolor{Dandelion!73} 0.73\scriptsize{(0.17)} & \cellcolor{Dandelion!86} 0.86\scriptsize{(0.09)} \\\cline{3-12}
    & & Knee & \cellcolor{Dandelion!84} 0.84\scriptsize{(0.11)} & \cellcolor{Dandelion!93} 0.93\scriptsize{(0.07)} & \cellcolor{Dandelion!93} 0.93\scriptsize{(0.06)} & \cellcolor{Dandelion!58} 0.58\scriptsize{(0.20)} & \cellcolor{Dandelion!88} 0.88\scriptsize{(0.12)} & \cellcolor{Dandelion!71} 0.71\scriptsize{(0.11)} & \cellcolor{Dandelion!73} 0.73\scriptsize{(0.17)} & \cellcolor{Dandelion!90} 0.90\scriptsize{(0.10)} & \cellcolor{Dandelion!83} 0.83\scriptsize{(0.12)} \\\cline{3-12}
    & & Shoulder & \cellcolor{Dandelion!83} 0.83\scriptsize{(0.10)} & \cellcolor{Dandelion!93} 0.93\scriptsize{(0.06)} & \cellcolor{Dandelion!97} 0.97\scriptsize{(0.03)} & \cellcolor{Dandelion!69} 0.69\scriptsize{(0.12)} & \cellcolor{Dandelion!71} 0.71\scriptsize{(0.11)} & \cellcolor{Dandelion!93} 0.93\scriptsize{(0.05)} & \cellcolor{Dandelion!86} 0.86\scriptsize{(0.09)} & \cellcolor{Dandelion!83} 0.83\scriptsize{(0.12)} & \cellcolor{Dandelion!95} 0.95\scriptsize{(0.05)} \\\hhline{|~|===========}
    
    & \multirow{3}{*}{\rotatebox[origin=c]{90}{\shortstack{DSL\\$+\mathcal{L}_{\text{SSC}}$}}} & Ankle & \cellcolor{Dandelion!84} 0.84\scriptsize{(0.17)} & \cellcolor{Dandelion!83} 0.83\scriptsize{(0.11)} & \cellcolor{Dandelion!84} 0.84\scriptsize{(0.08)} & \cellcolor{Dandelion!99} 0.99\scriptsize{(0.01)} & \cellcolor{Dandelion!25} 0.25\scriptsize{(0.02)} & \cellcolor{Dandelion!21} 0.21\scriptsize{(0.02)} & \cellcolor{Dandelion!90} 0.90\scriptsize{(0.11)} & \cellcolor{Dandelion!82} 0.82\scriptsize{(0.10)} & \cellcolor{Dandelion!83} 0.83\scriptsize{(0.08)} \\\cline{3-12}
    & & Knee & \cellcolor{Dandelion!83} 0.83\scriptsize{(0.11)} & \cellcolor{Dandelion!93} 0.93\scriptsize{(0.07)} & \cellcolor{Dandelion!93} 0.93\scriptsize{(0.05)} & \cellcolor{Dandelion!25} 0.25\scriptsize{(0.02)} & \cellcolor{Dandelion!99} 0.99\scriptsize{(0.01)} & \cellcolor{Dandelion!21} 0.21\scriptsize{(0.02)} & \cellcolor{Dandelion!82} 0.82\scriptsize{(0.10)} & \cellcolor{Dandelion!90} 0.90\scriptsize{(0.10)} & \cellcolor{Dandelion!85} 0.85\scriptsize{(0.08)} \\\cline{3-12}
    & & Shoulder & \cellcolor{Dandelion!84} 0.84\scriptsize{(0.08)} & \cellcolor{Dandelion!93} 0.93\scriptsize{(0.05)} & \cellcolor{Dandelion!98} 0.98\scriptsize{(0.02)} & \cellcolor{Dandelion!21} 0.21\scriptsize{(0.02)} & \cellcolor{Dandelion!21} 0.21\scriptsize{(0.02)} & \cellcolor{Dandelion!99} 0.99\scriptsize{(0.01)} & \cellcolor{Dandelion!83} 0.83\scriptsize{(0.08)} & \cellcolor{Dandelion!85} 0.85\scriptsize{(0.08)} & \cellcolor{Dandelion!97} 0.97\scriptsize{(0.04)} \\\hhline{|~|===========}
    
    & \multirow{3}{*}{\rotatebox[origin=c]{90}{\shortstack{DSL\\$+\mathcal{L}_{\text{MSC}}$}}} & Ankle & \cellcolor{Dandelion!99} 0.99\scriptsize{(0.01)} & \cellcolor{Dandelion!31} 0.31\scriptsize{(0.05)} & \cellcolor{Dandelion!29} 0.29\scriptsize{(0.04)} & \cellcolor{Dandelion!99} 0.99\scriptsize{(0.01)} & \cellcolor{Dandelion!29} 0.29\scriptsize{(0.02)} & \cellcolor{Dandelion!22} 0.22\scriptsize{(0.03)} & \cellcolor{Dandelion!98} 0.98\scriptsize{(0.02)} & \cellcolor{Dandelion!47} 0.47\scriptsize{(0.05)} & \cellcolor{Dandelion!39} 0.39\scriptsize{(0.03)} \\\cline{3-12}
    & & Knee & \cellcolor{Dandelion!31} 0.31\scriptsize{(0.05)} & \cellcolor{Dandelion!99} 0.99\scriptsize{(0.01)} & \cellcolor{Dandelion!34} 0.34\scriptsize{(0.03)} & \cellcolor{Dandelion!29} 0.29\scriptsize{(0.02)} & \cellcolor{Dandelion!99} 0.99\scriptsize{(0.01)} & \cellcolor{Dandelion!27} 0.27\scriptsize{(0.02)} & \cellcolor{Dandelion!47} 0.47\scriptsize{(0.05)} & \cellcolor{Dandelion!98} 0.98\scriptsize{(0.01)} & \cellcolor{Dandelion!47} 0.47\scriptsize{(0.05)} \\\cline{3-12}
    & & Shoulder & \cellcolor{Dandelion!29} 0.29\scriptsize{(0.04)} & \cellcolor{Dandelion!34} 0.34\scriptsize{(0.03)} & \cellcolor{Dandelion!99} 0.99\scriptsize{(0.01)} & \cellcolor{Dandelion!22} 0.22\scriptsize{(0.03)} & \cellcolor{Dandelion!27} 0.27\scriptsize{(0.02)} & \cellcolor{Dandelion!99} 0.99\scriptsize{(0.01)} & \cellcolor{Dandelion!39} 0.39\scriptsize{(0.03)} & \cellcolor{Dandelion!47} 0.47\scriptsize{(0.05)} & \cellcolor{Dandelion!99} 0.99\scriptsize{(0.01)} \\
    \hline
    
    \end{tabular}
\label{tab:quantitative_assessment_of_shared_representations}
\end{table*}

The quantitative evaluation (Table \ref{tab:quantitative_assessment_of_shared_representations}) further supported the visualizations obtained through the t-SNE algorithm (Fig. \ref{fig:visualization_contrastive_regularization}). Indeed, the shared Att-UNet representations presented inter-domain cosine similarity measures with high mean ($>0.58$) and standard deviation ($>0.06$) suggesting entangled domain representations with low cohesion. Moreover, as previously mentioned, the network bottleneck corresponding to the representation $z_{s_{5}}$ (Fig. \ref{fig:visualization_contrastive_regularization}) presented better domain disentanglement due to higher dimensionality. Additionally, the multi-scale contrastive regularization expectedly led to, at each scale, an increase in intra-domain similarity ($>0.98$) indicating more closely aligned representations from the same domain and a decrease in inter-domain similarity ($<0.47$) reflecting more discriminative (i.e. orthogonal) representations between different domains. However, we observed that domain representations were less disentangled at scale $s_{9}$ (inter-domain similarity greater than $0.39$) than at scales $s_{1}$ and $s_{5}$ (inter-domain similarity lower than $0.34$). Therefore, the effectiveness of contrastive learning to disentangle domain representations varies at each scale, as we observed a quantitative difference in the cosine similarity of the learnt representations. Finally, we also assessed the representations learnt with the single-scale contrastive regularization which only constrained the network bottleneck (i.e. encoder output or $z_{{s_5}}$) \citep{boutillon_multi-task_2021}. Compared with $\mathcal{L}_{\text{MSC}}$, only the representation associated with the $5^{th}$ scale was disentangled while $z_{s_{5}}$ and $z_{s_{9}}$ were not affected by the single-scale contrastive constraint. This further supported the necessity to employ a multi-scale contrastive $\mathcal{L}_{\text{MSC}}$ regularization to disentangle representations at each layer as opposed to $\mathcal{L}_{\text{SSC}}$ term.

\subsection{Benefits for clinical practice}
\label{sec:benefits_for_clinical_practice}

Current deep learning models are specific to anatomical region of interest and may suffer from the limited availability of imaging data, which is exacerbated in pediatric clinical workflows. Our approach demonstrated that designing a collaborative framework incorporating multi-anatomy datasets with close intensity domains and related segmentation tasks can lead to performance improvements on each dataset. In turn, this could lead to a more efficient utilization of imaging resources (pediatric or adult), most notably for the treatment of musculoskeletal disorders affecting different anatomical joints. Several patient cohorts impaired by distinct pathologies could be leveraged to optimize a single model with enhanced generalization capabilities, thus reducing the overall cost of medical image acquisition. More generally, our approach could be transposed to other sets of anatomical structures sharing common characteristics, such as blood vessels in brain, lungs, and retina images \citep{moccia_blood_2018}. Additionally, the multi-scale contrastive regularization could be integrated to enhance vascular segmentation by imposing domain-specific clusters in the embedded spaces of the shared neural network.

Similar to earlier studies employing highly compact multi-domain models \citep{karani_lifelong_2018, chang_domain-specific_2019, dou_unpaired_2020, liu_ms-net_2020}, our work demonstrated that deep neural networks can easily learn related segmentation tasks across multiple intensity domains. Specifically, this study further confirmed the usefulness of employing \texttt{DSBN} functions for multi-domain learning, which were previously successfully applied for multi-modal, multi-scanner, multi-center, or multi-protocol segmentation \citep{karani_lifelong_2018, chang_domain-specific_2019, dou_unpaired_2020, liu_ms-net_2020} and have now proven to be equally effective in a multi-anatomy scenario. Furthermore, when dealing with pediatric patients, it may be beneficial to define domains corresponding to different age groups, as anatomy is significantly modified during child development. However, in the current study, we were unable to explore such multi-age setting due to the limited amount of imaging resources per age group. Finally, as opposed to previous plain UNet models developed in \citep{karani_lifelong_2018, chang_domain-specific_2019, dou_unpaired_2020, liu_ms-net_2020}, our model relied on a more complex architecture based on an pre-trained \texttt{EfficientB3} encoder to achieve more accurate segmentation and integrated multi-domain spatial attention gates to improve its interpretability.

\subsection{Limitations}
\label{sec:limitations}

This study has certain limitations which are categorically listed in this section. First, although the coarse localization of the anatomical structures of interest computed by attention gates and the t-SNE visualizations of the learnt shared representations provide some interpretability of the network inference process, these approaches do not fully explain the features learnt by the segmentation model. Similarly, even though incorporating regularization through the loss function successfully constrains the network parameters and promotes the desired generalizable characteristics during training, the optimization procedure of deep neural networks remains difficult to analyse. Specifically, while the constraints computed by the multi-scale contrastive regularization are explicit, the interpretability of the multi-joint anatomical priors, on its part, is limited as it is based on a deep auto-encoder. It is thus essential to develop more interpretable models allowing a finer analysis of the internal behaviour of the framework during training and inference. In this direction, \citep{zhang_interpretable_2018} have proposed an interpretable CNN which provides a clear semantic representation by assigning to each filter a specific object part to explicitly memorize during the learning process. Such interpretable model could therefore be of great interest for medical image analysis applications, as it would allow a better analysis of the network failures.

Second, while a common hypothesis in machine learning is that the training and test data originate from the same data distribution, an emerging field (i.e. domain generalization or out-of-distribution generalization) has proposed to address the more challenging setting in which the goal is to learn a model that can generalize to an unseen test domain \citep{zhou_domain_2021, wang_generalizing_2021}. In the present study, although our model managed multiple domains, we only addressed plain generalizability within each (i.e. unseen test image from the same distribution as the training data). While the performance improvements obtained during the leave-one-out evaluation indicated better generalization abilities within each domain, our model is currently unable to generalize on new unseen domains (e.g. new modality or anatomical joint). In the context of life-long learning in which a single model continuously learns new domains, \citep{karani_lifelong_2018} has demonstrated that \texttt{DSBN} parameters could be fine tuned with limited amount of training data from the novel domain, while the convolutional filters remained fixed. We assumed that our model could be similarly fine tuned on a new domain without forgetting the knowledge learnt on the previous domains. However, this approach still requires access to labelled imaging data from the new domain, unlike domain generalization frameworks in which data from the new test domain is assumed to be unavailable. Out-of-distribution generalization is thus more generic than traditional domain adaption techniques or life-long learning schemes. Therefore, domain generalization appears crucial for medical image segmentation, where each anatomical region and acquisition protocol defines a new domain in which imaging resources are not necessarily available for network training or fine-tuning purposes. 

\subsection{Perspectives}
\label{sec:perspectives}

As our experiments were conducted on only one imaging modality (i.e. T1-weighted MR), we were unable to evaluate the genericity of our approach over multiple modalities (e.g. T2-weighted MR, CT, etc.) because of the lack of available data. However, previous studies have already demonstrated that a single neural network incorporating shared convolutional filters and \texttt{DSBN} functions can effectively process both CT and MR modalities simultaneously \citep{dou_unpaired_2020}. So, we assumed that our model could be easily extended to multiple modalities. Similarly, we limited our experiments to bone tissue segmentation without considering other musculoskeletal tissues such as muscle, ligaments, or cartilages due to the unavailability of annotations. We also hypothesised that our framework could be upgraded to multi-tissue segmentation since \citep{zhou_deep_2018, ambellan_automated_2019, conze_healthy_2020} have already demonstrated that deep learning models can effectively segment knee cartilages, knee muscles, and shoulder muscles, respectively. More generally, whereas methods developed on natural images employed up to ten domains \citep{rebuffi_efficient_2018}, our experiments involved only three imaging domains due to the scarcity of pediatric imaging resources and the lack of open access pediatric databases. Hence, future studies are aimed at incorporating supplementary MR imaging sequences to further promote generic features during optimization and segmenting additional tissues to provide a more complete description of the musculoskeletal system. Finally, in this direction of including an increasing number of imaging datasets, it may be beneficial to adapt our framework to federated learning scenarios similar to the ones developed by \citep{shen_multi-task_2021} for multi-task pancreas segmentation. This would allow optimization of a single model using training data from multiple institutions without centralizing imaging resources, thus allowing to prevent data privacy and security issues, which is crucial in medical workflows \citep{shen_multi-task_2021} .

Furthermore, we did not consider 3D architectures in our experiments due to their higher computational complexity and GPU memory consumption compared to their 2D counterparts \citep{milletari_v-net_2016}. Although our models did not integrate a third spatial dimension, we observed smooth delineations in all directions, indicating continuous segmentation predictions between adjacent 2D slices. Additionally, our experiments were performed using only four neural network architectures (Att-UNet, Inception-UNet, Dense-UNet, and Efficient-UNet), hence it would be beneficial to include supplementary comparisons based on additional deep learning models including Transformers-based ones \citep{hatamizadeh_unetr_2022} to further evaluate the genericity of our contributions. One could also consider an ensemble approach integrating all backbone architectures to combine the advantages of each model. Finally, with respect to the contrastive learning, we assumed that the temperature hyper-parameter $\tau$ should be constant at each scale, as the cosine similarity between representations was bounded in $[-1, 1]$ regardless of scale. However, we observed in Table \ref{tab:quantitative_assessment_of_shared_representations} that contrastive learning was less efficient at certain scales. Specifically, in the DSL$+\mathcal{L}_{\text{MSC}}$ learning scheme, the shared representation of the $9^{th}$ scale was less disentangled than in $1^{st}$ and $5^{th}$ scales. Hence, one could also propose to employ different temperatures at each scale and to learn such parameters during training, so that the contrastive metric be more sensitive at each scale and better disentangle representation between domains. Nevertheless, such a training procedure might be more challenging to optimize due the numerical instability associated with learnable temperature parameters.

\section{Conclusion}
\label{sec:conclusion}

Developing generalizable deep segmentation model is fundamental to provide accurate and reliable delineations on unseen images for clinical and morphological evaluation of the pediatric musculoskeletal system. This paper introduced a multi-task, multi-domain learning framework for pediatric bone segmentation in sparse MR imaging datasets acquired on separate anatomical joints. This multi-anatomy approach simultaneously benefited from robust shared representations and specialized layers that fitted to the domain-specific intensity distributions and task-specific segmentation label sets. Furthermore, the generalization capabilities of the segmentation model were enhanced by exploiting a multi-scale contrastive regularization to enforce domain clustering in the shared representations and multi-joint anatomical priors which encouraged anatomically consistent shape predictions. 

An important perspective from this study is that collaborative utilization of pediatric resources and intelligent design of deep learning models can improve the segmentation performance on small musculoskeletal imaging datasets. Nevertheless, our framework currently provides an incomplete description of the pediatric musculoskeletal system which solely encompass bone tissues. Hence, future work is aimed at improving our model to segment other anatomical structures (e.g. ankle cartilages, knee ligaments, or shoulder muscles). Thus, morphological and functional analysis will rely on a more complete modeling of the musculoskeletal system, towards a better management of pediatric disorders.

\section*{Compliance with ethical standards}
MR image acquisition on the pediatric cohorts used in this study were performed in line with the principles of the Declaration of Helsinki. Ethical approvals were respectively granted by the Ethics Committee (Comit\'e Protection de Personnes Ouest VI) of CHRU Brest (2015-A01409-40) and by the research ethics committee of the Children's Mercy Hospital, Kansas City, United States.

\section*{Acknowledgments} 
This work was funded by IMT, Fondation Mines-Télécom, and Institut Carnot TSN (Futur \& Ruptures program). Data were acquired with the support of Fondation motrice (2015/7), Fondation de l’Avenir (AP-RM-16-041), PHRC 2015 (POPB 282), and Innoveo (CHRU Brest). We would like to acknowledge Dr. Antonis Stylianou from the University of Missouri-Kansas City, Kansas City, United States and Dr. Donna Pacicca from Children's Mercy Hospital, Kansas City, United States for sharing the anonymized knee joint image dataset.

\bibliographystyle{abbrvnat}
\bibliography{refs}

\setcounter{table}{0}
\setcounter{section}{0}
\setcounter{figure}{0}
\setcounter{equation}{0}
\renewcommand{\thetable}{S\arabic{table}}
\renewcommand{\thesection}{S\arabic{section}}
\renewcommand{\thefigure}{S\arabic{figure}}
\renewcommand{\theequation}{S\arabic{equation}}

\begin{center}
\textbf{\large Supplementary Materials}
\end{center}

\section*{Metrics definition}
\label{sec:metric_definition}

Let $GT$ and $P$ be the ground truth and predicted 3D segmentation masks and let $S_{GT}$ and $S_P$ be the surface voxels of the corresponding sets. The metrics were defined as follows:
\begin{align}
    &\textnormal{Dice} = \dfrac{2 \vert GT \cdot P\vert}{\vert GT \vert + \vert P \vert}\\
    &\textnormal{Sensitivity} = \dfrac{\vert GT \cdot P \vert}{\vert GT \vert}\\
    &\textnormal{Specificity} = \dfrac{\vert \overline{GT} \cdot \overline{P} \vert}{\vert \overline{GT} \vert}\\
    &\textnormal{MSSD} = \max ( h(S_{GT}, S_P), h(S_P,S_{GT}) ) \\
    &\textnormal{with} \enspace h(S,S') = \max_{s \in S} \min_{s' \in S'} \norm{s-s'}_2 \nonumber\\
    &\textnormal{ASSD} = \dfrac{1}{\vert S_{GT} \vert + \vert S_P \vert} (\sum_{s \in S_{GT}} d(s,S_P) \nonumber\\
    & \hspace{8.9em} + \sum_{s \in S_{P}} d(s,S_{GT}) ) \\
    &\textnormal{with} \enspace d(s,S') = \min_{s' \in S'} \norm{s-s'}_2 \nonumber\\
    &\textnormal{RAVD} = \dfrac{\left| \vert GT \vert - \vert P \vert \right|}{\vert GT \vert}
\end{align}
\noindent Distance measures were transformed to millimeters using voxel size information extracted from DICOM metadata.

\end{document}